\documentclass[aps,prd,floatfix,showpacs,superscriptaddress,nofootinbib,10pt]{revtex4-2}
\usepackage{amsmath}
\usepackage{amssymb}
\usepackage{graphicx}
\usepackage{braket}
\usepackage{bbm}
\usepackage{xcolor}

\usepackage{subfigure}
\usepackage[export]{adjustbox}
\usepackage{siunitx,pslatex}
\usepackage{booktabs}

\begin{document}

\title{Interplay of magnetic field and chemical potential induced anisotropy and frame dependent chaos of a $Q\bar{Q}$ pair in holographic QCD}
\author{Bhaskar Shukla}\email{bhasker\_shukla@nitrkl.ac.in}\affiliation{Department of Physics and Astronomy, National Institute of Technology Rourkela, Rourkela - 769008, India}
\author{Jasper Nongmaithem}\email{524ph6005@nitrkl.ac.in}\affiliation{Department of Physics and Astronomy, National Institute of Technology Rourkela, Rourkela - 769008, India}
\author{David Dudal}\email{david.dudal@kuleuven.be}\affiliation{KU Leuven Campus Kortrijk -- Kulak, Department of Physics Etienne Sabbelaan 53 bus 7657, 8500 Kortrijk, Belgium}
\author{Subhash Mahapatra}\email{mahapatrasub@nitrkl.ac.in}\affiliation{Department of Physics and Astronomy, National Institute of Technology Rourkela, Rourkela - 769008, India}

\begin{abstract}
We investigate the role of both magnetic field and chemical potential on the emergence of chaotic dynamics in the QCD confining string from the holographic principle. An earlier developed bottom-up model of Einstein-Maxwell-dilaton gravity, which mimics QCD features quite well, is used. The qualitative information about the chaos is obtained using the Poincar\'{e} sections and Lyapunov exponents. We find signatures of chaos in energetically disfavored string configurations, that are closer to the horizon, whereas no chaos is observed in energetically favored string configurations that are away from the horizon. Our results depend quite strongly on the frame we consider in the analysis. In the string frame, the chemical potential and the magnetic field suppress the chaotic dynamics in both parallel and perpendicular orientations of the string with respect to the magnetic field. Meanwhile, in the Einstein frame, the magnetic field suppresses/enhances the chaotic dynamics when the string is orientated perpendicular/parallel to the magnetic field,  while the chemical potential enhances the chaotic dynamics for both orientations.  The reported Lyapunov exponents are consistent with a classical analogue of the MSS bound in the parameter space of the model and we find it to be always satisfied in both frames.
\end{abstract}

\maketitle

\section{Introduction}
\label{sec:intro}
Whether quantum chromodynamics (QCD) exhibits chaotic behavior at some level is a pivotal inquiry, holding significance for understanding QCD itself and the broader quest for a comprehensive understanding of interacting quantum field theories (QFT). Systems typically display chaotic characteristics in the natural world, except for some simple and isolated models. A parallel expectation is that QCD, a complex system, may inherently possess chaotic traits. Many researchers have worked on similar problems in recent years \cite{Pullirsch:1998ke, Asano:2015qwa, Pullirsch:1998wp, Bittner:2000nu, Hashimoto:2016wme, Ageev:2021poy, Yadav:2023hyg}. QCD is the fundamental QFT that underlies the precise understanding of subatomic particles such as quarks, gluons, and their interactions. Insights derived from weakly coupled approximations and out-of-time ordered correlators already indicate the possibility of chaos in QCD at high energies \cite{Muller1992Jun,biro1995lyapunov,tsukiji2016entropy,kunihiro2010chaotic}.

Notably, although lattice QCD is the most well-established approach for non-perturbative QCD studies, it does face limitations in capturing, e.g., time-dependent dynamics essential for analyzing chaos. Exploring alternative methodologies thus remains crucial in unraveling the delicate nature of chaos within the intricate realm of QCD. Since the introduction of the AdS/CFT correspondence \cite{Maldacena:1997re, Witten:1998qj, Gubser:1998bc}, it has grown into a versatile way for exploring various aspects of strongly coupled quantum systems, also for analyzing strongly coupled and thus non-perturbative QCD. 

Early inquiries into the chaotic dynamics of closed strings \cite{pando2010chaos} paved the way for a thorough examination of non-linear phenomena within the AdS/CFT framework, see for example \cite{Blake:2021hjj, Giombi:2022pas, Basu:2011dg, Basu:2011di, Stepanchuk:2012xi, Giataganas:2013dha, Bai:2014wpa, Panigrahi:2016zny, Basu:2016zkr, Asano:2016qsv, Ishii:2016rlk, Rigatos:2020hlq, Pal:2023kwc, Penin:2024rqb}. Over time, many diagnostic tools for chaos have been readily used, the important ones being Poincar\'{e} sections and Lyapunov exponents \cite{Shukla:2023pbp, Hashimoto:2018fkb, Colangelo:2020tpr, Colangelo:2021kmn, Akutagawa:2019awh, Giataganas:2017guj}, while out-of-time-order-correlators (OTOCs) have provided a holographic lens into quantifying quantum information scrambling~\cite{Shenker:2014cwa} and the saturation of the Maldacena-Shenker-Stanford (MSS) bound~\cite{maldacena2016bound}. Another important tool that is being prominently used in recent literature is pole-skipping~\cite{Blake:2017ris, Blake:2018leo, Baishya:2023ojl}, which establishes a direct connection between the early-time exponential growth of OTOCs and the near-horizon dynamics of black holes, as well as hydrodynamic collective modes. Level spacing distributions have emerged as a powerful diagnose of quantum chaos in holographic models. For instance, \cite{PandoZayas:2012ig} demonstrates that the spectrum of hadronic states in confining holographic models follows the Gaussian Orthogonal Ensemble (GOE), a hallmark of quantum chaos. Similarly, \cite{Basu:2013uva, Shukla:2024wsu} examine the transition from chaotic (Wigner GOE) to integrable (Poisson) regimes in the quantum spectrum of strings in an AdS soliton background, reinforcing the utility of level spacing analysis in holography. For a recent review on chaos in holography, see \cite{Jahnke:2018off}.

The holographic duality introduces a unique perspective, allowing for the exploration of quantum dynamical aspects in a dual, higher dimensional gravity (classical) system. In holography, classical strings are examined in higher dimensions, specifically in five dimensions, in the context of bottom-up or top-down holographic QCD models. Many external factors greatly influence QCD properties, particularly the chemical potential, magnetic field, and temperature. They are believed to be produced in central and/or non-central heavy ion collisions and are expected to be of the order of a hundred MeV \cite{Skokov:2009qp, Bzdak:2011yy, DElia:2010abb, Deng:2012pc, Tuchin:2013ie, Voronyuk:2011jd}. The dynamical and equilibrium properties of QCD are expected to change considerably depending upon the magnitudes of such factors. It is well known that various intricate phases in QCD appear in the parameter space of chemical potential, magnetic field, and temperature. However, due to various analytical and numerical issues, a complete understanding of QCD in the parameter space of such factors is still far from fully understood. For instance, studying the impact of chemical potential in lattice QCD poses challenges owing to the sign problem. The intricacies associated with this issue make it difficult to explore the effects comprehensively. As mentioned above, the situation worsens when one tries to address dynamic problems.

Much work has been done in recent years to study QCD-like physics from holography. By now various top-down and bottom-up holographic models have been constructed whose dual boundary field theories try to mimic real QCD as closely as possible in all the parameter space of temperature, chemical potential, and magnetic field. This includes simple holographic models having temperature as the only relevant parameter \cite{Gubser:2008ny, Gursoy:2007cb, Gursoy:2010fj}, models with both temperature and chemical potential \cite{DeWolfe:2010he, Dudal:2017max, Cai:2012xh, He:2013qq} or temperature and magnetic field \cite{bohra2021chiral, Dudal:2021jav, Rougemont:2015oea, Finazzo:2016mhm, Arefeva:2024xmg}, and models having all three parameters \cite{Bohra:2019ebj, Gursoy:2017wzz, Arefeva:2020vae, Arefeva:2024mtl}. Using these holographic models, a lot of new insights into the characteristic structure of QCD at strong coupling have been obtained, and in some cases, new and interesting predictions have also been made. More discussion on holographic QCD in the presence of magnetic field and chemical potential can be found in \cite{Dudal:2018ztm, Mahapatra:2019uql, Dudal:2016joz, Rougemont:2014efa, Fuini:2015hba, Cartwright:2019opv, Fukushima:2021got, Ballon-Bayona:2022uyy, Rodrigues:2017cha, Arefeva:2023jjh, Arefeva:2021jpa, Jena:2022nzw, Jain:2022hxl, Arefeva:2022avn, Chen:2021gop, Braga:2020hhs, Zhou:2020ssi, Ballon-Bayona:2020xtf, Zhao:2021ogc, Dudal:2018rki, Jena:2024cqs, Ballon-Bayona:2024twa, Braga:2021fey}, see also \cite{Rougemont:2023gfz, Hoyos:2021uff, Jarvinen:2021jbd} for reviews of this exciting subject. Since holography provides a unique mathematical tool to investigate QCD at strong couplings, therefore it would be interesting to analyse whether this duality can also provide new insights into the chaotic dynamics of the suspended open string, corresponding to the dynamics of the Wilson loop or the quark-antiquark pair, when all interesting parameters such as the temperature, chemical potential, and magnetic field are simultaneously present in QCD background. In this work, we undertake such an analysis.

We use a recently proposed dynamical holographic QCD model  \cite{Bohra:2019ebj}, which includes both the magnetic field and the chemical potential to dynamically investigate and find anisotropic imprints of chaotic behaviour in the open string. This is an extension of the work done in \cite{Hashimoto:2018fkb, Colangelo:2020tpr, Colangelo:2021kmn}, where
the chaotic behaviour of the open string was probed when parameters such as the temperature, chemical potential, or magnetic field were individually present in top-down holographic models. However, as is well known, these top-down models are not very reliable to model real QCD holographically. For instance, they do not exhibit a running coupling constant and, as opposed to real QCD, there is no self-consistent chemical potential or magnetic field dependent confined/deconfined phase transition in the dual field theory. These issues can be rectified in more dynamical bottom-up holographic QCD models. The main motive of this work is to fill this gap and analyse chaotic features of the suspended open string at finite temperature in a dynamical bottom-up holographic QCD model where magnetic field and chemical potential are simultaneously present. For this purpose, we consider the holographic QCD model of \cite{Bohra:2019ebj}. This particular model is not only analytically tractable but was also shown to be successful in capturing a variety of real QCD results from holography, including the inverse magnetisation phenomena, linear Regge trajectories for heavy vector mesons, area law for the Wilson loop in the confined phase, etc. We find that the dynamics of the energetically disfavored open string exhibit chaos in the deconfined phase and it depends nontrivially on the background chemical potential and magnetic field.

A key development in the study of chaos was the Maldacena-Shenker-Stanford (MSS) bound \cite{maldacena2016bound}, which imposes an upper limit on the Lyapunov exponent \(\lambda\) for quantum systems at finite temperature $T$, i.e., $\lambda \leq 2\pi T \,$. This bound arises from the analysis of out-of-time-ordered correlators (OTOCs) in thermal quantum field theories and reflects a fundamental constraint on quantum chaos. However, recent studies of classical string and particle motions in black hole backgrounds have identified a ``classical analog" of this bound in specific symmetric spacetimes \cite{Hashimoto:2016dfz, Giataganas:2021ghs}. While the classical Lyapunov exponent \(\lambda_L\) for strings or particles near horizons often saturates \(2\pi T\) in certain setups, this behavior is distinct from the quantum MSS bound and not universally guaranteed, as it may depend on the symmetry of the background and the nature of the probe~\cite{Giataganas:2021ghs, Djukic:2023dgk}. For instance, for geometries with broken symmetries, \(\lambda_L\) can deviate from \(2\pi T\) \cite{Giataganas:2021ghs}. 

In this work, we focus on the classical Lyapunov exponent of suspended strings near black hole horizons. The near-horizon region is critical because gravitational instabilities dominate here, and prior studies \cite{Hashimoto:2016dfz, Ishii:2015wua} have shown that probes in this regime exhibit universal chaotic features tied to the horizon’s surface gravity. Our analysis demonstrates that \(\lambda_L\) for the suspended string is less than \(2\pi T_H\) (where \(T_H\) is the Hawking temperature) in the Einstein-Maxwell-Dilaton (EMD) black hole background. This result aligns with the classical extrapolation of the MSS-like scaling observed in holographic setups \cite{Djukic:2023dgk}, where the string’s Lyapunov exponent corresponds to thermal decay rates or quasinormal mode spectra rather than to quantum chaos. The suspended string’s dual interpretation as a Wilson loop in the boundary theory further bridges the gravitational dynamics to thermal gauge theory observables, reinforcing the consistency of this classical bound in strongly coupled systems. In this work, without losing any generality, we will use the symbol $\lambda_L=\lambda$ for the string Lyapunov exponent.

Another crucial observation in gravity research, in general, is that sometimes significant differences can arise between the string and Einstein frames' physics. In general, the debate over which choice complies with the correct physical processes is still ongoing~\cite{faraoni1999einstein, cho1992reinterpretation, sk2017scalar, capozziello2010physical, corda2011gravitational, quiros2013conformal, macias2001jordan, galaverni2022jordan, nojiri2001string, jarv2007scalar}. From the holographic QCD perspective, usually, the string frame is advocated when discussing
the Wilson loop/string tension in relation to ``holographic confinement''. To the best of our knowledge, there is no strict proof as to why one should always use the string frame in this context, see for example \cite{Girardello:1999hj} where the Wilson loop has been computed explicitly in the Einstein frame, also exhibiting the area law. Here, for completeness, we will thus investigate chaos in the open string configuration in both frames.

Specializing to our holographic setup, in \cite{Shukla:2023pbp}, it was observed that the magnetic field-dependent Poincar\'{e} sections and Lyapunov exponents of the perturbed string dynamics get substantially modified in different frames. In particular, the anisotropic features and magnitudes of chaotic observables changed considerably. Therefore, in this paper, we shall further investigate, in both frames, the chaotic nature of an open string using the QCD-gravity dual mentioned above, which includes both the magnetic field and the chemical potential. For simplicity, we work in the usual units $G=\hbar=c=k_{B}=1$.  

The paper is structured into five sections. In section~\ref{Methodology}, we discuss the methodology, focusing on the bottom-up magnetized EMD gravity model, setting the stage with the necessary theoretical background. In section~\ref{sec:stringframe}, we examine the dynamics in the string frame, starting with the derivation of the string equation of motion in section~\ref{subsec:stringmodel}, followed by the introduction of perturbations in section~\ref{subsec:stringperturbation}. This section also includes the Poincar\'{e} section analysis to visualize chaotic trajectories, section~\ref{StringPoincare} and the Lyapunov exponent analysis to quantify chaos, section~\ref{StringLyapunov}. To complete our study, we also include the analysis of saddle points in section~\ref{SaddleMSS}. In section~\ref{sec:conclusion}, we conclude our paper by highlighting the main results with some discussion. In Appendix~\ref{Einsteinframe}, we discuss the results of the string dynamics within the Einstein frame. This includes Poincar\'{e} analysis in section~\ref{EinsteinPoincare}, Lyapunov exponents in section~\ref{EinsteinLyapunov}, and MSS bound analysis in section~\ref{einsteinsaddle}. 

\section{Magnetised EMD gravity with chemical potential}\label{Methodology}
In this section, we briefly discuss the holographic QCD model of \cite{Bohra:2019ebj,bohra2021chiral}. 
This will set the stage for later sections for investigating chemical potential and magnetic field effects on chaotic string dynamics.  We start with a five-dimensional gravity system,
\begin{equation}
S=-\frac{1}{16 \pi  G_5}\int d^5x\sqrt{-g}\left[R-\frac{1}{4} f_1(\phi ) F_{(1) {MN}} F^{{MN}} 
		-\frac{1}{4} f_2(\phi ) F_{(2) {MN}} F^{MN}-\frac{1}{2}\partial _M\phi\partial ^M\phi -V(\phi )\right],
 \label{eq:OriginalAction}
\end{equation}
where $F_{(1)MN}$ and $F_{(2)MN}$ are two $U(1)$ field strength tensors used to introduce a background chemical potential ($\mu$) and magnetic field ($B$) respectively, $\phi$ is the dilaton field, $f_1$($\phi$) and $f_2$($\phi$) are the coupling functions between the dilaton field and the two $U(1)$ gauge fields, $V(\phi)$ is the dilaton potential, and $G_5$ is Newton's gravitational constant in five dimensions.

By varying the action (\ref{eq:OriginalAction}), we get the Einstein, gauge fields, and dilaton equations of motion. To determine solutions of these equations of motion, we take the following Ans\"atze
\begin{eqnarray}
& & ds^2=\frac{L_{AdS}^2~e^{2 A(z)}}{z^2}\left[-g(z)dt^2 +\frac{dz^2}{g(z)}+dx_1^2+e^{B^2 z^2}\left(dx_2^2+dx_3^2\right)\right]\,, \nonumber\\
& & \phi =\phi (z),~~A_{(1) M}=A_t(z) \delta _M^t,~~F_{(2) MN}=B dx_2\land dx_3\,,
    \label{eq:Ansatzmetric}
\end{eqnarray}
where $z$ is the usual holographic radial coordinate, $z=1/r$, and $L_{AdS}$ is the AdS length scale which we will set to ``1'' from now on. The above Ans\"atze gives us four Einstein, one gauge field (coming from the first gauge field $A_{(1) M}$), and one dilaton equation of motion. The gauge field equation of motion coming from the second gauge field $A_{(2) M}$ is trivially satisfied. Therefore, overall, there are six equations of motion. However, only five of them are independent. Interestingly, taking the dilaton equation as a constraint equation and using the potential reconstruction technique \cite{Dudal:2017max, Mahapatra:2018gig, Arefeva:2018hyo, Arefeva:2020byn, Alanen:2009xs, Mahapatra:2020wym, Priyadarshinee:2021rch, Priyadarshinee:2023cmi, Daripa:2024ksg}, complete closed-form solutions to the gravity equations can be found. One particular solution corresponding to a black hole is given by
\begin{equation}\label{asol}
A(z)= -a z^2 \,,
\end{equation}

\begin{equation}\label{f1sol}
f_{1}(z)=e^{-A(z)-B^2 z^2-c z^2} \,,    
\end{equation}

\begin{equation}\label{gsol}
\begin{split}
g(z) &= 1+\frac{c  \left(e^{z^2 \left(3 a-B^2+c\right)} \left(z^2 \left(3 a-B^2+c\right)-1\right)+1\right)\mu ^2}{ \left(e^{c z_{h}^2}-1\right)^2 \left(3 a-B^2+c\right)^2} \\
& -\frac{\left(e^{z^2 \left(3 a-B^2\right)} \left(z^2 \left(3 a-B^2\right)-1\right)+1\right) \left(\frac{c  \left(e^{z_{h}^2 \left(3 a-B^2+c\right)} \left(z_{h}^2 \left(3 a-B^2+c\right)-1\right)+1\right)\mu ^2}{L^2 \left(e^{c z_{h}^2}-1\right)^2 \left(3 a-B^2+c\right)^2}+1\right)}{e^{z_{h}^2 \left(3 a-B^2\right)} \left(z_{h}^2 \left(3 a-B^2\right)-1\right)+1}
\end{split}
\end{equation}

\begin{equation}\label{phisol}
\begin{split}
\phi(z) &= \frac{\left(9 a-B^2\right) \log \left(\sqrt{6 a^2-B^4} \sqrt{z^2 \left(6 a^2-B^4\right)+9 a-B^2}+6 a^2
   z -B^4 z \right)}{\sqrt{6 a^2-B^4}} \\
  & + z \sqrt{z^2 \left(6 a^2-B^4\right)+9 a-B^2} -\frac{\left(9 a-B^2\right) \log \left(\sqrt{9 a-B^2} \sqrt{6 a^2-B^4}\right)}{\sqrt{6 a^2-B^4}}\,,
\end{split}
\end{equation}

\begin{equation}\label{fsol}
f_2(z)=g(z)e^{2 A(z)+2 B^2 z^2} \left(-\frac{6 A'(z)}{z}-4 B^2+\frac{4}{z^2}\right)-\frac{2 e^{2 A(z)+2 B^2 z^2} g'(z)}{z} \,,
\end{equation}

\begin{equation}\label{Vsol}
\begin{split}
V(z)&=\frac{1}{2} e^{-2 A(z)} (z \left(g'(z) \left(9-9 z A'(z)-4 B^2 z^2\right)-z g''(z)\right) \nonumber\\
& -2 g(z) \left(3 z \left(z A''(z)+3 A'(z) \left(z A'(z)+B^2 z^2-2\right)\right)+2 \left(B^4 z^4-4 B^2 z^2+6\right)\right))\,.
\end{split}
\end{equation}
This solution is obtained by imposing the boundary conditions
\begin{eqnarray}
\begin{aligned}
& & g(0)=1\,,~~ g\left(z_h\right)=0\,, \\
& & A_t(0)=\mu\,,~~ A_t\left(z_h\right)=0\,,\\
& & A(0)=0\,,\\
\end{aligned}
\end{eqnarray}
where $z_{h}$ is the radius of the black hole horizon. Hence, the $z$-coordinate runs from $z=0$ to $z=z_h$ (or $r=r_h$ to $r=\infty$), where $z=0$ (or $r=\infty$) is the asymptotic boundary. These boundary conditions are chosen to ensure that the spacetime asymptotes to AdS at the asymptotic boundary $z\rightarrow0$ and that it has a well-defined horizon at $z=z_h$. Apart from these boundary conditions, we have further demanded that the dilaton field $\phi$ remains real throughout the bulk and goes to zero at the asymptotic boundary $\phi(0)=0$. Notice the presence of the magnetic field in the Ans\"atze metric, see Eq.~(\ref{eq:Ansatzmetric}), partially breaks the boundary $SO(3)$ symmetry, but recovers it as soon as $B\rightarrow0$. Note that $B$ is the 5-dimensional magnetic field, so it carries dimension of GeV.

The above solution explicitly depends on $\mu$ and $B$, with $a$ and $c$ being the only arbitrary and free parameters. The value of $a$ and $c$ is usually fixed by taking inputs from the dual boundary QCD theory. In \cite{Bohra:2019ebj,he2013phase}, $a=0.15~\text{GeV}^2$ was fixed by demanding the confinement/deconfinement transition temperature to be around $270~\text{MeV}$ at zero chemical potential and magnetic field. Similarly, the magnitude of $c$ was determined by aligning the holographic meson mass spectrum with that of the lowest-lying heavy meson states. This fixes $c=1.16~\text{GeV}{^2}$. Moreover, for the dilaton field to be real, we must impose $B^4\le 6a^2$, further putting a constraint on the maximally allowed magnetic field at $B\simeq0.6~\text{GeV}$ \cite{bohra2021chiral}. We want to emphasize that we are not considering varying magnetic field scenarios. We will consider it a constant throughout spacetime.

There are some features of this particular form of $A(z)$ that we have chosen in Eq.~(\ref{asol}). As said, our metric is now asymptotically AdS, i.e., at $z\rightarrow0$. This can also be checked from our potential form $V(z)$ in Eq.~(\ref{Vsol}). As $z\rightarrow0$, $V(z)$ reduces to the value of the cosmological constant expected in five-dimensional AdS spacetime. The form of our scalar potential satisfies the Gubser criterion \cite{gubser2000curvature}, as it is bounded from above. Moreover, the mass of the scalar field also satisfies the Breitenl\"ohner-Freedman bound \cite{Breitenlohner:1982bm}. Another solution to the gravity system also exists, one corresponding to thermal-AdS, which has no horizon. It can be obtained from the black hole solution by taking the limit $z_h\rightarrow\infty$. The black hole and thermal-AdS solutions correspond to the deconfined and confined phases of the dual boundary QCD, respectively, and can undergo magnetic field and chemical potential dependent Hawking/Page type phase transitions \cite{Jena:2022nzw}.

Since the chaotic nature of the string is intimately tied to the presence of the horizon, the thermal-AdS solution will not play any part in the rest of the discussion, and we will concentrate mainly on the black hole solution. 

The metric function that we have solved is in the Einstein frame. In our previous analysis \cite{Shukla:2023pbp}, it was observed that different frames had a significant impact on the chaotic nature of the string. In particular, the string frame metric takes a different form than the Einstein frame metric in the presence of a non-trivial dilaton field, thereby leading to different profiles of the Poincar\'{e} section and Lyapunov exponent in these two frames.  We can thus expect to find further significant differences in the chaotic dynamics of the string when the chemical potential is also turned on. 

Now, let us note the metric expression in the string frame. The standard way to go from the Einstein to the string frame metric involves the transformation $(g_s)_{MN}=e^{\sqrt{2/3}\phi} g_{MN}$ \cite{Bohra:2019ebj}, where the subscript ``s'' stands for string frame. The metric solution (\ref{gsol}) in the string frame then reads
\begin{equation}\label{4s}
	ds^2=L_{AdS}^2 r^2 e^{2 A_s(r)} \left[-g(r) dt^2+\frac{dr^2}{r^4 g(r)}+dx_1^2+e^{\frac{B^2}{r^2}} \left(dx_2^2+dx_3^2\right)\right]\,,
\end{equation}
where $A_s(r) = A(r) + \sqrt{1/6} \phi(r)$.
Writing down the above Einstein and string frame metrics in the following general form
\begin{equation}\label{7}
	ds^2=g_{tt} dt^2+g_{11} dx_1^2+g_{22} dx_2^2+g_{33} dx_3^2+g_{rr} dr^2 \,,
\end{equation}
will further allow us to write various expressions related to the string dynamics in a unified manner and we can minimise repetition of various expressions. Note that in the string frame, we have:
\begin{eqnarray}\label{eq:Aforstringframe}
g_{tt}=-r^{2}e^{2A_s(r)}g(r),~~g_{11}=r^{2}e^{2A_s(r)}h(r),~~g_{22}=g_{33}=r^{2}e^{2A_s(r)}q(r),~~g_{rr}=\frac{e^{2A_s(r)}}{r^2g(r)} \,,
\end{eqnarray}
with
\begin{equation}\label{10}
h(r) = 1,~~q(r) = e^{\frac{B^{2}}{r^{2}}} \,,
\end{equation}
with similar expressions existing in the Einstein frame. With this complete picture of our five-dimensional metric, we can now move on to model our string system on it. In the next section, we will investigate the string dynamics in the string frame.

\section{Chaotic dynamics in the string frame}\label{sec:stringframe}
\subsection{String model}\label{subsec:stringmodel}
In this section, we investigate the string dynamics and its chaotic behaviour in the presence of chemical potential and
magnetic field at finite temperature. We use an infinitely thin string that traces out a world sheet as it moves through time. Its motion is determined by the Nambu-Goto action
\begin{equation}
	S=-\frac{1}{2 \pi \alpha'}\int \sqrt{-h} d\ell dt \,,
\end{equation}
where $\alpha'$ is the string tension, $h$ is the determinant of the induced metric $h_{\alpha\beta}$ on the world sheet 
\begin{equation}
	h_{\alpha\beta}=g_{(s)MN}\frac{\partial X^M}{\partial \xi ^\alpha}\frac{\partial X^N}{\partial \xi ^\beta}\,.
\end{equation}
Here $g_{(s)}$ is the metric tensor in the string frame, and $\xi ^i$ are the coordinates on the string world sheet, i.e., $i=(t,\ell)$.

Let us first investigate the static string configuration. In the presence of a background magnetic field, we will consider two special cases of string configuration, either parallel or perpendicular to the magnetic field. Accordingly, there are two ways to parametrize the string. In the static case, the location of the string is specified by $r(\ell)$ and $x_{i}(\ell)$, with $i=1$ when the magnetic field $B$ is parallel to the string endpoints and $i=3$ when perpendicular. Here, $\ell$ corresponds to a proper length measured along the string with endpoints of the string located at $x_{i}=\pm L/2$.  

Following the above notation, the Nambu-Goto action of the static string looks 
\begin{equation}
	S=-\frac{1}{2 \pi \alpha' }\int dtd\ell \sqrt{\left|g_{rr} g_{tt} \left(r'\right)^2+g_{ii} g_{tt} \left(x_i'\right){}^2\right|} \,,
\end{equation}
where the prime $'$ denotes the derivative with respect to $\ell$. Notice that in the above Lagrangian, the coordinate $x_i$ is cyclic. This gives us a conserved quantity
\begin{equation}
	\frac{\partial \mathcal{L}}{\partial x_i'}=-\frac{1}{2 \pi \alpha' }\frac{\left|g_{tt}\right|g_{ii} x_i'}{\sqrt{\left|g_{tt}\right|g_{ii} \left(x_i'\right){}^2+|g_{tt}|g_{rr} \left(r'\right)^2}}\,.
\end{equation}
At the tip of the string $r(\ell=0)=r_0$, we have $\left.\frac{dr}{dx_{i}}\right\vert_{\ell=0}=0$. This gives 
\begin{equation}
	\frac{g_{ii} \sqrt{|g_{tt}|} x_i'}{\sqrt{g_{rr} \left(r'\right)^2+g_{ii} \left(x_i'\right){}^2}}=\left.\sqrt{g_{ii} |g_{tt}|}\right\vert_{\ell=0} \,.
\end{equation}
The symmetry of the string configuration ensures that the tip $r_0$ lies at the point $x_i=0$ (or $\ell=0$). Using the elementary length element set by
\begin{equation}
d\ell^2=g_{rr} dr^2 +g_{ii} dx_i^2 \,,
\end{equation}
we get the following static profile of the string
\begin{eqnarray}
x' & = &\pm\frac{\sqrt{-g_{ii}\left(r_0\right) g_{tt}\left(r_0\right)}}{g_{ii} \sqrt{-g_{tt}}}\,,
 \label{eq:stringprof01} \\
 r' & = &\pm\frac{\sqrt{g_{ii}\left(r_0\right) g_{tt}\left(r_0\right)-g_{ii} g_{tt}}}{\sqrt{-g_{tt} g_{rr} g_{ii}}} \,.
 \label{eq:stringprof02}
\end{eqnarray}
Using the boundary condition that the endpoints of the string lie on the asymptotic boundary at $x_i=\pm L/2$, we can further determine the string length $L$ as a function of $r_0$
\begin{equation}
	L=\int_{r_0}^{\infty } 2 \sqrt{\frac{g_{rr}(r)}{g_{ii}(r)} \left(\frac{g_{ii}(r_{0}) g_{tt}(r_{0})}{ g_{ii}(r) g_{tt}(r)-g_{ii}(r_{0}) g_{tt}(r_{0})}\right)} \, dr \,.
 \label{eq:StringLength}
\end{equation}

\begin{figure}[htbp]
\centering
\includegraphics[width=.47\textwidth]{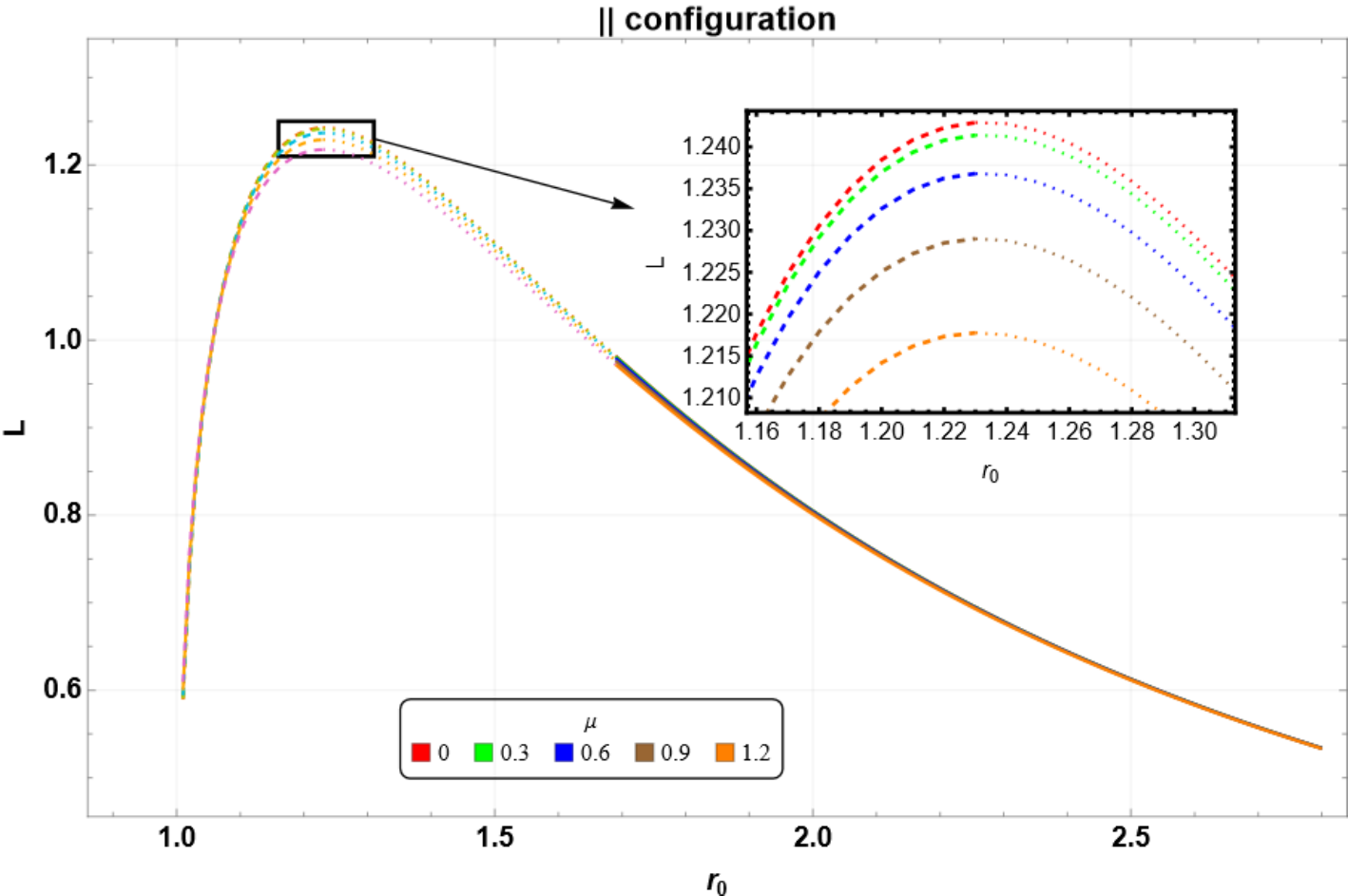}
\qquad
\includegraphics[width=.47\textwidth]{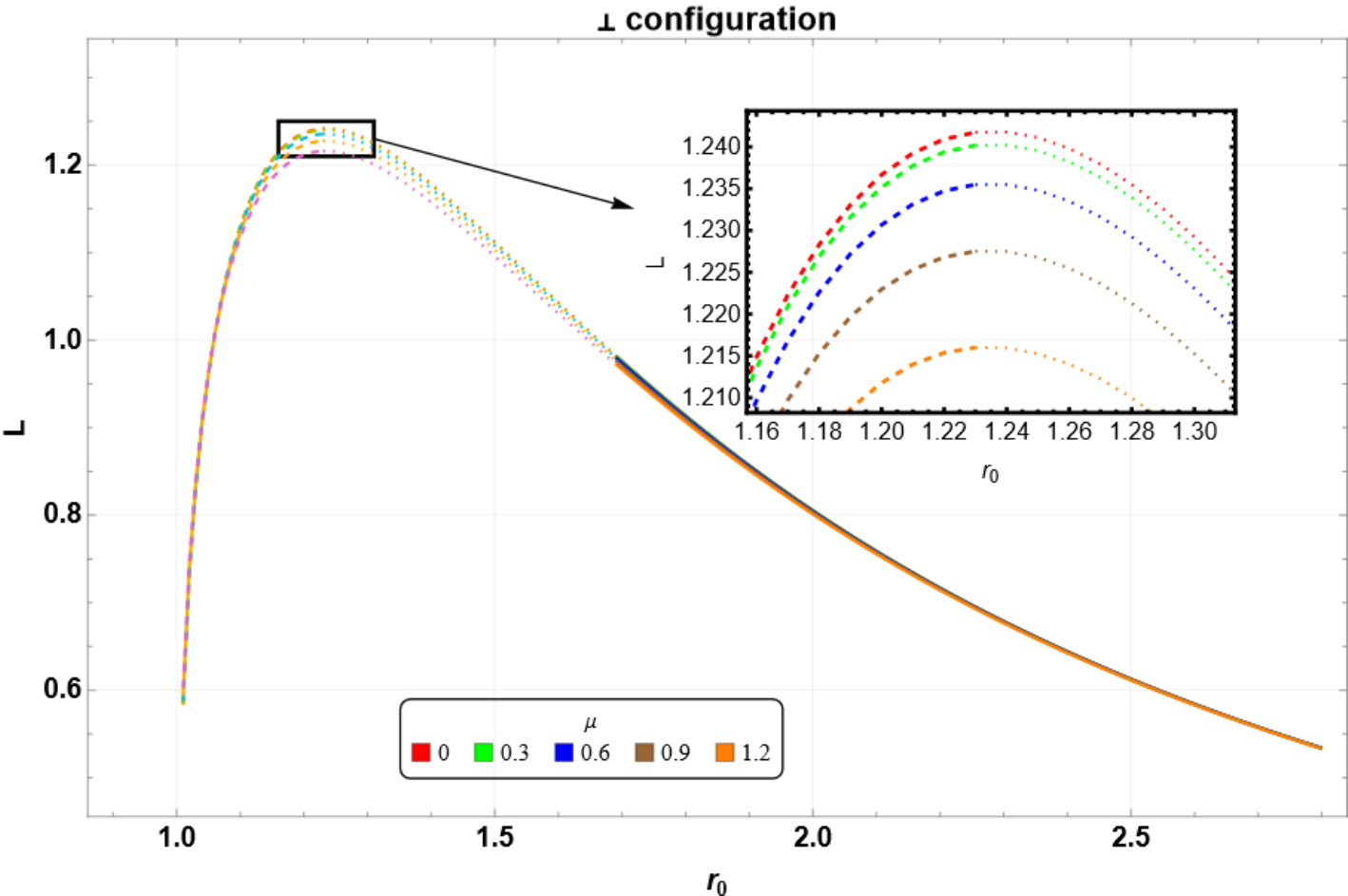}

\caption{The behaviour of the string length $L$ as a function of $r_0$ is depicted for both parallel (left) and perpendicular (right) orientations of the string, in relation to the varying chemical potential $\mu$, with the magnetic field fixed at $B = 0.2$ and horizon radius $r_h=1$. All quantities are in units of GeV.   
\label{fig:StringLvsr0}}
\end{figure}

In Fig.~\ref{fig:StringLvsr0}, the behaviour of the string length $L$ as a function of $r_0$ for different values of the chemical potential at a fixed magnetic field $B=0.2$ aligned both in the parallel and perpendicular direction is shown. Here, we have taken $r_h=1$ for illustration,
but analogous results appear for other $r_h$ values as well. There appears a maximal string length $L_{max}$, beyond which no connected solution exists, i.e., no solution to Eq.~(\ref{eq:StringLength}) exists for $L>L_{max}$. This behaviour is true irrespective of the values of $\mu$ and $B$. The precise value of this $L_{max}$, however, depends non-trivially on $\mu$ and $B$. In particular, it decreases with $\mu$ and $B$ in parallel and perpendicular orientations. Moreover, below $L_{max}$, we can identify two types of connecting string solutions for each value of $L$, one where the tip of the string is near the black hole horizon (represented by dashed lines) and one where the tip of the string is away from the horizon (represented by dotted and solid lines).  

\begin{figure}[htbp!]
	\centering
	\includegraphics[width=0.35\linewidth]{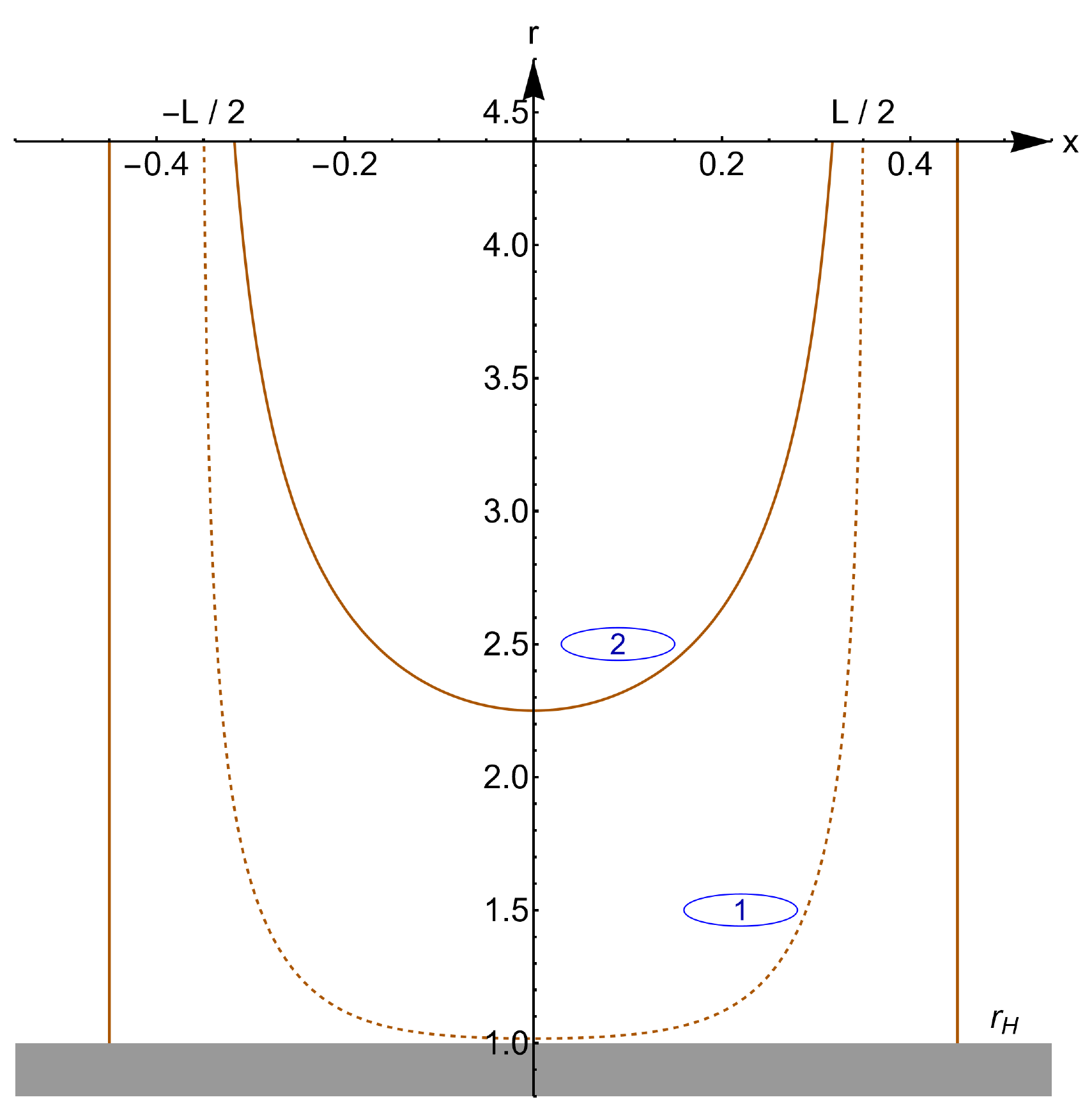}
	\caption{The figure illustrates three distinct configurations of static suspended strings. The solid (2) and dashed (1) lines represent connected strings, corresponding to the local minima and maxima of the energy, respectively. The two solid, straight, parallel lines denote disconnected strings extending from the asymptotic boundary to the horizon. The parameters used are $B=0.1$, $\mu=0.2$, $r_{h}=1$, and $L=0.7$ for the parallel magnetic field. All quantities are expressed in units of GeV.}
	\label{stringprofileBPt1muPt2}
\end{figure}

We further need to probe their free energy profile to describe the stability of the connected string solutions. This can be computed from the on-shell Nambu-Goto action. For the string solutions in Eq.~(\ref{eq:StringLength}), the free energy is given by
\begin{equation}
    F=-\frac{2}{2 \pi  \alpha' } \int_{r_0}^{\infty } \frac{\sqrt{-g_{tt}(r)g_{rr}(r)}}{\sqrt{\frac{g_{ii}\left(r_0\right) g_{tt}\left(r_0\right)}{g_{ii}(r) g_{tt}(r)} \left(\frac{g_{ii}(r) g_{tt}(r)}{g_{ii}\left(r_0\right) g_{tt}\left(r_0\right)}-1\right)}} \, dr\,.
\end{equation}
Observe that in the limit $r\rightarrow \infty$, $F$ contains a UV divergence. Generally, this can be taken care of by subtracting the free energy of the two disconnected strings that are separated by a distance $L$ and extended from the boundary to the horizon. Since the disconnected strings are also solutions of the string equations of motion (EOMs), one may consider these disconnected strings as the third string solution. Accordingly, there exist three string profiles (two connected and one disconnected) for $L<L_{max}$, whereas only the disconnected profile appears for $L>L_{max}$. These different string profiles are shown in Fig.~\ref{stringprofileBPt1muPt2}, where two connected profiles are indicated by the solid line (corresponding to large $r_0$ solution), marked as (2) and the dashed line (corresponding to small $r_0$ solution) marked as (1). In contrast, two parallel solid vertical lines indicate the disconnected profile. The free energy of the disconnected string profile is given by
\begin{equation}
    F_{dis}=-\frac{2}{2 \pi  \alpha ' }\int_r^{\infty } \sqrt{-g_{tt}(r)g_{rr}(r) } \, dr\,.
\end{equation} 
Since the UV divergence structure of $F$ and $F_{dis}$ is the same, it allows us to regularise $F$ in a minimalistic way. 

\begin{figure}[htbp!]
\centering
\includegraphics[width=.47\textwidth]{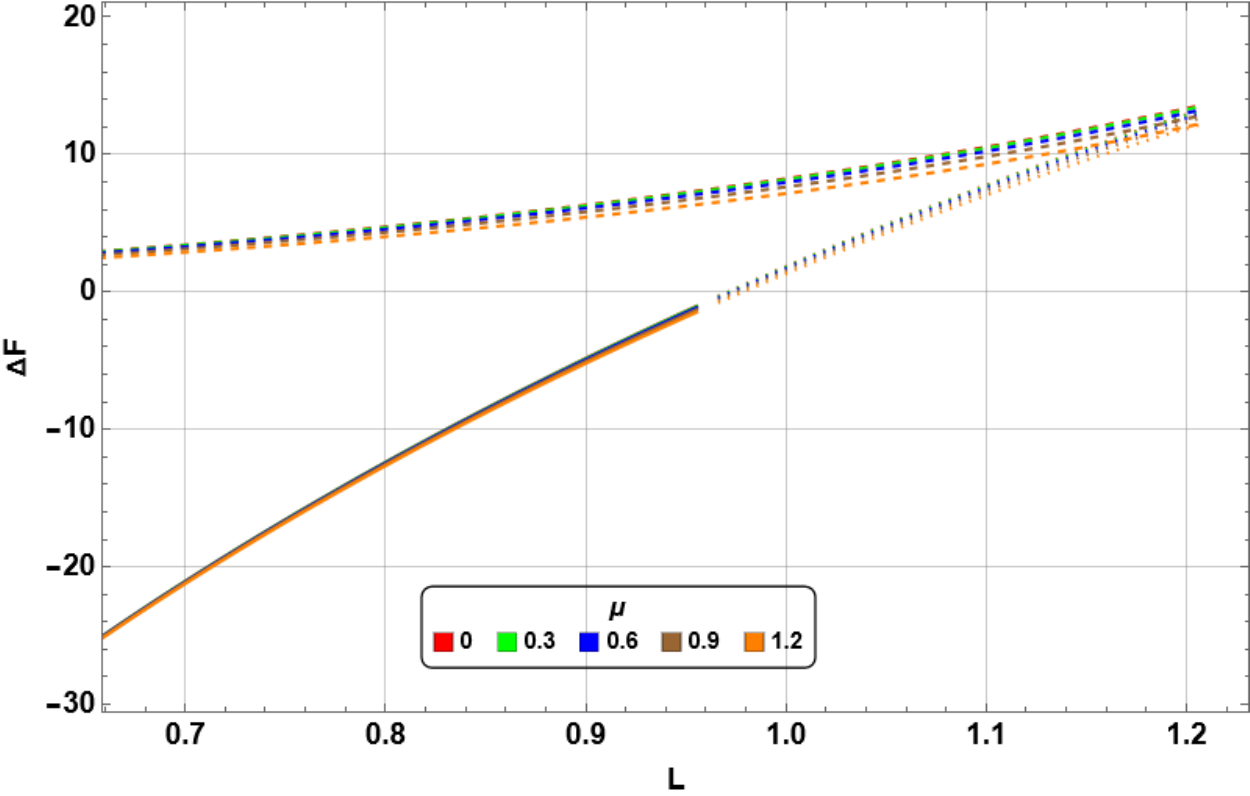}
\qquad
\includegraphics[width=.47\textwidth]{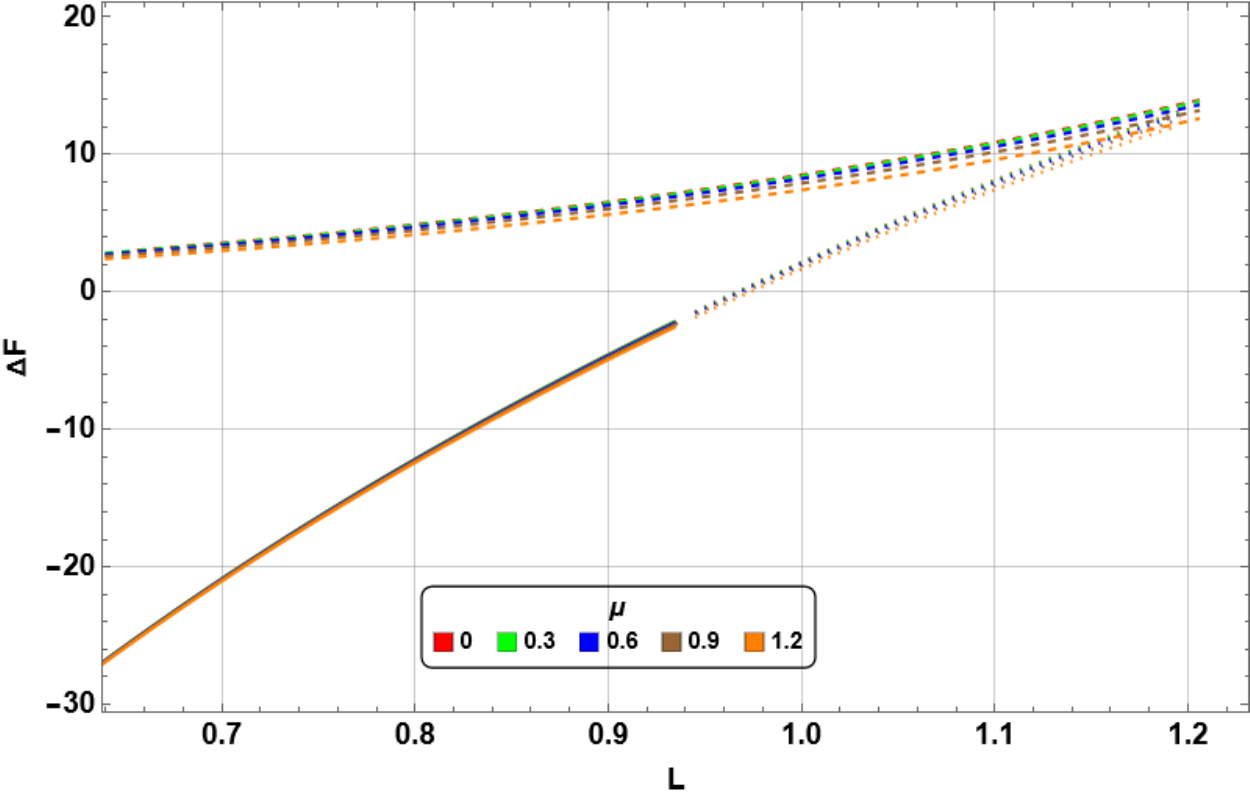}
\caption{The free energy difference is presented for the string in the $x_1$ $(\parallel)$ configuration on the left and the $x_3$ $(\perp)$ configuration on the right, at a magnetic field strength of $0.2$ GeV, across different values of the chemical potential $\mu$.   \label{fig:StringFreeEnergy}}
\end{figure}

The free energy behaviour is illustrated in Fig.~\ref{fig:StringFreeEnergy}, where the difference $\Delta F=F-F_{dis}$ is plotted. The dashed and solid lines again correspond to the small and large $r_0$ solutions, respectively. Analysing the free energy behaviour reveals a clear distinction between the large $r_0$ and small $r_0$ solutions. For the large $r_0$ configuration, it is consistently observed that this solution exhibits lower free energy compared to the small $r_0$ solution. This behaviour indicates that the true minimum energy state, and thus the most stable configuration, is associated with the large $r_0$ solution. Interestingly, the free energy of the large $r_0$ solution can exchange dominance relative to $F_{dis}$ depending on the length $L$ of the string.  Specifically, there is a critical length $L_{crit}$ at which $\Delta F =0$, and the free energy is minimised by the large $r_0$ solution only for $L\le L_{crit}$.  This string configuration gives the stable phase of the string configuration for small $L$. Beyond this critical length, for $L_{crit}\le L\le L_{max}$, the reference-free energy $F_{dis}$ becomes minimum, indicating a metastable phase of the string configuration. Dotted lines in Figs.~\ref{fig:StringLvsr0} and~\ref{fig:StringFreeEnergy} represent this metastable phase. On the other hand, the small $r_0$ solution consistently has free energy greater than $F_{dis}$, suggesting a higher and less stable energy state. The local maxima of the energy profile for the small $r_0$ solution is notably close to the horizon, emphasising its instability relative to the large $r_0$ and parallel string configurations. This unstable phase of the string configuration is represented by dashed lines in Figs.~\ref{fig:StringLvsr0} and~\ref{fig:StringFreeEnergy}. This behaviour of the string's free energy is typical for both parallel and perpendicular magnetic field configurations for all possible chemical potential values.

To explore the chaotic dynamics of the string, we focus on the small $r_0$ solution, where the free energy is comparatively large. The small $r_0$ solution consistently exhibits higher free energy than the large $r_0$ solution, indicating a less stable and more energetically excited state - as we have observed earlier from its free energy. Additionally, one could expect that black holes, being thermal objects, might enhance or induce chaotic dynamics of the string as its tip approaches the horizon.

\subsection{Perturbing the string }\label{subsec:stringperturbation}
Probing chaotic dynamics of any system requires that we introduce slight perturbations to the initial condition of the system, giving us information on how the dynamics change w.r.t. the other initial conditions. In particular, we consider a time-dependent perturbation around the static string solution and expand the Nambu-Goto action order by order in this perturbation. 

\begin{figure}[t]
	\centering
	\includegraphics[width=0.4\linewidth]{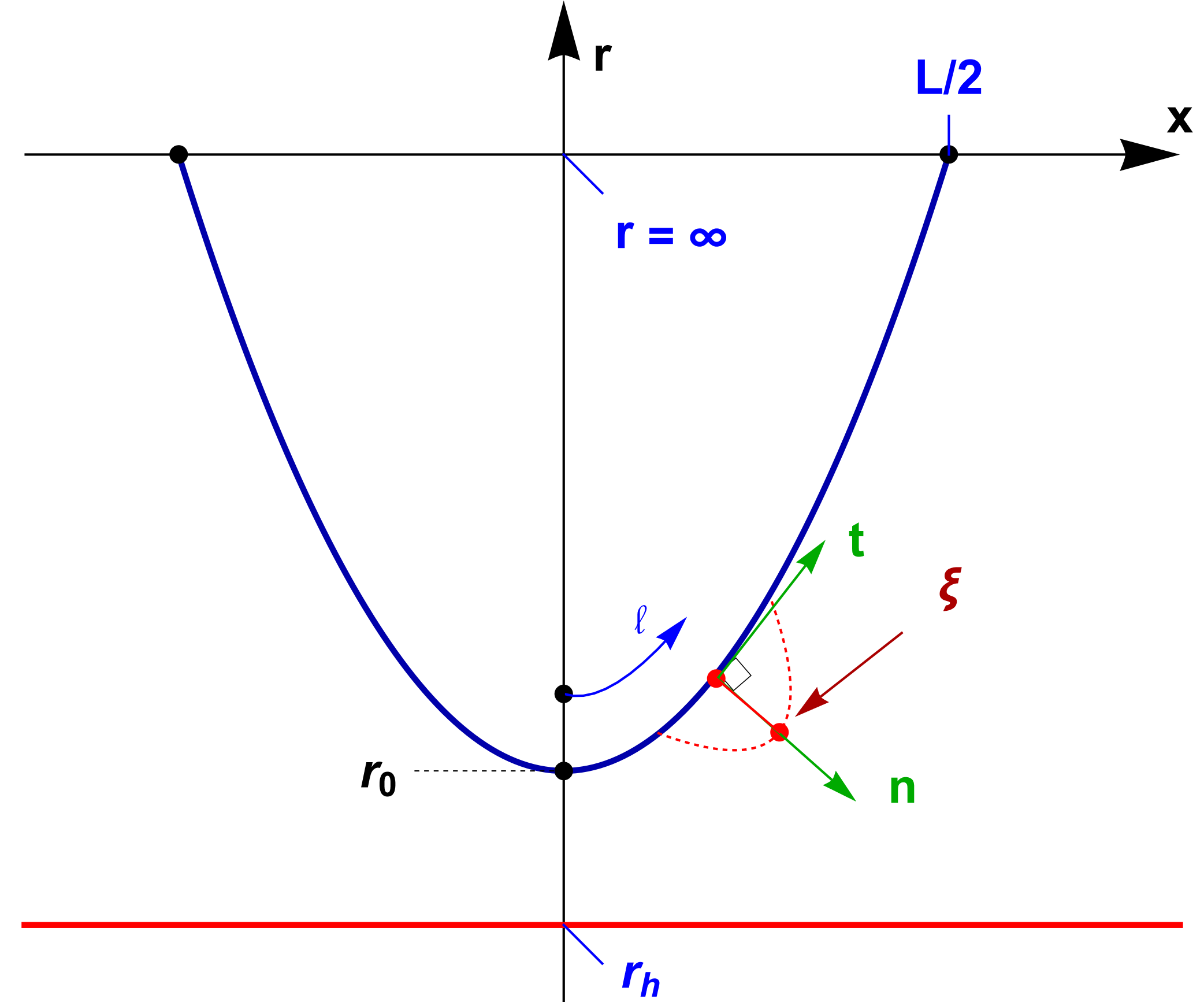}
	\caption{An illustration of the static string profile and its normal perturbation.}
	\label{fig:string-profile}
\end{figure}

To begin such analysis, we choose a specific value of the string length $L$ that lies in the unstable area. As we have seen from the previous demonstration of the free energy diagram (Fig.~\ref{fig:StringFreeEnergy}) and the length solution diagram  (Fig.~\ref{fig:StringLvsr0}), the unstable area corresponds to the small $r_0$ case, i.e., $r_0$ near the black hole horizon. Introducing the perturbation as
\begin{equation}
	r(t,\ell)=n^r(\ell) \xi (t,\ell)+r_b(\ell)\,,
 \label{eq:stringperturb01}
\end{equation}
and
\begin{equation}
	x(t,\ell)=n^x(\ell) \xi (t,\ell)+x_b(\ell)\,,
 \label{eq:stringperturb02}
\end{equation}
where $r_{b}(\ell)$ and $x_{b}(\ell)$ correspond to the static position, $\xi(t,\ell)$ is the small perturbation that carries our desired time-dependent part, and $x$ is either $x_{1}$ (parallel) or $x_{3}$ (perpendicular) depending upon the orientation of the string with respect to the magnetic field. The normal vector $n^{M} = (0,n^{x},0,0,n^{r})$ is orthogonal to tangent vector $t^{M}$ and makes sure the perturbation is perpendicular to the string profile. Accordingly, 
\begin{eqnarray}
& & 	\left(n^x\right)^2 g_{xx}(r)+\left(n^r\right)^2 g_{rr}(r)=1\,, \nonumber\\
& & n^x g_{xx}(r) x'(\ell)+n^r g_{rr}(r) r'(\ell)=0\,.
\end{eqnarray}
Next, we solve for the normal vectors $n_{x}$ and $n_{r}$ using the above equations. For an outward perturbation, as in Fig.~\ref{fig:string-profile}, these are given by
\begin{eqnarray}
& & n^x(\ell)=\sqrt{\frac{g_{rr}}{g_{xx}}} r'(\ell)\,,\\
& & n^r(\ell)=-\sqrt{\frac{g_{xx}}{g_{rr}}} x'(\ell)\,.
 \label{eq:normalvector02}
\end{eqnarray}
With Eqs.~(\ref{eq:stringperturb01}) and (\ref{eq:stringperturb02}), we can expand our action around the static configuration $r_{b}$ and $x_{b}$ up to any order in the perturbation parameter. For our analysis, we will be taking only up to the next-to-leading order term. The leading order generates an oscillatory term and the next-to-leading order produces a trapping potential. The expanded action contains quadratic order and cubic order terms in $\xi$.
The quadratic part of the action has the form
\begin{equation}
	S^{(2)}=\frac{1}{2 \pi \alpha' }\int dt \int _{-\infty }^{\infty }d\ell \left(C_{\ell\ell}^{x_i} \left(\xi '\right)^2+C_{00}^{x_i} \xi ^2+\dot{\xi }^2 C_{tt}^{x_i}\right)\,,
 \label{eq:secondorderaction}
\end{equation}
where the prime ($'$) denotes a derivative w.r.t.~$\ell$ and the dot ($^{.}$) w.r.t.~$t$. The form of the coefficients $C_{tt}$, $C_{\ell\ell}$, and $C_{00}$ are as follows
\begin{equation}\label{eq:coefficients}
\begin{aligned}
C_{tt}^{x_i}={} & \frac{e^{-A_s(r_{b}(\ell))}}{2 r_{b}(\ell) \sqrt{g(r_{b}(\ell))}} \,, \\
C_{\ell\ell}^{x_i}={} & -\frac{1}{2} r_{b}(\ell) \sqrt{g(r_{b}(\ell))} e^{A_s(r_{b}(\ell))}\,,\\
C_{00}^{x_1}={} &\frac{e^{-5 A_s(r_b(\ell))}}{8 r_b(\ell)^3 g(r_b(\ell))^{3/2} h(r_b(\ell))^2} \\ 
              &(r_0^4 g(r_0) h(r_0)e^{4A_s(r_0)}(r_b(\ell) g(r_b(\ell)) (2 h(r_b(\ell)) (g'(r_b(\ell)) \left(3 r_b(\ell) A_s'(r_b(\ell))+1\right)- \\ 
              &r_b(\ell) g''(r_b(\ell)))+r_b(\ell) g'(r_b(\ell)) h'(r_b(\ell)))+ \\ 
              &2 g(r_b(\ell))^2 (r_b(\ell) h'(r_b(\ell)) \left(r_b(\ell) A_s'(r_b(\ell))+1\right)+ \\ 
              &h(r_b(\ell)) \left(r_b(\ell)^2 \left(6 A_s'(r_b(\ell)){}^2-2 A_s''(r_b(\ell))\right)+8 r_b(\ell) A_s'(r_b(\ell))+4\right))+ \\ 
	          &2 r_b(\ell)^2 h(r_b(\ell)) g'(r_b(\ell))^2)-r_b(\ell)^4 g(r_b(\ell))^2 e^{4 A_s(r_b(\ell))} (2r_b(\ell) h(r_b(\ell)) g'(r_b(\ell)) \\ 
	          &\left(2 h(r_b(\ell)) \left(r_b(\ell) A_s'(r_b(\ell))+1\right)+r_b(\ell) h'(r_b(\ell))\right)+	g(r_b(\ell)) (2 r_b(\ell) \\ 
              &h(r_b(\ell)) (2 h'(r_b(\ell)) \left(r_b(\ell) A_s'(r_b(\ell))+2\right)+r_b(\ell) \\ 
              &h''(r_b(\ell)))+
	           4 h(r_b(\ell))^2 \left(r_b(\ell)^2 \left(A_s'(r_b(\ell)){}^2+A_s''(r_b(\ell))\right)+
	           4 r_b(\ell) A_s'(r_b(\ell))+2\right)- \\ 
              &r_b(\ell)^2 h(r_b(\ell))^2)))\,.
\end{aligned}
\end{equation}
where $i=1$ and $i=3$ again correspond to the parallel and perpendicular magnetic field orientations, respectively. The expression of $C_{00}^{x_3}$ is identical to $C_{00}^{x_1}$, except $h(r)$ is to be replaced by $q(r)$. We determine the equation of motion from the perturbed action to analyse the dynamics. The equation of motion is given by
\begin{equation}
	C_{tt}^{x_{i}}\ddot{\xi }+\partial_{\ell} \left(C_{\ell\ell}^{x_{i}} \xi '\right)-C_{00}^{x_{i}} \xi =0\,,
\end{equation}
using the factorization $\xi(t,\ell)=\xi(\ell)e^{i\omega t}$, the above equation can be re-written into a Sturm-Liouville form
\begin{equation}
	\partial_{\ell} \left(C_{\ell\ell}^{x_{i}} \xi '\right)-C_{00}^{x_{i}} \xi =  \omega ^2 \xi C_{tt}^{x_{i}}\,,
 \label{eq:Strum-Liouville}
\end{equation}
where $W(\ell) = -C_{tt}^{x_{i}}(\ell)$ is the weight function, with inner product
\begin{equation}\label{28}
	(\xi,\zeta) \equiv \int_{-\infty}^{\infty} W(\ell)\xi(\ell)\zeta(\ell) d\ell\,.
\end{equation}
To solve Eq.~(\ref{eq:Strum-Liouville}) numerically, we impose the boundary condition $\xi(\ell)\xrightarrow{\ell\longrightarrow  \pm\infty}0$ and set the total length of the string $L=1.1$. Fixing $L$ to a particular value makes $r_0$ a $\mu$ and $B$ dependent quantity. Note that $L=1.1$ is arbitrary but as long as it lies in the unstable region, most of our results remain qualitatively the same for different values of $L$.

For $L=1.1$, we have collected some $r_0$ values for the unstable string at different magnetic field and chemical potential values in Table~\ref{tab:StringLvsr0}. The value of $r_0$ increases if we increase either the magnetic field or the chemical potential. This implies that the tip of the string moves away from the black hole horizon when we increase these parameters. There have been suggestions that the black hole can act as a source of chaos \cite{Hashimoto:2018fkb, Shukla:2023pbp}.  If indeed this is true, we can also expect a decrease in chaos with the chemical potential and magnetic field as the tip of the string moves away from the horizon for higher values of these parameters. This expectation turns out to be true.

\begin{table}[!htb]
    \setlength\arraycolsep{4pt} 
    \renewcommand{\arraystretch}{1.25}
    
    $\begin{array}[t]{@{}
    r S[table-format=3.5] S[table-format=-3.5]
    @{\hspace{12pt}} !{\vrule width 0.2pt} @{\hspace{12pt}}
    r S[table-format=3.5] S[table-format=-3.5]
    @{\hspace{12pt}} !{\vrule width 0.2pt} @{\hspace{12pt}}
    r S[table-format=3.5] S[table-format=-3.5]
    @{}}
     \toprule
     \mu & {r_{0}~(||)} & {r_{0}~(\perp)} & \mu & {r_{0}~(||)} & {r_{0}~(\perp)} & \mu & {r_{0}~(||)} & {r_{0}~(\perp)} \\
     \midrule
    
     \multicolumn{9}{c}{\mathbf{B=0.0} \hspace{4cm} \mathbf{B=0.1} \hspace{4cm} \mathbf{B=0.2}}\\
     0.0 & 1.08318 & 1.08318 & 0.0 & 1.08382 & 1.08429 & 0.0 & 1.08581 & 1.08768\\
     0.3 & 1.08334 & 1.08334 & 0.3 & 1.08398 & 1.08445 & 0.3 & 1.08596 & 1.08786\\
     0.6 & 1.08386 & 1.08386 & 0.6 & 1.08449 & 1.08499 & 0.6 & 1.08647 & 1.08843\\
     0.9 & 1.08485 & 1.08485 & 0.9 & 1.08547 & 1.08600 & 0.9 & 1.08742 & 1.08951\\
     1.2 & 1.08657 & 1.08657 & 1.2 & 1.08718 & 1.08776 & 1.2 & 1.08908 & 1.09161\\
     \addlinespace
     \multicolumn{9}{c}{\mathbf{B=0.3} \hspace{4cm} \mathbf{B=0.4} \hspace{4cm} \mathbf{B=0.5}}\\
     0.0 & 1.08940 & 1.09417 & 0.0 & 1.09598 & 1.10519 & 0.0 & 1.10728 & 1.12473\\
     0.3 & 1.08955 & 1.09441 & 0.3 & 1.09615 & 1.10551 & 0.3 & 1.10750 & 1.12524\\
     0.6 & 1.09004 & 1.09516 & 0.6 & 1.09671 & 1.10652 & 0.6 & 1.10818 & 1.12686\\
     0.9 & 1.09113 & 1.09657 & 0.9 & 1.09778 & 1.10840 & 0.9 & 1.10947 & 1.12985\\
     1.2 & 1.09302 & 1.09895 & 1.2 & 1.09963 & 1.11177 & 1.2 & 1.11202 & 1.13549\\
    
     \bottomrule
     \end{array}$ 
\caption{\label{tab:StringLvsr0}The magnitude of $r_0$ for the unstable string configuration is shown here for various magnetic field and chemical potential values for both parallel and perpendicular orientations. The parameter $L = 1.1$ is used, with all quantities in units of GeV.}
\end{table}

\begin{figure}[htb!]
	\centering
	\subfigure[Parallel configuration]{\label{fig:eigen_states_x1_stringframe}	\includegraphics[width=0.47\linewidth]{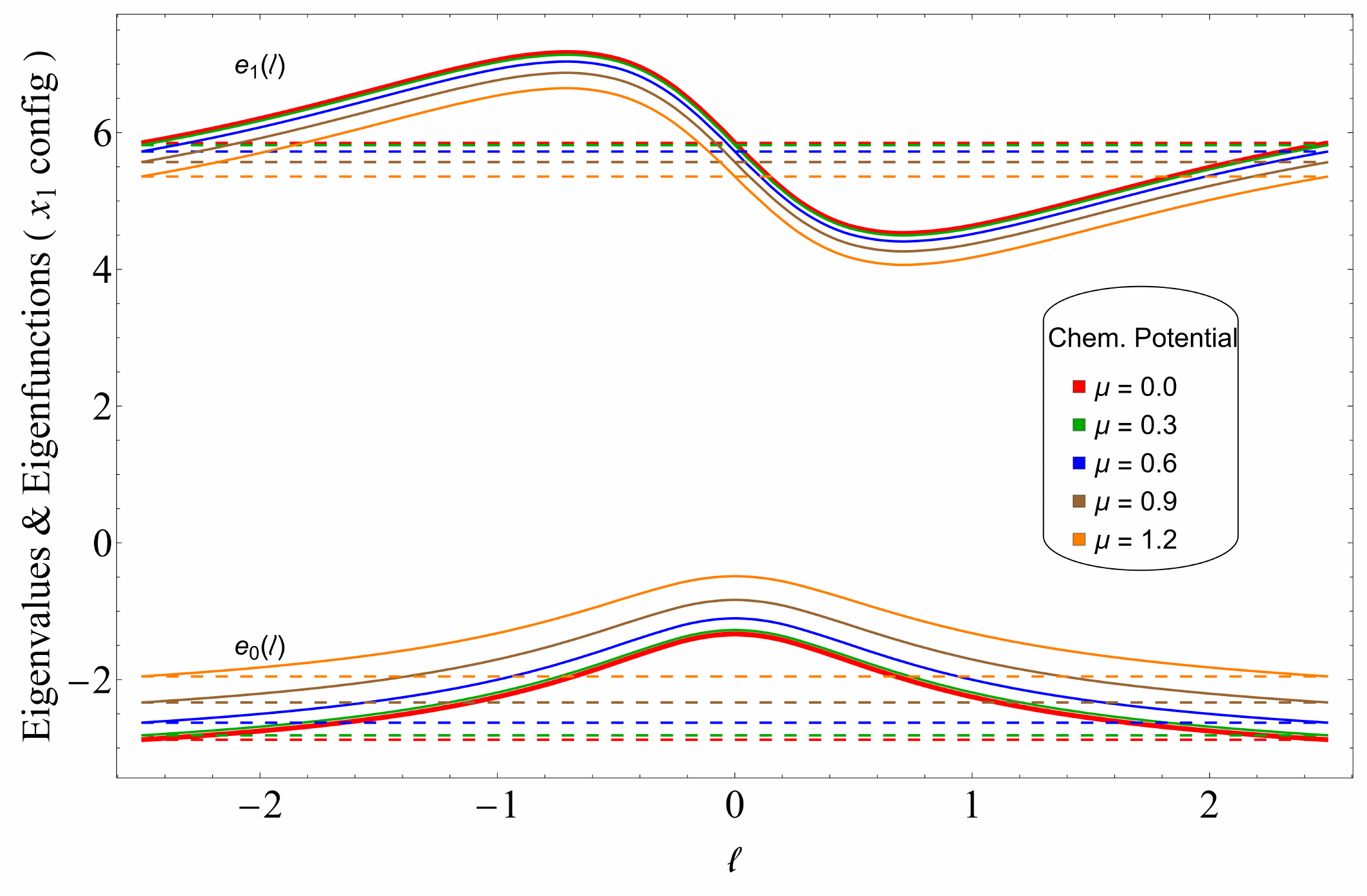}}
    \subfigure[Perpendicular configuration]{\label{fig:eigen_states_x3_stringframe}
	\includegraphics[width=0.47\linewidth]{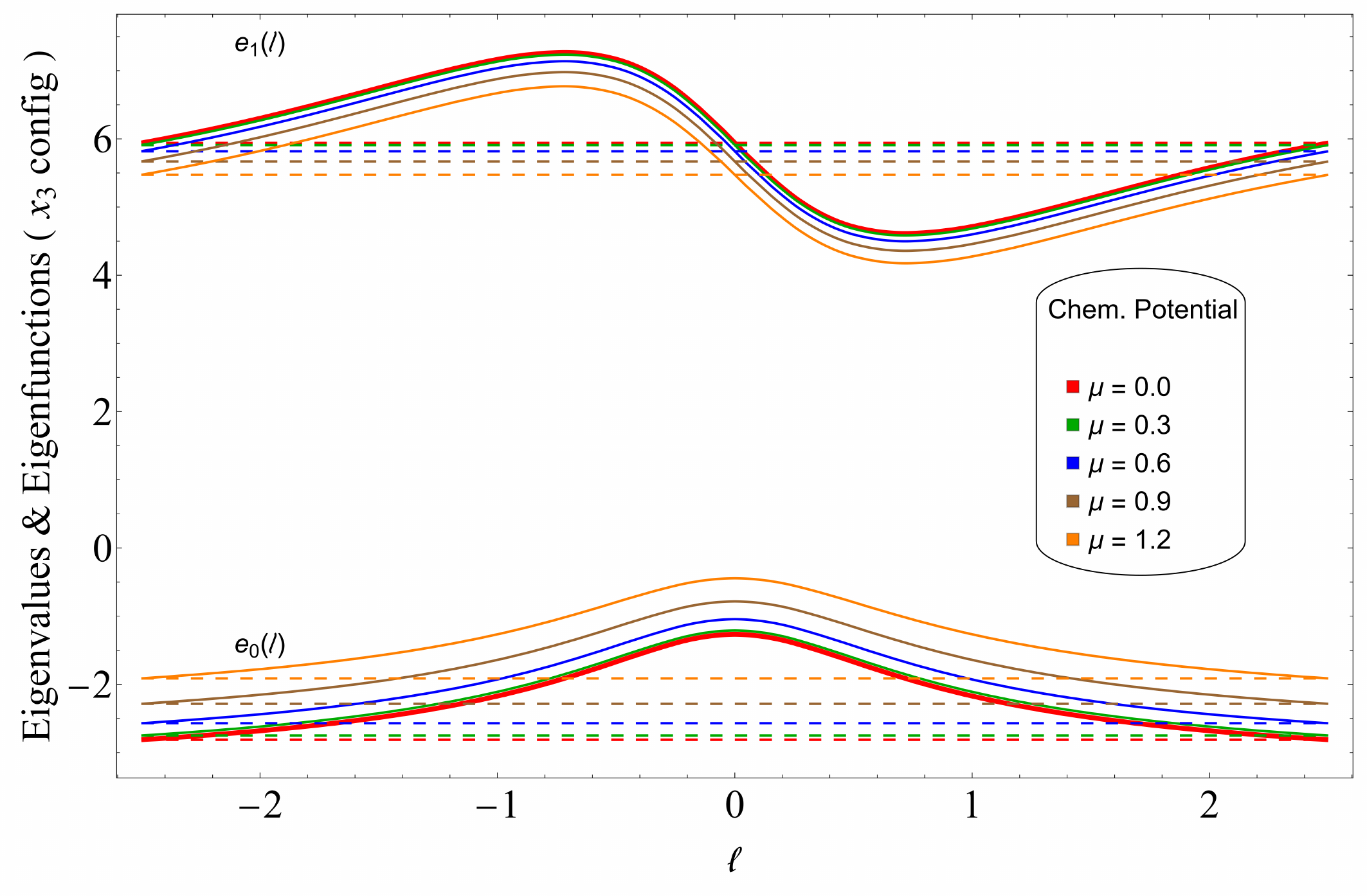}}
	\caption{\label{fig3}The eigenfunctions $e_{0}(\ell)$ and $e_{1}(\ell)$ from Eq.~(\ref{eq:Strum-Liouville}) are shown as a function of $\ell$ for different values of the chemical potential for both parallel and perpendicular string configurations. The values $B = 0.2$ and $L = 1.1$ are used, with all quantities in units of GeV.}
\end{figure}

Both eigenvalues and eigenfunctions of Eq.~(\ref{eq:Strum-Liouville}) can be obtained numerically. We collected the two lowest eigenvalues ($\omega_0^2$, $\omega_1^2$) for different values of $B$ and $\mu$, for both parallel and perpendicular cases, in Table~\ref{tab:StringEigenValues}. The corresponding eigenfunctions $\xi(\ell)=e_{0}(\ell)$ and $\xi(\ell)=e_{1}(\ell)$ are even and odd functions of $\ell$, respectively, and are illustrated in Fig.~\ref{fig3}. The occurrence of negative eigenvalues is a common feature of unstable systems. That is indeed the case for the unstable string, where we can see that the lowest eigenvalue is negative for all values of $B$ and $\mu$. This observation suggests instability of the string under perturbation. Moreover, the value of our lowest eigenvalue $\omega_{0}^2$ increases if we increase either $B$ or $\mu$. This is true for both parallel and perpendicular string configurations. Hence, we can expect our system to stabilise if we increase these parameters. Moreover, for a fixed value of $B$ or $\mu$, the magnitude of $\omega_{0}^2$ is more negative for the parallel case than the perpendicular one. This suggests that the magnetic field induces more instability in the string configuration in the former case compared to the latter case. Later, via the Poincar\'{e} section and Lyapunov exponent,  we will see that these expectations indeed turn out to be true. Let us mention here that our results with $B$ are different from the results of \cite{Colangelo:2021kmn}, where more stability was found for the parallel case compared to the perpendicular one. 

\begin{table}[!htb]
    \setlength\arraycolsep{4pt} 
    \renewcommand{\arraystretch}{1.25}
    
    $\begin{array}[t]{@{}
    r S[table-format=3.5] S[table-format=-3.5]
    @{\hspace{12pt}} !{\vrule width 0.2pt} @{\hspace{12pt}}
    r S[table-format=3.5] S[table-format=-3.5]
    @{}}
     \toprule
     \mu & {\omega_{0}^{2}~(||)} & {\omega_{1}^{2}~(||)} & \mu & {\omega_{0}^{2}~(\perp)} & {\omega_{1}^{2}~(\perp)} \\
     \midrule
    
     \multicolumn{6}{c}{\mathbf{B=0.0}}\\
     0.0 & -2.9503 & 5.8881 & 0.0 & -2.9503 & 5.8881\\
     0.3 & -2.8844 & 5.8561 & 0.3 & -2.8844 & 5.8561\\
     0.6 & -2.6911 & 5.7604 & 0.6 & -2.6911 & 5.7604\\
     0.9 & -2.3847 & 5.6024 & 0.9 & -2.3847 & 5.6024\\
     1.2 & -1.9891 & 5.3863 & 1.2 & -1.9891 & 5.3863\\
     \addlinespace
     \multicolumn{6}{c}{\mathbf{B=0.1}} \\  
     0.0 & -2.9165 & 5.9011 & 0.0 & -2.9155 & 5.9006\\
     0.3 & -2.8517 & 5.8691 & 0.3 & -2.8508 & 5.8688\\
     0.6 & -2.6617 & 5.7736 & 0.6 & -2.6607 & 5.7746\\
     0.9 & -2.4110 & 5.5798 & 0.9 & -2.3598 & 5.6186\\
     1.2 & -2.0498 & 5.3305 & 1.2 & -1.9710 & 5.4058 \\
     \addlinespace
     \multicolumn{6}{c}{\mathbf{B=0.2}} \\     
     0.0 & -2.8151 & 5.9427 & 0.0 & -2.8121 & 5.9394\\
     0.3 & -2.7537 & 5.9108 & 0.3 & -2.7506 & 5.9090\\
     0.6 & -2.5730 & 5.8165 & 0.6 & -2.5707 & 5.8181\\
     0.9 & -2.2864 & 5.6604 & 0.9 & -2.2852 & 5.6683\\
     1.2 & -1.9155 & 5.4464 & 1.2 & -1.9127 & 5.4728\\
     
     \bottomrule
     \end{array}$ 
     \hfill       
     $\begin{array}[t]{@{}
    r S[table-format=3.5] S[table-format=3.5]
    @{\hspace{12pt}} !{\vrule width 0.2pt} @{\hspace{12pt}}
    r S[table-format=3.5] S[table-format=3.5]
    @{}}
     \toprule
     \mu & {\omega_{0}^{2}~(||)} & {\omega_{1}^{2}~(||)} & \mu & {\omega_{0}^{2}~(\perp)} & {\omega_{1}^{2}~(\perp)} \\
     \midrule
    
     \multicolumn{6}{c}{\mathbf{B=0.3}}\\
     0.0 & -2.6448 & 6.0230 & 0.0 & -2.6291 & 6.0307\\
     0.3 & -2.5884 & 5.9917 & 0.3 & -2.5724 & 6.0034\\
     0.6 & -2.4229 & 5.8987 & 0.6 & -2.4068 & 5.9221\\
     0.9 & -2.2036 & 5.7089 & 0.9 & -2.1439 & 5.7893\\
     1.2 & -1.8119 & 5.5503 & 1.2 & -1.8042 & 5.6114\\
     \addlinespace
     \multicolumn{6}{c}{\mathbf{B=0.4}} \\  
     0.0 & -2.3834 & 6.1918 & 0.0 & -2.3512 & 6.2170\\
     0.3 & -2.3385 & 6.1621 & 0.3 & -2.3015 & 6.1938\\
     0.6 & -2.1879 & 6.0742 & 0.6 & -2.1558 & 6.1251\\
     0.9 & -1.9554 & 5.9299 & 0.9 & -1.9246 & 6.0417\\
     1.2 & -1.6527 & 5.7349 & 1.2 & -1.6221 & 5.8790\\
     \addlinespace
     \multicolumn{6}{c}{\mathbf{B=0.5}} \\     
     0.0 & -2.0131 & 6.5095 & 0.0 & -1.9275 & 6.6160\\
     0.3 & -1.9716 & 6.4825 & 0.3 & -1.8855 & 6.6001\\
     0.6 & -1.8502 & 6.4016 & 0.6 & -1.7621 & 6.5549\\
     0.9 & -1.6563 & 6.2700 & 0.9 & -1.5661 & 6.4866\\
     1.2 & -1.3980 & 6.1060 & 1.2 & -1.3018 & 6.4314\\
     
     \bottomrule
     \end{array}$ 
\caption{\label{tab:StringEigenValues}The first two eigenvalues, $\omega_{0}^{2}$ and $\omega_{1}^{2}$, for the unstable string configuration are presented here for various magnetic field and chemical potential values for both parallel and perpendicular orientations. The parameter $L = 1.1$ is used, with all values in units of GeV.}
\end{table}

Now we expand the string action up to cubic order in the perturbation. Considering only the two lowest eigenfunctions,
\begin{equation}
    \xi(t,\ell) = c_0(t)e_0(\ell)+c_1(t)e_1(\ell)\,,
\end{equation}
the cubic order term of the expanded action, up to the surface terms, takes the form 
\begin{equation}
	S^{(3)}=\frac{1}{2 \pi \alpha' }\int dt \int _{-\infty }^{\infty } d\ell \left(D_0^{x_i} \xi ^3+D_1^{x_i} \xi \left(\xi '\right)^2+D_2^{x_i} \dot{\xi }^2 \xi \right)\,,
\end{equation}
where $c_{0}(t)$ and $c_{1}(t)$ encode the perturbation's time dependence and $D_{0,1,2}^{x_{i}}$ are functions of $\ell$.  In terms of $e_0$ and $e_1$, the above action becomes
\begin{eqnarray}
&&S^{(3)}=\frac{1}{2\pi \alpha' }\int dt\int _{-\infty }^{\infty }d\ell[\left(e_0^3 D_0^{x_i}+D_1^{x_i}e_0\acute{e}_0{}^2\right)c_0^3+
    3 e_0 e_1^2 c_0c_1^2 D_0^{x_i}\\ \nonumber
&& +\left(2e_0e_1\acute{e}_1+e_0\acute{e}_1{}^2\right)c_0c_1^2 D_1^{x_i}+\left(c_0 \dot{c}_0^2 e_0^3+c_0 \dot{c}_1^2 e_1^2 e_0+2 c_1 \dot{c}_0 \dot{c}_1 e_1^2 e_0\right) D_2^{x_i}]\,.
    \label{eq:thirdorderaction}
\end{eqnarray}
Summing $S^{(2)}$ + $S^{(3)}$, and integrating out $\ell$ from the equation, we get
\begin{equation}
        S^{(3)}+S^{(2)}=\frac{1}{2\pi \alpha' }\int dt[\sum _{n=0,1} \left(\dot{c}_n^2-c_n^2 \omega _n^2\right)+c_0^3 K_1^{x_i}+  
        c_0 c_1^2 K_2^{x_i}+c_0 \dot{c}_0^2 K_3^{x_i}+c_0 \dot{c}_1^2 K_4^{x_i}+\dot{c}_0 c_1 \dot{c}_1 K_5^{x_i}]. 
        \label{eq:S2S3}
\end{equation}
The above action can be used to set the equation of motion for $c_0(t)$ and $c_1(t)$. The coefficients $K_{1,\ldots,5}^{x_i}$ can be obtained numerically. Their values depend on the magnitude of $L$, $B$, and $\mu$. For $L=1.1$, the values of $K_{1,\ldots,5}^{x_i}$ are collected in the Table~\ref{tab:StringKValues} for different values of $B$ and $\mu$.

\begin{table}[!htb]
    \setlength\arraycolsep{4pt} 
    \renewcommand{\arraystretch}{1.25}
    
    $\begin{array}[t]{@{}
    r S[table-format=3.3] S[table-format=3.3] S[table-format=3.3] S[table-format=3.3] S[table-format=3.3]
    @{\hspace{10pt}} !{\vrule width 0.2pt} @{\hspace{10pt}}
    r S[table-format=3.3] S[table-format=3.3] S[table-format=3.3] S[table-format=3.3] S[table-format=3.3]
    @{}}
     \toprule
     \mu & {K_{1}~(||)} & {K_{2}~(||)} & {K_{3}~(||)} & {K_{4}~(||)} & {K_{5}~(||)} & {K_{1}~(\perp)} & {K_{2}~(\perp)} & {K_{3}~(\perp)} & {K_{4}~(\perp)} & {K_{5}~(\perp)}\\
     \midrule
    
     \multicolumn{11}{c}{\mathbf{B=0.0}}\\
     0.0 & 9.873 & 12.432 & 5.896 & 2.197 & 4.394 & 9.873 & 12.432 & 5.896 & 2.197 & 4.394\\
     0.3 & 9.577 & 12.174 & 5.866 & 2.197 & 4.395 & 9.577 & 12.174 & 5.866 & 2.197 & 4.395\\
     0.6 & 8.725 & 11.417 & 5.777 & 2.197 & 4.395 & 8.725 & 11.417 & 5.777 & 2.197 & 4.395\\
     0.9 & 7.423 & 10.223 & 5.631 & 2.197 & 4.395 & 7.423 & 10.223 & 5.631 & 2.197 & 4.395\\
     1.2 & 5.835 & 8.694 & 5.432 & 2.197 & 4.394 & 5.835 & 8.694 & 5.432 & 2.197 & 4.394\\
     \addlinespace
     \multicolumn{11}{c}{\mathbf{B=0.1}} \\  
     0.0 & 9.782 & 12.352 & 5.885 & 2.196 & 4.391 & 9.771 & 12.340 & 5.872 & 2.197 & 4.395\\
     0.3 & 9.491 & 12.098 & 5.855 & 2.196 & 4.392 & 9.481 & 12.086 & 5.843 & 2.197 & 4.395\\
     0.6 & 8.653 & 11.352 & 5.768 & 2.196 & 4.392 & 8.644 & 11.340 & 5.755 & 2.198 & 4.395\\
     0.9 & 7.371 & 10.176 & 5.625 & 2.196 & 4.392 & 7.365 & 10.165 & 5.612 & 2.198 & 4.395\\
     1.2 & 5.805 & 8.666 & 5.429 & 2.196 & 4.392 & 5.803 & 8.658 & 5.417 & 2.197 & 4.394\\
     \addlinespace
     \multicolumn{11}{c}{\mathbf{B=0.2}} \\  
     0.0 & 9.511 & 12.115 & 5.852 & 2.191 & 4.382 & 9.476 & 12.069 & 5.804 & 2.198 & 4.397\\
     0.3 & 9.234 & 11.871 & 5.824 & 2.191 & 4.382 & 9.201 & 11.826 & 5.776 & 2.198 & 4.397\\
     0.6 & 8.545 & 11.240 & 5.753 & 2.191 & 4.382 & 8.409 & 11.116 & 5.694 & 2.199 & 4.398\\
     0.9 & 7.429 & 10.195 & 5.629 & 2.191 & 4.382 & 7.195 & 9.993 & 5.559 & 2.199 & 4.398\\
     1.2 & 5.714 & 8.587 & 5.418 & 2.193 & 4.386 & 5.703 & 8.547 & 5.370 & 2.198 & 4.395\\
     \addlinespace
     \multicolumn{11}{c}{\mathbf{B=0.3}} \\  
     0.0 & 9.066 & 11.734 & 5.797 & 2.182 & 4.364 & 8.982 & 11.611 & 5.684 & 2.199 & 4.397\\
     0.3 & 8.812 & 11.509 & 5.772 & 2.182 & 4.365 & 8.730 & 11.385 & 5.658 & 2.199 & 4.397\\
     0.6 & 8.078 & 10.848 & 5.695 & 2.184 & 4.367 & 8.005 & 10.726 & 5.581 & 2.199 & 4.397\\
     0.9 & 6.942 & 9.795 & 5.566 & 2.185 & 4.369 & 6.893 & 9.686 & 5.456 & 2.198 & 4.397\\
     1.2 & 5.545 & 8.440 & 5.389 & 2.185 & 4.371 & 5.527 & 8.354 & 5.287 & 2.198 & 4.396\\
     \addlinespace
     \multicolumn{11}{c}{\mathbf{B=0.4}} \\  
     0.0 & 8.414 & 11.182 & 5.697 & 2.165 & 4.329 & 8.293 & 10.961 & 5.510 & 2.197 & 4.393\\
     0.3 & 8.189 & 10.980 & 5.675 & 2.165 & 4.330 & 8.073 & 10.760 & 5.487 & 2.197 & 4.393\\
     0.6 & 7.540 & 10.390 & 5.607 & 2.167 & 4.334 & 7.440 & 10.175 & 5.420 & 2.197 & 4.393\\
     0.9 & 6.536 & 9.452 & 5.494 & 2.169 & 4.339 & 6.467 & 9.251 & 5.310 & 2.196 & 4.393\\
     1.2 & 5.292 & 8.238 & 5.341 & 2.172 & 4.345 & 5.262 & 8.062 & 5.159 & 2.195 & 4.391\\
     \addlinespace
     \multicolumn{11}{c}{\mathbf{B=0.5}} \\  
     0.0 & 7.552 & 10.485 & 5.543 & 2.134 & 4.268 & 7.378 & 10.074 & 5.261 & 2.187 & 4.373\\
     0.3 & 7.364 & 10.314 & 5.524 & 2.135 & 4.270 & 7.196 & 9.905 & 5.242 & 2.186 & 4.373\\
     0.6 & 6.819 & 9.810 & 5.467 & 2.137 & 4.275 & 6.674 & 9.410 & 5.185 & 2.186 & 4.373\\
     0.9 & 5.972 & 9.008 & 5.374 & 2.141 & 4.283 & 5.869 & 8.629 & 5.094 & 2.186 & 4.372\\
     1.2 & 4.907 & 7.960 & 5.241 & 2.145 & 4.290 & 4.864 & 7.621 & 4.964 & 2.184 & 4.368\\
     
     \bottomrule
     \end{array}$ 
\caption{\label{tab:StringKValues} The $K$ values are depicted for various $\mu$ and $B$ values for both parallel and perpendicular string orientations within the string frame. The parameter $L = 1.1$ is used, with all quantities expressed in units of GeV.}
\end{table}

Interestingly, the above action (\ref{eq:S2S3}) contains a trapping potential for the unstable string configuration. Analysing the dynamics of $c_0$ and $c_1$ around the trapping potential is crucial to establish the stability of the string motion. However, in some parts of the potential region, the kinetic term of $c_0$ and $c_1$ in action (\ref{eq:S2S3}) becomes negative. In order to avoid such a situation and make the kinetic term positive definite, we transform $c_{0,1}\rightarrow\Tilde{c}_{0,1}$, where $c_0=\Tilde{c}_{0} + \alpha_{1}\Tilde{c}_{0}^{2} + \alpha_{2}\Tilde{c}_{1}^{2}$ and $c_{1}=\Tilde{c}_{1}+\alpha_{3}\Tilde{c}_{0}\Tilde{c}_{1}$, following the strategy adopted in \cite{Hashimoto:2018fkb, Shukla:2023pbp, Colangelo:2020tpr}. Now, the values of $\alpha_{1}$, $\alpha_{2}$, and $\alpha_{3}$ can be chosen to get rid of the negative kinetic situation. One such choice for $\alpha$'s are $\alpha_{1}=-1.35$, $\alpha_{2}=-0.5$, and $\alpha_{3}=-1$. Using the modified values, our action becomes
\begin{equation}
        S^{(\text{modified})}=\frac{1}{2\pi \alpha' }\int dt[\sum _{n=0,1} \left(\dot{\Tilde{c}}_{n}^{2}-\Tilde{c}_{n}^{2} \omega _n^2\right)+\Tilde{c}_{0}^{3} \Tilde{K}_{1}^{x_i}+  
        \Tilde{c}_{0} \Tilde{c}_{1}^{2} \Tilde{K}_{2}^{x_i}+\Tilde{c}_{0} \dot{\Tilde{c}}_{0}^{2} \Tilde{K}_{3}^{x_i}+\Tilde{c}_{0} \dot{\Tilde{c}}_{1}^{2} \Tilde{K}_{4}^{x_i}+\dot{\Tilde{c}}_{0} \Tilde{c}_{1} \dot{\Tilde{c}}_{1} \Tilde{K}_{5}^{x_i}].
    \label{eq:ActionModified}
\end{equation}
The above transformation makes the time evolution of $c_0$ and $c_1$ well-posed without affecting
their dynamics. In addition to that the chaotic structure also shows up in the modified action. From the modified action we can further find out the equations of motion of $\tilde{c_0}$ and $\tilde{c_1}$
\begin{equation}
\begin{aligned}
    \ddot{\Tilde{c}}_{0}={} &-\frac{1}{4 \left(\Tilde{c}_{0} \left(4 \alpha _1+K_{3}\right)+1\right) \left(\Tilde{c}_{0} \left(2 \alpha _3+K_{4}\right)+1\right)-\Tilde{c}_{1}^{2} \left(4 \alpha _2+2 \alpha _3+K_{5}\right)^2} \\ 
    &\times \left(2 \Tilde{c}_{0} \left(2 \alpha _3+K_{4}\right)+2\right) (4 \alpha _1 \dot{\Tilde{c}}_{0}^{2}+4 \alpha _2 \dot{\Tilde{c}}_{1}^{2}-3 \Tilde{c}_{0}^{2} \left(K_{1}-2 \alpha _1 \omega _0^2\right) \\ 
    &+\Tilde{c}_{1}^{2} \left(2 \alpha _2 \omega _0^2+2 \alpha _3 \omega _1^2-K_{2}\right)+\dot{\Tilde{c}}_{0}^{2} K_{3}-\dot{\Tilde{c}}_{1}^{2} K_{4}+\dot{\Tilde{c}}_{1}^{2} K_{5}+2 \Tilde{c}_{0} \omega _0^2) \\ 
    &-2 \Tilde{c}_{1} \left(4 \alpha _2+2 \alpha _3+K_{5}\right) (\Tilde{c}_{0} \Tilde{c}_{1} \left(2 \alpha _2 \omega _0^2+2 \alpha _3 \omega _1^2-K_{2}\right)+\dot{\Tilde{c}}_{0} \dot{\Tilde{c}}_{1} \left(2 \alpha _3+K_{4}\right)+\Tilde{c}_{1} \omega _1^2)\,,
\end{aligned}
\end{equation}
\begin{equation}
\begin{aligned}
    \ddot{\Tilde{c}}_1={} &\frac{1}{4 \Tilde{c}_{0}^{2} \left(4 \alpha                       _1+K_{3}\right) \left(2 \alpha                             _3+K_{4}\right)+4 \Tilde{c}_{0}                                \left(4 \alpha _1+2 \alpha                                       _3+K_{3}+K_{4}\right)-                             \Tilde{c}_{1}^{2} \left(4 \alpha _2+2                                \alpha _3+K_{5}\right)^2+4} \\ 
                          &(\Tilde{c}_{1}^{3} \left(4 \alpha _2+2 \alpha _3+K_{5}\right) \left(2 \alpha _2 \omega _0^2+2 \alpha _3 \omega _1^2-K_{2}\right)-\Tilde{c}_{0}^{2} \Tilde{c}_{1} (8 \alpha _1 \alpha _2 \omega _0^2-12 \alpha _1 \alpha _3 \omega _0^2+32 \alpha _1 \alpha _3 \omega _1^2\\ 
                          &-6 \alpha _{1} K_{5} \omega _0^2+8 \alpha _2 K_{3} \omega _0^2+8 \alpha _3 K_{3} \omega _1^2-4 K_{2} \left(4 \alpha _1+K_{3}\right)+3 K_{1} \left(4 \alpha _2+2 \alpha _3+K_{5}\right))\\ 
                          &+\Tilde{c}_{1} (\dot{\Tilde{c}}_{0}^{2} \left(4 \alpha _1+K_{3}\right) \left(4 \alpha _2+2 \alpha _3+K_{5}\right)-\dot{\Tilde{c}}_{1}^{2} \left(-4 \alpha _2+K_{4}-K_{5}\right)\\ 
                          &\times \left(4 \alpha _2+2 \alpha _3+K_{5}\right)-4 \omega _1^2)+2 \Tilde{c}_{0} (\Tilde{c}_{1} \left(2 \alpha _3 \omega _0^2-8 \alpha _1 \omega _1^2-4 \alpha _3 \omega _1^2+K_{5} \omega _0^2-2 K_{3} \omega _1^2+2 K_{2}\right)\\ 
                          &-2 \dot{\Tilde{c}}_0 \dot{\Tilde{c}}_1 \left(4 \alpha _1+K_{3}\right) \left(2 \alpha _3+K_{4}\right))-4 \dot{\Tilde{c}}_{0} \dot{\Tilde{c}}_{1} \left(2 \alpha _3+K_{4}\right))\,.
\end{aligned}
\end{equation}
We have written down these equations of motion explicitly as they will be useful later on in the analysis of the Lyapunov exponent at the unstable fixed point.

\subsection{Poincar\'{e} sections} \label{StringPoincare}
Poincar\'{e} sections are used to better observe the qualitative nature of chaotic dynamics. Although we can not use it to quantify the chaos of string dynamics (which can be done using the Lyapunov exponents, see the next subsection~\ref{StringLyapunov}), we can get information about where the chaotic behavior of the string emerges via the Poincar\'{e} section. We take our equation of motion and trace out in the phase space with time as the parameter and construct the Poincar\'{e} section for bound orbits with the section identified by $\tilde{c}_{1}(t)=0$ and $\dot{\tilde{c}}_{1}(t)\ge0$ in the phase space. The conditions $\{\tilde{c}_{1}(t)=0, \dot{\tilde{c}}_{1}(t)\ge0 \}$ correspond to bound orbits within the trapping potential. To obtain the Poincar\'{e} sections, we start with various initial conditions with fixed energy $E=10^{-5}$ and time interval  $0<t<15000$. This energy is chosen for illustrative purposes. 

\begin{figure}[htbp!]
	\centering
    \begin{tabular}{c c c}
		\textbf{$\mu$ value} & \textbf{Parallel Configuration} & \textbf{Perpendicular Configuration} \\
		\textbf{$\mu=0.0$} & \includegraphics[scale=0.2,valign=c]{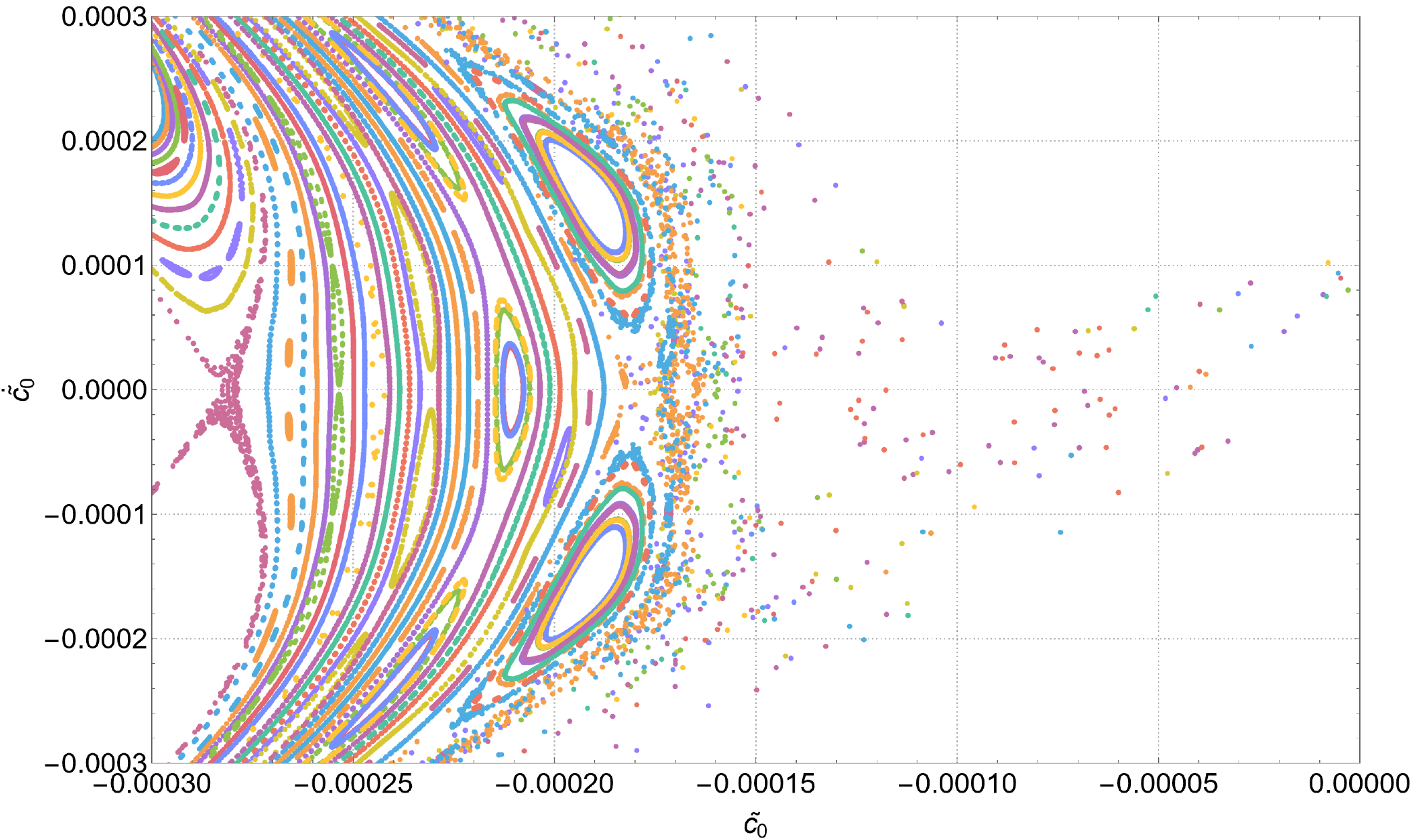} & \includegraphics[scale=0.2,valign=c]{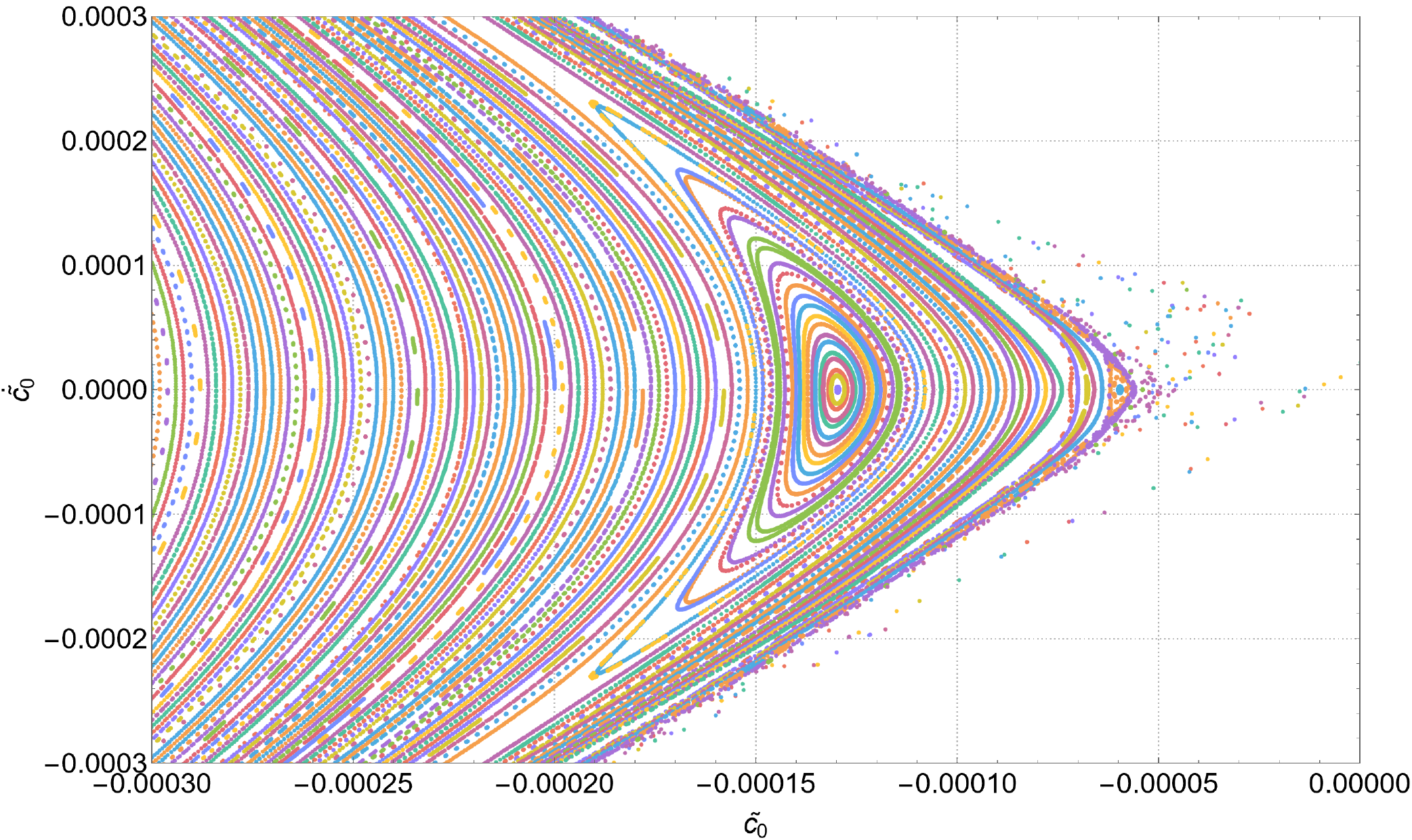} \\
		\textbf{$\mu=0.3$} & \includegraphics[scale=0.2,valign=c]{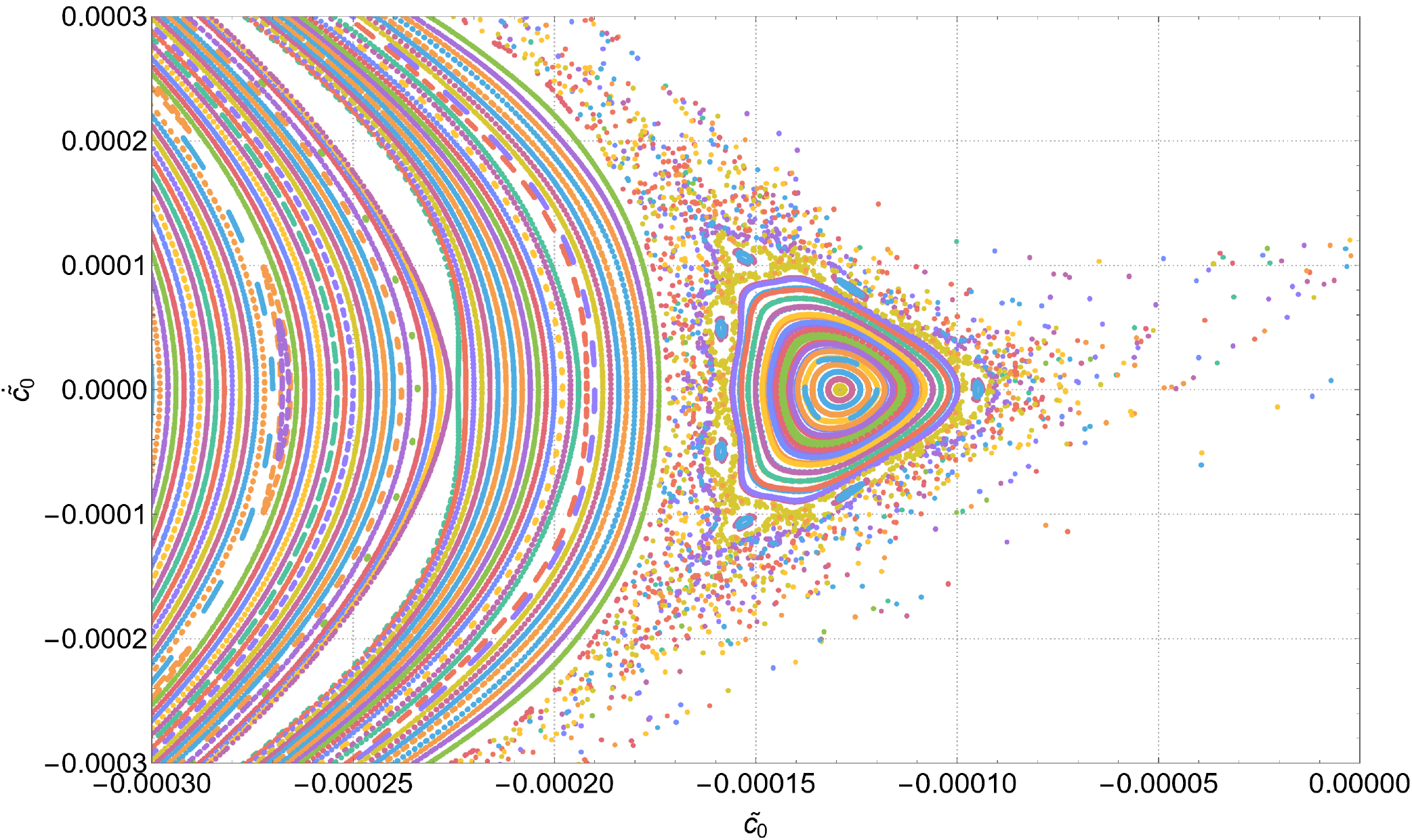} & \includegraphics[scale=0.2,valign=c]{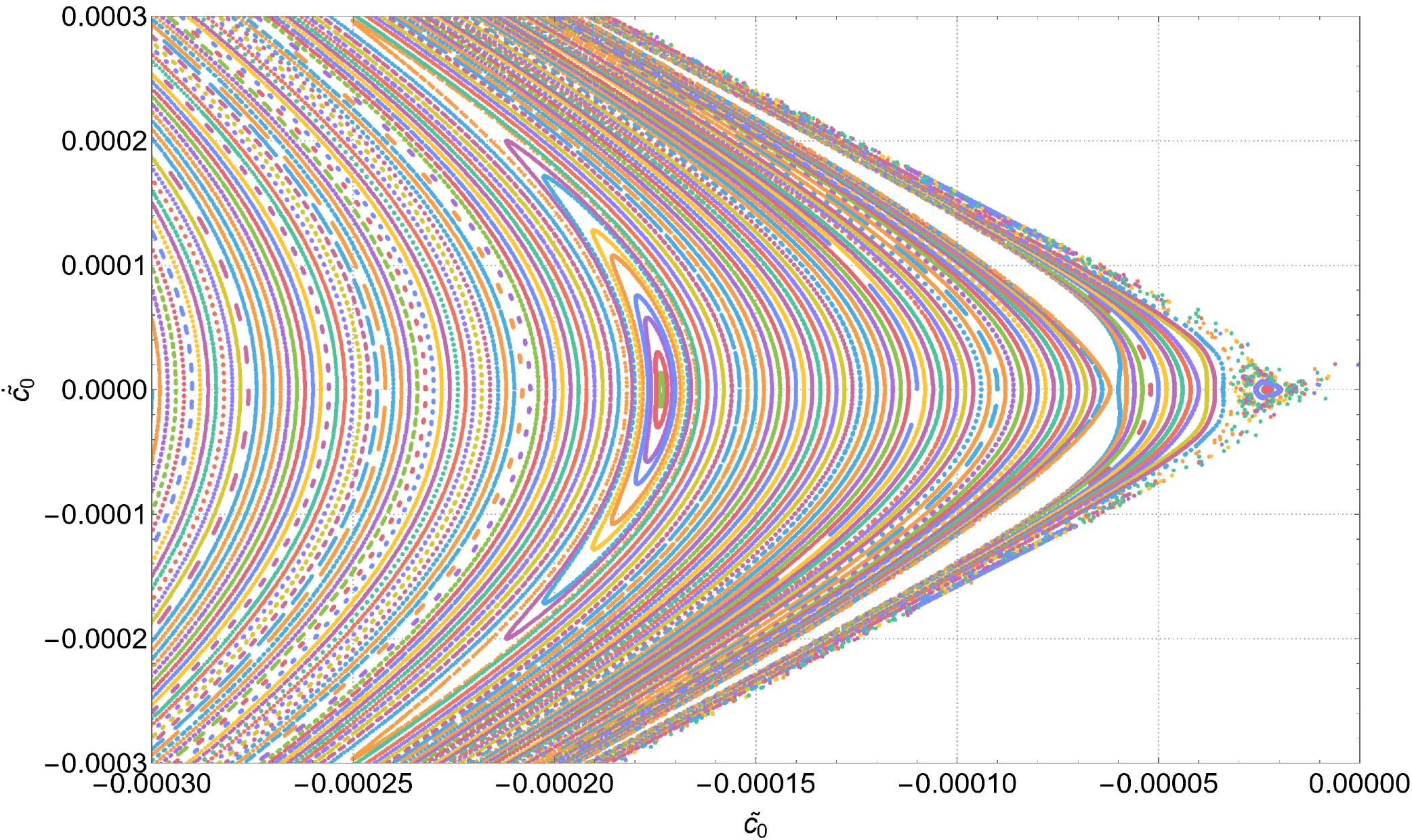} \\
		\textbf{$\mu=0.6$} & \includegraphics[scale=0.2,valign=c]{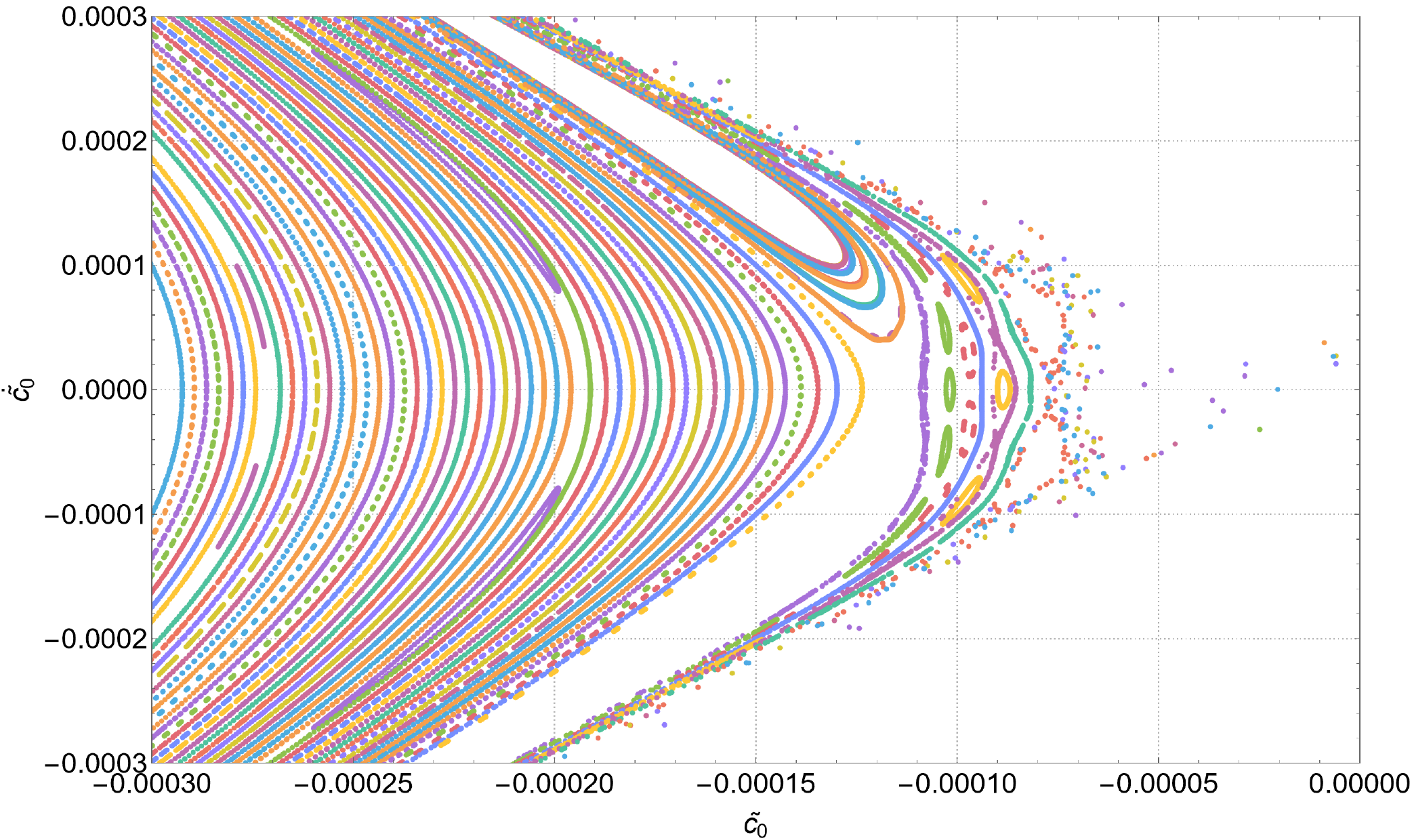} & \includegraphics[scale=0.2,valign=c]{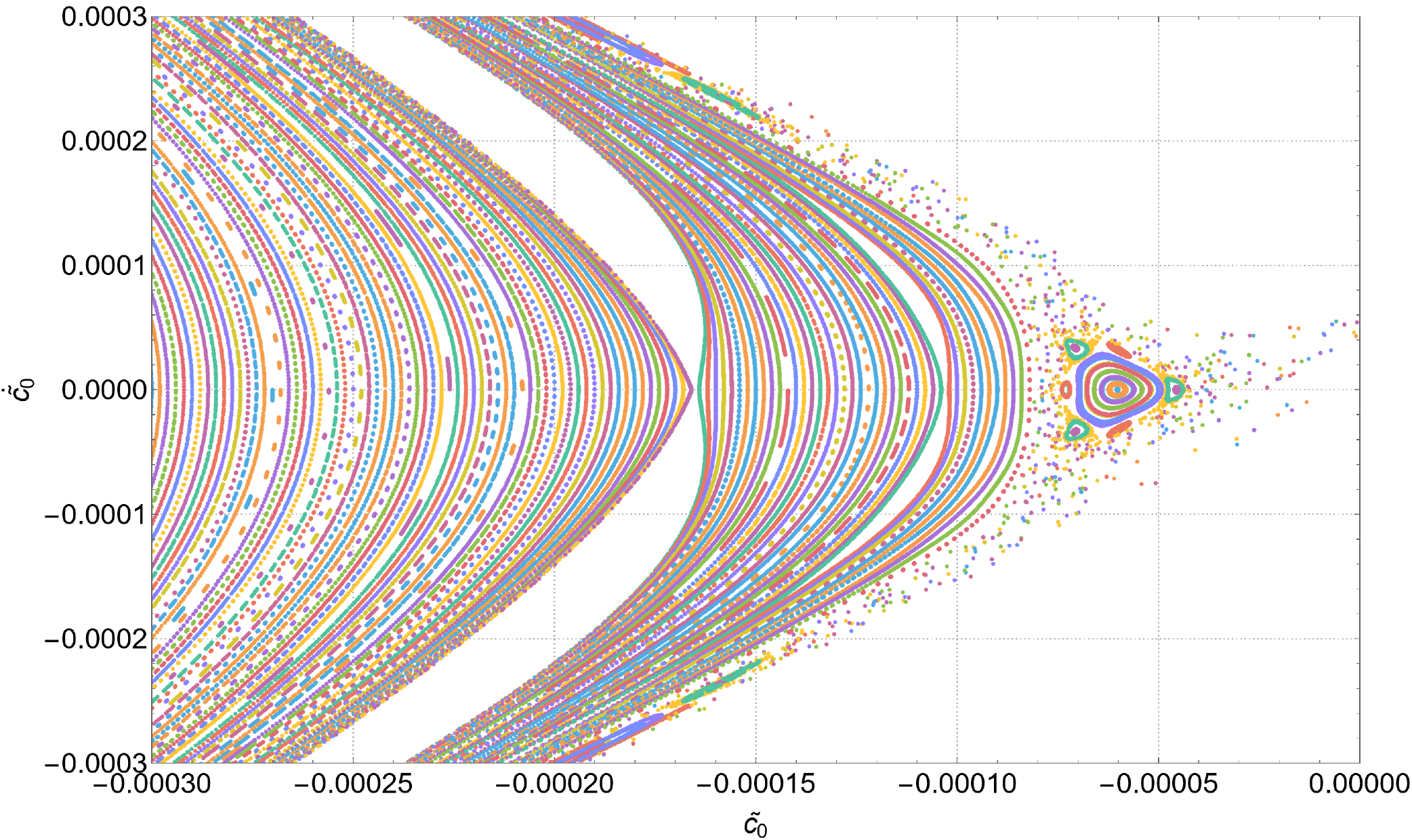} \\
        \textbf{$\mu=0.9$} & \includegraphics[scale=0.2,valign=c]{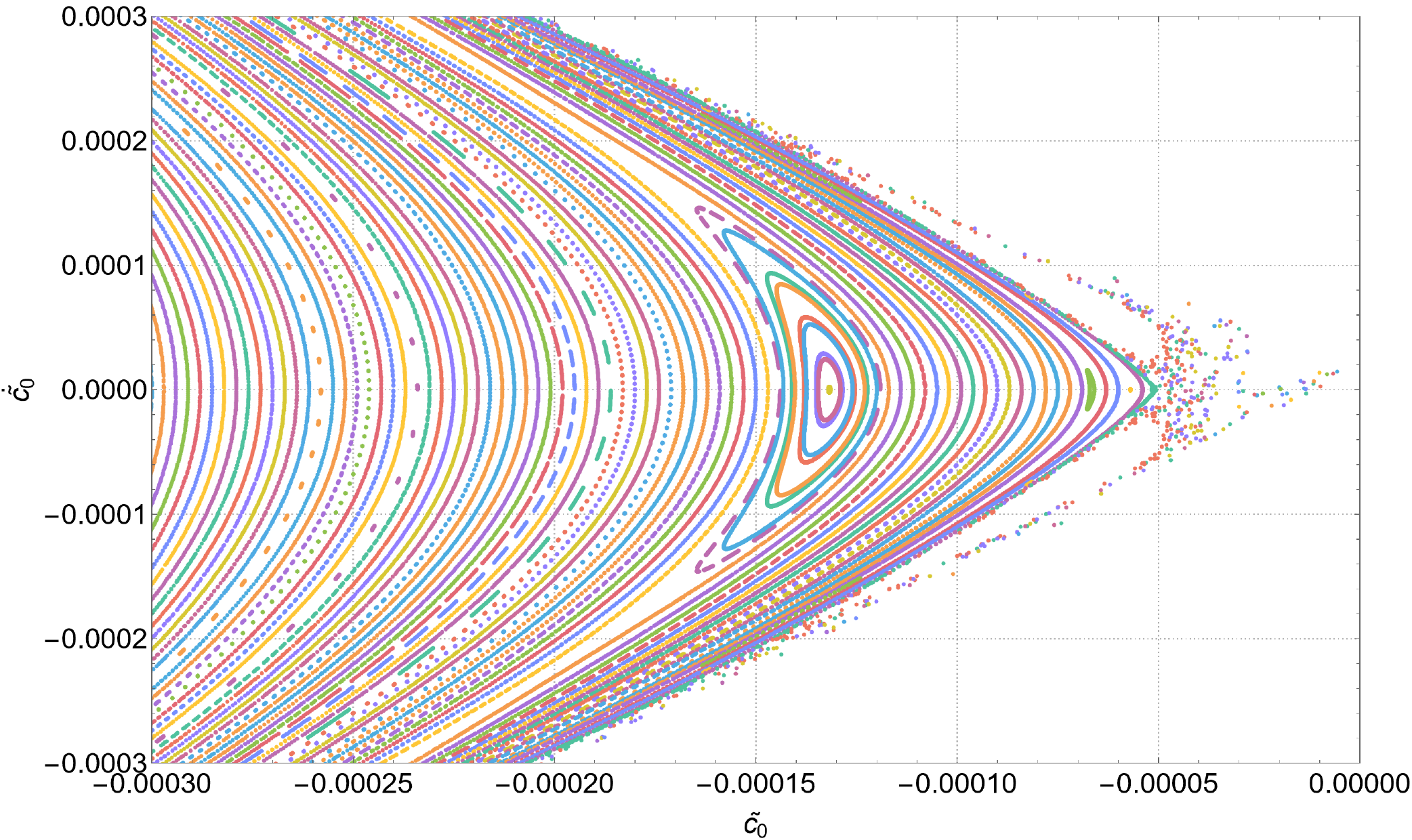} & \includegraphics[scale=0.2,valign=c]{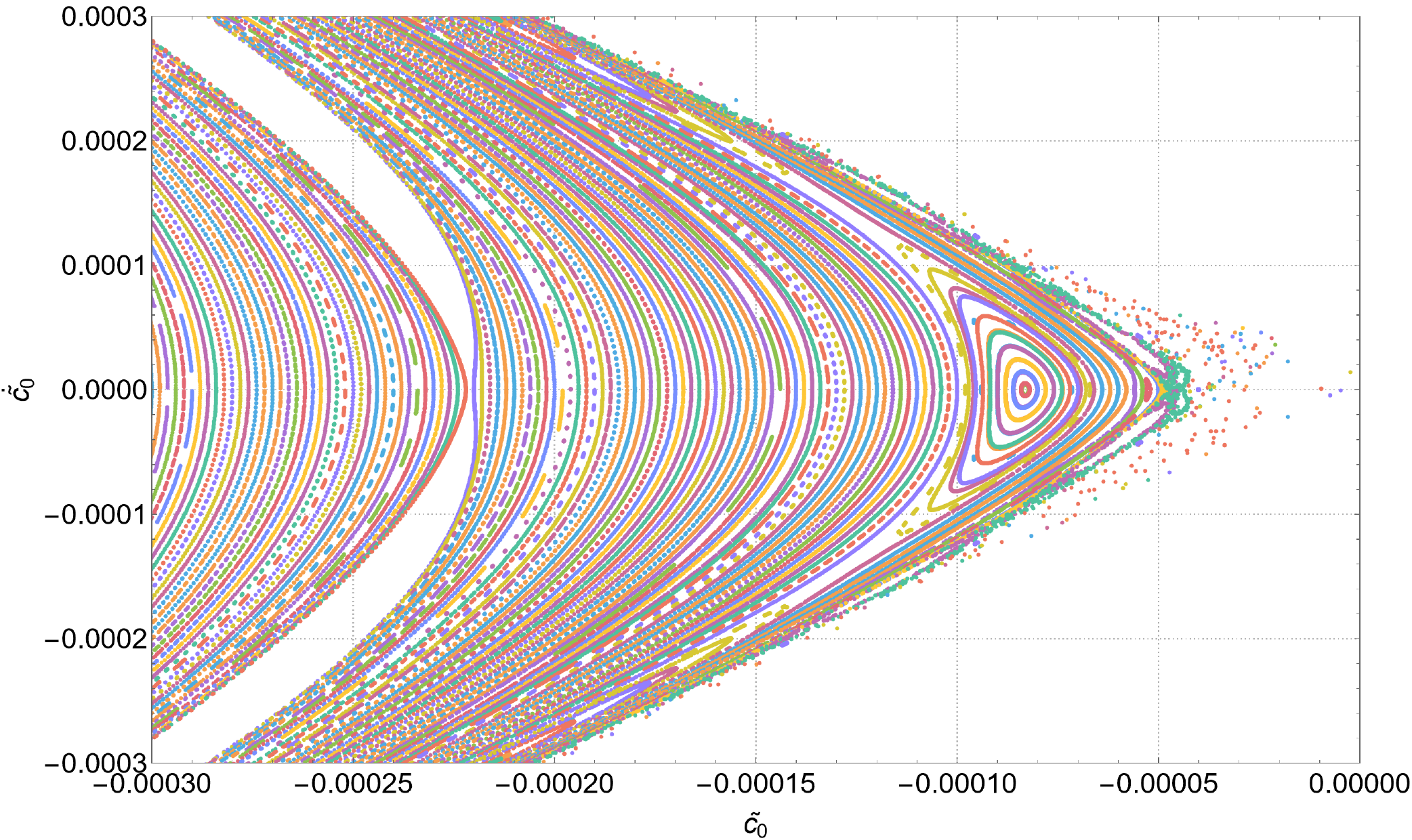} \\
        \textbf{$\mu=1.2$} & \includegraphics[scale=0.2,valign=c]{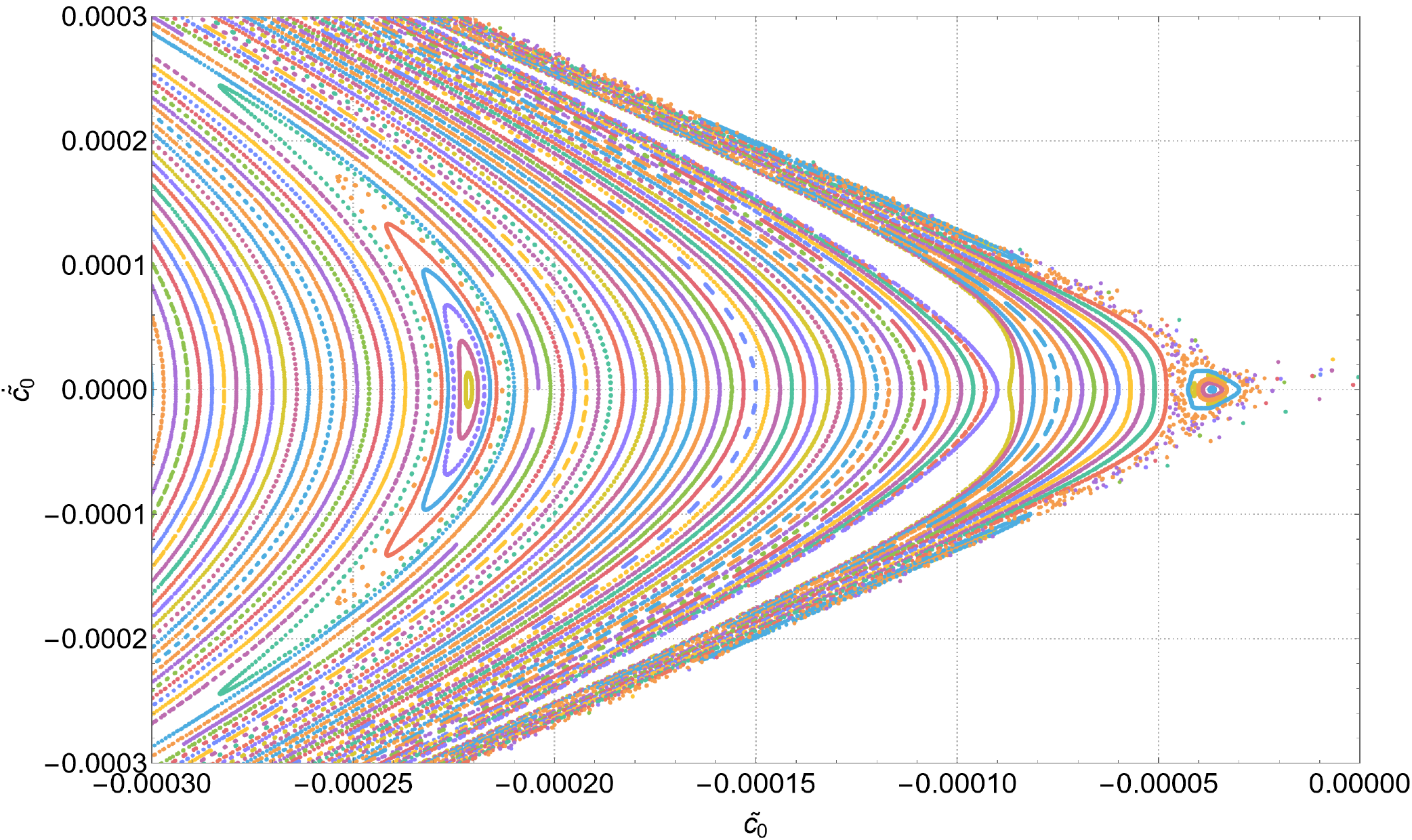} & \includegraphics[scale=0.2,valign=c]{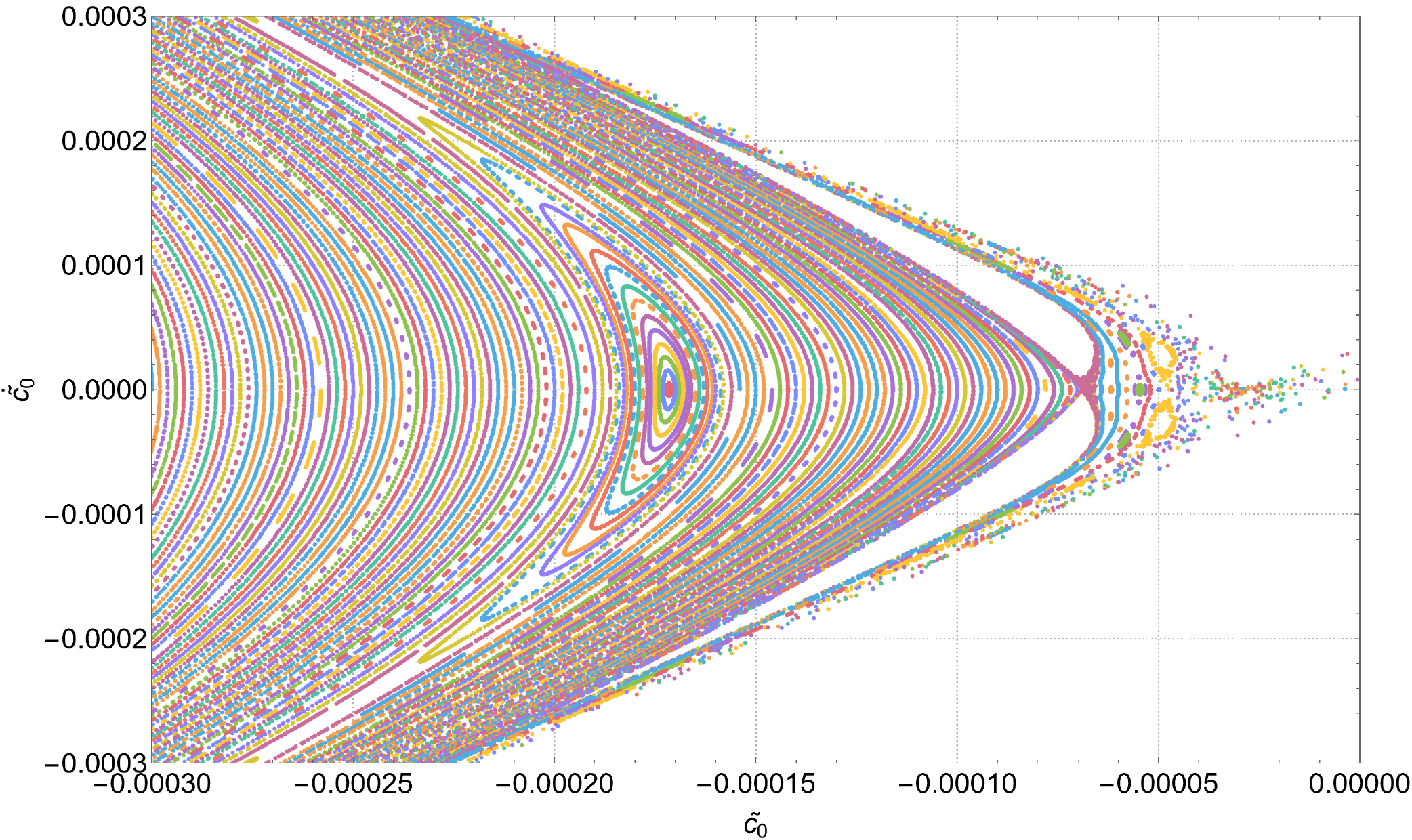} \\
	\end{tabular}
	\caption{The Poincar\'{e} sections are shown for different values of $\mu$, with the left figure representing the $x_1$ (parallel) string configuration and the right figure representing the $x_3$ (perpendicular) configuration. Each section is identified by $\tilde{c}_{1}(t)=0$ and $\dot{\tilde{c}}_{1}(t)\ge0$ with $E=10^{-5}$. The parameters $L = 1.1$ and $B = 0.2$ are fixed, with all quantities in units of GeV.
\label{fig:StringPoincare}}
\end{figure}

Fig.~\ref{fig:StringPoincare} shows a close-up profile of the Poincar\'{e} section near the origin, where most of the interesting things happen, both for parallel and perpendicular string configurations. Here $B=0.2$ and $\mu= 0$, $0.3$, $0.6$, $0.9$, and $1.2$ are used for illustrative purposes. The choice of $B=0.2$ is arbitrary, in the sense that it is chosen just to demonstrate the anisotropic chaotic nature of the string as we vary the value of $\mu$. The data points in different colors correspond to numerical data of orbits for different starting conditions. 

We find from the Poincar\'{e} sections that for both $x_1$ ($\parallel$) and $x_3$ ($\perp$) cases the chaos shows up only near the black hole event horizon.  We observe scattered points concentrating near the origin $\tilde{c_{0}}\simeq 0$, showing strong dependence on initial conditions. These scattered points convert into regular paths for higher values of $B$ and $\mu$. This suggests that the effect of the magnetic field and chemical potential is to reduce the chaotic behaviour. This is true for both parallel and perpendicular configurations of the string. Referring to Table \ref{tab:StringLvsr0}, we can see that as the chemical potential increases, the tip of the unstable string $r_0$ moves away from the horizon, although the change in $r_{0}$ with the chemical potential is small as compared to the changes by the magnetic field.  This implies that the chaos's source is the black hole's event horizon. The same behaviour was observed in \cite{Shukla:2023pbp}.

We can also observe that the nature of the chaos is quite different between the $x_1$ and $x_3$ configurations of the string.  The chaos in the perpendicular configuration is less compared to the parallel configuration. This analysis makes sense if we look at Table \ref{tab:StringLvsr0} again. We can see that at the same magnetic field value, the tip of the string is farther away from the horizon for the perpendicular than for the parallel configuration. Hence less chaos is expected in the former case compared to the latter. We would like to again point out that the chaos is significant only when the unstable string is near the horizon irrespective of the magnetic configuration. For instance, for the stable string (i.e., the large $r_0$ solution), which is relatively far from the horizon, there are only stable orbits without scattered points, suggesting no chaos for the stable string configuration for any values of $B$ and $\mu$.

Although the Poincar\'{e} section is quite helpful in identifying the chaotic ``position'' in the phase space, the inability to give a quantitative result makes it difficult for us to compare the chaos between different configurations. For this reason, we will move on to the analysis of our system using the Lyapunov exponents.

\subsection{Lyapunov exponents}\label{StringLyapunov}
\begin{figure}[htbp!]
	\centering
	\begin{tabular}{c c c}
		\textbf{$\mu$ value} & \textbf{Parallel Configuration} & \textbf{Perpendicular Configuration} \\
		\textbf{$\mu=0.0$} & \includegraphics[scale=0.16,valign=c]{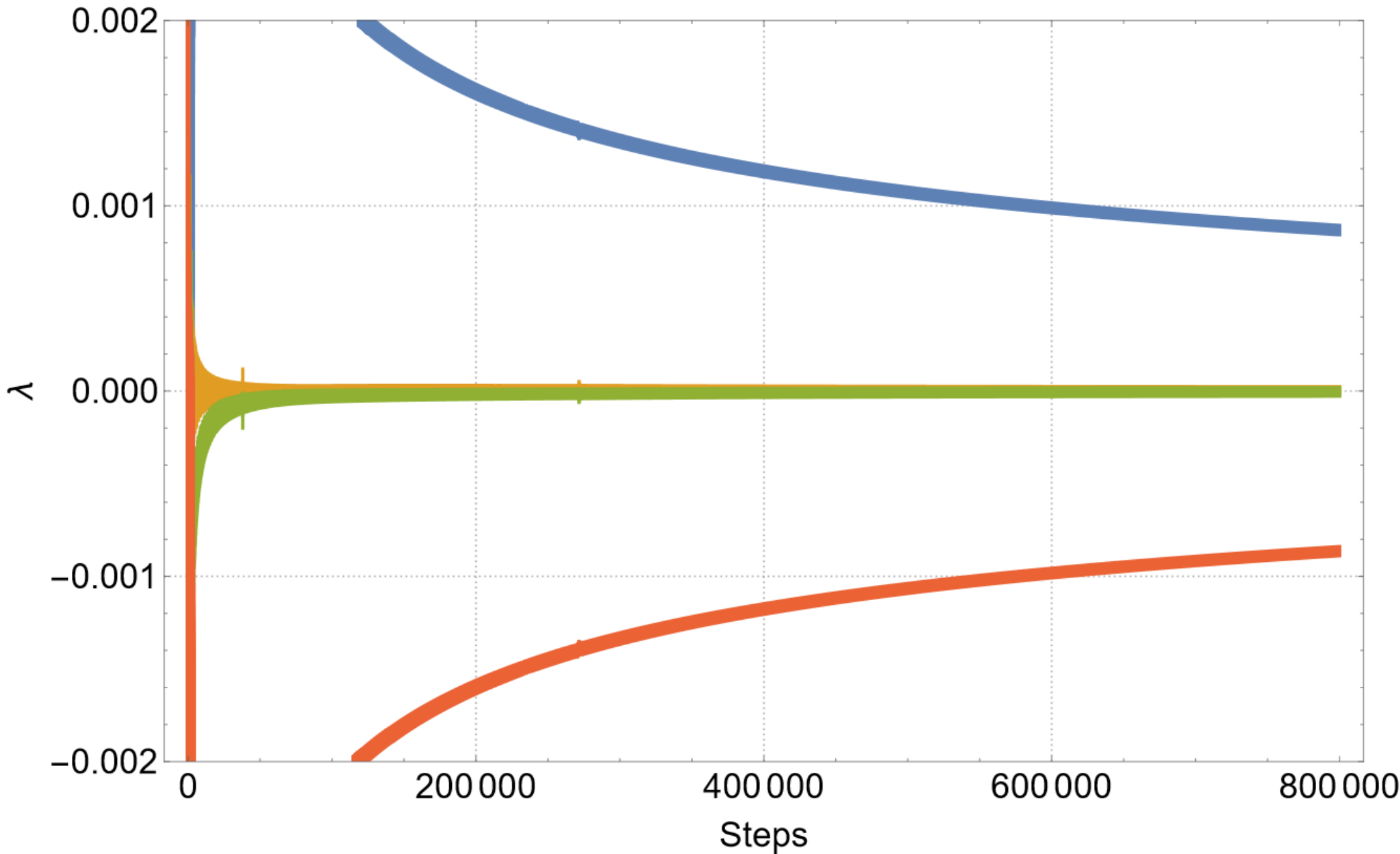} & \includegraphics[scale=0.16,valign=c]{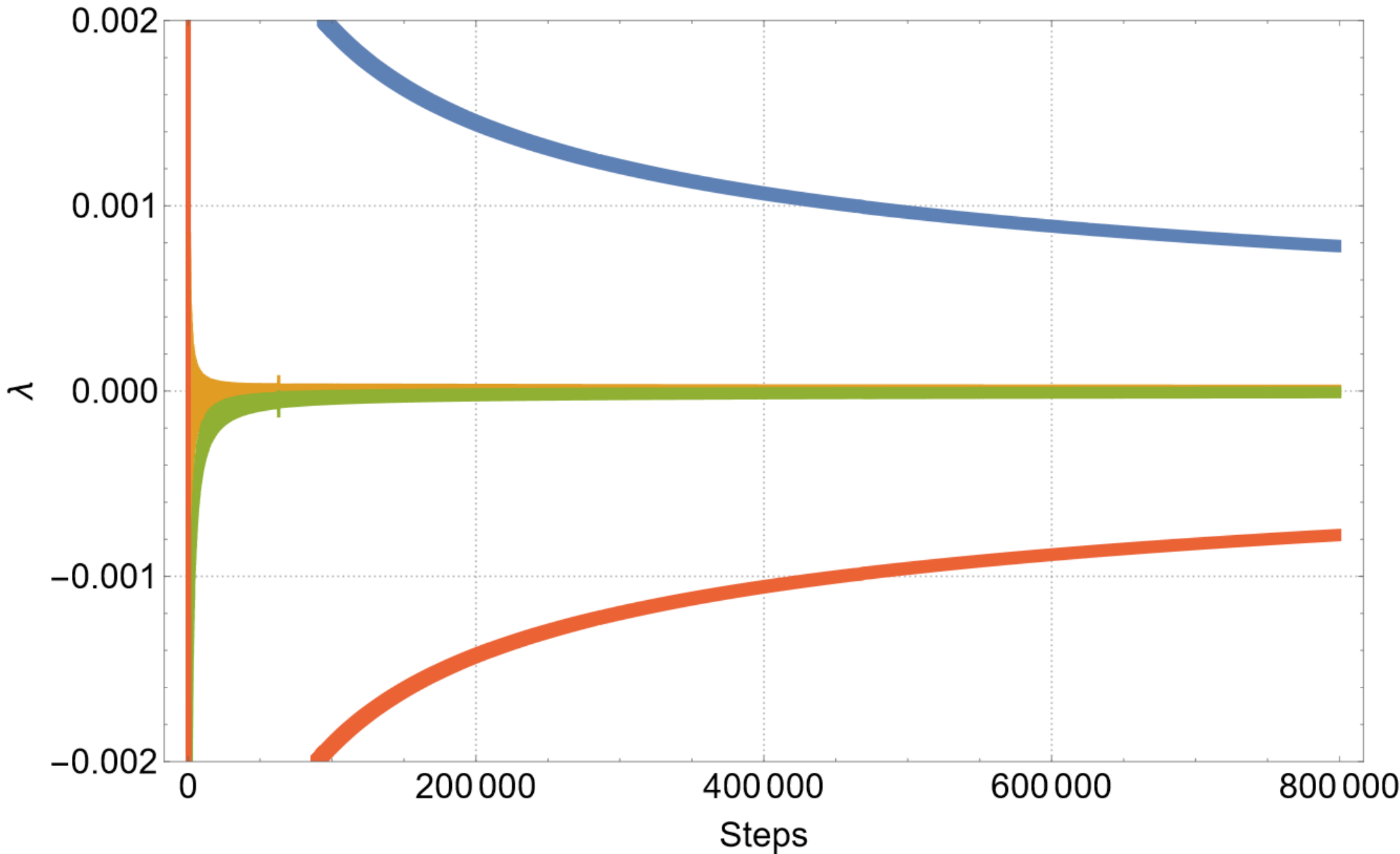} \\
		\textbf{$\mu=0.3$} & \includegraphics[scale=0.16,valign=c]{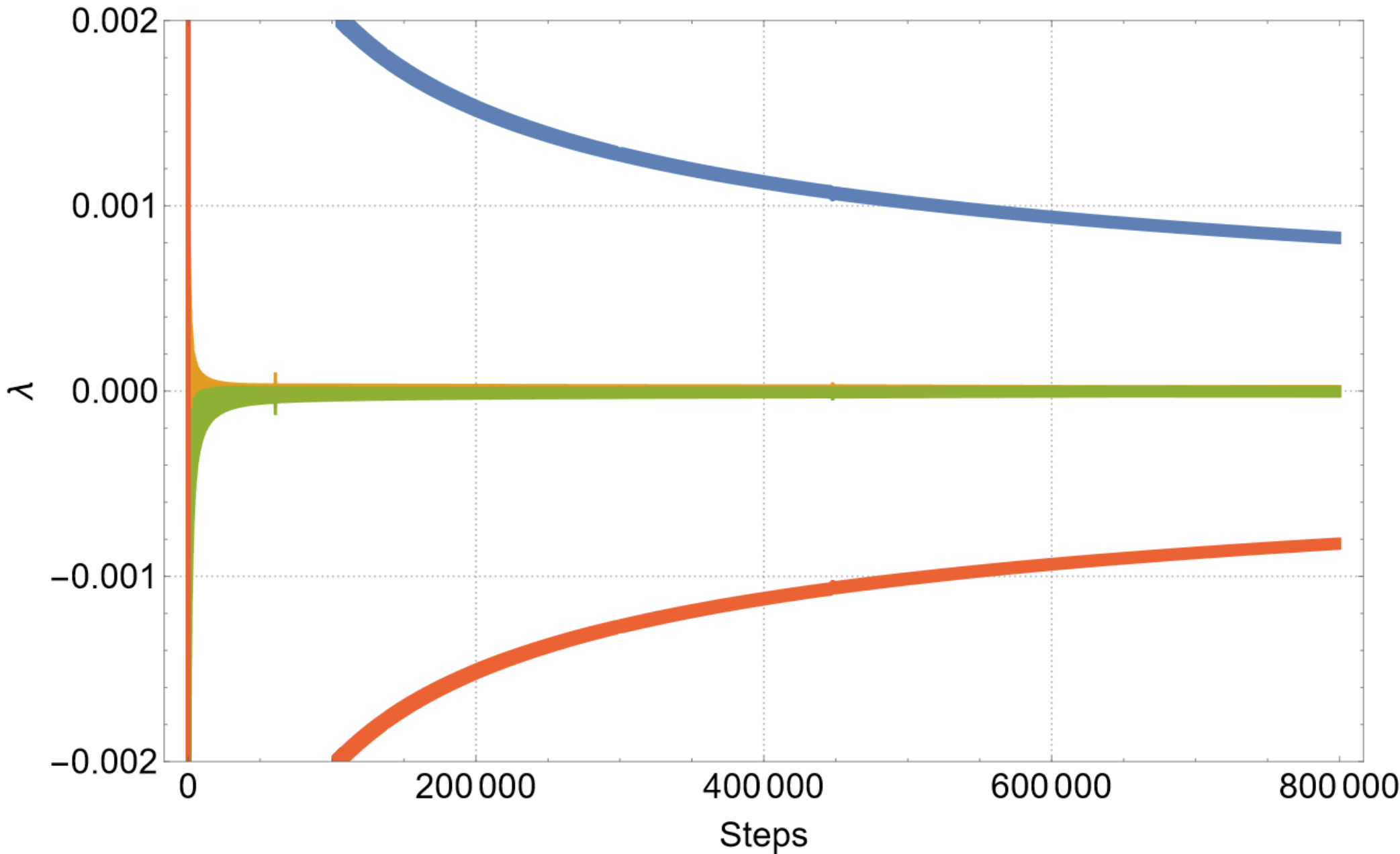} & \includegraphics[scale=0.16,valign=c]{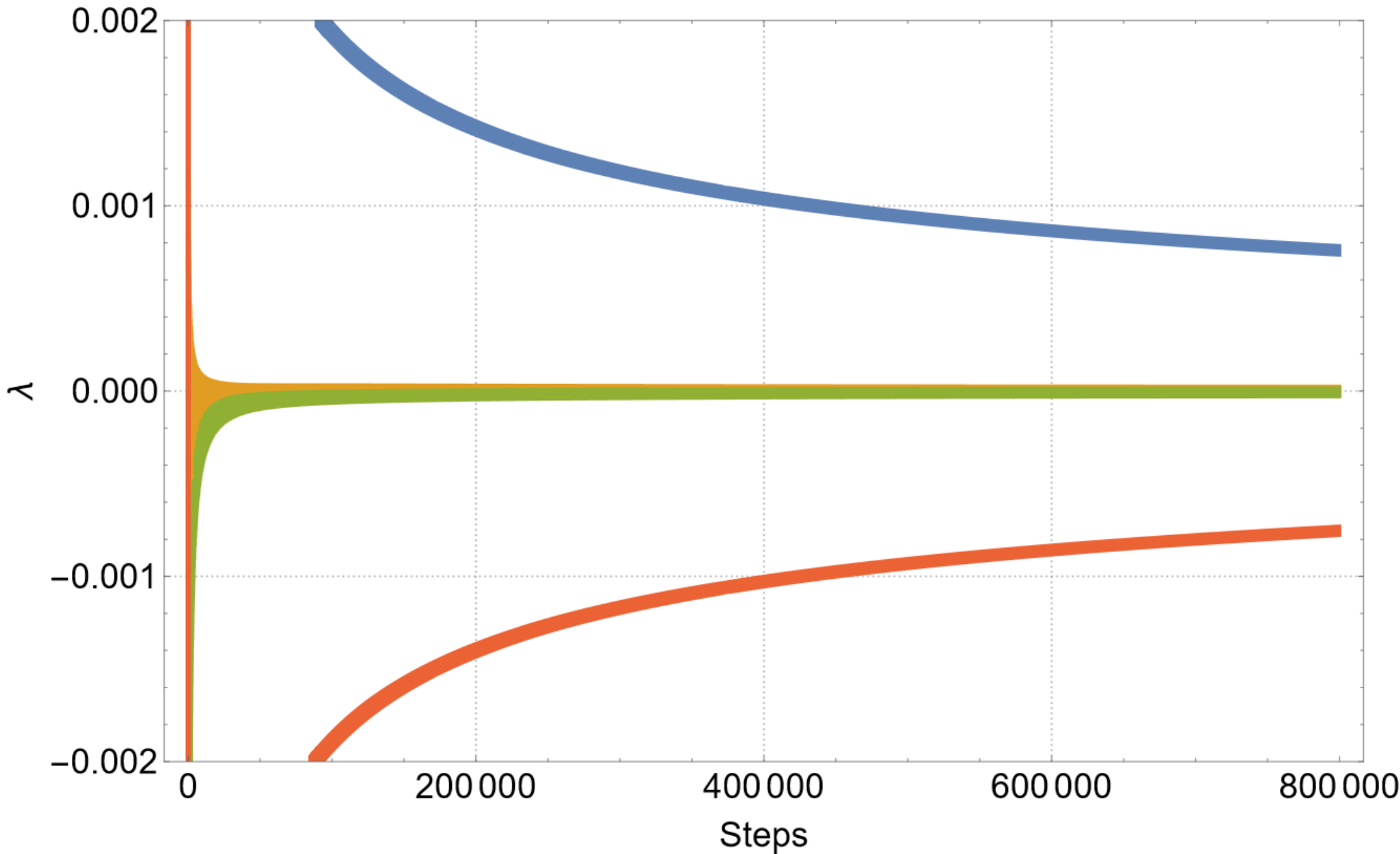} \\
		\textbf{$\mu=0.6$} & \includegraphics[scale=0.16,valign=c]{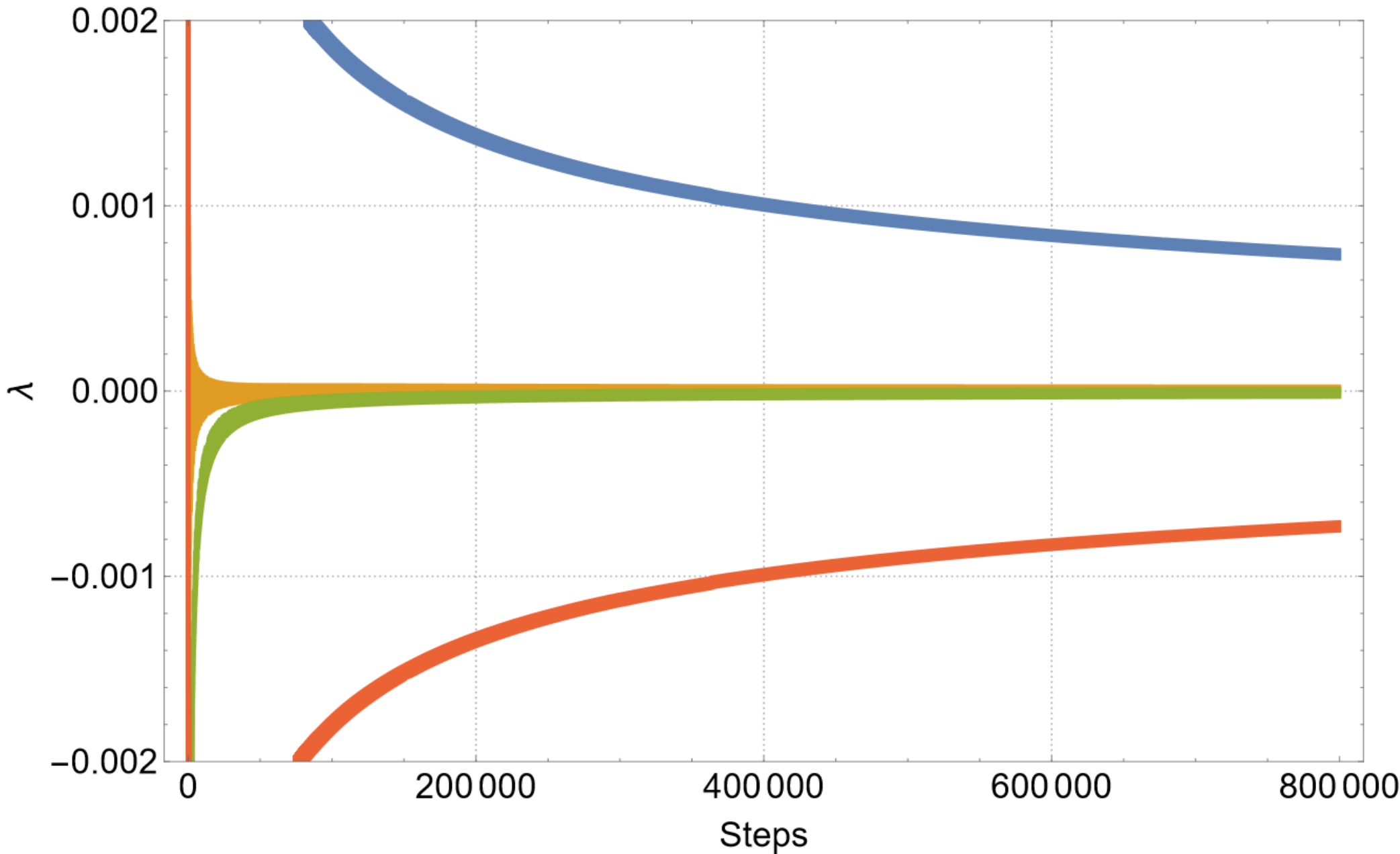} & \includegraphics[scale=0.16,valign=c]{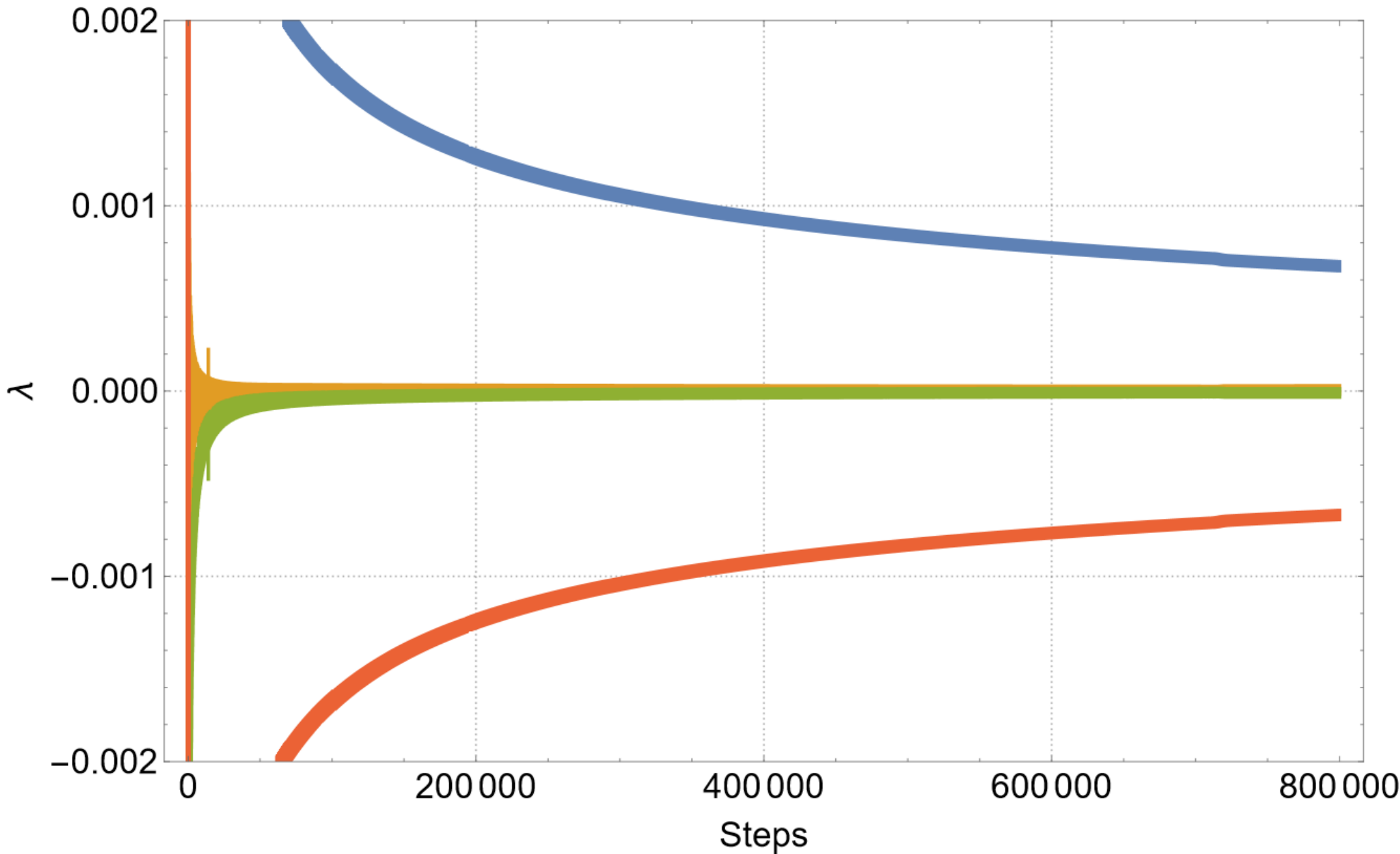} \\
        \textbf{$\mu=0.9$} & \includegraphics[scale=0.16,valign=c]{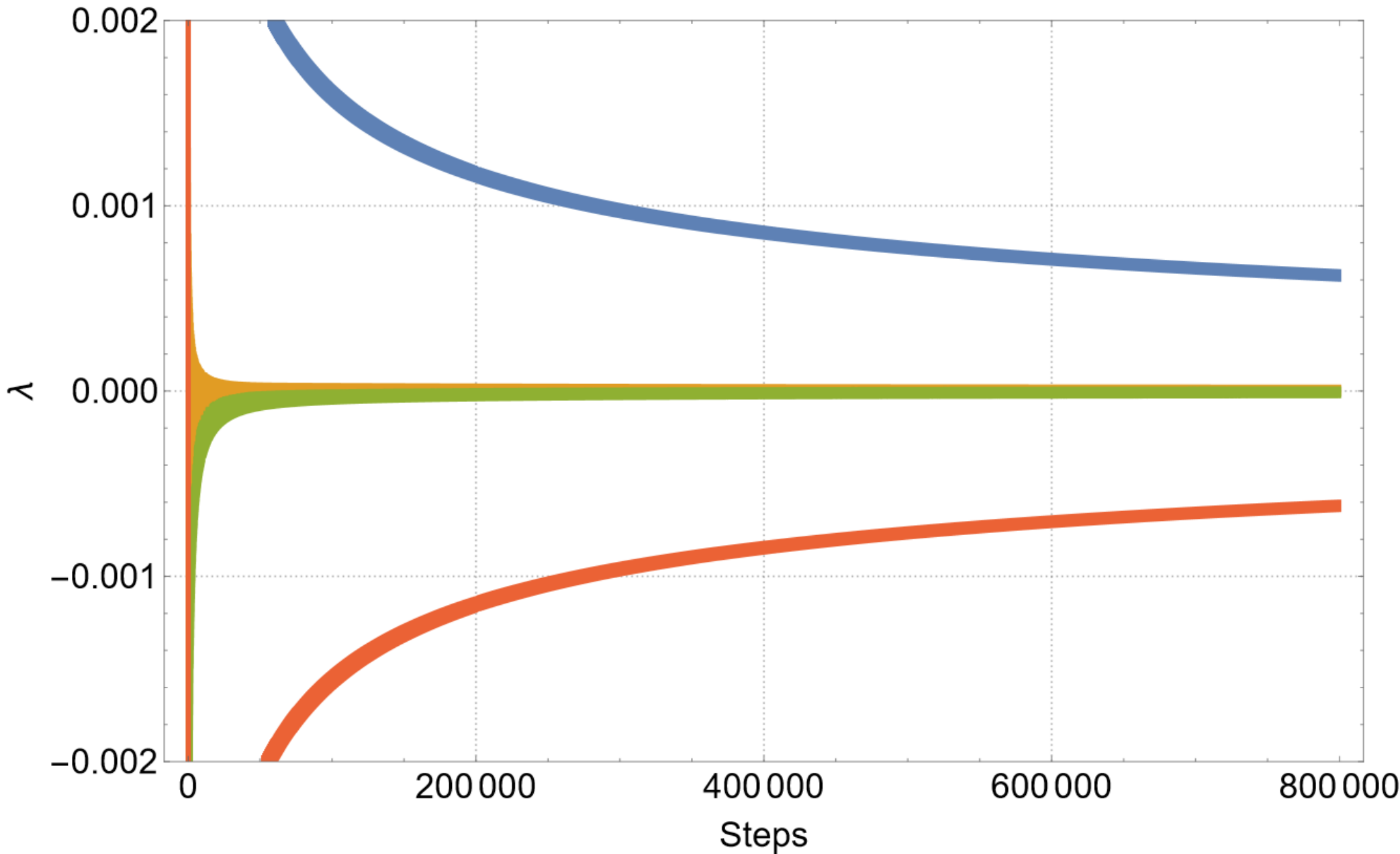} & \includegraphics[scale=0.16,valign=c]{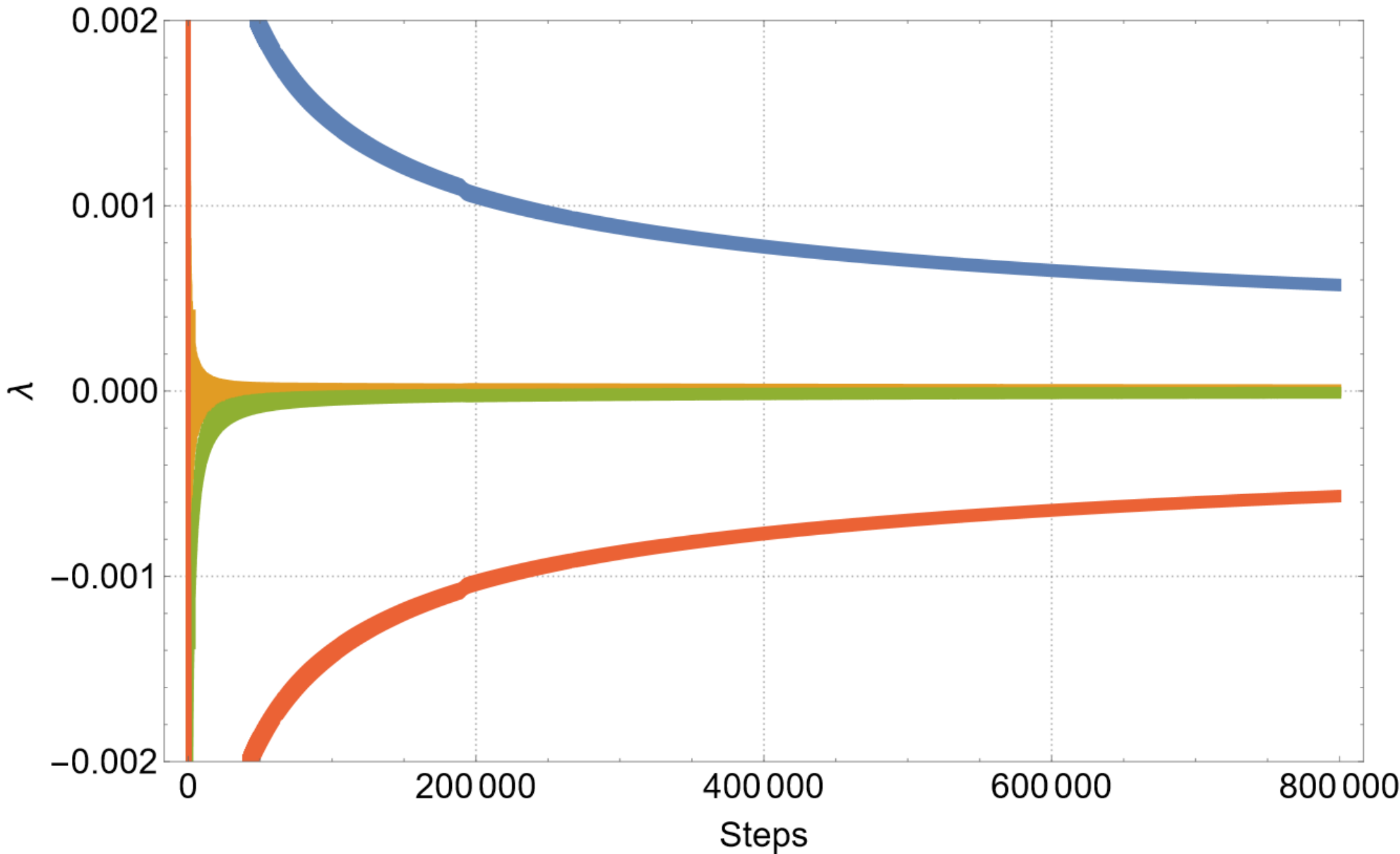} \\
        \textbf{$\mu=1.2$} & \includegraphics[scale=0.16,valign=c]{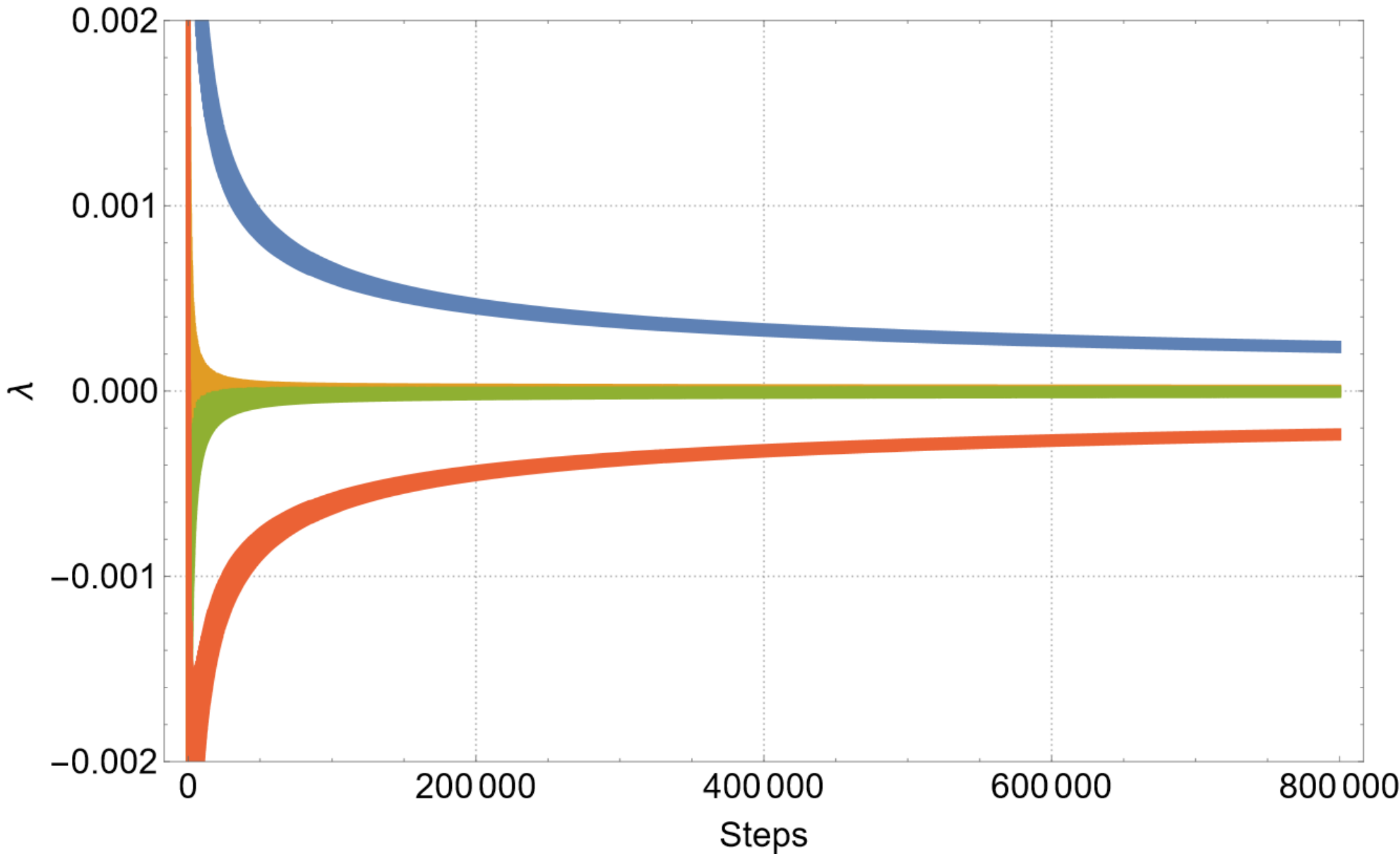} & \includegraphics[scale=0.16,valign=c]{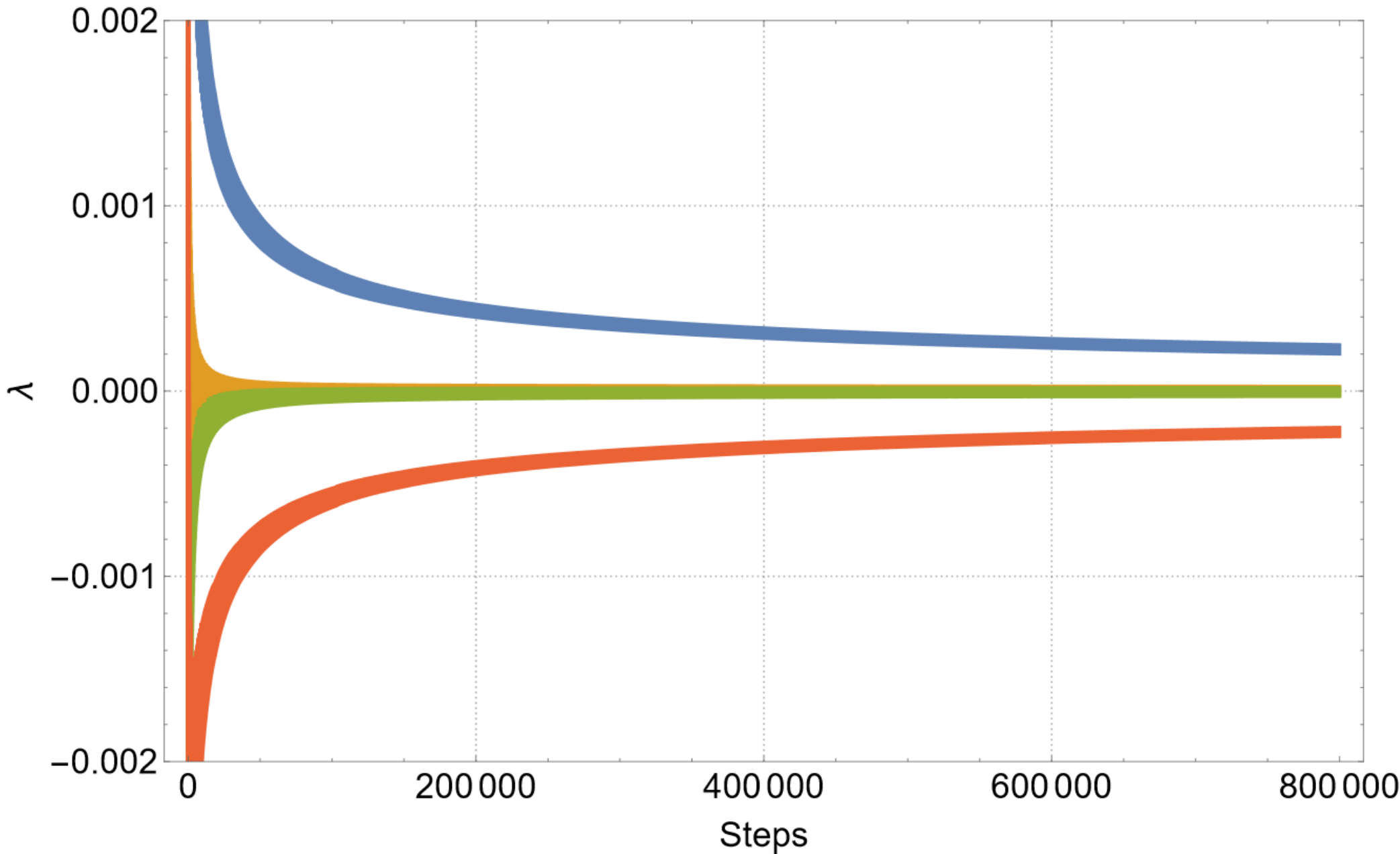} \\
        \textbf{Sum of $\lambda$} & \includegraphics[scale=0.16,valign=c]{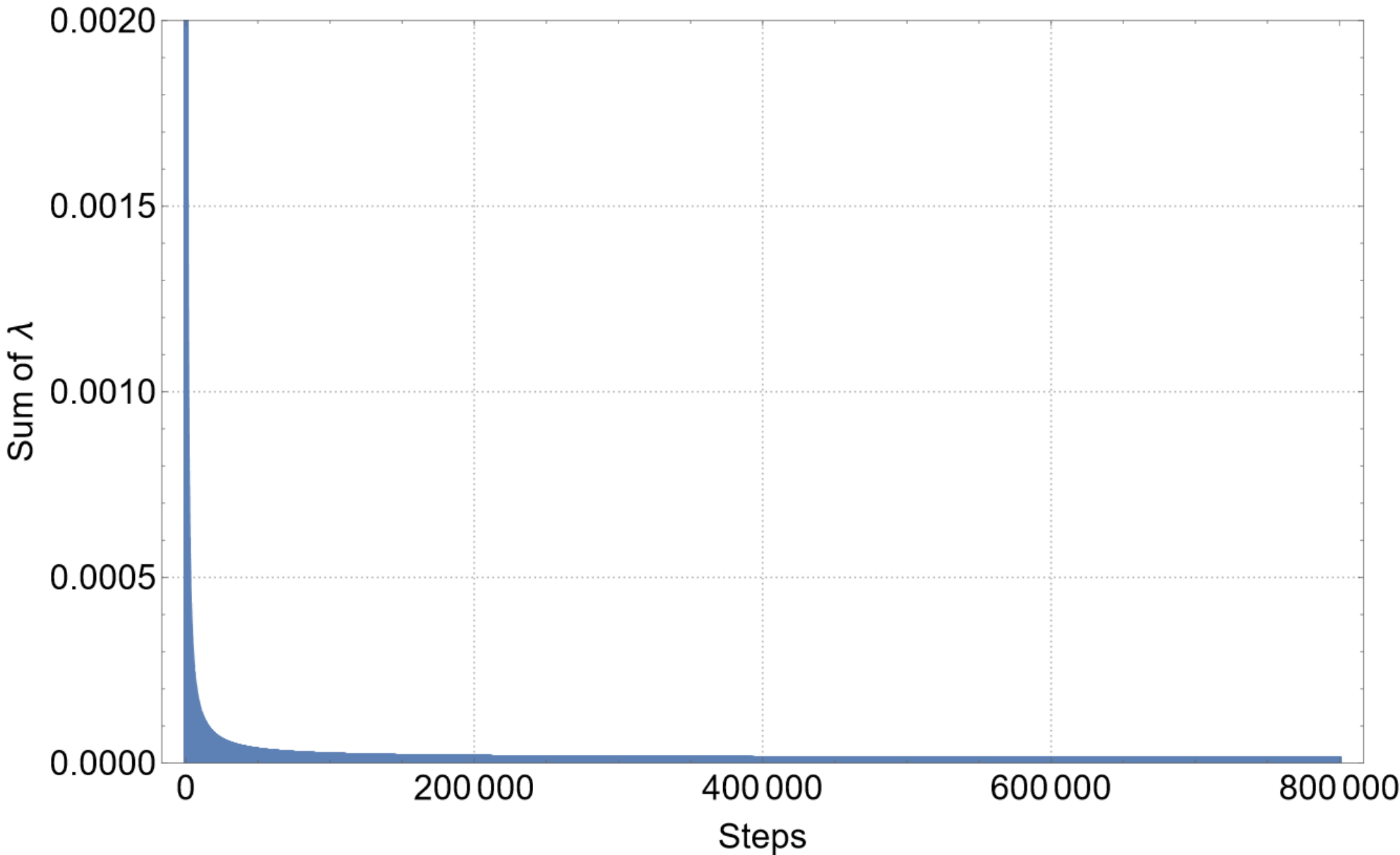} & \includegraphics[scale=0.16,valign=c]{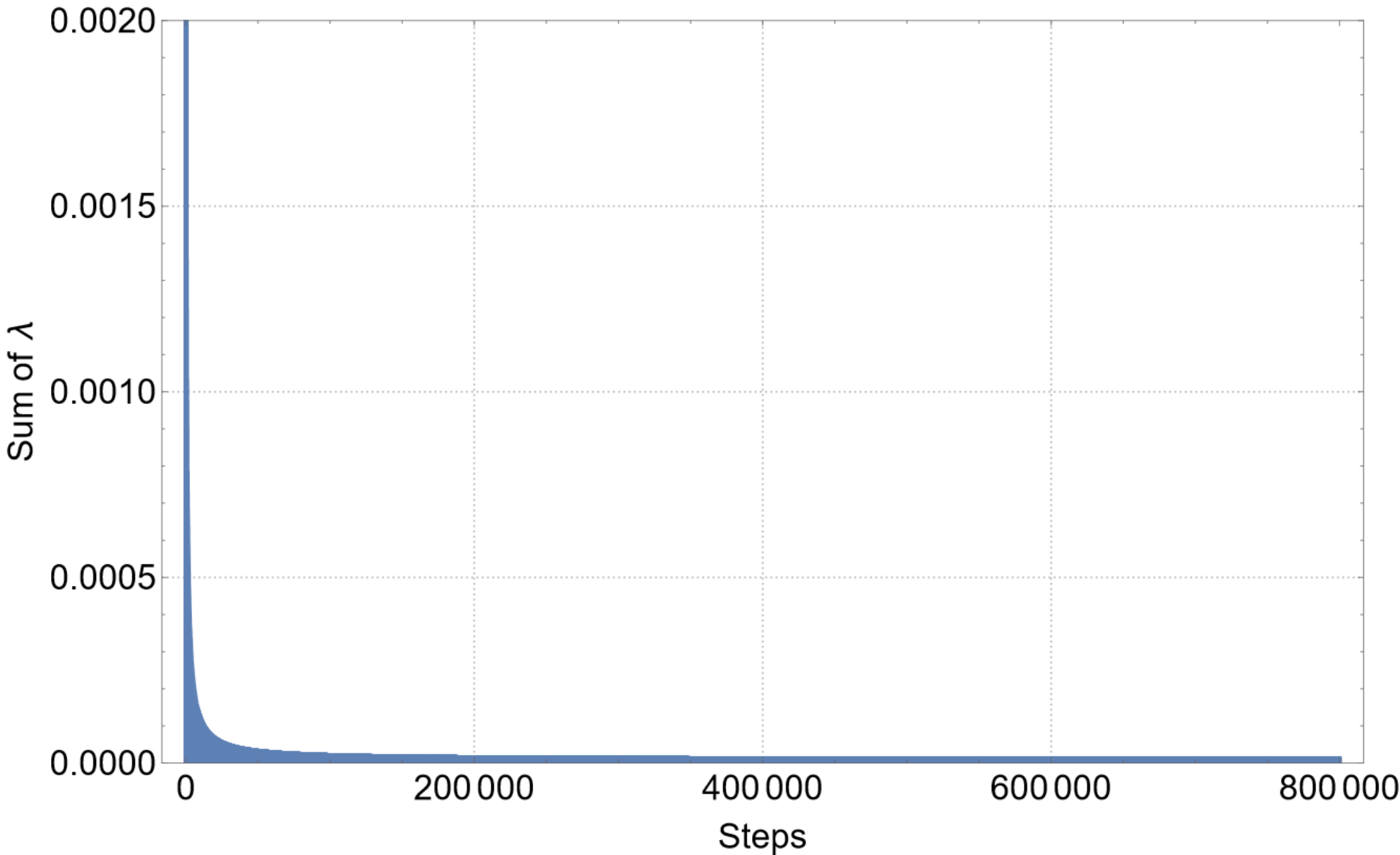}
        \end{tabular}
\caption{Convergence plots of the four Lyapunov exponents are displayed for different values of $\mu$, with the string oriented parallel (left column) and perpendicular (right column) to the magnetic field. The last row shows the sum of the Lyapunov exponents for $\mu=0.6$ and $B=0.2$. Similar converging behaviour is observed for all values of the magnetic field and chemical potential. Here the initial conditions $\Tilde{c}_{0}=-0.0003$, $\dot{\tilde{c}}_{1}(t)=0.00001$, and $\Tilde{c}_{1}=0.0008$ are used for both the configurations. The parameters $L = 1.1$, $E=10^{-5}$, and $B = 0.2$ are fixed, with all quantities in units of GeV.   \label{fig:StringLyapunov}}
\end{figure}

Lyapunov exponents can be used to quantitatively analyze chaotic dynamics in classical mechanics.\footnote{The Lyapunov exponent was also employed more recently to demonstrate the phase transition of AdS black holes \cite{Shukla:2024tkw, guo2022probing, yang2023lyapunov, lyu2023probing, kumara2024lyapunov, Du:2024uhd, Gogoi:2024akv}.} We numerically compute the Lyapunov exponent in the four-dimensional phase space $(\Tilde{c}_{0}, \Tilde{c}_{1})$ using methods given in \cite{sandri1996numerical,wolf1985determining}. As before, we focus on the system with $L=1.1$ and energy $E=10^{-5}$ for both the $x_1$ and $x_3$ orientations of the string. Fig.~\ref{fig:StringLyapunov}, shows the convergence diagram of the four Lyapunov exponents. The behavior is that of a damped oscillation, although we cannot make that observation from our figure due to the high-resolution nature of the plots. In the previous subsection~\ref{StringPoincare}, we discussed the results of the Poincar\'{e} section analysis (Fig.~\ref{fig:StringPoincare}) for various $\mu$ values and fixed $B=0.2$. To create the necessary data on the quantitative nature of the chaos, we also extract the Lyapunov exponents for similar values of $B$ and $\mu$ in Fig.~\ref{fig:StringLyapunov}. In our calculation, we took $8\times10^5$ time steps with step size $0.001$, making our analysis numerically
sufficiently accurate. Also, we have taken the coefficients $K_{1,\ldots,5}^{x_i}$ up to
a $10^{-3}$ level of accuracy. We did not find any substantial effect when enforcing greater accuracy.

From our action (\ref{eq:ActionModified}), we can already see that there will be four Lyapunov exponents, which we represent in different colours in each graph. We have computed the convergency plots of these four Lyapunov coefficients and their sum. The behaviour of each exponent is that of a damped oscillation. The sum of the Lyapunov exponents converges to zero at later times, suggesting the conservative nature of the system. This can be explicitly observed from the last row of Fig.~\ref{fig:StringLyapunov}. The same structure is present for all values of $B$ and $\mu$, irrespective of the orientation of the string concerning the magnetic field. 

\begin{figure}[htbp]
\centering
\includegraphics[width=0.65\textwidth]{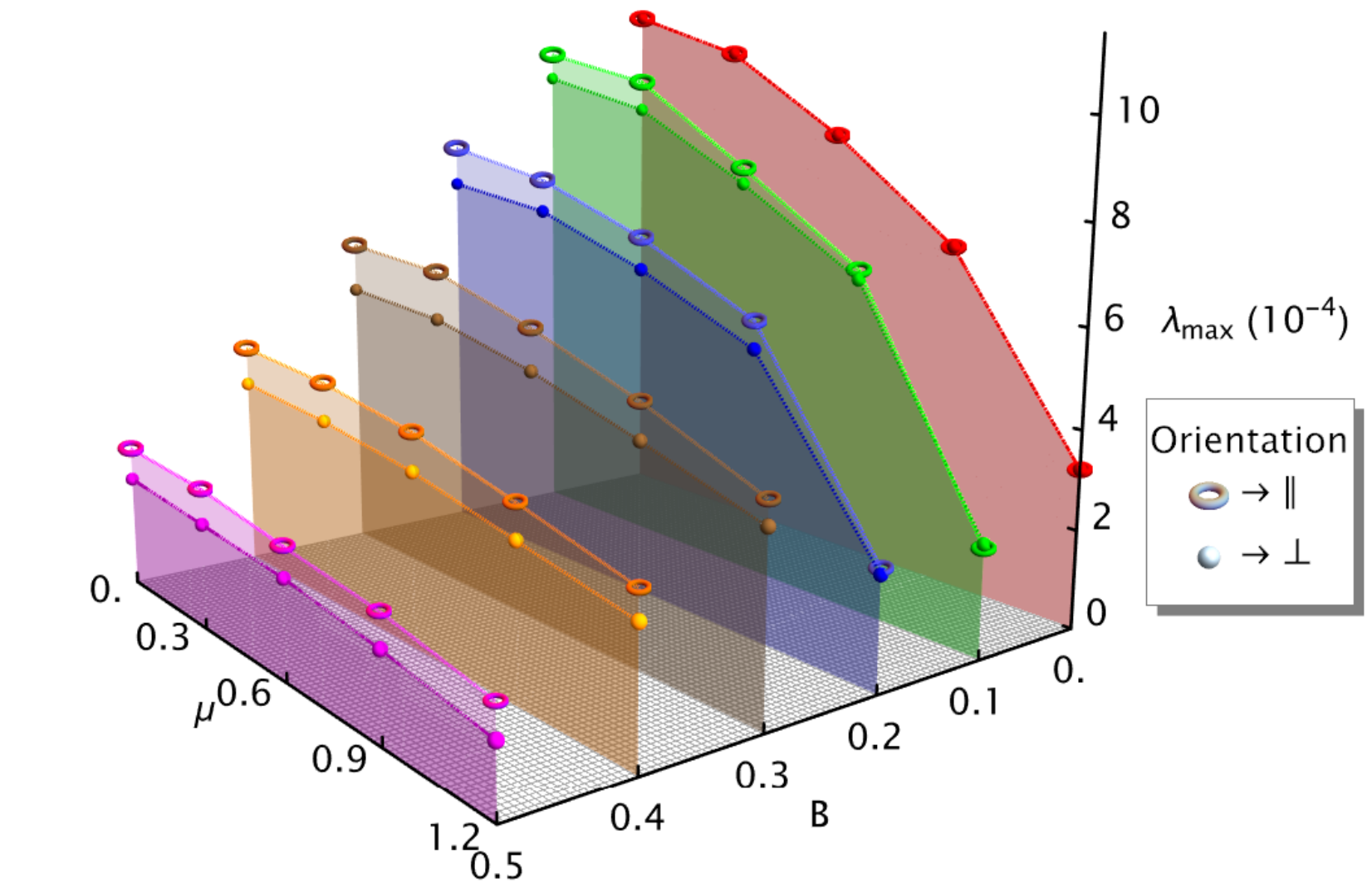}
\caption{The maximum Lyapunov exponent $\lambda_{\text{max}}$ is compared for the $x_1$ $(\parallel)$ and $x_3$ $(\perp)$ orientations of the string at various values of $B$ and $\mu$, close to the unstable saddle point. Data points are colour-coded as follows: red, green, blue, brown, orange, and magenta colours represent $B = 0.0$, $0.1$, $0.2$, $0.3$, $0.4$, and $0.5$, respectively. The string length is fixed at $L = 1.1$, and $E = 10^{-5}$.}
\label{fig:StringLmax}
\end{figure}

The maximal Lyapunov exponent, denoted as $\lambda_{max}$, can be extrapolated from various configurations by considering sufficient time steps and fitting the maximum in each oscillation. The obtained $\lambda_{max}$ results are shown in Fig.~\ref{fig:StringLmax} for parallel and perpendicular cases for different values of the chemical potential and magnetic field. Here the string length $L=1.1$ is kept fixed for all cases.  It is observed that $\lambda_{max}$ decreases for both orientations of the string as $\mu$ or $B$ increases. Additionally, for fixed values of $B$ and $\mu$, $\lambda_{max}$ is smaller in the perpendicular case compared to the parallel one. The Poincar\'{e} section analysis supported such observation already in the previous section. We further observe that the Lyapunov exponent decreases sharply as $\mu$ increases starting from $0.6$ for lower magnetic field values, i.e., $B=0.0$, $0.1$ and $0.2$ for both the $x_1$ and $x_3$ configurations. This indicates a rapid stabilisation of the system by the chemical potential at a lowering of the magnetic field. But this rapid stabilisation vanishes as the magnetic field $B$ increases. Our overall  Lyapunov exponent analysis confirms more qualitatively that the string dynamics become less chaotic when the chemical potential or magnetic field increases for both parallel and perpendicular string orientations.

Our choice of the string length $L=1.1$ is still arbitrary. As long as we stick to the unstable area of the system, i.e., $r_0$ near the horizon, we have already seen that the chaotic nature is only significant near the black hole horizon. 

\subsection{Saddle point analysis and MSS bound}\label{SaddleMSS}
\begin{figure}[htbp!]
	\centering
	\subfigure[Parallel configuration]{\label{fig:potentialstringframex1}	\includegraphics[width=0.4\linewidth]{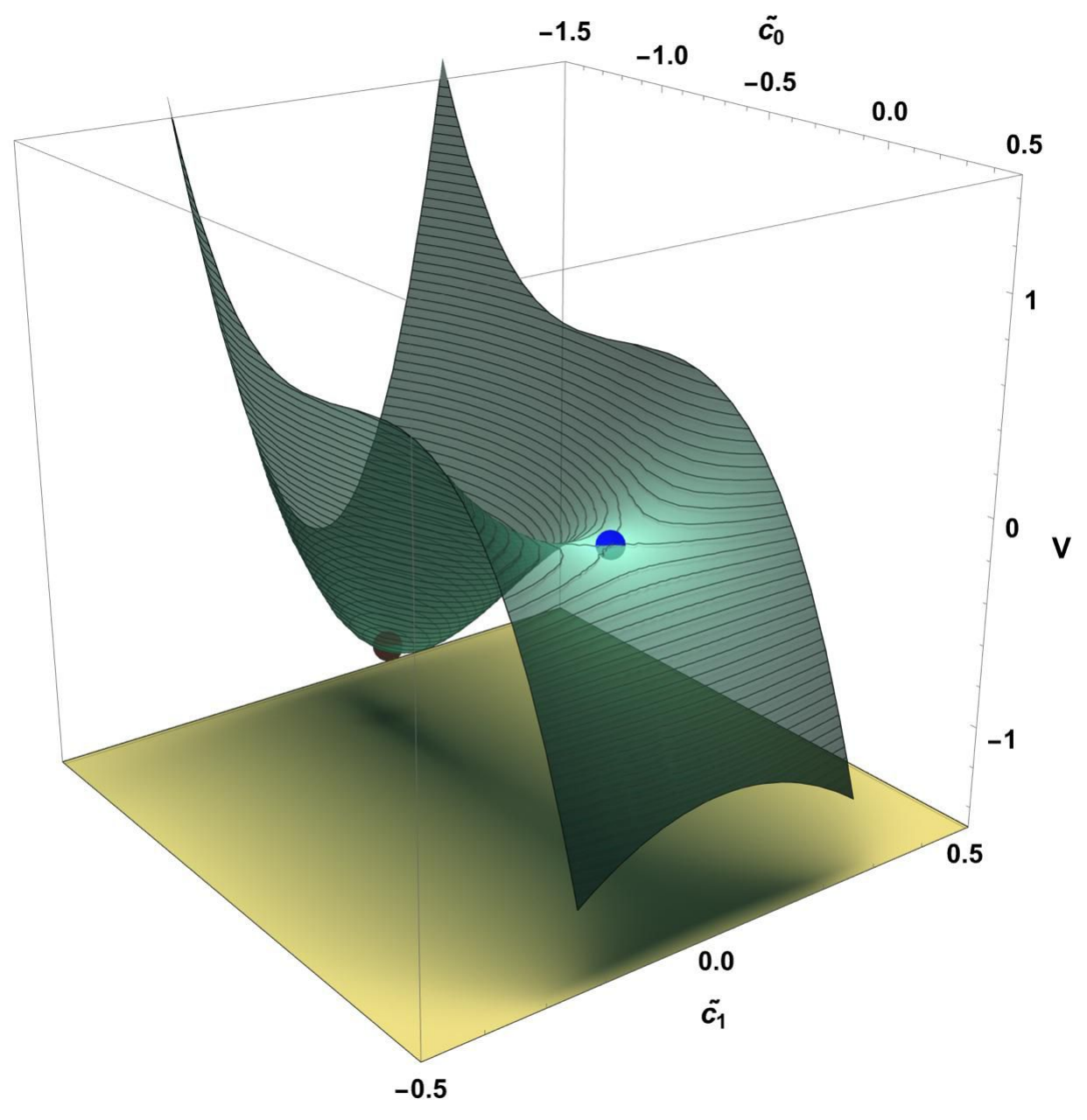}}
    \subfigure[Perpendicular configuration]{\label{fig:potentialstringframex3}
	\includegraphics[width=0.4\linewidth]{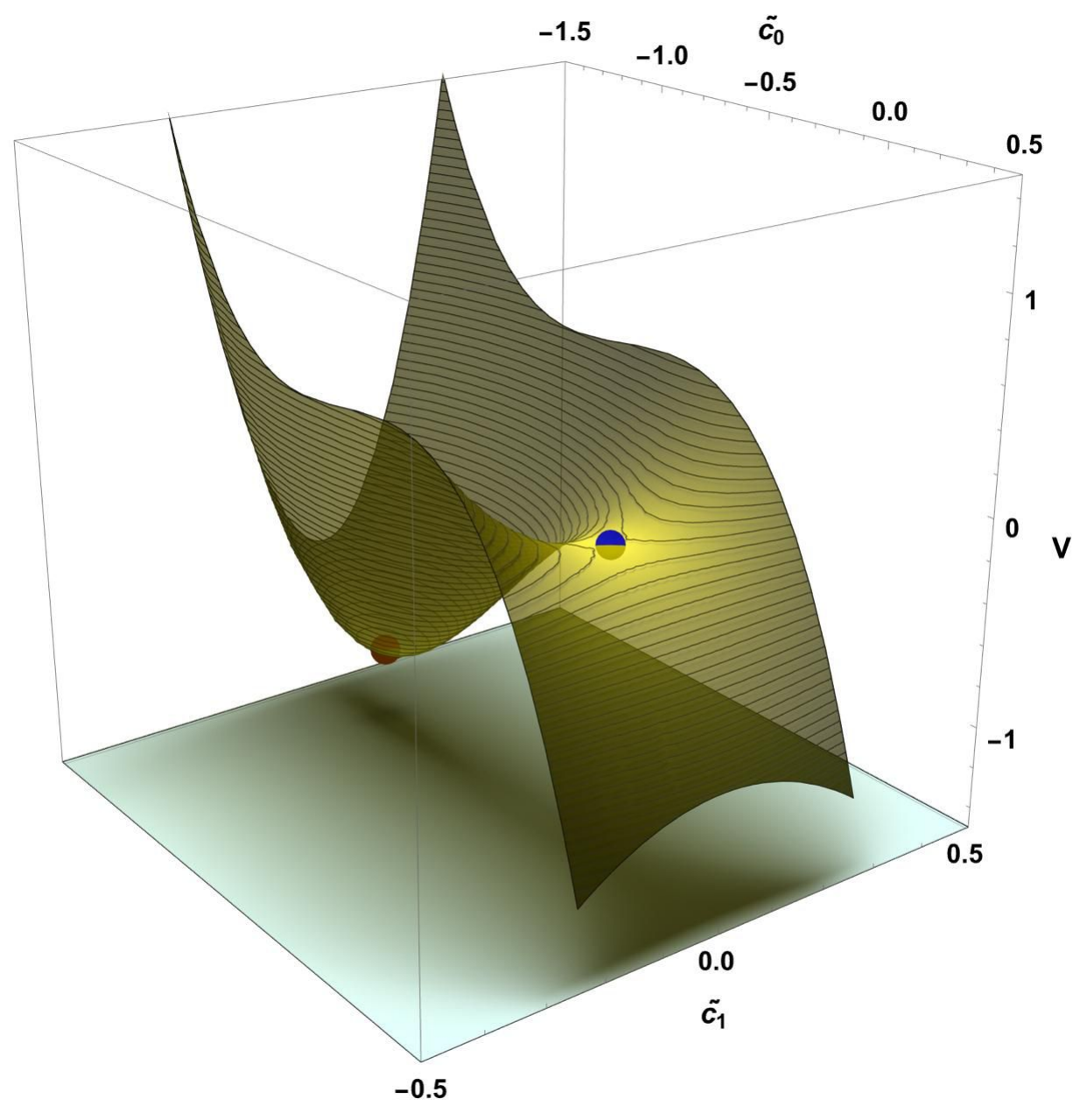}}
	\caption{\label{fig:potentialstringframe}A three-dimensional plot of the potential is shown, with the red dot indicating the local minima and the blue dot representing the saddle point. The parameters used are $B = 0.2$, $\mu = 0.6$, and $L = 1.1$, with all values expressed in units of GeV.}
\end{figure}

\begin{table}[!htb]
    \setlength\arraycolsep{5pt} 
    \renewcommand{\arraystretch}{1.25}
    
    $\begin{array}[t]{@{}
    r S[table-format=3.5] S[table-format=-3.5]
    @{\hspace{12pt}} !{\vrule width 0.2pt} @{\hspace{12pt}}
    r S[table-format=3.5] S[table-format=-3.5]
    @{\hspace{12pt}} !{\vrule width 0.2pt} @{\hspace{12pt}}
    r S[table-format=3.5] S[table-format=-3.5]
    @{}}
     \toprule
     \mu & {\tilde{c_{0}}~(||)} & {\tilde{c_{0}}~(\perp)} & \mu & {\tilde{c_{0}}~(||)} & {\tilde{c_{0}}~(\perp)} & \mu & {\tilde{c_{0}}~(||)} & {\tilde{c_{0}}~(\perp)} \\
     \midrule
    
     \multicolumn{9}{c}{\mathbf{B=0.0} \hspace{4cm} \mathbf{B=0.1} \hspace{4cm} \mathbf{B=0.2}}\\
     0.0 & -1.0313 & -1.0313 & 0.0 & -1.0196 & -1.0231 & 0.0 & -0.9826 & -0.9952\\
     0.3 & -1.0746 & -1.0746 & 0.3 & -1.0614 & -1.0652 & 0.3 & -1.0202 & -1.0332\\
     0.6 & -1.2293 & -1.2293 & 0.6 & -1.2101 & -1.2145 & 0.6 & -1.1519 & -1.1672\\
     0.9 & -1.6141 & -1.6141 & 0.9 & -1.5765 & -1.5827 & 0.9 & -1.4654 & -1.4860\\
     1.2 & -2.8563 & -2.8563 & 1.2 & -2.7184 & -2.7301 & 1.2 & -2.3554 & -2.3682\\
     \addlinespace
     \multicolumn{9}{c}{\mathbf{B=0.3} \hspace{4cm} \mathbf{B=0.4} \hspace{4cm} \mathbf{B=0.5}}\\
     0.0 & -0.9158 & -0.9307 & 0.0 & -0.8030 & -0.8061 & 0.0 & -0.6339 & -0.5912\\
     0.3 & -0.9465 & -0.9610 & 0.3 & -0.8240 & -0.8252 & 0.3 & -0.6441 & -0.5969\\
     0.6 & -1.0516 & -1.0646 & 0.6 & -0.8936 & -0.8873 & 0.6 & -0.6766 & -0.6130\\
     0.9 & -1.2852 & -1.2941 & 0.9 & -1.0374 & -1.0100 & 0.9 & -0.7363 & -0.6364\\
     1.2 & -1.8500 & -1.8337 & 1.2 & -1.3285 & -1.2258 & 1.2 & -0.8230 & -0.6432\\
    
     \bottomrule
     \end{array}$ 
\caption{\label{tab:StringMinima} The location of the minima point $(\tilde{c_{0}}, \tilde{c_{1}}=0)$ is shown for various values of the magnetic field and chemical potential for both parallel and perpendicular orientations. The parameter $L = 1.1$ is used, with all quantities expressed in units of GeV.}
\end{table}

It is clear from the previous analysis that the dynamics of the string produces positive values for the largest Lyapunov exponents, suggesting chaos in its dynamics. In order to make our discussion complete, in this section, we will analyse the Lyapunov exponent at the unstable fixed points and make a comparison with the MSS bound. The potential in the action~(\ref{eq:ActionModified}) gives rise to two fixed points, a stable and an unstable one. The former corresponds to the minimum of the potential, while the latter to the saddle point. These fixed points are explicitly shown in Fig.~\ref{fig:potentialstringframe}, where the red dot corresponds to the local minima and the blue dot corresponds to the saddle point of the potential. These two fixed points are similar in positioning and appearance for both $x_1$ and $x_3$ configurations. 

The entire evolution of the system given in~(\ref{eq:ActionModified}) is governed by  $\Dot{\overrightarrow{y}}=\overrightarrow{F}$, where $\overrightarrow{y}=\left(\Tilde{c_0},\Dot{\Tilde{c_0}},\Tilde{c_1},\Dot{\Tilde{c_1}} \right)$. The two fixed points correspond to $\overrightarrow{F}=0$. For energy $E=0$, the saddle point is found at the location $\overrightarrow{y}=\left(0,0,0,0\right)$ for the unstable string. Interestingly, the saddle point is always located at $\overrightarrow{y}=\left(0,0,0,0\right)$ for all values of $B$ and $\mu$. Whereas the position of the local minimum changes with $B$ and $\mu$. In particular, the location of the minimum, specified by $(\tilde{c_{0}},\tilde{c_{1}}=0)$, moves further away from the saddle point as $B$ and $\mu$ increases. The precise location of the minimum is specified in Table~\ref{tab:StringMinima} for different $B$ and $\mu$ values. 

We numerically obtain the Lyapunov exponents at the saddle point $\overrightarrow{y}=\left(0,0,0,0\right)$
in a similar way as in the last section and find that they asymptotically converge to $\left(\sqrt{-\omega_0^2},-\sqrt{-\omega_0^2},0,0\right)$, where the eigenvalues $\omega_0^2$ are given in Table~\ref{tab:StringEigenValues}. Therefore, the largest Lyapunov exponent at the saddle point is $\lambda_{max}=\sqrt{-\omega_0^2}$. Accordingly, the largest Lyapunov exponent decreases with $B$ and $\mu$ for both parallel and perpendicular string orientation. 

Interestingly, we can also obtain the Lyapunov exponents analytically at the saddle point. In particular, at the saddle point, the Lyapunov exponents are given by the real part of the eigenvalues of the Jacobian matrix of $\overrightarrow{F}$ \cite{sandri1996numerical}. At the saddle point, the Jacobian matrix ($J$) is of the form
\begin{equation}    
J=
    \begin{pmatrix}
        0    & -\omega_0^{2}       & 0      & 0      \\
        1      & 0 & 0 & 0      \\
        0 & 0  & 0 & -\omega_1^{2} \\
        0      & 0       & 1 & 0
    \end{pmatrix}\,,
\label{eq:jacobianmatrix}
\end{equation}
and its eigenvalues are $\left(-i \sqrt{\omega_0^2},i \sqrt{\omega_0^2},-i \sqrt{\omega_1^2},i \sqrt{\omega_1^2}\right)$. Since $\omega_0^2 <0$ and $\omega_1^2 >0$, we have two non-zero Lyapunov exponent $\left(\sqrt{-\omega_0^2},-\sqrt{-\omega_0^2},0,0\right)$, with largest Lyapunov exponent being $\lambda_{max}=\sqrt{-\omega_0^2}$. This is the same expression that we obtained earlier at the saddle point. This further advocates for the correctness of the numerical routine, and hence the corresponding
numerical results, considered in this work.

\begin{figure}[htbp]
\centering
\includegraphics[width=0.65\textwidth]{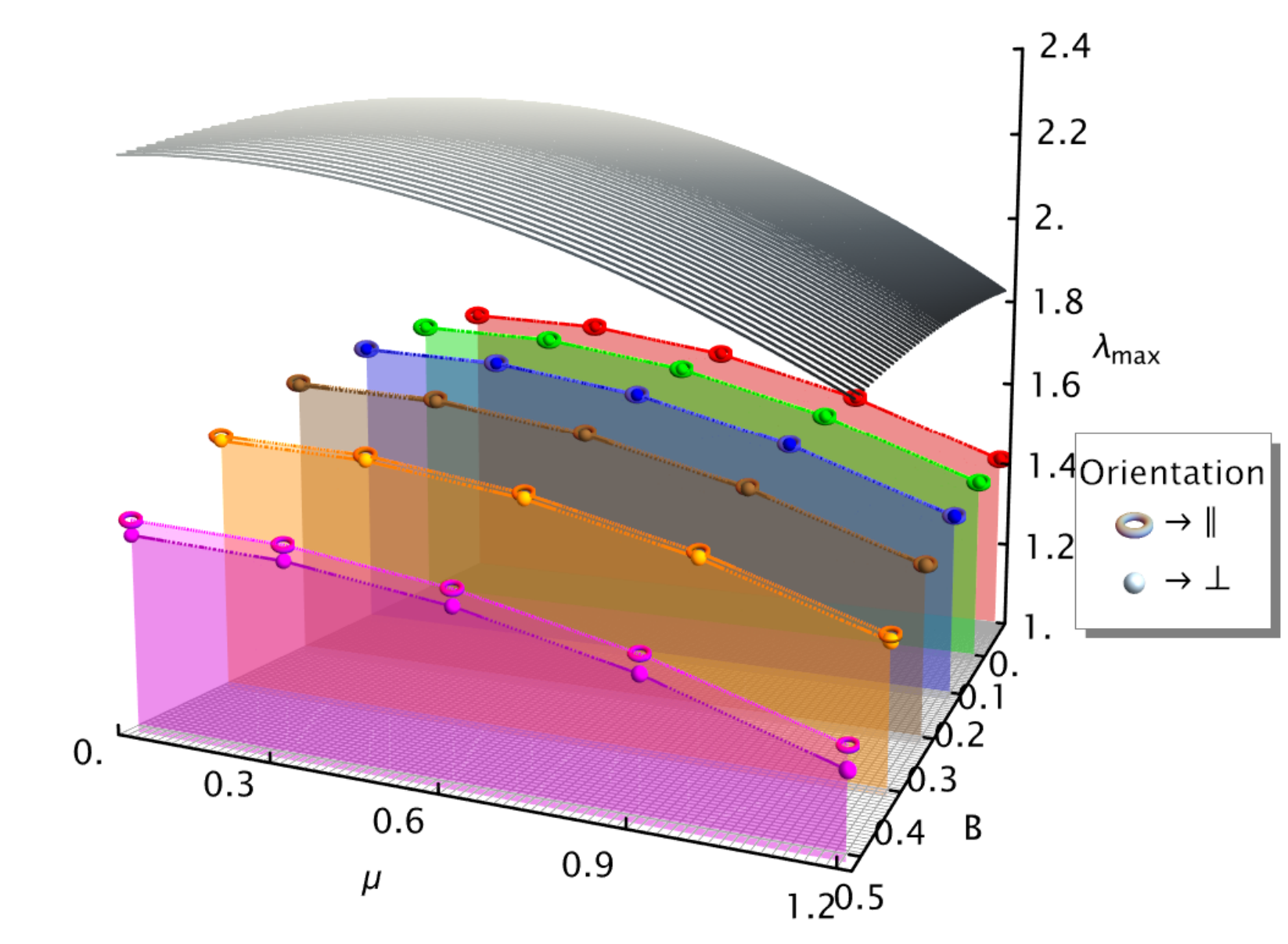}
\caption{The comparison of the maximum Lyapunov exponent $\lambda_{\text{max}}$ and the MSS bound is presented for the $x_1$ $(\parallel)$ and $x_3$ $(\perp)$ orientations at the unstable saddle point. Data points are colour-coded as follows: red, green, blue, brown, orange, and magenta colours represent $B = 0.0$, $0.1$, $0.2$, $0.3$, $0.4$, and $0.5$, respectively. The MSS bound is indicated by the grey surface above the data points.}
\label{fig:StringLmaxMSS}
\end{figure}

It is also important to compare the obtained Lyapunov exponents with the MSS bound. The MSS
bound, from Eq.~(\ref{gsol}) is given by
\begin{equation}\label{35}
	\lambda_{MSS} = \frac{{r_h}^2 g'({r_h})}{2}\,.
\end{equation}
We find that the largest Lyapunov exponent at the saddle point is always smaller than $\lambda_{MSS}$ for all $B$ and $\mu$ values, suggesting that it always satisfies the MSS bound for all values of chemical potential and magnetic field irrespective of its orientation with respect to the string. The comparison between $\lambda_{max}$ and $\lambda_{MSS}$ is shown in Fig.~\ref{fig:StringLmaxMSS}. 

As mentioned in the introduction, the above bound is essentially a classical analog of the famous MSS bound. That being said, it is also important to mention that one can in principle formulate the Lyapunov exponent of the suspended string in terms of OTOCs, with the Wilson loop appearing as one of the operators in OTOCs. The fact that the fluctuation of the Wilson loop expectation value grows exponentially does suggest a relationship between the obtained string Lyapunov exponent and OTOC. A qualitative discussion on this can be found in \cite{Hashimoto:2018fkb}. It would be interesting to try to explicitly establish this relation in the future.

In summary, our investigation of the string frame suggests that chaos is produced in the proximity of the black hole horizon, and the role of the chemical potential and magnetic field is to weaken the dependence of the system dynamics on the initial condition, thereby making the system less chaotic. In the context of holographic QCD, this indicates that less chaos is expected in the bound state of the quark-antiquark pair at higher chemical potential and magnetic field values. This might be related to having a less easy deconfinement, see also \cite{Hashimoto:2018fkb} where chaos was speculated to have a connection to the onset of deconfinement. In return, this could be related to the issue of inverse magnetic catalysis in QCD, which received a lot of attentions since the seminal lattice estimates of \cite{Bali:2011qj}. Although our result is true for both orientations of the string relative to the magnetic field, the string's chaotic dynamics get softened more along the transverse than the parallel direction. This could be related to the anisotropies seen in the confining properties of lattice QCD as well, see \cite{DElia:2021tfb}.

\section{Conclusion}\label{sec:conclusion}
In this paper, we have numerically analyzed the effects of chemical potential and magnetic field on the dynamics of the confining string based on the EMD model of \cite{Bohra:2019ebj,bohra2021chiral} that includes two $U(1)$ gauge fields to incorporate both chemical potential and background magnetic field in the dual boundary QCD system. The model has been shown elsewhere to capture several lattice-supported QCD features. Here, we analysed the system in both the string and Einstein frames. Our investigations reveal that the effects strongly depend on the considered gravitational frame. 

To be more precise, in the string frame we find that, as we increase the chemical potential the tip of the string moves away from the horizon both in parallel and perpendicular directions of the magnetic field. Similar behaviour is also observed with increasing magnetic field in both directions. The Poincar\'{e} section exhibits a more structured picture with a decrease in the number of scatter points as the value of the chemical potential and magnetic field increases. This behaviour is common for both magnetic field directions. The largest Lyapunov exponent decreases as we increase the value of the chemical potential and magnetic field,  also irrespective of the orientation of the magnetic field.  The Lyapunov exponent is, however, smaller in the perpendicular direction when compared to the parallel one. The largest Lyapunov exponent always remains below the MSS upper bound both at the stable and unstable fixed points. These results confirm the role of chemical potential and magnetic field as stabilisers of the system,  with a stronger stabilisation in the perpendicular direction. Chaos only shows up for the unstable string configuration, which is near the horizon, whereas no chaos appears in the stable string configuration, which is relatively far away from the horizon, highlighting the horizon as the source of the chaos.

On the other hand, in the Einstein frame, as we increase the chemical potential the tip of the string moves towards the horizon both in parallel and perpendicular directions of the magnetic field. With increasing magnetic field, the tip of the string moves towards/away from the horizon for the parallel/perpendicular directions. The Poincar\'{e} section becomes less structured around the origin as we increase the chemical potential. In fact, the Poincar\'{e} section becomes less/more structured when the magnetic field is increased in the parallel/perpendicular direction. The largest Lyapunov exponent increases as we increase the chemical potential, irrespective of the orientation of the magnetic field. Similarly, the largest Lyapunov exponent increases/decreases with the magnetic field for parallel/perpendicular cases. These results are quite different from the string frame case. From the Poincar\'{e} section and the Lyapunov exponents, we conclude that the effect of the chemical potential is to destabilise the system as it increases (as it is evident from Figure~\ref{fig:EinsteinLmax}). The largest Lyapunov exponent again always remains below the MSS bound both at the stable and unstable fixed points. Interestingly, as opposed to the string frame case, the largest Lyapunov exponent exhibits distinct chemical potential and magnetic field-related behaviour near the unstable and stable fixed points for both parallel and perpendicular orientations.

In summary, our results clearly distinguish between the string and Einstein frames, also in the context of holographic QCD. The choice of frame considerably alters the dynamical nature of the system. Specifically, in the string frame, the dependence of the confined quark-antiquark pair's dynamics on the initial conditions is shown to weaken as the chemical potential increases. Conversely, in the Einstein frame, the dynamics of the quark-antiquark pair were found to become more sensitive to its initial conditions with increasing chemical potential.

We end this paper by pointing out potential sources of chaos in the quark-antiquark pair in holographic confined and deconfined phases and possible directions for future research. One can also analyze the dynamics of the open string in the AdS-soliton background, corresponding to analysing chaos in the dual confined phase. In the AdS soliton background, the Nambu-Goto action and its corresponding equations of motion exhibit only one string profile for a fixed string length at the boundary. This profile is always stable. In \cite{Akutagawa:2019awh}, it was observed that by giving fluctuations/or perturbations at the end-points of the string, i.e., giving a pulse force (quench) to the quarks, the stable string also exhibits chaos for large enough quench. Perturbing the endpoints of the sting corresponds to giving a small deformation of the Wilson line along the time direction in the dual field theory side. A similar result also appeared while analysing the closed string dynamics in \cite{Basu:2011dg, Shukla:2024wsu}, where it was found that chaos appears in the closed string dynamics only in the high energy regime.

In our work, with a black hole geometry, there are two connected string profiles: a stable one (away from the horizon) and an unstable or saddle (near the horizon). We analyzed chaos in both stable and saddle string configurations and found that chaos is present only in the higher energy (or saddle) string configuration, that is, the one close to the black hole horizon, which we thus identify as the source of the chaos, see also \cite{Hashimoto:2018fkb}, where it was pointed out that the unstable string should play the role of the instanton amplitude relevant for string breaking, that is, the transition from confining to deconfining solution, thereby highlighting the potential physical relevance of the unstable string. 

In our study, we did not introduce any quench at the endpoints of the open string, which are completely fixed from the outset. 

The fact that the horizon does not act as a source of chaos for the stable open string opens up interesting possibilities of analysing chaos in it from other potential sources. Following \cite{Akutagawa:2019awh}, the obvious thing would be to introduce a quench at the endpoints of the stable string. This has been done for AdS-Schwarzschild background in \cite{Hashimoto:2018fkb}, and it was observed that the dynamics of the stable string become chaotic for large quench amplitude. It would be interesting to investigate this chaotic scenario of the stable open string with quench in our more sophisticated bottom-up holographic EMD model. We plan to undertake such an investigation in the future.

\section*{Acknowledgements}
The work of S.M.~is supported by the core research grant from the Science and Engineering Research Board, a statutory body under the Department of Science and Technology, Government of India, under grant agreement number CRG/2023/007670. 

\appendix

\section{Chaotic dynamics in the Einstein frame}\label{Einsteinframe}
We now look at how the chaotic string dynamics work out in the Einstein frame. Eq.~(\ref{eq:Ansatzmetric}) gives the corresponding black hole geometry. The dynamical analysis here is quite similar to the string frame calculation, so we can be more brief. We just need to replace the factor $A_{s}(r)$ by its counterpart $A(r)$.  As before, our analytical framework hinges on Eqs.~(\ref{7}) and (\ref{eq:Aforstringframe}), albeit with $A_{s} (r)\rightarrow A(r)$. Through the application of Eqs.~(\ref{eq:stringprof01}) and (\ref{eq:stringprof02}), we can extract the static string profile within the Einstein frame. Moreover, Eq.~(\ref{eq:StringLength}) gives us a means to quantify the total length of the string for both parallel and perpendicular orientations of the string.

\begin{figure}[htbp!]
\centering
\includegraphics[width=.47\textwidth]{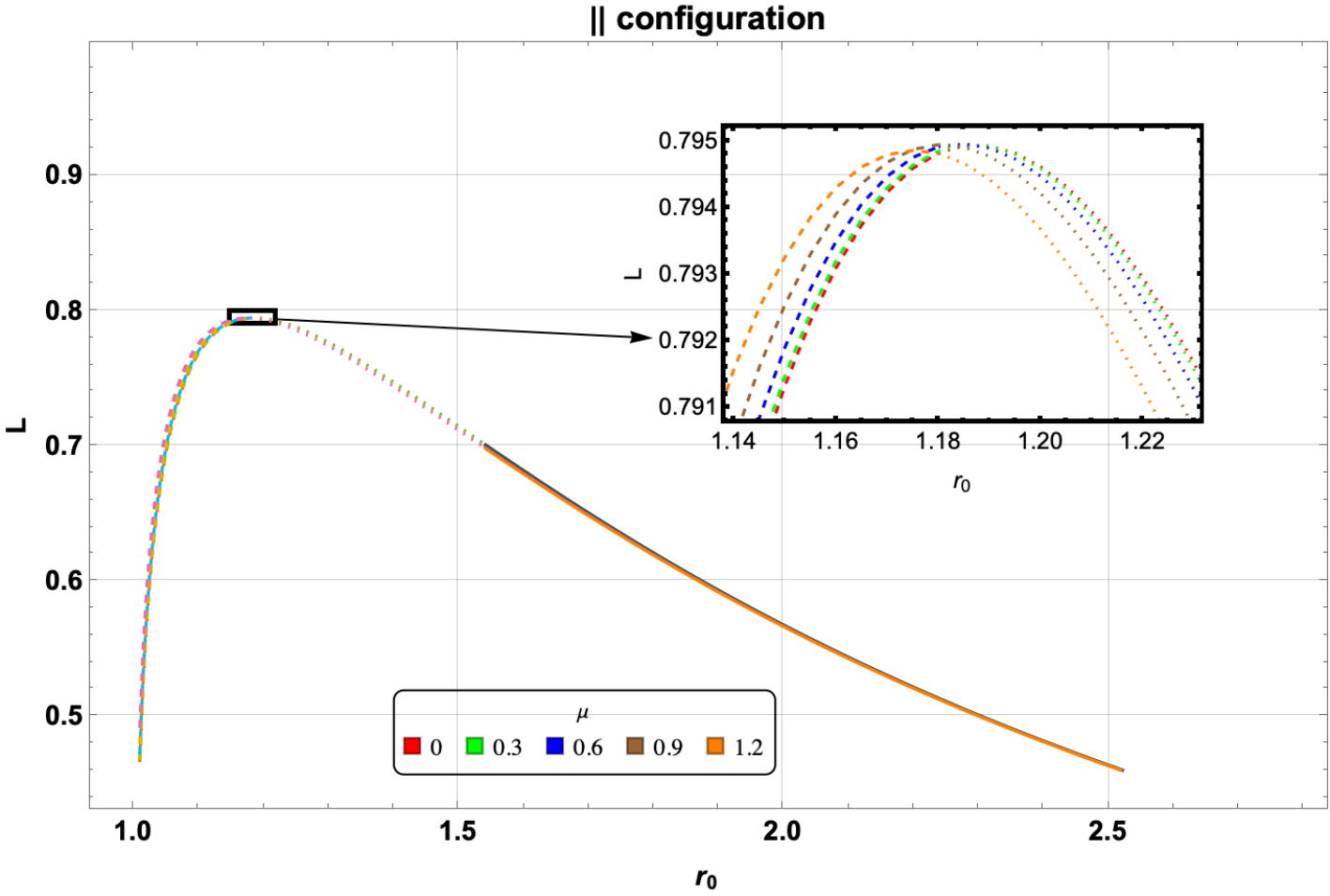}
\qquad
\includegraphics[width=.47\textwidth]{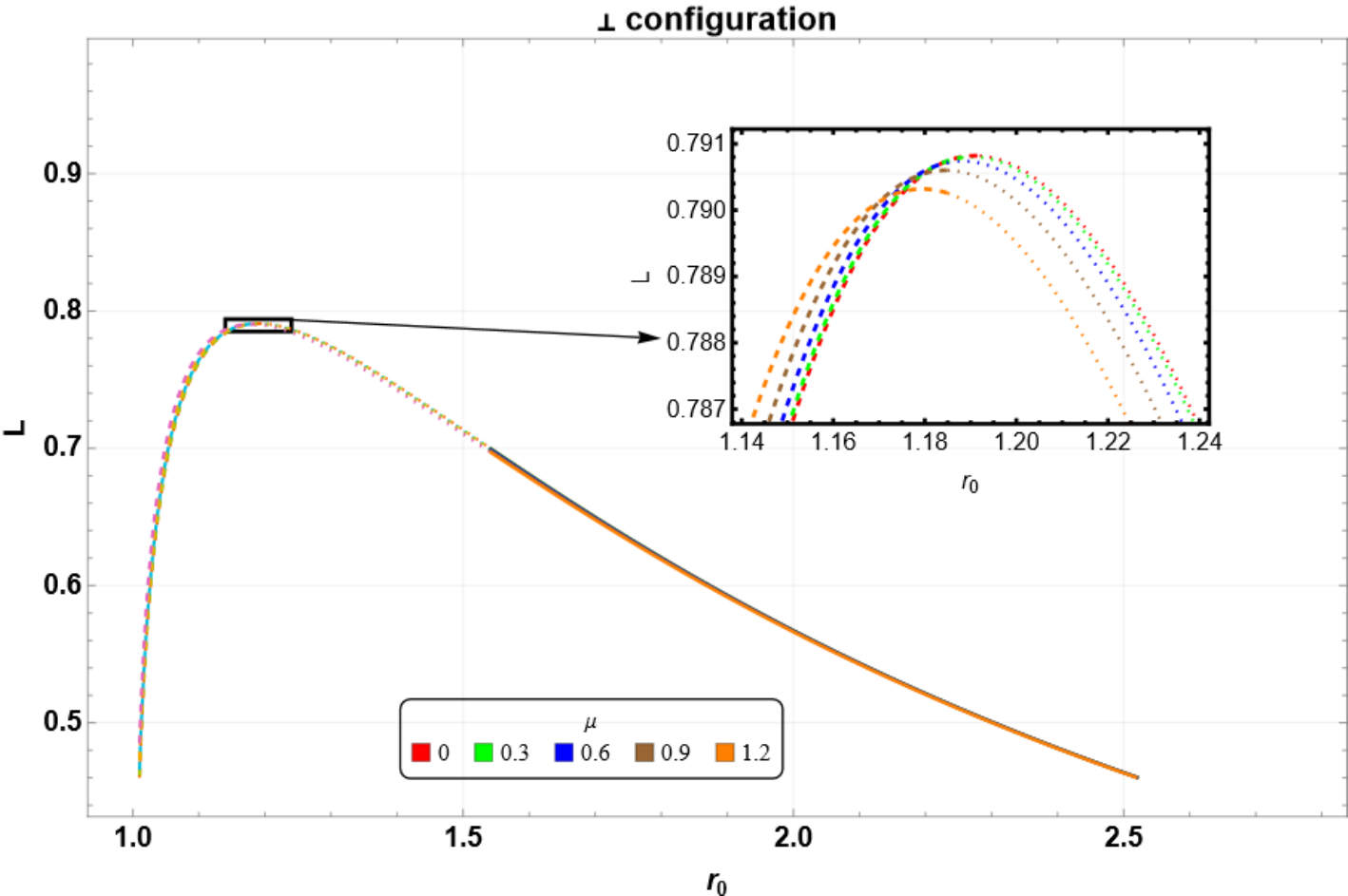}
\caption{Behavior of the string length $L$ as a function of $r_0$ for parallel (left) and perpendicular (right) orientations of the string, with varying chemical potential $\mu$ and a fixed magnetic field $B = 0.2$. All quantities are expressed in units of GeV.}
\label{fig:EinsteinLvsr0}
\end{figure}

\begin{table}[!htb]
    \setlength\arraycolsep{4pt} 
    \renewcommand{\arraystretch}{1.25}
    
    $\begin{array}[t]{@{}
    r S[table-format=3.5] S[table-format=-3.5]
    @{\hspace{12pt}} !{\vrule width 0.2pt} @{\hspace{12pt}}
    r S[table-format=3.5] S[table-format=-3.5]
    @{\hspace{12pt}} !{\vrule width 0.2pt} @{\hspace{12pt}}
    r S[table-format=3.5] S[table-format=-3.5]
    @{}}
     \toprule
     \mu & {r_{0}~(||)} & {r_{0}~(\perp)} & \mu & {r_{0}~(||)} & {r_{0}~(\perp)} & \mu & {r_{0}~(||)} & {r_{0}~(\perp)} \\
     \midrule
    
     \multicolumn{9}{c}{\mathbf{B=0.0} \hspace{4cm} \mathbf{B=0.1} \hspace{4cm} \mathbf{B=0.2}}\\
     0.0 & 1.08235 & 1.08235 & 0.0 & 1.08214 & 1.08351 & 0.0 & 1.08149 & 1.08693\\
     0.3 & 1.08178 & 1.08178 & 0.3 & 1.08157 & 1.08296 & 0.3 & 1.08091 & 1.08643\\
     0.6 & 1.08003 & 1.08003 & 0.6 & 1.07984 & 1.08126 & 0.6 & 1.07929 & 1.08487\\
     0.9 & 1.07752 & 1.07752 & 0.9 & 1.07732 & 1.07857 & 0.9 & 1.07672 & 1.08209\\
     1.2 & 1.07358 & 1.07358 & 1.2 & 1.07336 & 1.07475 & 1.2 & 1.07267 & 1.07818\\
     \addlinespace
     \multicolumn{9}{c}{\mathbf{B=0.3} \hspace{4cm} \mathbf{B=0.4} \hspace{4cm} \mathbf{B=0.5}}\\
     0.0 & 1.08037 & 1.09303 & 0.0 & 1.07896 & 1.10255 & 0.0 & 1.07710 & 1.11701\\
     0.3 & 1.07981 & 1.09250 & 0.3 & 1.07846 & 1.10204 & 0.3 & 1.07657 & 1.11658\\
     0.6 & 1.07833 & 1.09085 & 0.6 & 1.07692 & 1.10045 & 0.6 & 1.07495 & 1.11528\\
     0.9 & 1.07568 & 1.08831 & 0.9 & 1.07415 & 1.09809 & 0.9 & 1.07204 & 1.11300\\
     1.2 & 1.07150 & 1.08464 & 1.2 & 1.06984 & 1.09461 & 1.2 & 1.06803 & 1.10964\\
    
     \bottomrule
     \end{array}$ 
\caption{\label{tab:EinsteinLvsr0}The parameter $r_0$ of the unstable string configuration for various magnetic field strengths and chemical potentials, for both parallel and perpendicular string orientations. The string length is set to $L = 0.75$. All values are given in units of GeV.}
\end{table}

The variation of the string length $L(r_0)$ for different values of $\mu$ and $B=0.2$ for parallel and perpendicular configurations is
shown in Fig.~\ref{fig:EinsteinLvsr0}. The dashed line in the figure gives the unstable string profile, solid lines give the stable profile and the dotted line denotes the meta-stable profile. Similar to what we observed in the string frame, here in the Einstein frame, we also find a maximum length for the string, denoted as $L_{max}$, beyond which no connected solution of the string exists. Below $L_{max}$, there are two distinct types of connected string solutions: one closer to the black hole horizon (shown by dashed lines) and another further away from the horizon (represented by dotted and solid lines). The small $r_0$ solution again corresponds to the local maximum of the energy, whereas the large $r_0$ solution corresponds to the local minimum. The free energy of the string is observed to be similar to that of Fig.~\ref{fig:StringFreeEnergy}, i.e., there is a phase transition from the large $r_0$ connected solution to the parallel disconnected strings as the length of the string is varied, with the connected (large $r_0$) solution always corresponding to the stable and the disconnected one (small $r_0$) to the unstable configuration. 

An important quantity of interest is again the tip ($r_0$) of the unstable string. For fixed string length $L=0.75$, the variation of $r_0$ for different $B$ and $\mu$ is given in Table~\ref{tab:EinsteinLvsr0}. In the subsequent analysis, we will be maintaining this fixed string length. Here, we can already appreciate some subtle differences with the string frame case. Indeed, as we increase the chemical potential, the $r_0$ value decreases for both the parallel ($x_1$) and perpendicular ($x_3$) orientations of the string, i.e., the tip of the string moves toward the horizon, in contrast to the string frame findings. Similarly, for a fixed $\mu$, the value of $r_{0}$ decreases as $B$ increases in the $x_1$ direction while it increases as $B$ increases in the $x_3$ direction. This already suggests that not only substantial changes might appear in the chaotic dynamics of the string in terms of the chemical potential, but also magnetic field-dependent anisotropy may appear in the chaotic string dynamics in the Einstein frame. In particular, the chaotic structure might now lessen/enhance with magnetic fields in the perpendicular/parallel directions.

\subsection{Perturbing the string}
We now move on to analyse the perturbed string dynamics in the Einstein frame following the same procedure as in the string frame. Again, we are only interested in the unstable string (given by the dashed line in Fig.~\ref{fig:EinsteinLvsr0}), where $r_0$ is closer to the horizon, as the stable string profiles do not show any chaos. 

\begin{figure}[htb!]
	\centering
	\subfigure[Parallel configuration]{\label{fig:eigen_states_x1_einsteinframe}	\includegraphics[width=0.47\linewidth]{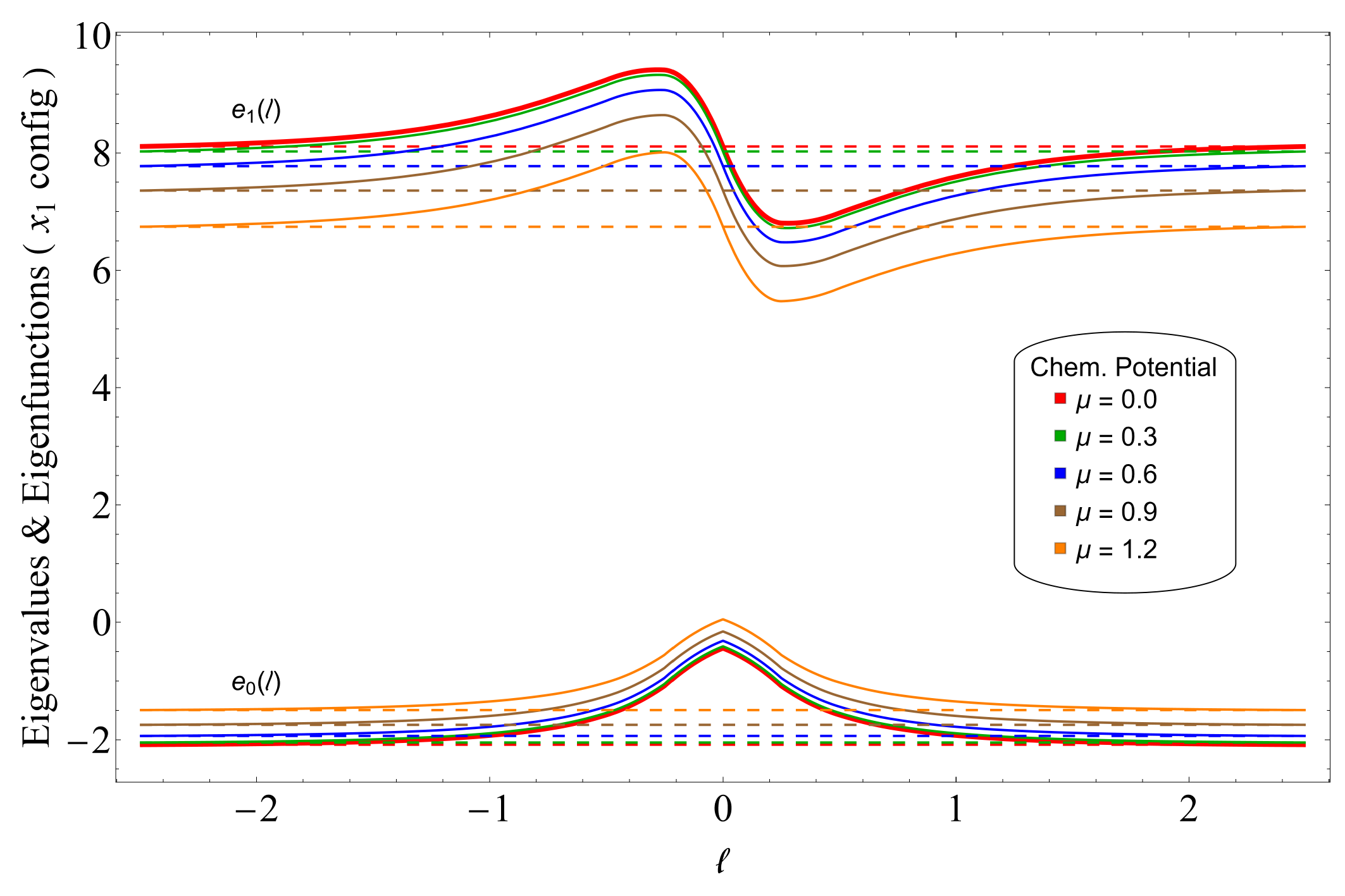}}
    \subfigure[Perpendicular configuration]{\label{fig:eigen_states_x3_einsteinframe}
	\includegraphics[width=0.47\linewidth]{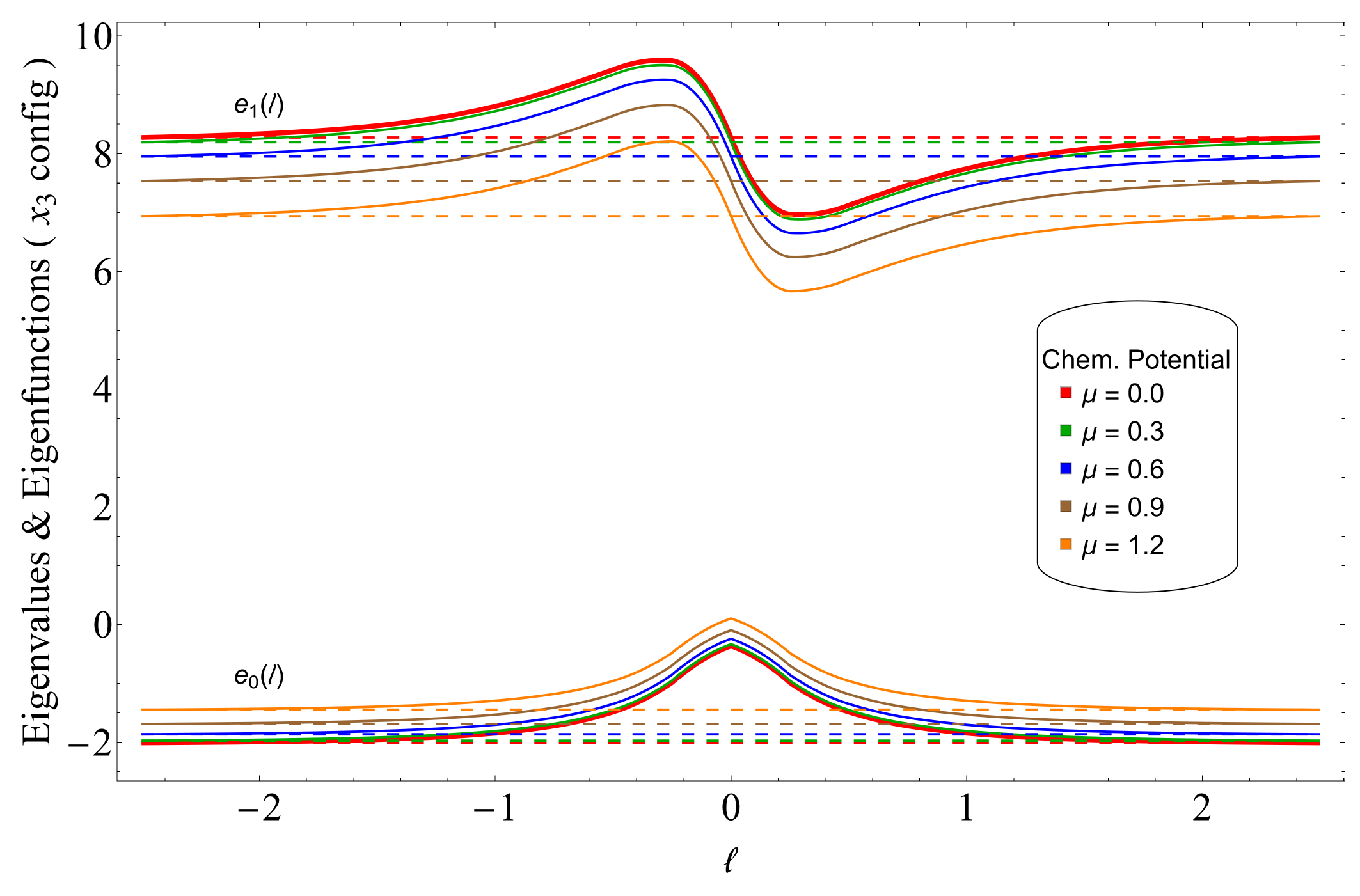}}
	\caption{\label{fig:eigenstates-einsteinframe}The eigenfunctions $e_{0}(\ell)$ and $e_{1}(\ell)$ as a function of $\ell$ for different values of the chemical potential for both parallel and perpendicular string configurations. The magnetic field is set at $B = 0.2$ and the string length is set at $L = 0.75$. All values in GeV units.}
\end{figure}

The relevant equations take the same form as in Eq.~(\ref{eq:stringperturb01}) and (\ref{eq:stringperturb02}), thence the analysis will be very similar. We expand the action up to the second and third order in the perturbation parameter. The dynamics can then be obtained from the second-order action [see Eq.~(\ref{eq:secondorderaction})], which is again of the Sturm-Liouville form [given by Eq.~(\ref{eq:Strum-Liouville})], with coefficients $C_{\ell\ell}^{x_i}$, $C_{00}^{x_i}$ and $C_{tt}^{x_i}$. After factorizing the equation of motion, we can determine the two lowest eigenvalues and the eigenfunction of the Sturm-Liouville equation. Table \ref{tab:EinsteinEigenValues} gives us the two lowest eigenvalues for both $x_1$ and $x_3$ orientations. The lowest eigenvalue $\omega_0^2$ is again negative for all values of $\mu$ and $B$, suggesting chaos in the Einstein frame as well. This is true for both $x_1$ and $x_3$ orientations of the string. Moreover, the nature of $\omega_0^2$ is again quite similar to the string frame case. In particular, it decreases with the magnetic field and is stronger in the perpendicular direction. A similar trend can also be observed with the chemical potential. The two lowest eigenfunctions are shown in Fig.~\ref{fig:eigenstates-einsteinframe}, which again exhibits qualitatively similar features as in the string frame case.  

The perturbed action up to the third order (see Eq.~\eqref{eq:thirdorderaction}) can be used to determine the form of potential and the dynamics of $c_0$ and $c_1$. This term also gives us a trapping potential in the Einstein frame. The values of the coefficient $K^{x_i}_{1,\ldots,5}$ are given in the Table \ref{tab:EinsteinKValues}. Although the values are recorded only up to the third decimal, higher accurate values have been used in numerical computations. Similar to the string frame, to avoid a negative kinetic term in some parts of the parameter space, we transformed the parameters $c_{0,1}\rightarrow\Tilde{c}_{0,1}$, where $c_0=\Tilde{c}_{0} + \alpha_{1}\Tilde{c}_{0}^{2} + \alpha_{2}\Tilde{c}_{1}^{2}$ and $c_{1}=\Tilde{c}_{1}+\alpha_{3}\Tilde{c}_{0}\Tilde{c}_{1}$ following similar methods adopted in \cite{Hashimoto:2018fkb, Colangelo:2021kmn, Shukla:2023pbp}, where for the current case e.g.~$\alpha_{1}=-3$, $\alpha_{2}=-1$, and $\alpha_{3}=-1.5$ are good choices. 

\begin{table}[!htb]
    \setlength\arraycolsep{4pt} 
    \renewcommand{\arraystretch}{1.25}
    
    $\begin{array}[t]{@{}
    r S[table-format=3.5] S[table-format=-3.5]
    @{\hspace{12pt}} !{\vrule width 0.2pt} @{\hspace{12pt}}
    r S[table-format=3.5] S[table-format=-3.5]
    @{}}
     \toprule
     \mu & {\omega_{0}^{2}~(||)} & {\omega_{1}^{2}~(||)} & \mu & {\omega_{0}^{2}~(\perp)} & {\omega_{1}^{2}~(\perp)} \\
     \midrule
    
     \multicolumn{6}{c}{\mathbf{B=0.0}}\\
     0.0  &   -2.1182 & 8.2144  & 0.0   &   -2.1182 &  8.2144 \\
     0.3 & -2.0799 & 8.1291 & 0.3  &  -2.0799 & 8.1291 \\
     0.6 &  -1.9660 &7.8695 & 0.6 & -1.9660 &7.8695 \\
     0.9 & -1.7688 &7.4501 & 0.9 & -1.7688 &7.4501 \\
     1.2 & -1.5063 &6.8297 & 1.2 & -1.5063 &6.8297 \\
     \addlinespace
     \multicolumn{6}{c}{\mathbf{B=0.1}} \\  
     0.0  &  -2.1099 & 8.1884 & 0.0   &   -2.0905 & 8.2306 \\
     0.3  & -2.0719& 8.1034 & 0.3  &  -2.0525 & 8.1469 \\
     0.6  & -1.9586 & 7.8455 & 0.6 & -1.9400 &7.8918 \\
     0.9 & -1.7634& 7.4271 & 0.9 & -1.7506 &7.4673 \\
     1.2 &-1.5035& 6.8080 & 1.2 & -1.4911 &6.8581 \\
     \addlinespace
     \multicolumn{6}{c}{\mathbf{B=0.2}} \\     
     0.0   &  -2.0855& 8.1091 & 0.0   &   -2.0114 & 8.2745 \\
     0.3  & -2.0767& 8.0564 & 0.3  &  -1.9746 & 8.1951 \\
     0.6  & -1.9360& 7.7742 & 0.6 & -1.8662 &7.9526 \\
     0.9 &-1.7470& 7.3577 & 0.9 & -1.6905 &7.5345 \\
     1.2 &-1.4953& 6.7409  & 1.2 & -1.4481 &6.9364 \\
     
     \bottomrule
     \end{array}$ 
     \hfill       
     $\begin{array}[t]{@{}
    r S[table-format=3.5] S[table-format=-3.5]
    @{\hspace{12pt}} !{\vrule width 0.2pt} @{\hspace{12pt}}
    r S[table-format=3.5] S[table-format=-3.5]
    @{}}
     \toprule
     \mu & {\omega_{0}^{2}~(||)} & {\omega_{1}^{2}~(||)} & \mu & {\omega_{0}^{2}~(\perp)} & {\omega_{1}^{2}~(\perp)} \\
     \midrule
    
     \multicolumn{6}{c}{\mathbf{B=0.3}}\\
     0.0  &   -2.0454& 7.9744 & 0.0   &   -1.8794 & 8.3604 \\
     0.3 &  -2.0099& 7.8920 & 0.3  &  -1.8472 & 8.2828 \\
     0.6 &  -1.8992& 7.6524 & 0.6 & -1.7518 &8.0458 \\
     0.9 & -1.7201& 7.2393 & 0.9 & -1.5898 &7.6552 \\
     1.2 & -1.4814& 6.6270 & 1.2 & -1.3689 &7.0929 \\
     \addlinespace
     \multicolumn{6}{c}{\mathbf{B=0.4}} \\  
     0.0  &  -1.9856& 7.7916 & 0.0   &   -1.6924 & 8.5113 \\
     0.3  & -1.9511& 7.7138 & 0.3  &  -1.6648 & 8.4387 \\
     0.6  & -1.8486& 7.4770 & 0.6 & -1.5833 &8.2169 \\
     0.9 &-1.6831& 7.0680 & 0.9 & -1.4430 &7.8557 \\
     1.2 &-1.4610& 6.4642 & 1.2 & -1.2531 &7.3336 \\
     \addlinespace
     \multicolumn{6}{c}{\mathbf{B=0.5}} \\     
     0.0   &  -1.9098 &  7.5516 & 0.0   &  -1.4408 & 8.7752 \\
     0.3  & -1.8788& 7.4746 & 0.3  &  -1.4179 & 8.7107 \\
     0.6  & -1.7863& 7.2411 & 0.6 & -1.3496 &8.5158 \\
     0.9 &-1.6368& 6.8376 & 0.9 & -1.2389 &8.1830 \\
     1.2 &-1.4277& 6.2673 & 1.2 & -1.0888 &7.7030 \\
     
     \bottomrule
     \end{array}$ 
\caption{\label{tab:EinsteinEigenValues}The first two eigenvalues, $\omega_{0}^{2}$ and $\omega_{1}^{2}$, of the unstable string configuration as a function of magnetic field strength and chemical potential for both parallel and perpendicular string orientations. The string length is fixed at $L = 0.75$. All quantities are in units of GeV.}
\end{table}

\begin{table}[!htb]
    \setlength\arraycolsep{4pt} 
    \renewcommand{\arraystretch}{1.25}
    
    $\begin{array}[t]{@{}
    r S[table-format=3.3] S[table-format=3.3] S[table-format=3.3] S[table-format=3.3] S[table-format=3.3]
    @{\hspace{10pt}} !{\vrule width 0.2pt} @{\hspace{10pt}}
    r S[table-format=3.3] S[table-format=3.3] S[table-format=3.3] S[table-format=3.3] S[table-format=3.3]
    @{}}
     \toprule
     \mu & {K_{1}~(||)} & {K_{2}~(||)} & {K_{3}~(||)} & {K_{4}~(||)} & {K_{5}~(||)} & {K_{1}~(\perp)} & {K_{2}~(\perp)} & {K_{3}~(\perp)} & {K_{4}~(\perp)} & {K_{5}~(\perp)}\\
     \midrule
    
     \multicolumn{11}{c}{\mathbf{B=0.0}}\\
     0.0 & 19.199 & 33.873 & 13.720 & 3.783 & 7.566 & 19.199 & 33.873 & 13.720 & 3.783 & 7.566\\
     0.3 & 18.728 & 33.420 & 13.724 & 3.808 & 7.616 & 18.728 & 33.420 & 13.724 & 3.808 & 7.616\\
     0.6 & 17.357 & 32.062 & 13.733 & 3.883 & 7.766 & 17.357 & 32.062 & 13.733 & 3.883 & 7.766\\
     0.9 & 15.104 & 29.719 & 13.696 & 4.002 & 8.005 & 15.104 & 29.719 & 13.696 & 4.002 & 8.005\\
     1.2 & 12.220 & 26.420 & 13.647 & 4.184 & 8.368 & 12.220 & 26.420 & 13.647 & 4.184 & 8.368\\
     \addlinespace
     \multicolumn{11}{c}{\mathbf{B=0.1}} \\  
     0.0 & 19.093 & 33.764 & 13.725 & 3.790 & 7.580 & 18.904 & 33.615 & 13.603 & 3.768 & 7.535\\
     0.3 & 18.628 & 33.311 & 13.728 & 3.815 & 7.629 & 18.440 & 33.168 & 13.604 & 3.791 & 7.583\\
     0.6 & 17.270 & 31.955 & 13.736 & 3.889 & 7.779 & 17.093 & 31.821 & 13.610 & 3.865 & 7.729\\
     0.9 & 15.042 & 29.629 & 13.700 & 4.008 & 8.017 & 14.914 & 29.530 & 13.592 & 3.986 & 7.972\\
     1.2 & 12.188 & 26.355 & 13.652 & 4.189 & 8.379 & 12.076 & 26.279 & 13.536 & 4.162 & 8.323\\
     \addlinespace
     \multicolumn{11}{c}{\mathbf{B=0.2}} \\  
     0.0 & 18.784 & 33.423 & 13.740 & 3.812 & 7.624 & 18.076 & 32.875 & 13.271 & 3.725 & 7.449\\
     0.3 & 18.636 & 33.263 & 13.772 & 3.830 & 7.659 & 17.633 & 32.444 & 13.270 & 3.746 & 7.492\\
     0.6 & 17.004 & 31.640 & 13.742 & 3.908 & 7.815 & 16.354 & 31.142 & 13.267 & 3.813 & 7.626\\
     0.9 & 14.852 & 29.364 & 13.711 & 4.026 & 8.052 & 14.345 & 28.975 & 13.262 & 3.931 & 7.862\\
     1.2 & 12.091 & 26.154 & 13.671 & 4.207 & 8.413 & 11.671 & 25.877 & 13.226 & 4.101 & 8.201\\
     \addlinespace
     \multicolumn{11}{c}{\mathbf{B=0.3}} \\  
     0.0 & 18.283 & 32.860 & 13.767 & 3.848 & 7.697 & 16.770 & 31.704 & 12.739 & 3.652 & 7.305\\
     0.3 & 17.855 & 32.433 & 13.769 & 3.872 & 7.744 & 16.379 & 31.306 & 12.742 & 3.673 & 7.346\\
     0.6 & 16.576 & 31.117 & 13.754 & 3.939 & 7.878 & 15.244 & 30.116 & 12.750 & 3.737 & 7.474\\
     0.9 & 14.543 & 28.918 & 13.730 & 4.056 & 8.113 & 13.400 & 28.089 & 12.738 & 3.841 & 7.682\\
     1.2 & 11.929 & 25.811 & 13.704 & 4.236 & 8.472 & 10.995 & 25.229 & 12.703 & 3.993 & 7.986\\
     \addlinespace
     \multicolumn{11}{c}{\mathbf{B=0.4}} \\  
     0.0 & 17.159 & 31.632 & 13.782 & 3.918 & 7.836 & 15.091 & 30.197 & 12.032 & 3.551 & 7.102\\
     0.3 & 17.218 & 31.664 & 13.789 & 3.918 & 7.835 & 14.762 & 29.838 & 12.035 & 3.569 & 7.139\\
     0.6 & 15.980 & 30.378 & 13.773 & 3.984 & 7.968 & 13.783 & 28.765 & 12.045 & 3.625 & 7.251\\
     0.9 & 14.110 & 28.283 & 13.762 & 4.101 & 8.202 & 12.191 & 26.956 & 12.036 & 3.716 & 7.431\\
     1.2 & 11.689 & 25.317 & 13.749 & 4.278 & 8.557 & 10.115 & 24.408 & 12.016 & 3.848 & 7.697\\
     \addlinespace
     \multicolumn{11}{c}{\mathbf{B=0.5}} \\  
     0.0 & 16.669 & 30.990 & 13.808 & 3.958 & 7.917 & 13.161 & 28.490 & 11.170 & 3.419 & 6.837\\
     0.3 & 16.305 & 30.603 & 13.808 & 3.980 & 7.960 & 12.885 & 28.183 & 11.171 & 3.433 & 6.867\\
     0.6 & 15.250 & 29.425 & 13.806 & 4.046 & 8.091 & 12.072 & 27.253 & 11.174 & 3.478 & 6.956\\
     0.9 & 13.569 & 27.455 & 13.810 & 4.162 & 8.323 & 10.777 & 25.703 & 11.180 & 3.555 & 7.110\\
     1.2 & 11.334 & 24.636 & 13.774 & 4.325 & 8.650 & 9.080 & 23.524 & 11.183 & 3.668 & 7.336\\
     
     \bottomrule
     \end{array}$ 
\caption{\label{tab:EinsteinKValues}$K$ values for various chemical potentials $\mu$ and magnetic field strengths $B$ for both parallel and perpendicular string orientations in the Einstein frame. The string length is set to $L = 0.75$. All quantities are expressed in units of GeV.}
\end{table}

\subsection{Poincar\'{e} sections}\label{EinsteinPoincare}
\begin{figure}[htbp!]
	\centering
    \begin{tabular}{c c c}
		\textbf{$\mu$ value} & \textbf{Parallel Configuration} & \textbf{Perpendicular Configuration} \\
		\textbf{$\mu=0.0$} & \includegraphics[scale=0.22,valign=c]{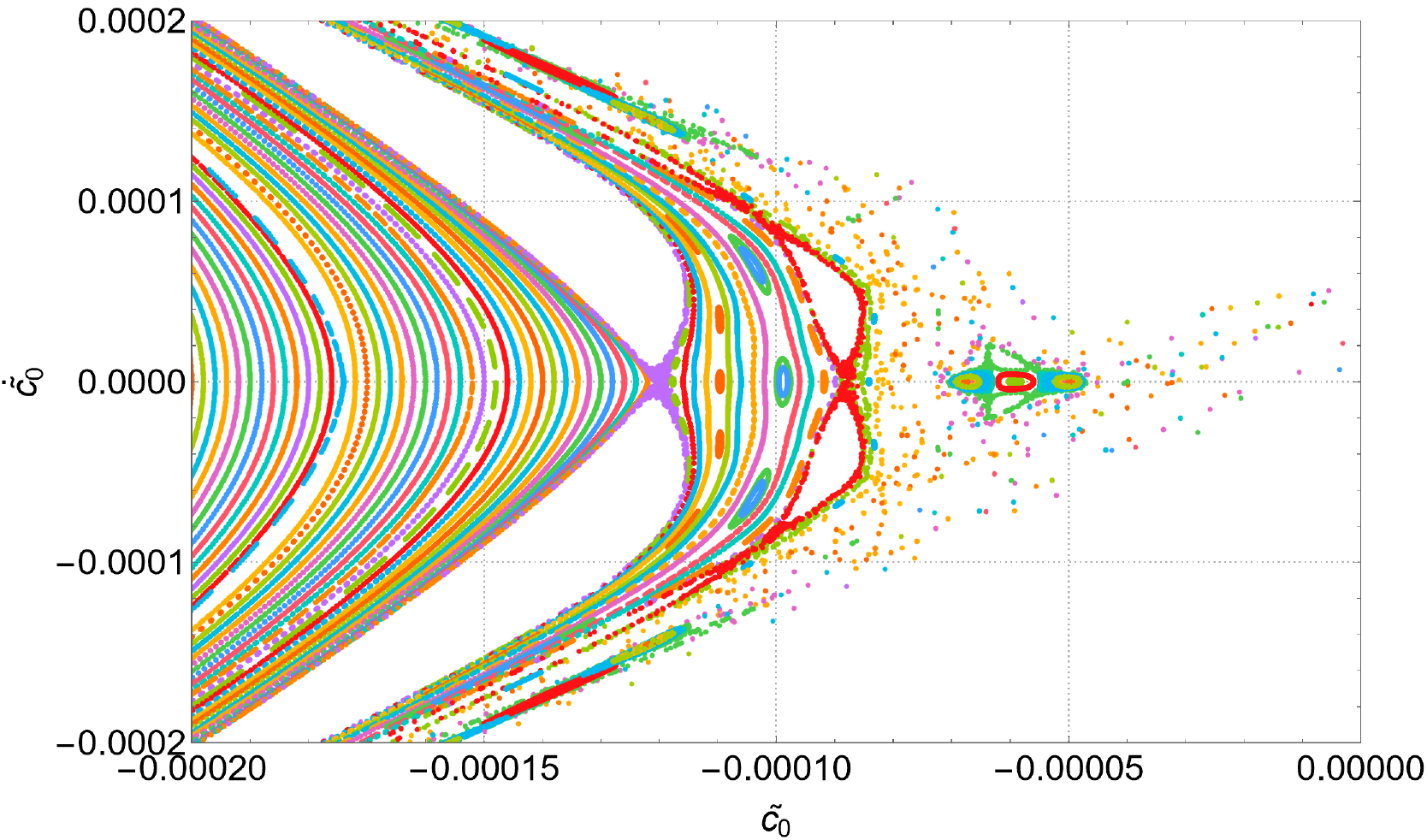} & \includegraphics[scale=0.22,valign=c]{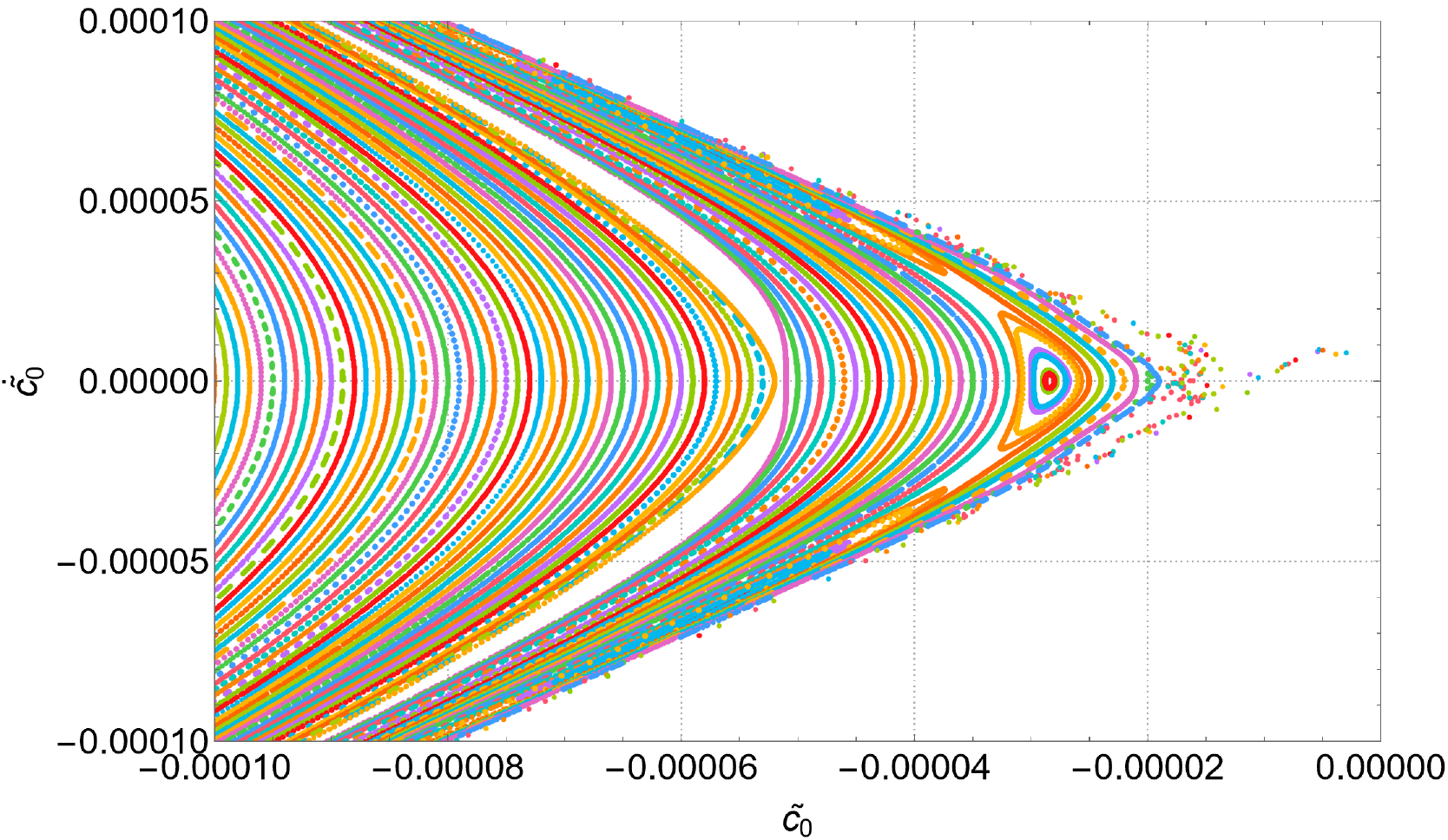} \\
		\textbf{$\mu=0.3$} & \includegraphics[scale=0.22,valign=c]{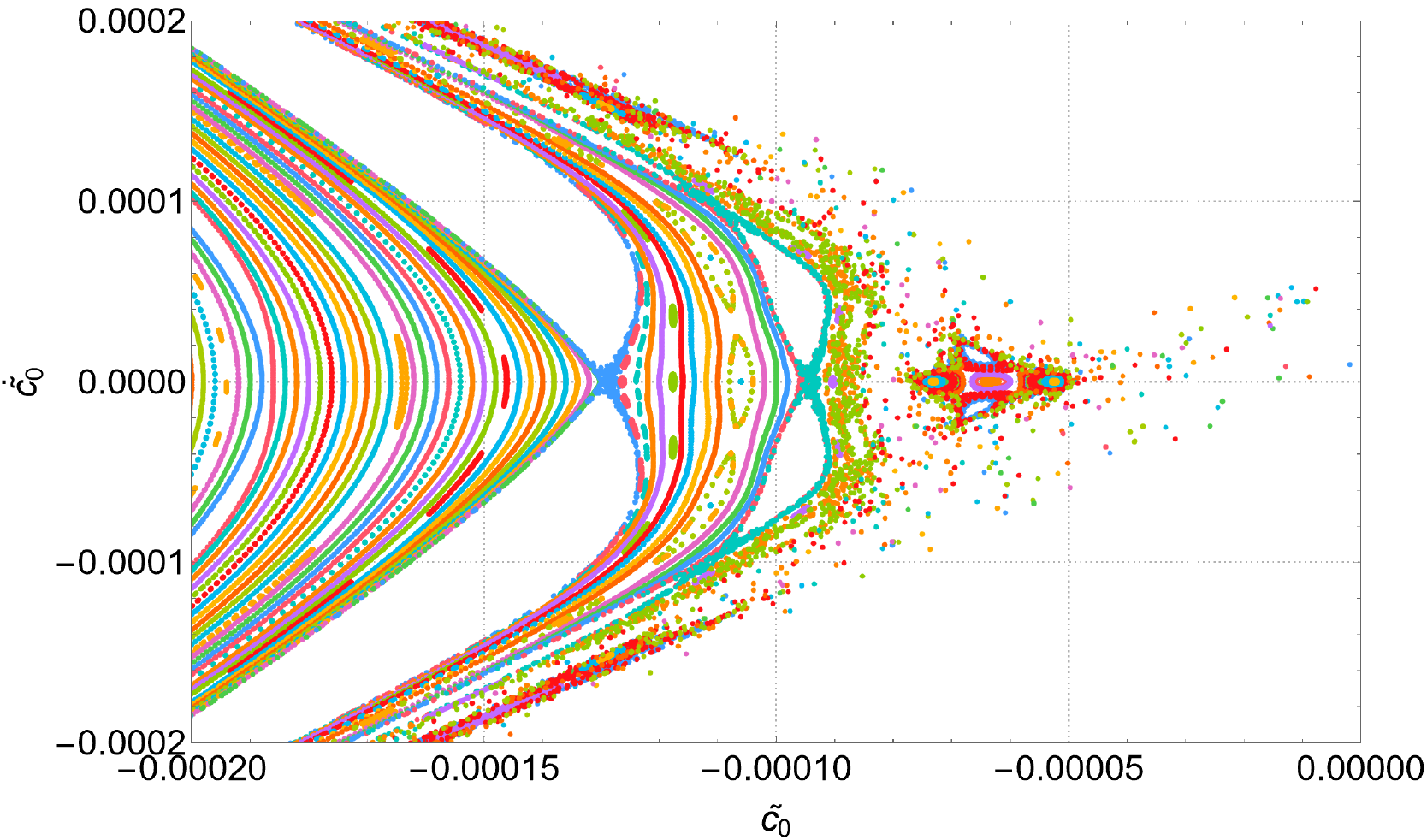} & \includegraphics[scale=0.22,valign=c]{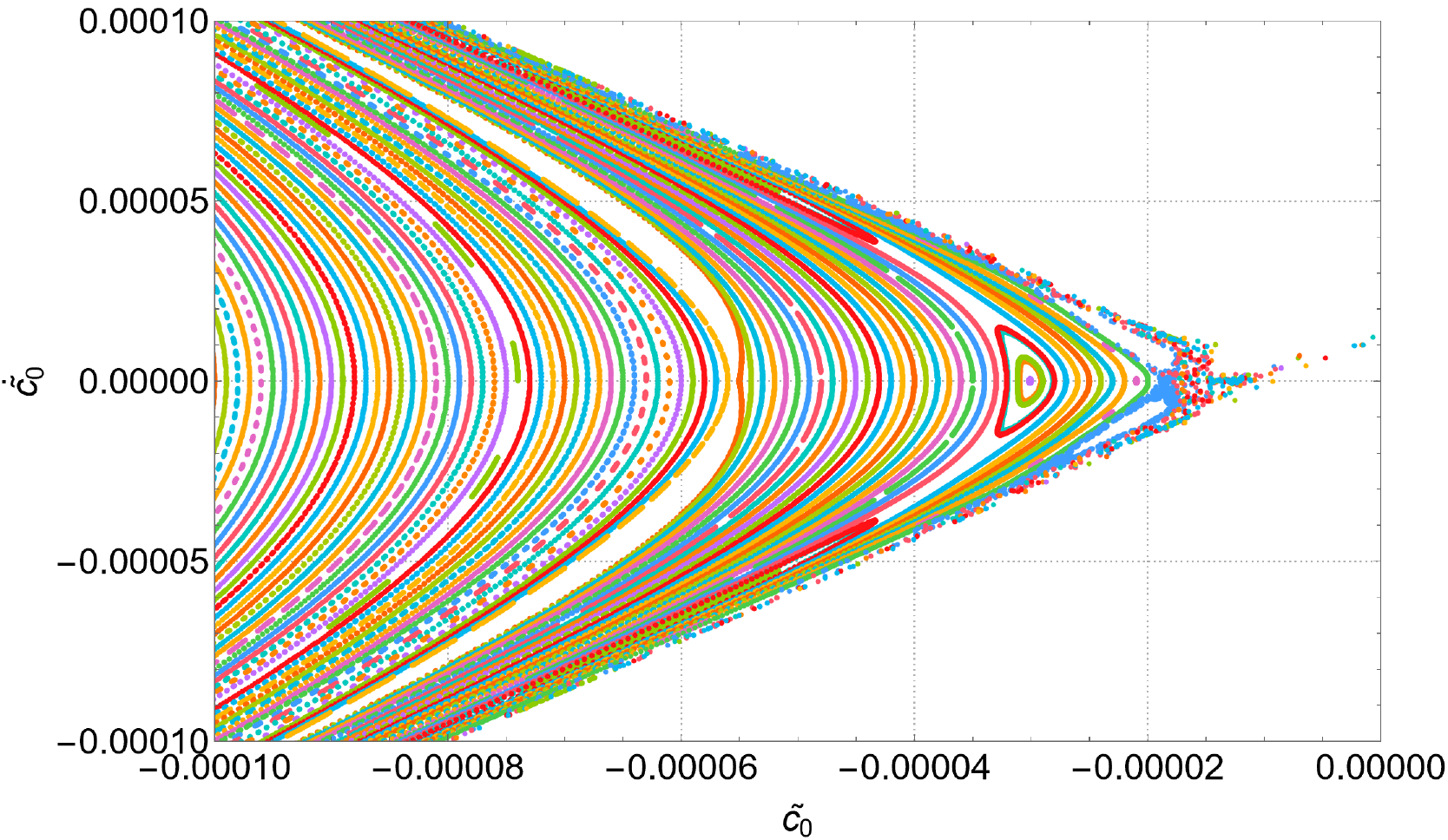} \\
		\textbf{$\mu=0.6$} & \includegraphics[scale=0.22,valign=c]{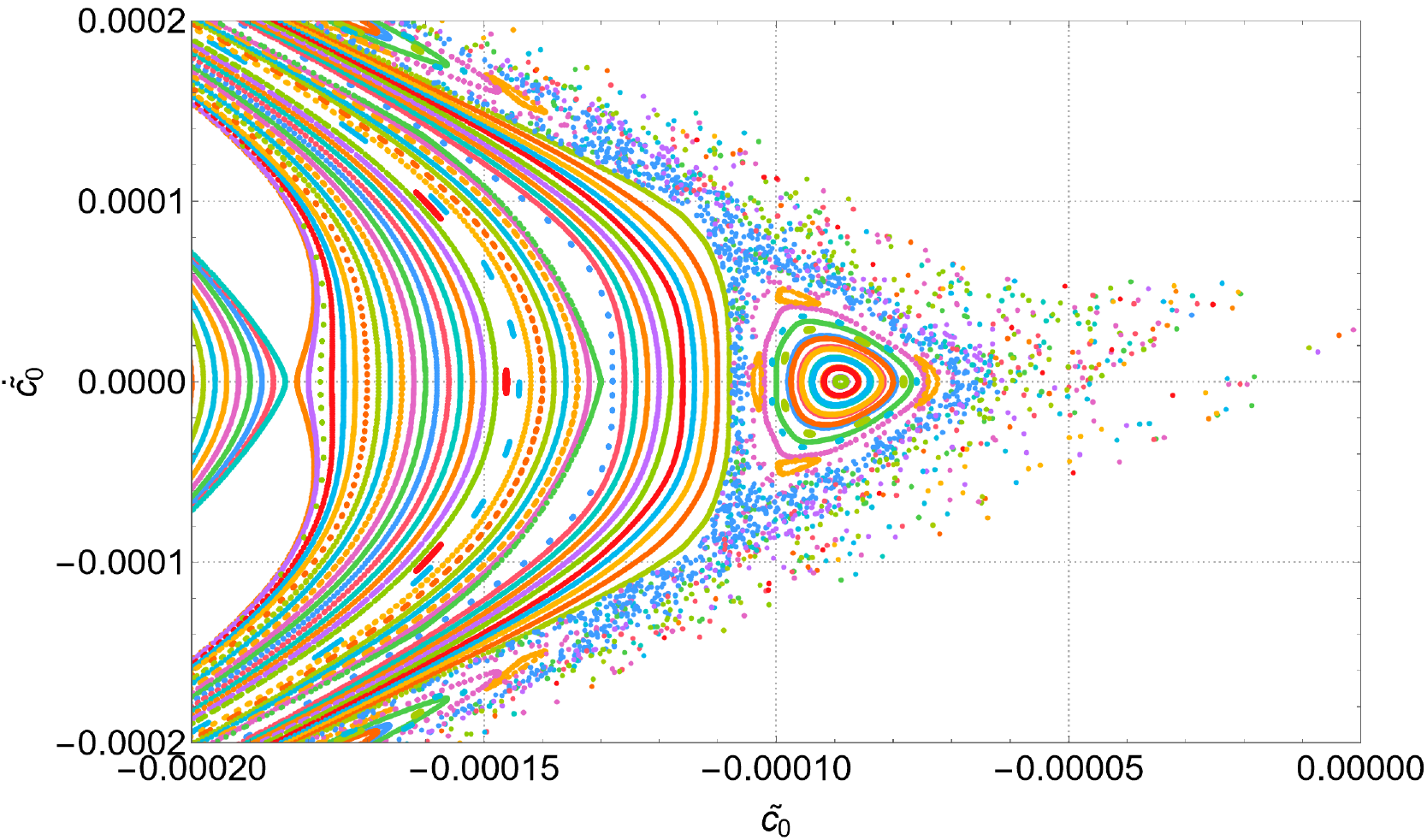} & \includegraphics[scale=0.22,valign=c]{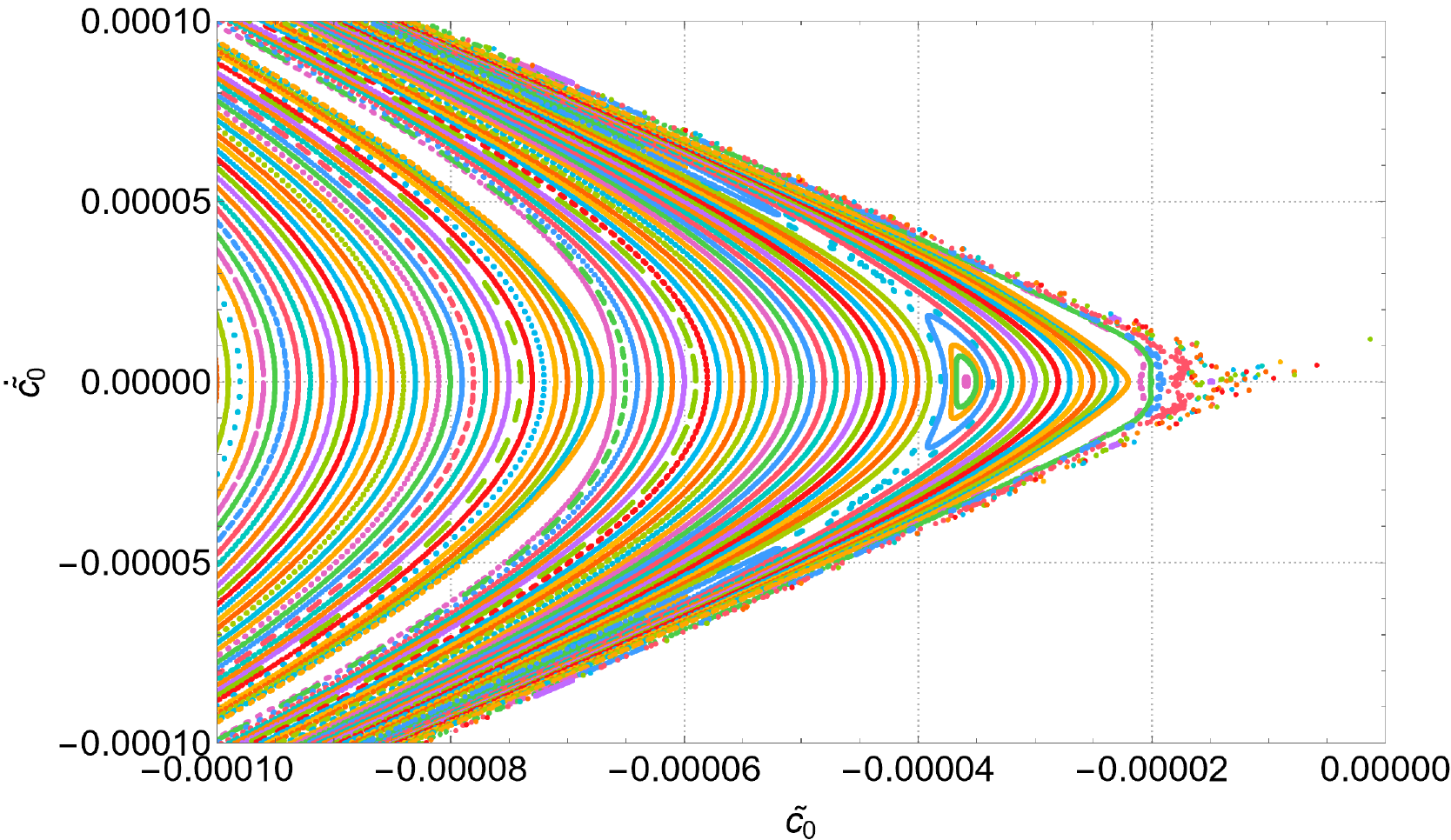} \\
        \textbf{$\mu=0.9$} & \includegraphics[scale=0.22,valign=c]{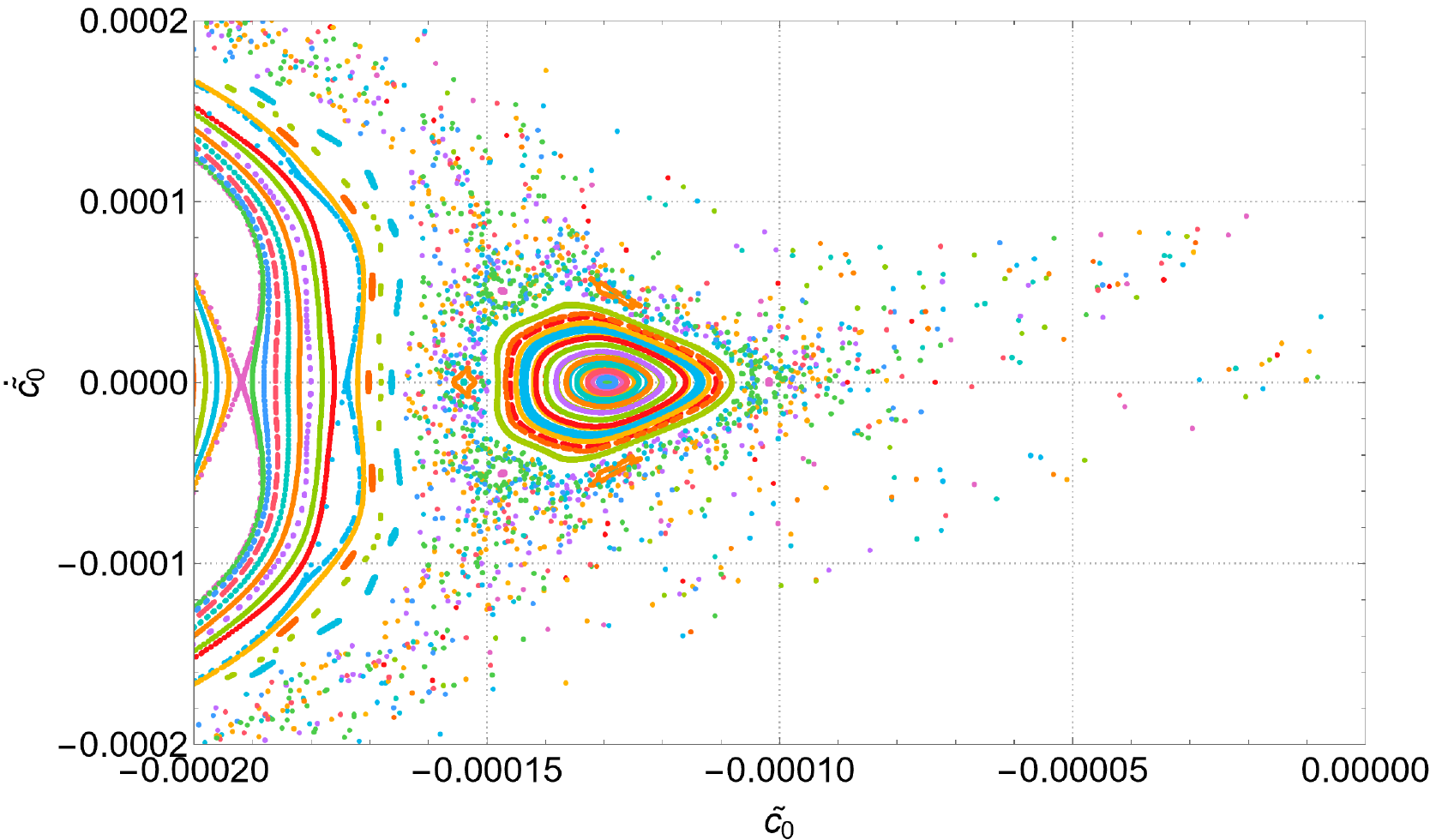} & \includegraphics[scale=0.22,valign=c]{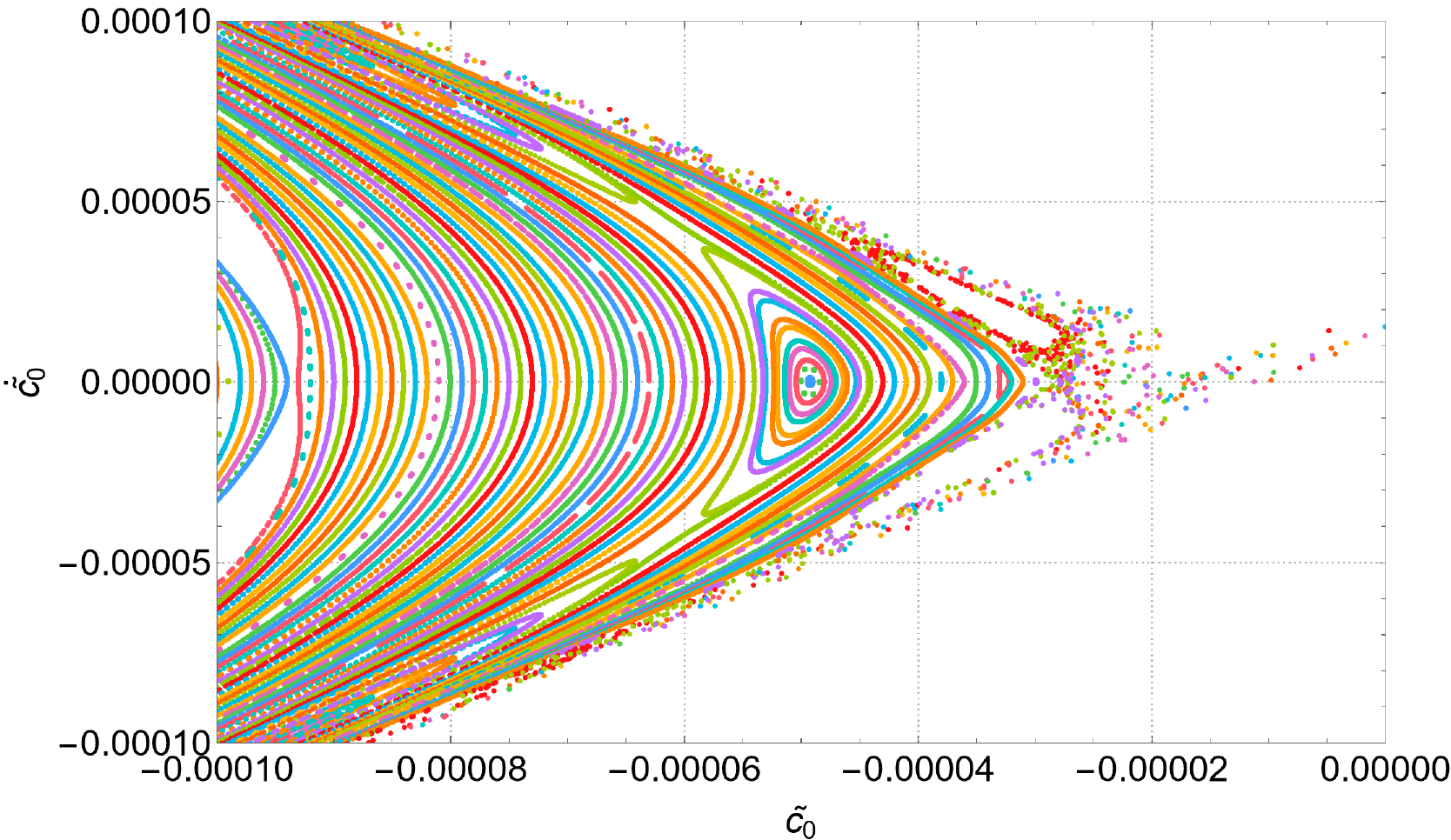} \\
        \textbf{$\mu=1.2$} & \includegraphics[scale=0.22,valign=c]{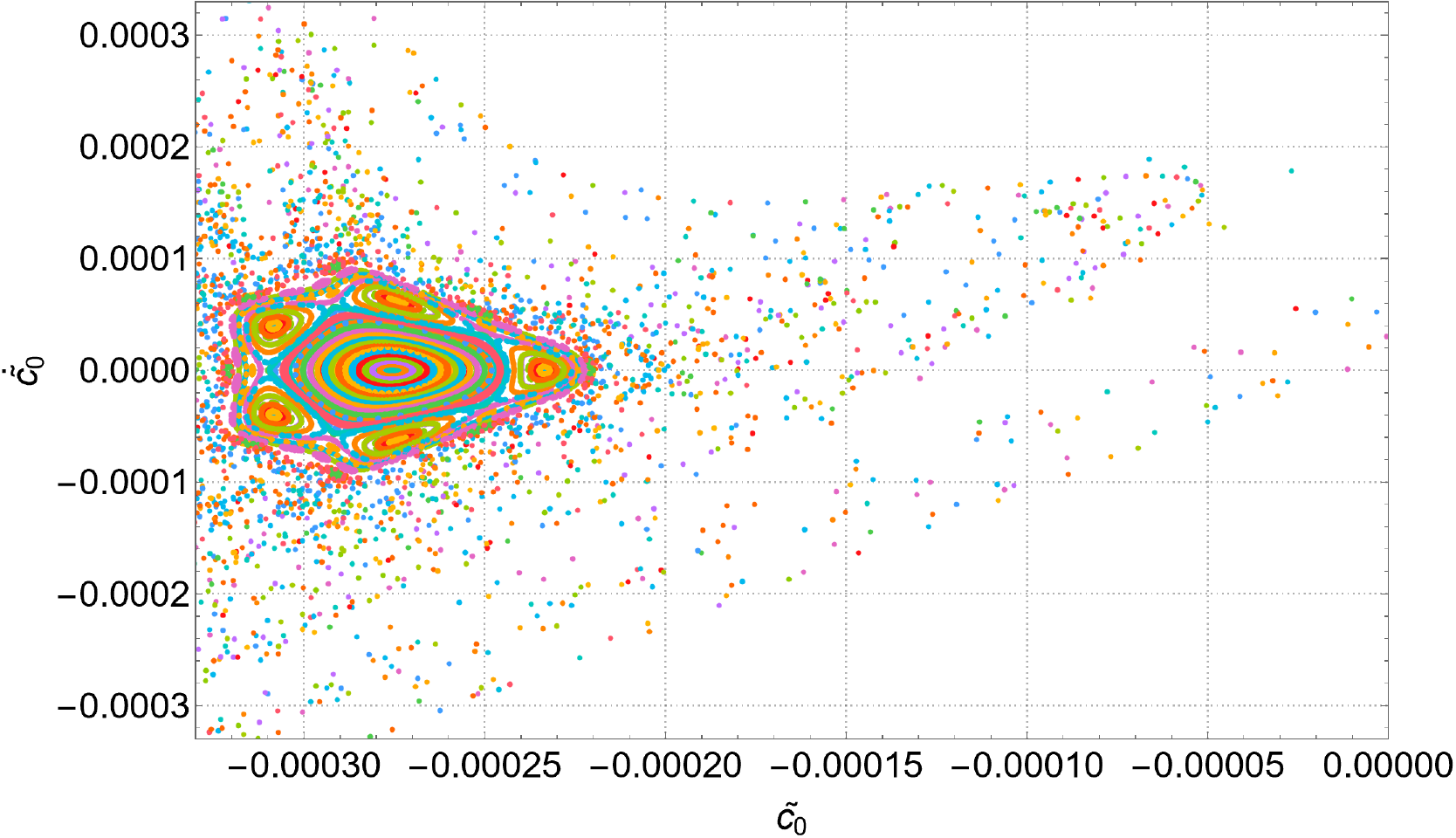} & \includegraphics[scale=0.22,valign=c]{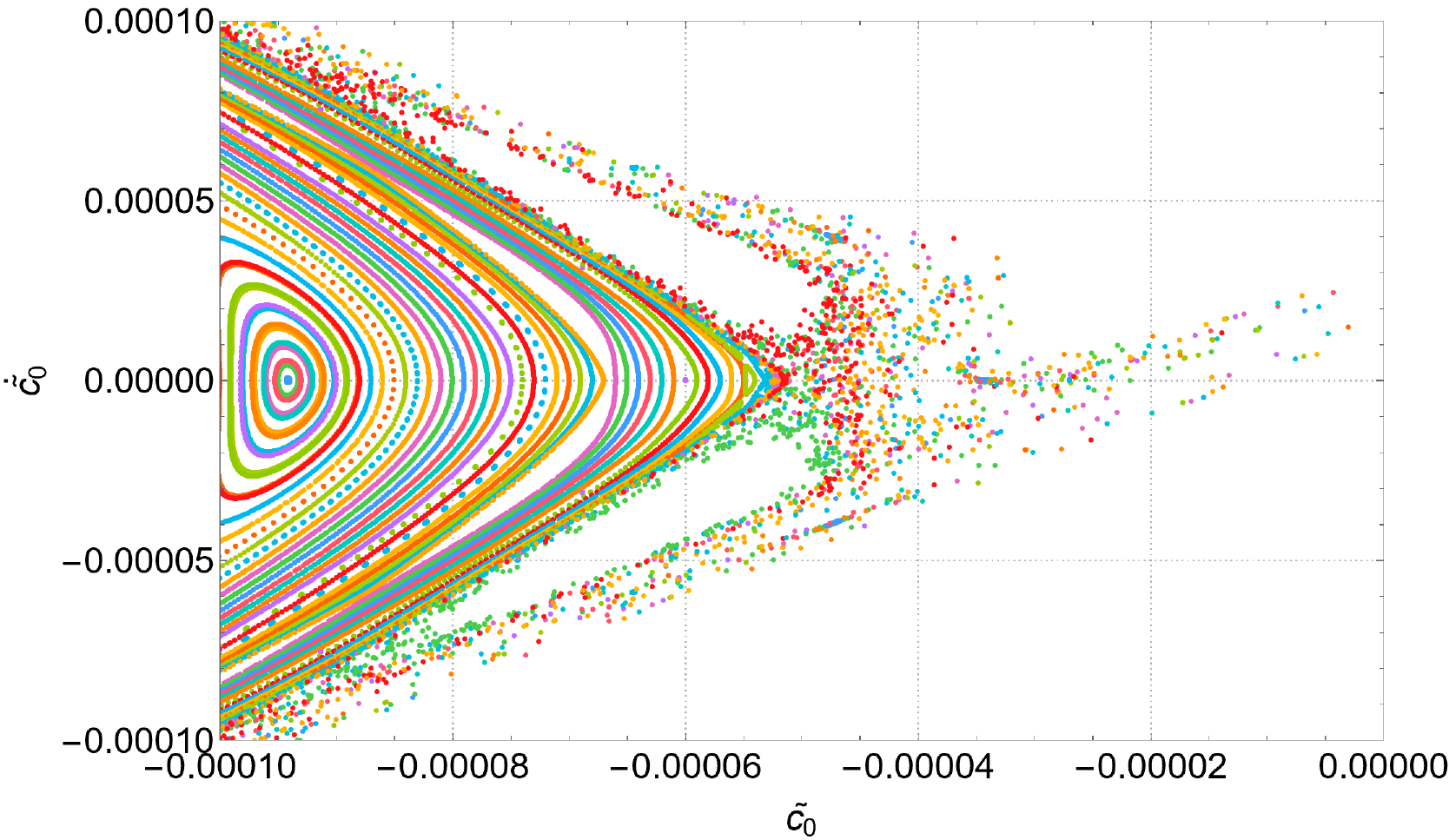} \\
	\end{tabular}
	\caption{The Poincar\'{e} sections for different values of $\mu$. The left panel is for $x_1$ orientation (parallel), while the right panel is for $x_3$ orientation (perpendicular). The section is identified by $\tilde{c}_{1}(t)=0$ and $\dot{\tilde{c}}_{1}(t)\ge0$ with $E=10^{-5}$. The string length is fixed at $L = 0.75$, and the magnetic field is set to $B = 0.2$. All values in GeV units.
\label{fig:EinsteinPoincare}}
\end{figure}

Next, we study the Poincar\'{e} section in the Einstein frame for bound orbits with the section identified by $\tilde{c}_{1}(t)=0$ and $\dot{\tilde{c}}_{1}(t)\ge0$ in the phase space. We utilize our equation of motion and trace trajectories in the phase space using time as the parameter. Our numerical results for the Poincar\'{e} sections near the origin for different values of the chemical potential are shown in Fig.~\ref{fig:EinsteinPoincare} for both parallel and perpendicular configurations of the string. Here we have used a fixed string length $L=0.75$, magnetic field $B=0.2$, and energy $E=10^{-5}$ with $t\le 15,000$. In the figure, different colors denote orbits with different initial conditions.

From Fig.~\ref{fig:EinsteinPoincare}, we can observe some interesting behaviour of the Poincar\'{e} sections as the chemical potential increases. In particular, there is an increase in scatter points near the origin $\Tilde{c_0}$ for higher values of $\mu$, indicating strong dependence on the initial conditions. This is again true for both string orientations, suggesting that turning on the chemical potential increases the chaotic behaviour. This result should be contrasted with the string frame result, where an increased chemical potential resulted in a decreased chaos for both string orientations. Similarly, though not explicitly presented here for brevity, we find that for higher values of the magnetic field, the number of scatter points amplify in the system for the parallel configuration, whereas the points order themselves into regular paths for the perpendicular configuration. This suggests that the effect of switching on the magnetic field is to aggravate the chaotic behaviour in the parallel configuration, whereas it reduces the chaotic behaviour in the perpendicular configuration. Therefore, unlike in the string frame case, the magnetic field introduces a manifest anisotropy in the chaotic behaviour of the string in the Einstein frame. This observation also differentiates the role of chemical potential from that of the magnetic field. Nonetheless, the increase/decrease of chaos with the magnetic field correlates with the tip of the string moving towards/away from the horizon for parallel/perpendicular configuration. It again confirms that the horizon is the source of the chaos. The latter result can be further advocated by the lack of chaos in the stable string dynamics, which is relatively far away from the horizon. 

\subsection{Lyapunov exponents}\label{EinsteinLyapunov}
\begin{figure}[htbp!]
	\centering
	\begin{tabular}{c c c}
		\textbf{$\mu$ value} & \textbf{Parallel Configuration} & \textbf{Perpendicular Configuration} \\
		\textbf{$\mu=0.0$} & \includegraphics[scale=0.16,valign=c]{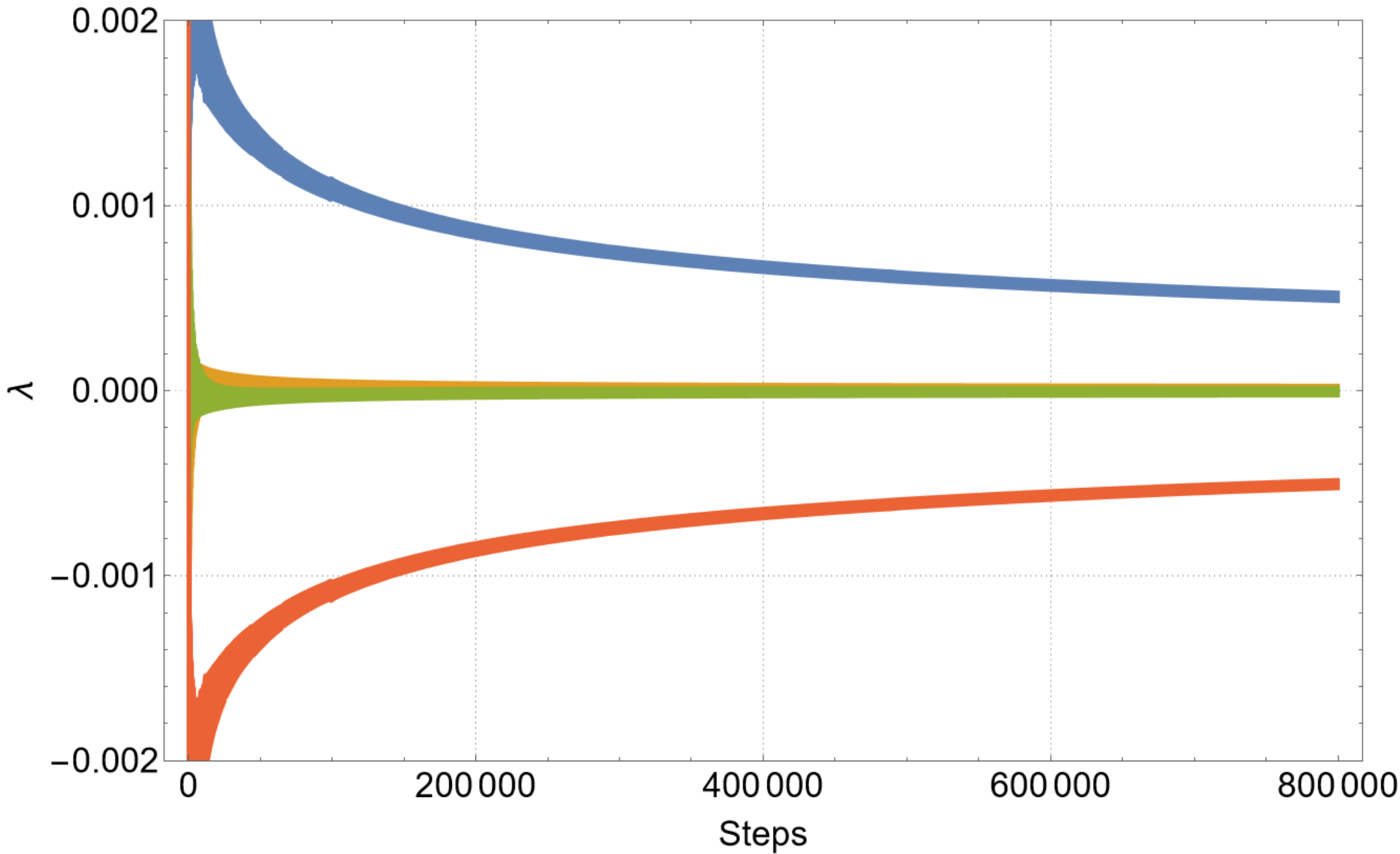} & \includegraphics[scale=0.16,valign=c]{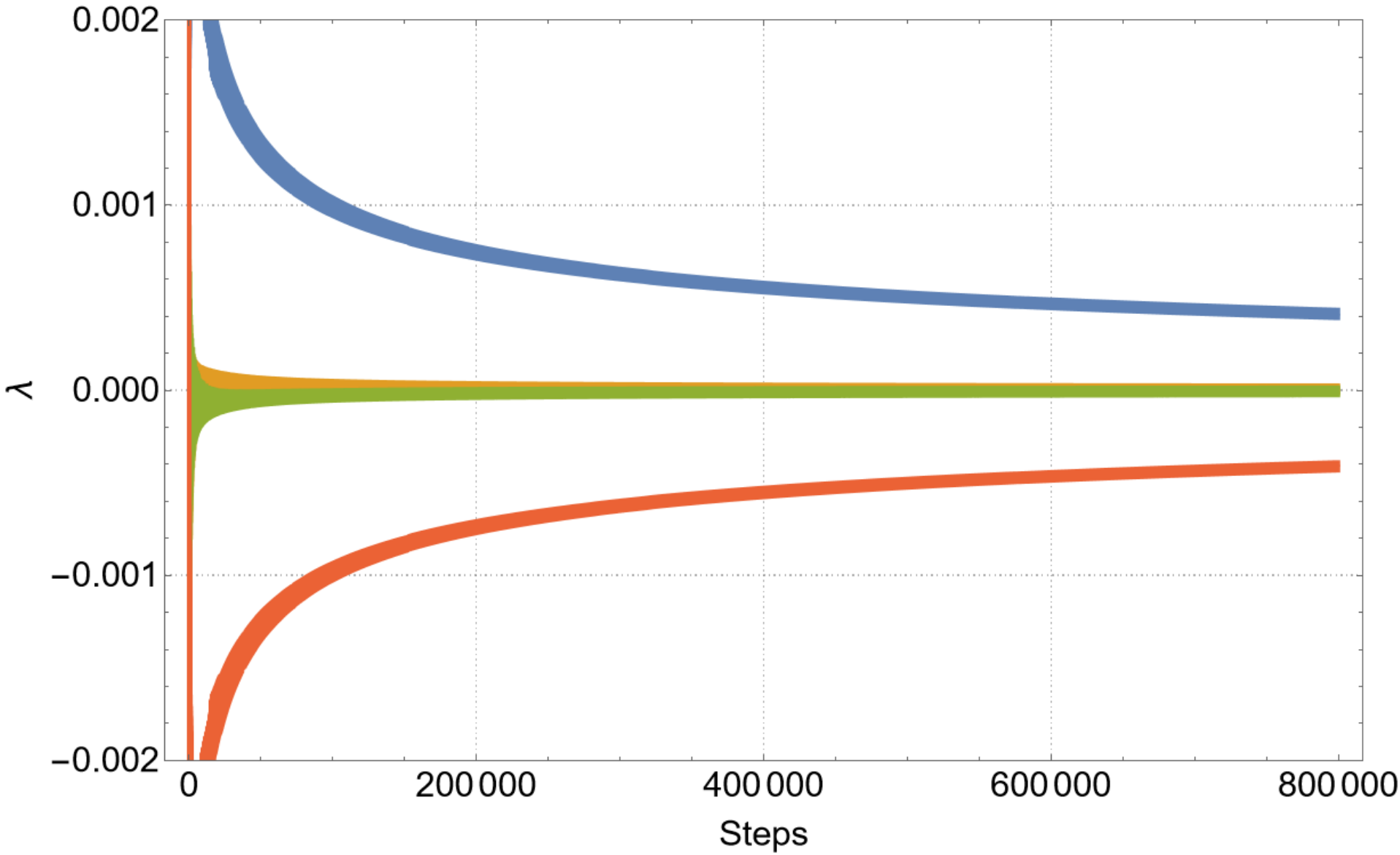} \\
		\textbf{$\mu=0.3$} & \includegraphics[scale=0.16,valign=c]{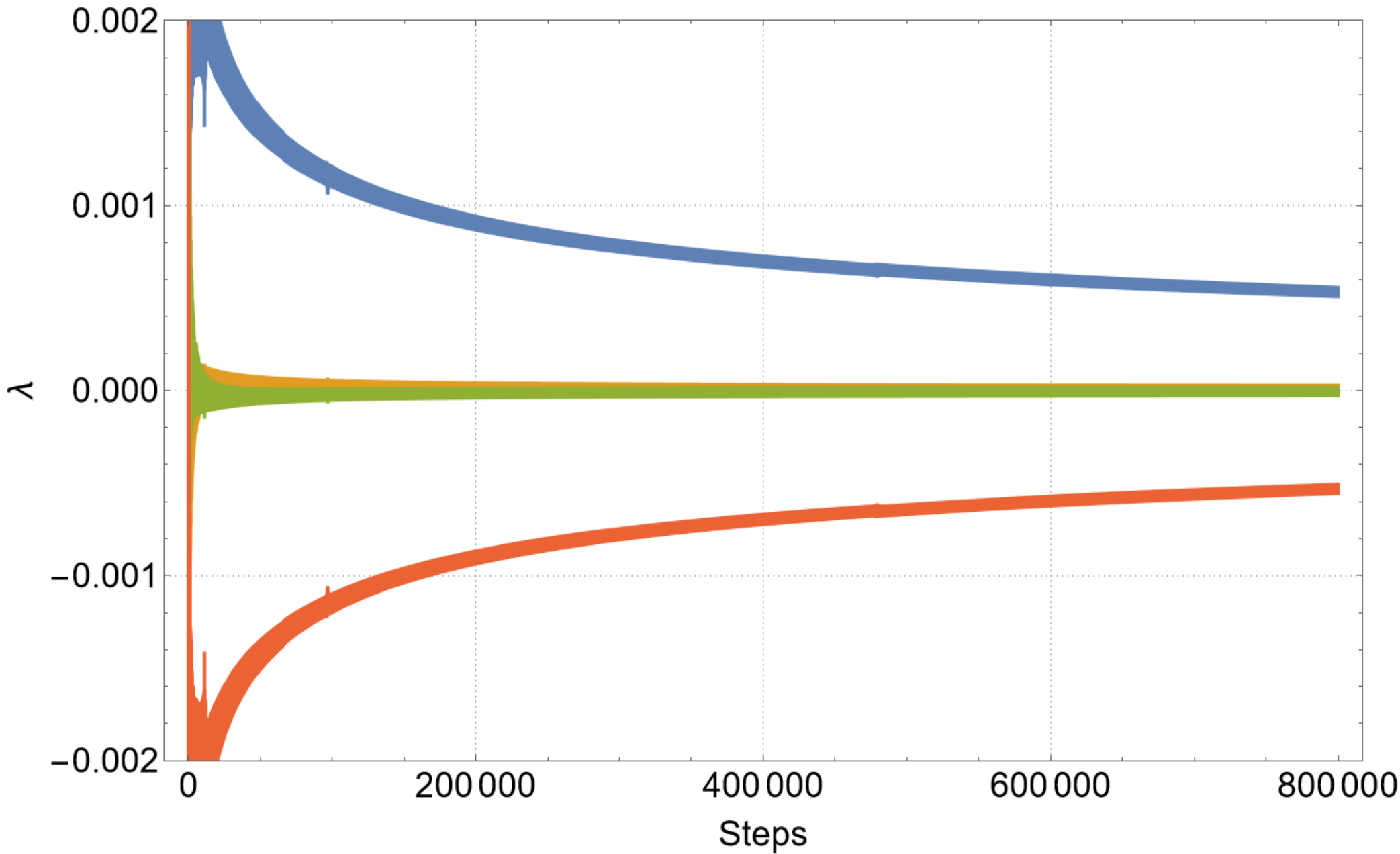} & \includegraphics[scale=0.16,valign=c]{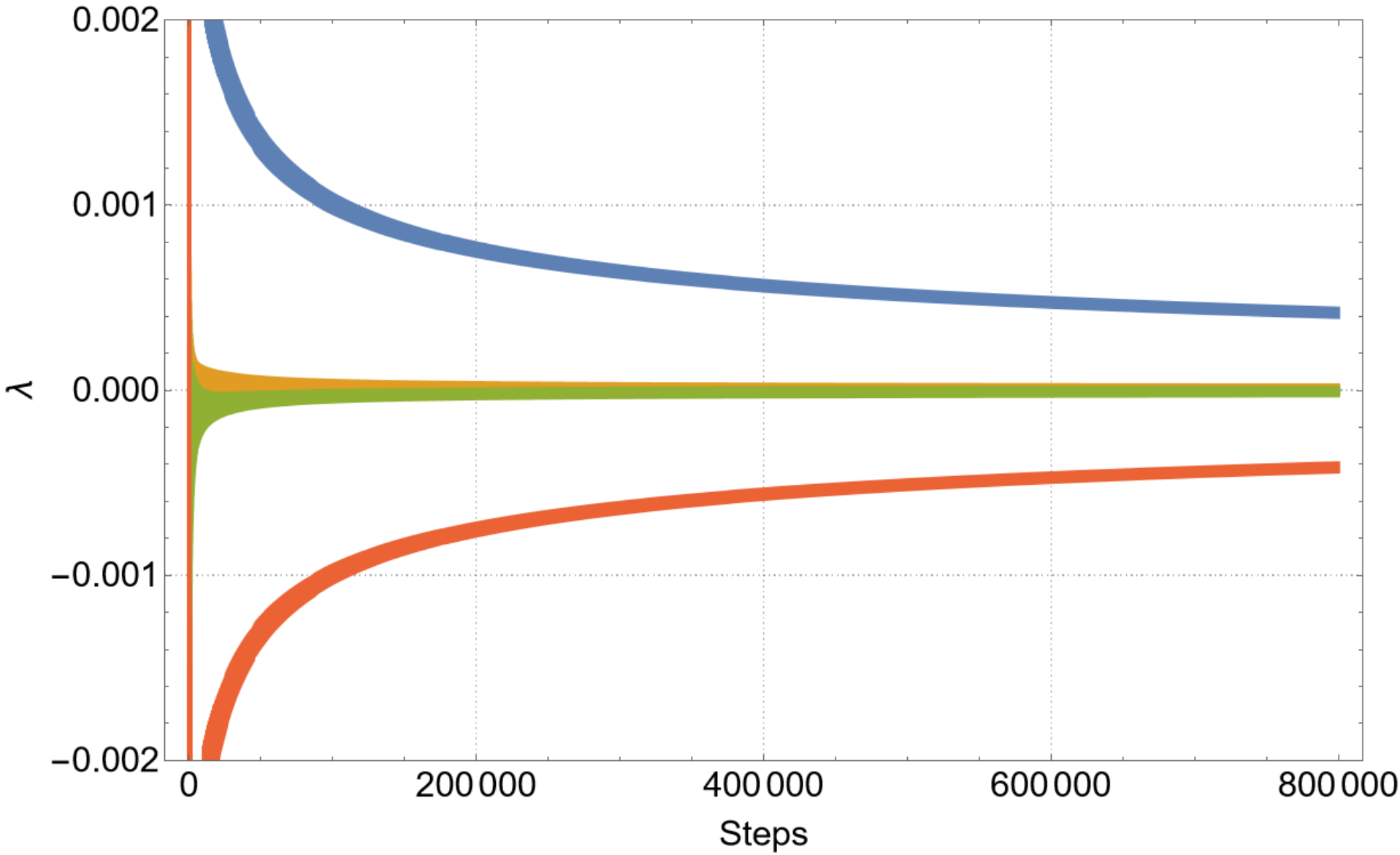} \\
		\textbf{$\mu=0.6$} & \includegraphics[scale=0.16,valign=c]{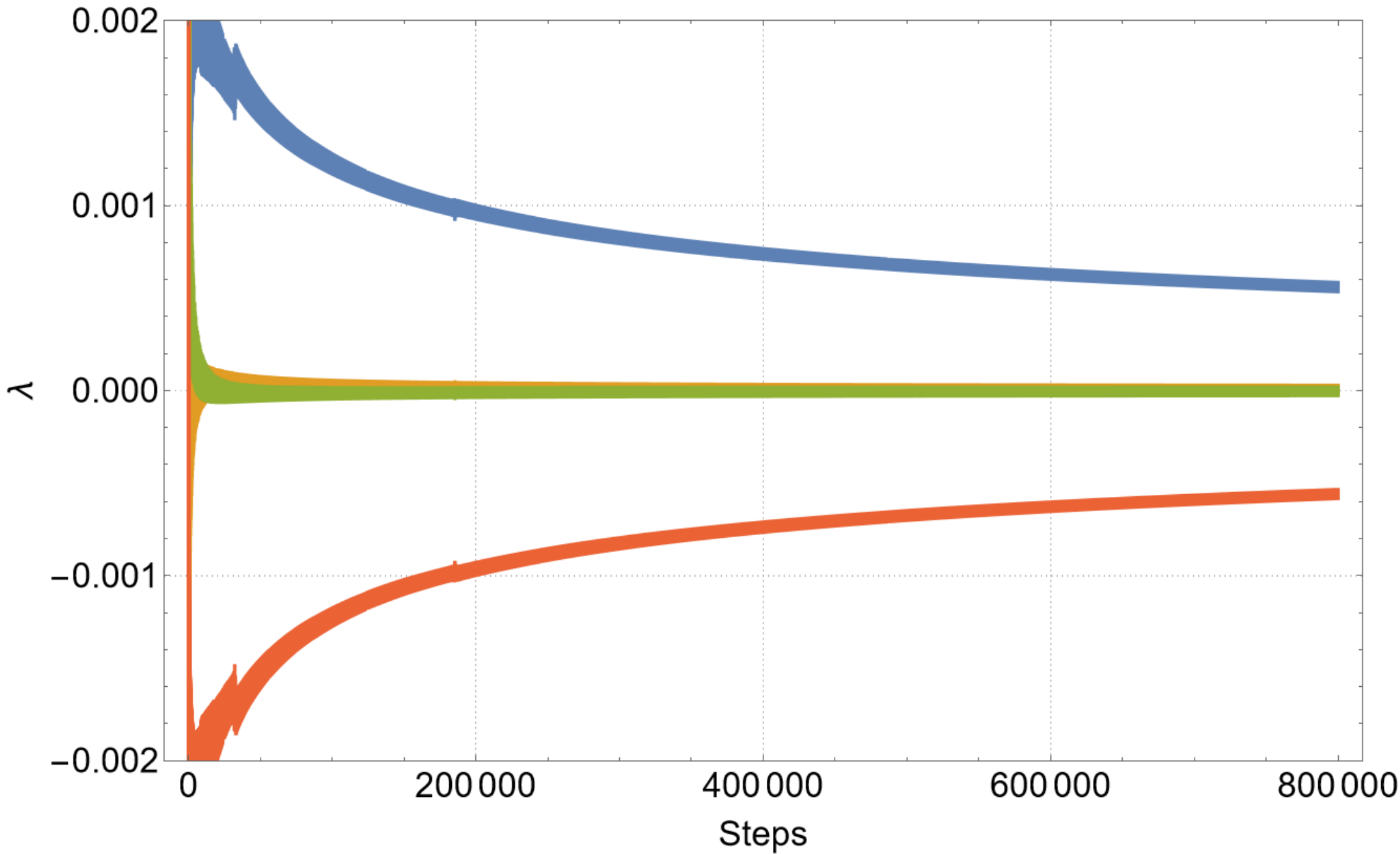} & \includegraphics[scale=0.16,valign=c]{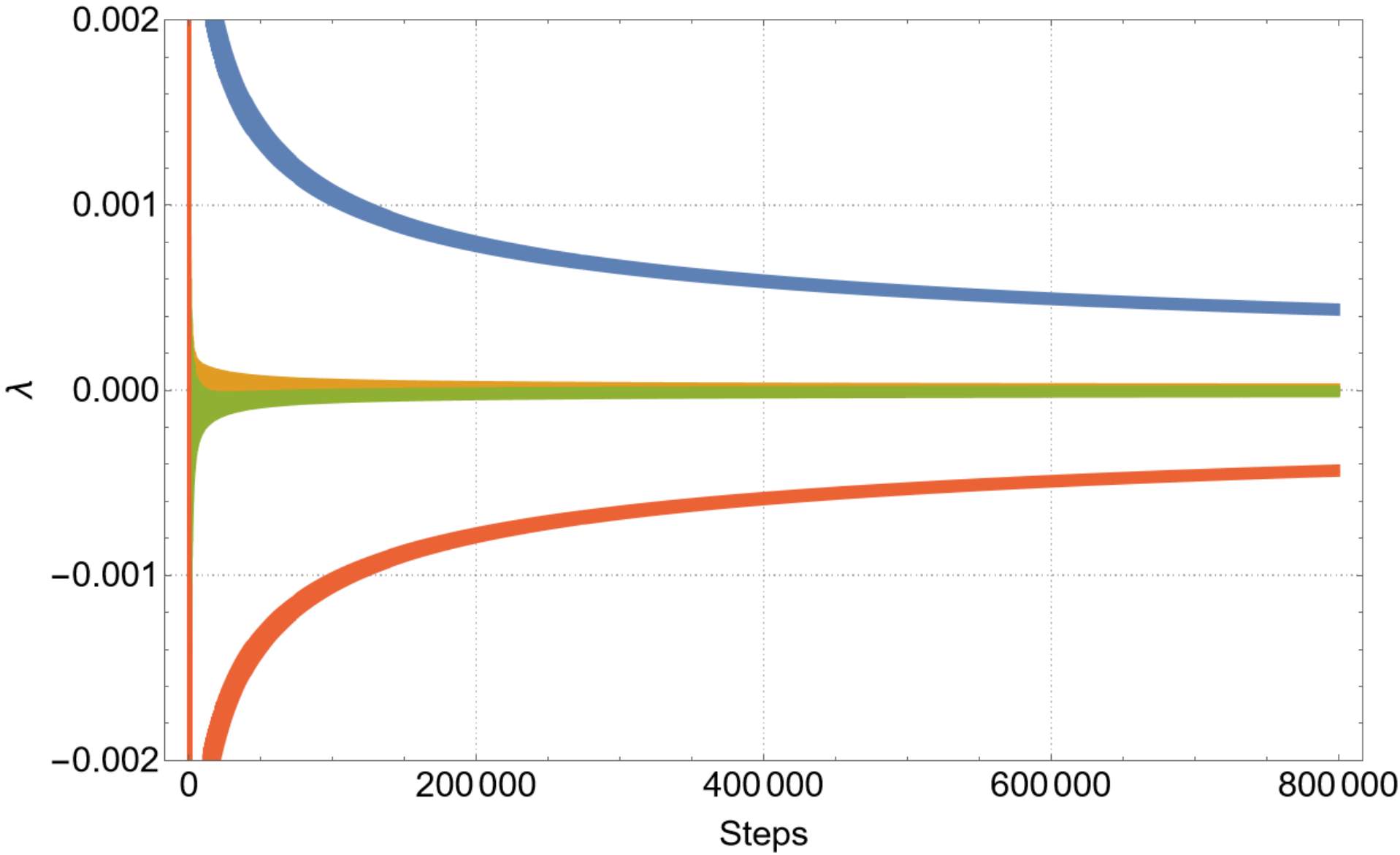} \\
        \textbf{$\mu=0.9$} & \includegraphics[scale=0.16,valign=c]{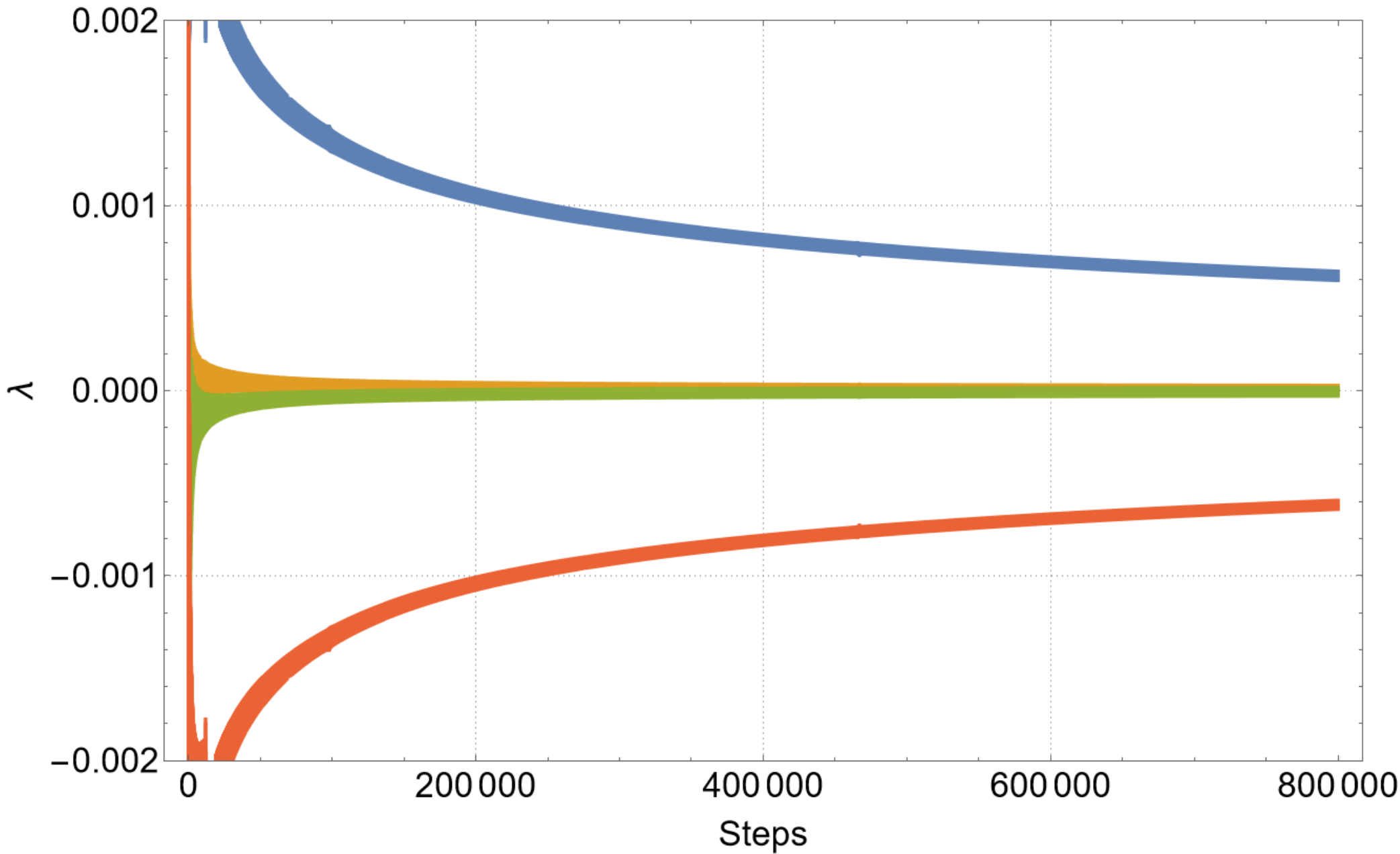} & \includegraphics[scale=0.16,valign=c]{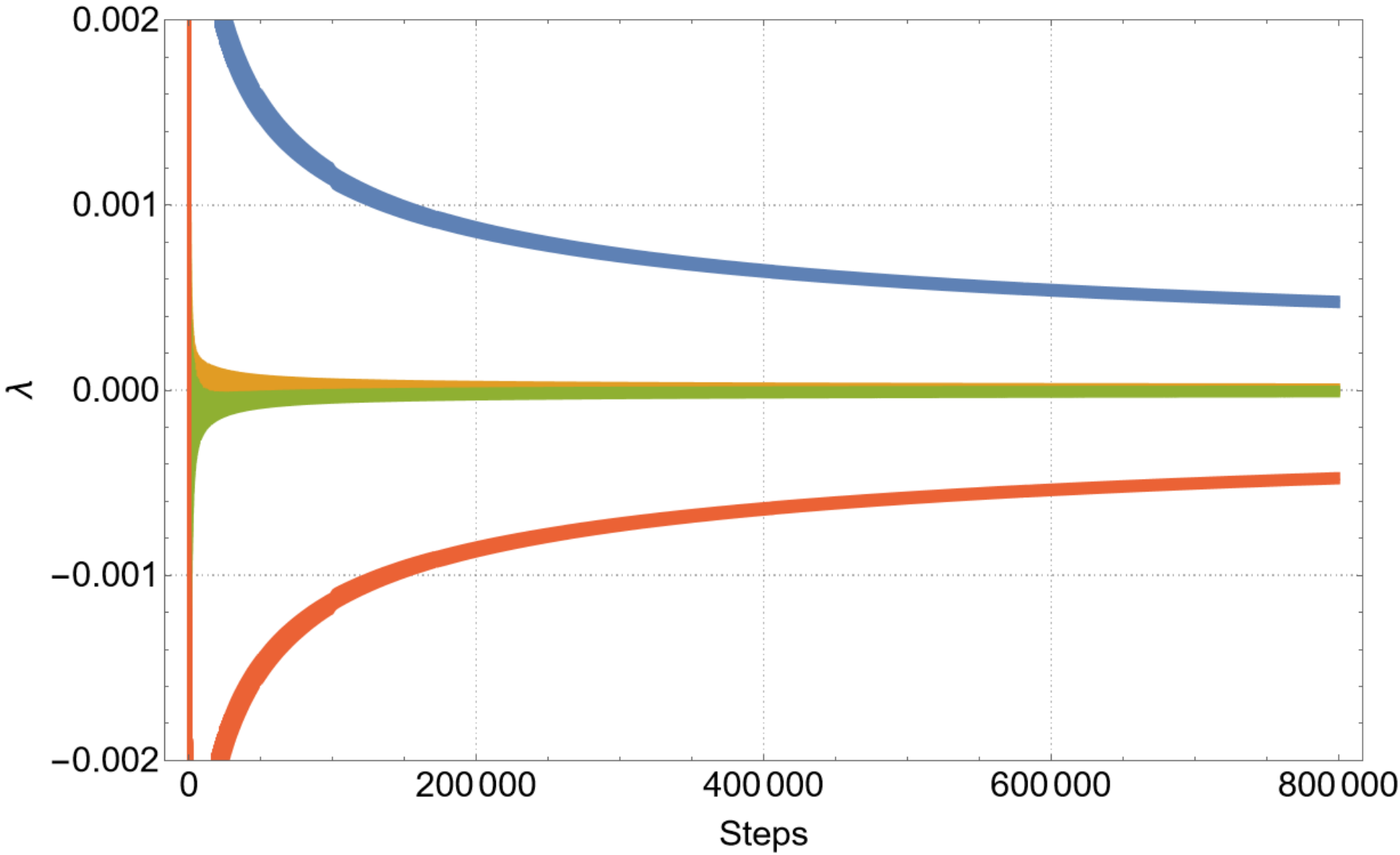} \\
        \textbf{$\mu=1.2$} & \includegraphics[scale=0.16,valign=c]{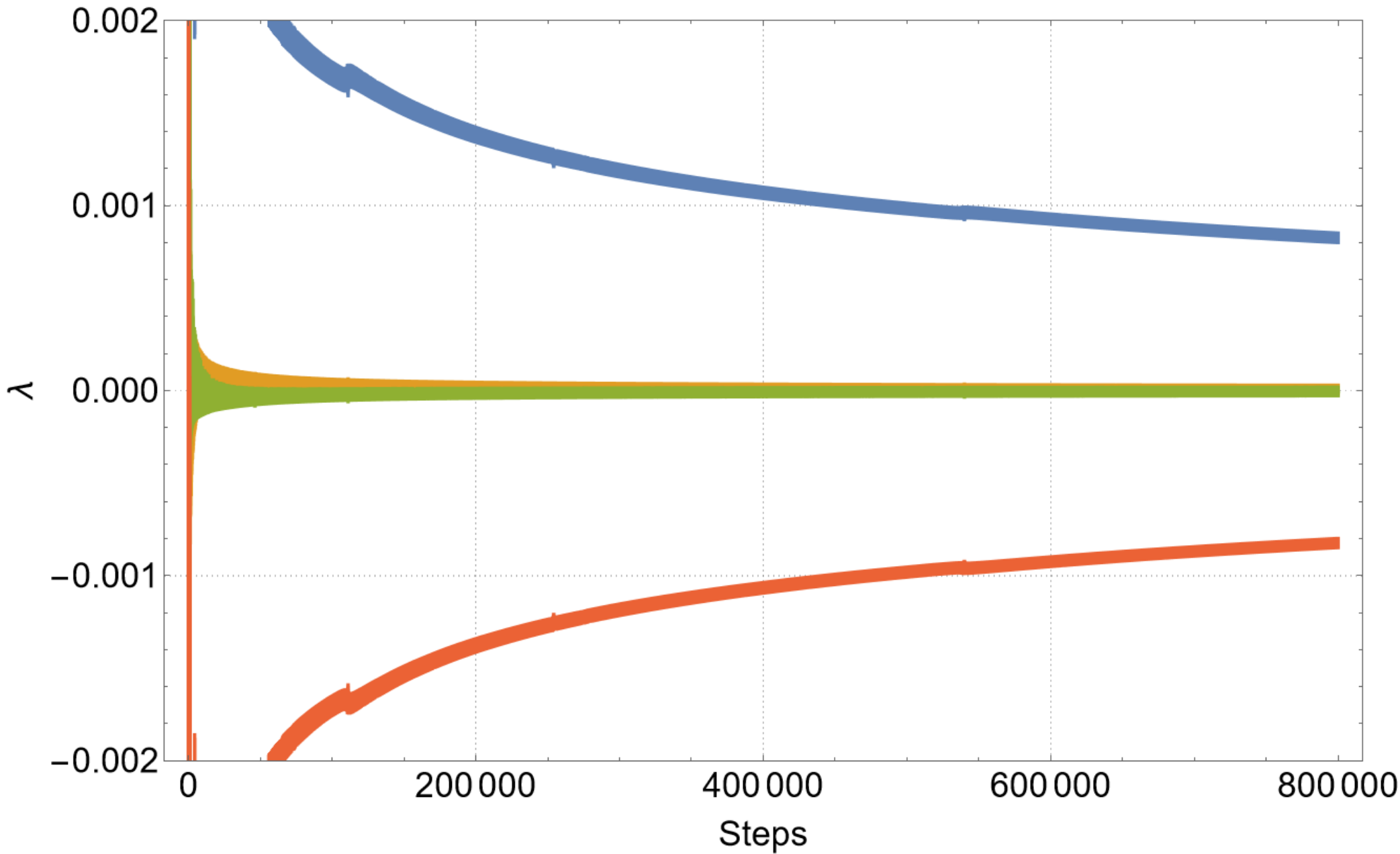} & \includegraphics[scale=0.16,valign=c]{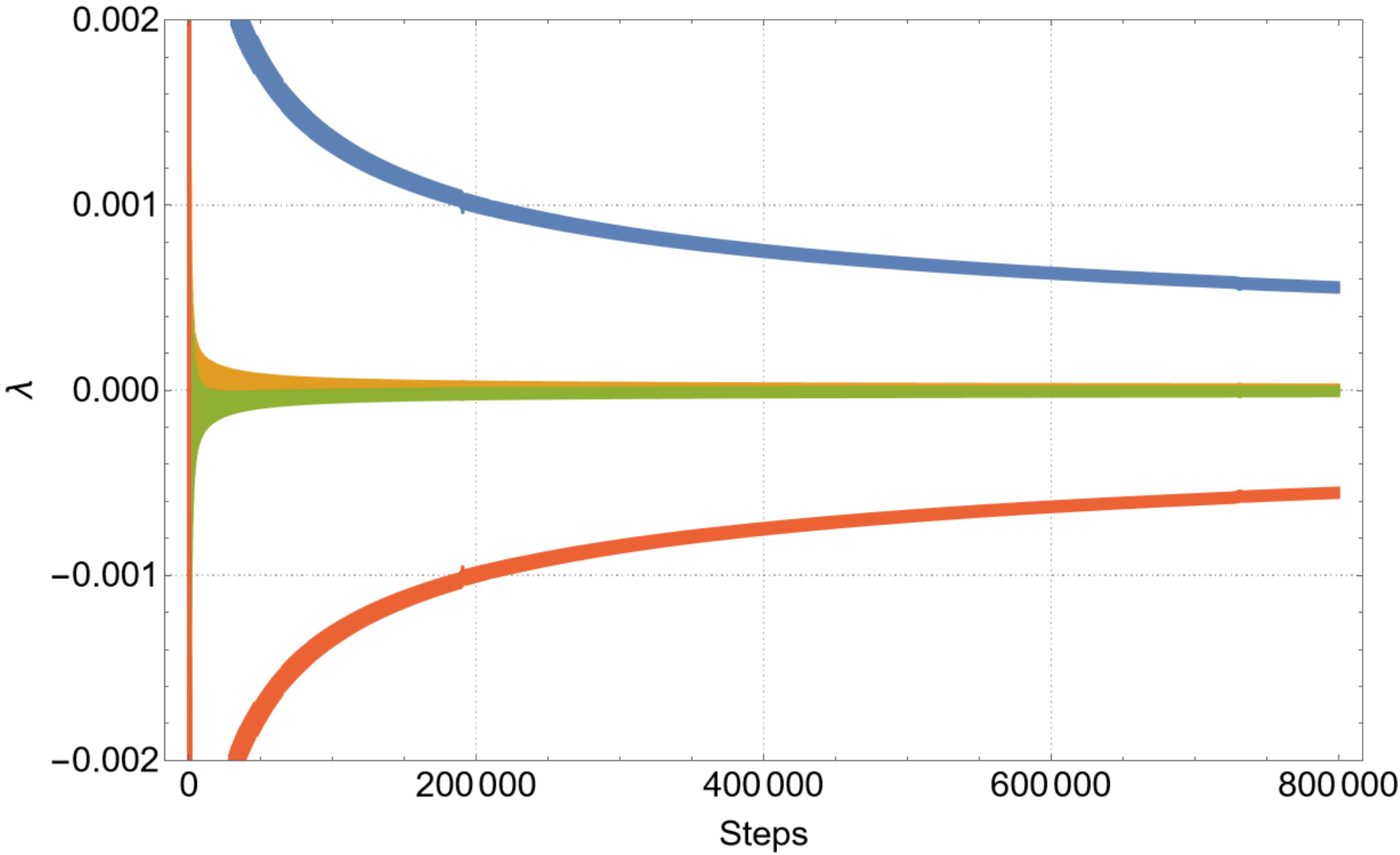} \\
        \textbf{Sum of $\lambda$} & \includegraphics[scale=0.16,valign=c]{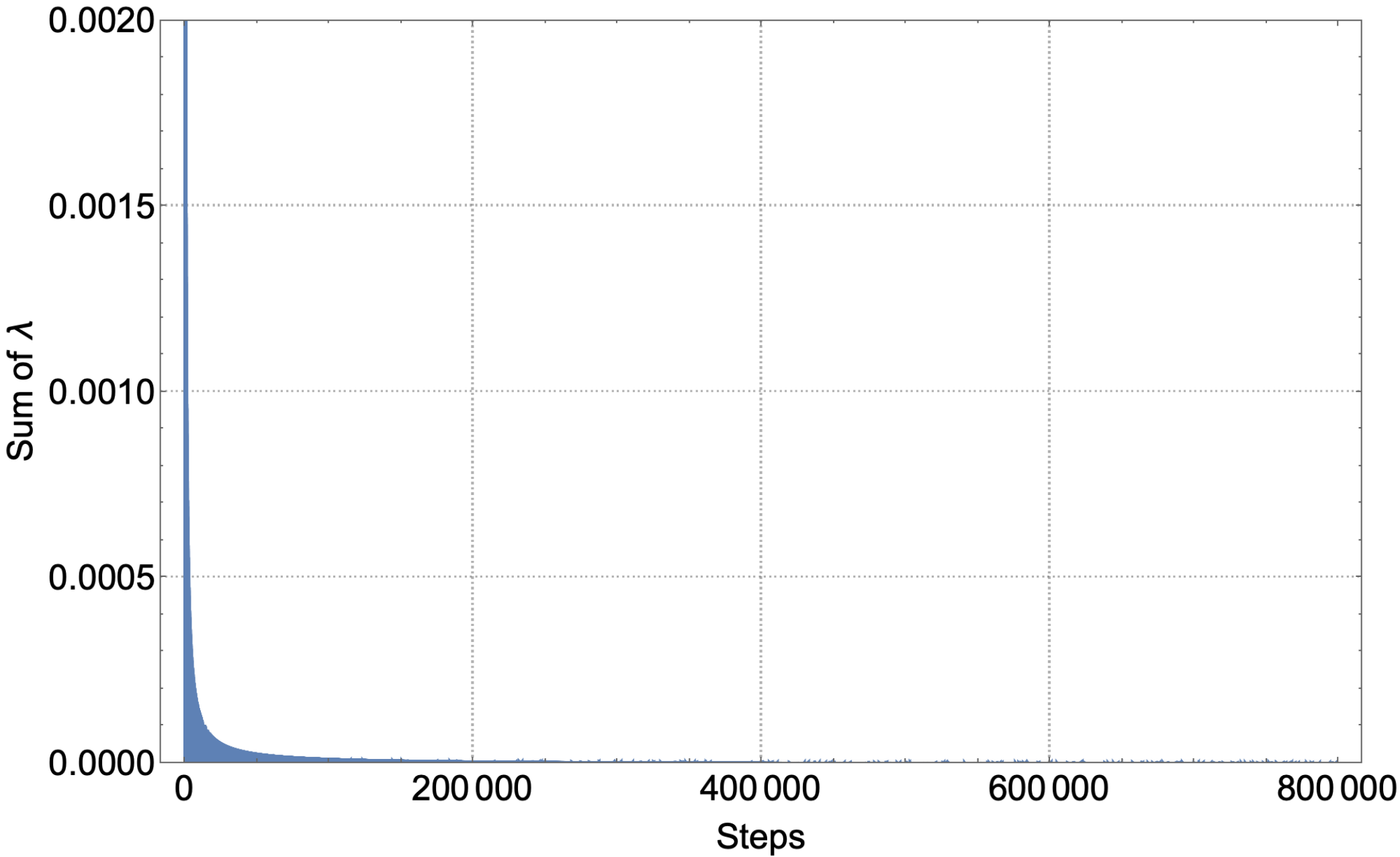} & \includegraphics[scale=0.16,valign=c]{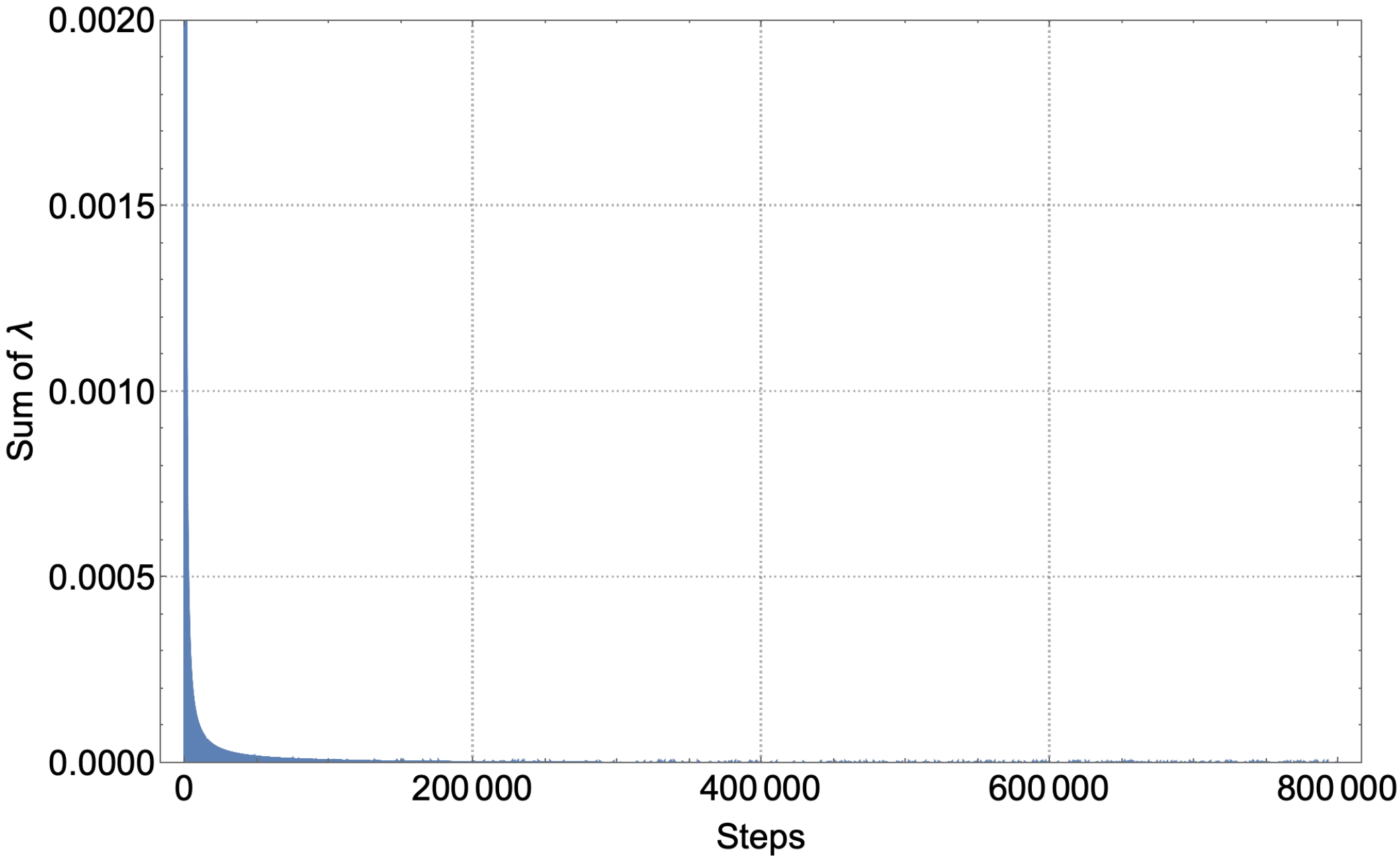}
        \end{tabular}
	\caption{Convergence plots of the four Lyapunov exponents for different values of $\mu$ for a string oriented parallel (left column) and perpendicular (right column) to the magnetic field. Here the initial conditions $(\Tilde{c}_{0}, \dot{\tilde{c}}_{0}, \Tilde{c}_{1})=(-0.002, 0, 0.001)$ are taken and fixed $L = 0.75$,  $E=10^{-5}$, and $B = 0.2$ are used. The sum of the Lyapunov exponents is displayed in the bottom row for $\mu=0.6$ and $B=0.2$. Similar converging behaviour is observed for all values of the magnetic field and chemical potential. All quantities are expressed in units of GeV.   \label{fig:EinsteinLyapunovB0pt4}}
\end{figure}

\begin{figure}[htbp!]
\centering
\includegraphics[width=0.65\textwidth]{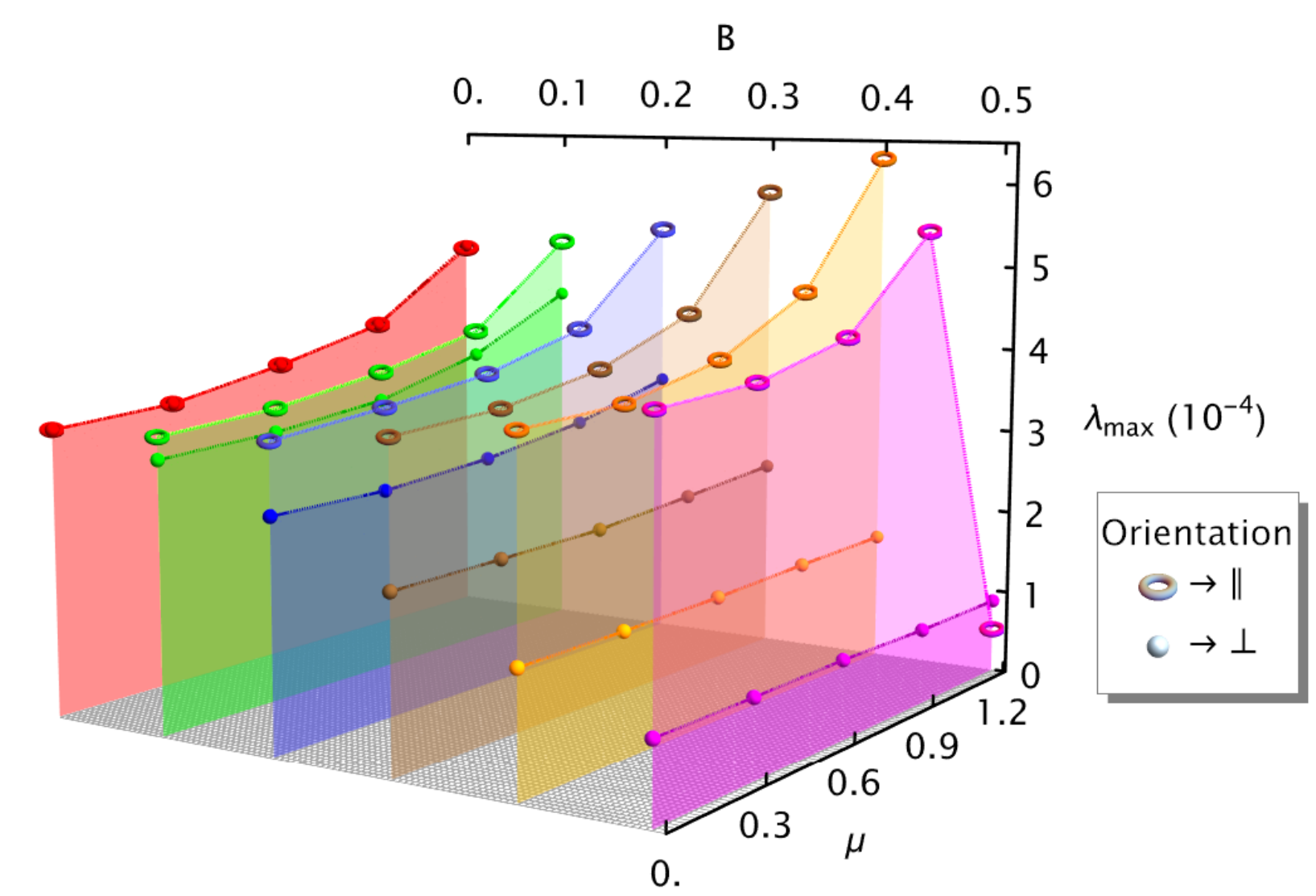}
\caption{Comparison of the maximum Lyapunov exponent $\lambda_{\text{max}}$ for the $x_1$ (parallel) and $x_3$ (perpendicular) magnetic field orientations for various values of $B$ and $\mu$ near the unstable saddle point. Data points are colour-coded as follows: red for $B = 0.0$, green for $B = 0.1$, blue for $B = 0.2$, brown for $B = 0.3$, orange for $B = 0.4$, and magenta for $B = 0.5$. Here, we use the initial conditions $(\Tilde{c}_{0}, \dot{\tilde{c}}_{0}, \Tilde{c}_{1}) = (-0.002, 0, 0.001)$.  The string length is fixed at $L = 0.75$ and $E = 10^{-5}$.}
\label{fig:EinsteinLmax}
\end{figure}

To conclude, we also analyse the Lyapunov exponents in the four-dimension phase space $(\Tilde{c}_{0},\Tilde{c}_{1})$. Here, we focus on the system with fixed length $L = 0.75$ and energy $E = 10^{-5}$ for the string's $x_1$ and $x_3$ orientations. As for the string frame calculations, we have taken $8\times10^5$ time steps with step size $0.001$. The numerical results for the convergence of the Lyapunov exponents and their sum are illustrated in Fig.~\ref{fig:EinsteinLyapunovB0pt4}. The maximum Lyapunov exponent is recorded in Fig.~\ref{fig:EinsteinLmax} for $x_1$ and $x_3$ string configurations. As in the string frame, our convergence is similar to that of the damped oscillator. The total sum of the Lyapunov exponents converges to zero, suggesting that the system remains conservative. This result is true for all magnetic field and chemical potential values, irrespective of the relative orientation of the string and magnetic field.

From the Poincar\'{e} plots (Fig.~\ref{fig:EinsteinPoincare}), we already observe that the scatter points were concentrated near the origin of the phase space. This observation suggests the dynamics of the string becomes more chaotic as its tip approaches the horizon. This result is now also supported by the largest Lyapunov exponents $\lambda_{max}$ for both parallel and perpendicular string orientation, as shown in figure~\ref{fig:EinsteinLmax}.

\begin{figure}[htbp!]
	\centering
	\subfigure[Poincar\'{e} Section for $B=0.5$ and $\mu=1.2$]{\label{fig:EinsteinPoincare_x1_B0pt5mu1pt2-a}	\includegraphics[width=0.44\linewidth]{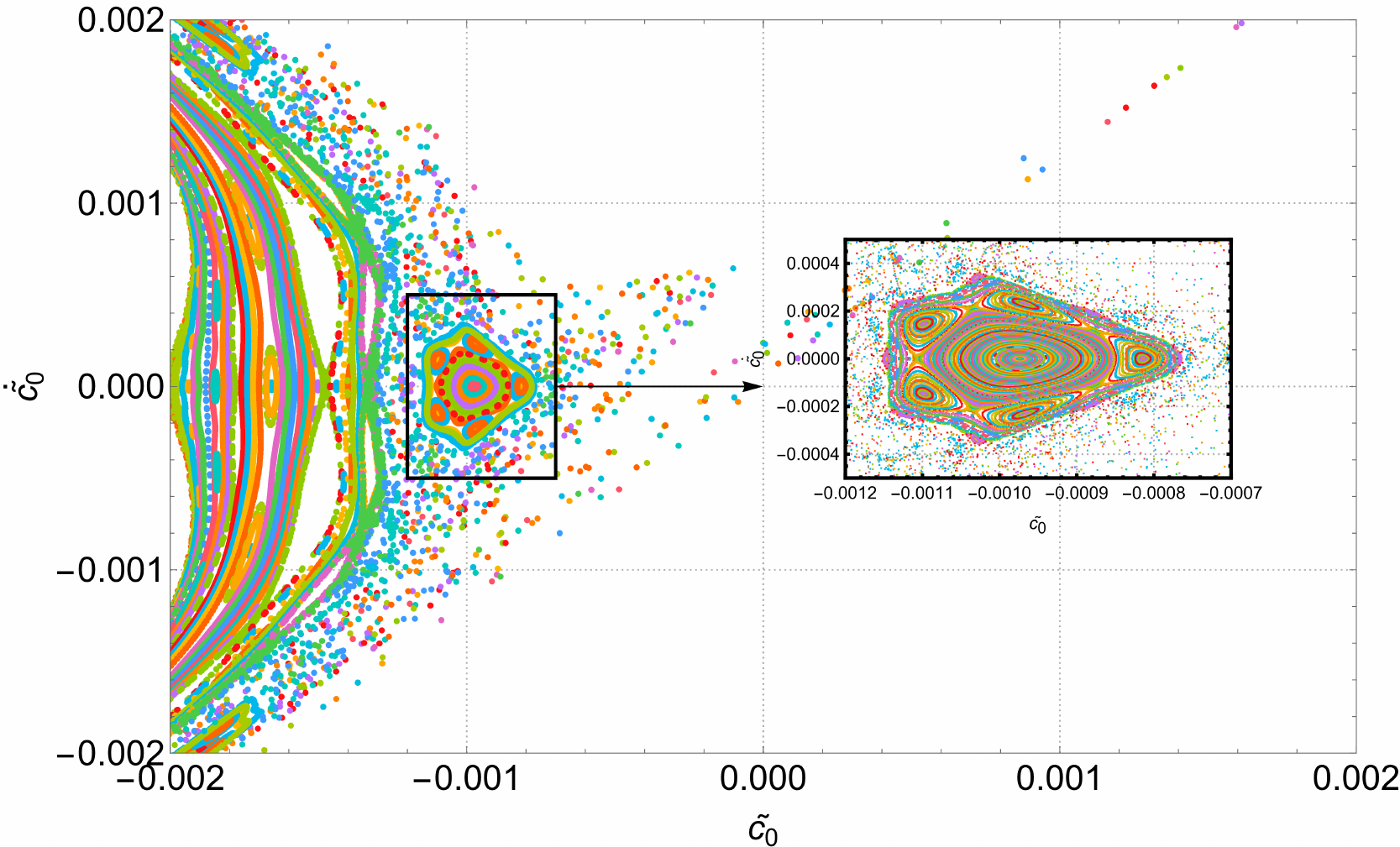}}
    \subfigure[Close up of (a)]{\label{fig:EinsteinPoincare_x1_B0pt5mu1pt2-b}
	\includegraphics[width=0.44\linewidth]{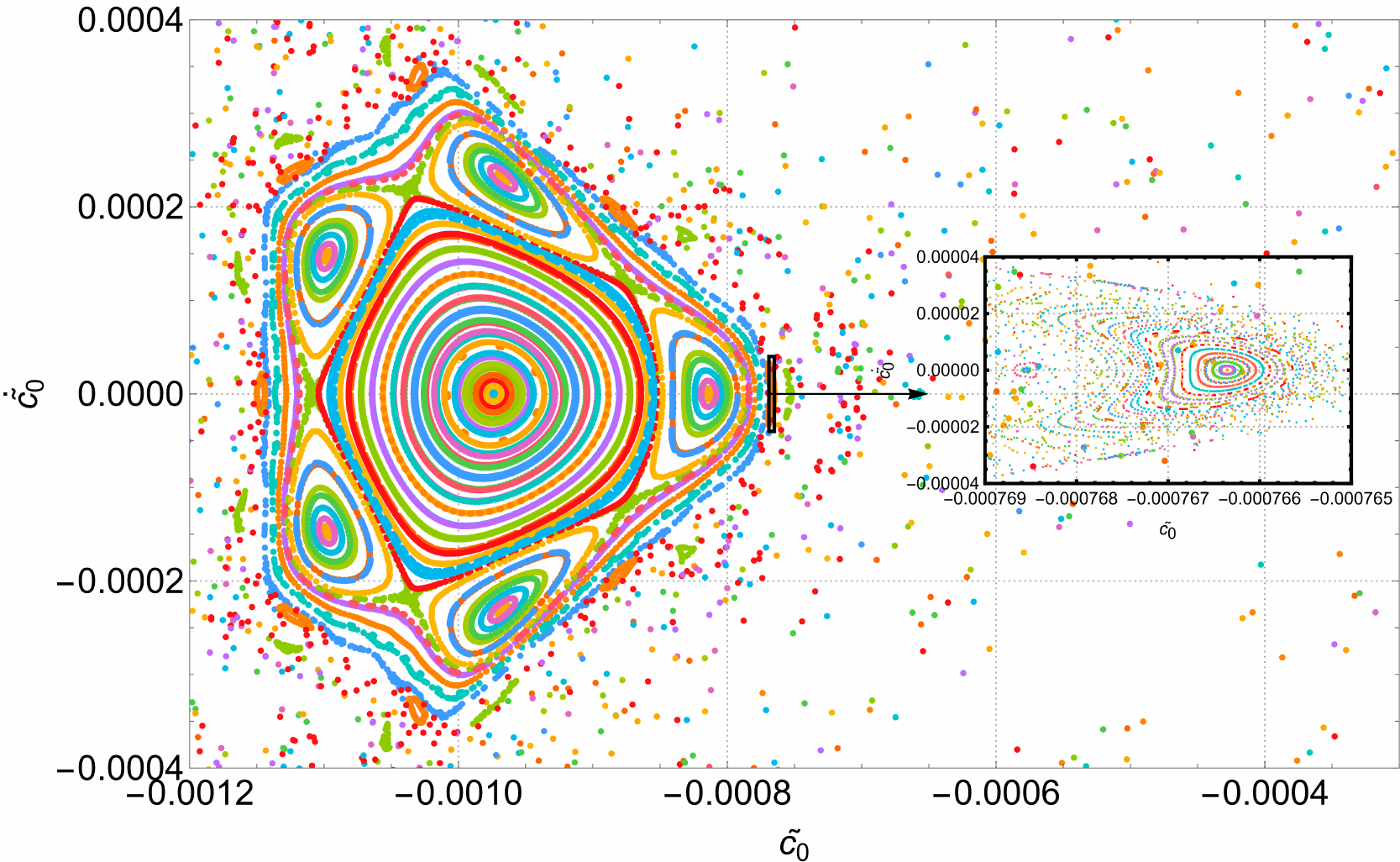}}
	\caption{\label{fig:EinsteinPoincare_x1_B0pt5mu1pt2}Poincar\'{e} section, identified by $\tilde{c}_{1}(t)=0$ and $\dot{\tilde{c}}_{1}(t)\ge0$ with $E = 10^{-5}$, for $\mu = 1.2$ with the magnetic field fixed at $B = 0.5$. The section reveals a fractal island near the origin, with the fractal nature further examined in an enlarged view. The string length is fixed at $L = 0.75$.}
\end{figure}
The maximal Lyapunov exponent is extracted for all magnetic field and chemical potential values; the data is then plotted for the $x_1$ and $x_3$ string orientation cases with the string length fixed. In contrast to the string frame case, we see that $\lambda_{max}$ increases as $\mu$ increases for both $x_1$ and $x_3$ string configurations when $B$ is small, i.e., the overall system exhibits an enhanced chaotic nature as $\mu$ increases, irrespective of its orientation relative to the magnetic field. If we look more closely, there is actually a difference in the behaviour of $\lambda_{max}$ between the $x_1$ and $x_3$ string orientations for large $B$ values. Indeed, in the case of the $x_1$ orientation, increasing the $\mu$ value significantly increases $\lambda_{max}$ while it remains almost constant for the $x_3$ orientation. This highlights one important aspect of the interplay of the magnetic field with the chemical potential. Since increasing the $B$ value increases the chaos in the $x_1$ orientation and decreases it in the $x_3$ orientation, the effect of increasing the chemical potential also differs in both orientations.

From Fig.~\ref{fig:EinsteinLmax}, we also see a small departure from the overall behaviour of $\lambda_{max}$ in the $x_1$ configuration. In particular, there is a sudden drop in $\lambda_{max}$ for $B=0.5$ as $\mu$ goes from $0.9$ to $1.2$. This differs from the Lyapunov exponent's overall nature as we increase $\mu$, keeping the $B$ value fixed. The reason for such difference is what appears to be the presence of fractal islands in the vicinity of origin, which are like small non-chaotic pockets in a sea of chaos. This can be explicitly observed in the Poincar\'{e} section profile for $\mu=1.2$,  shown in Fig.~\ref{fig:EinsteinPoincare_x1_B0pt5mu1pt2}. The occurrence of these islands is related to KAM (Kolmogorov–Arnold–Moser) theory and transition to chaos~\cite{Gutzwiller1990}. The KAM theorem is one of the most important results in the study of dynamical systems. Oversimplifying and under certain assumptions, if a tiny Hamiltonian perturbation is applied to a system, some of the so-called tori---to which the phase space trajectories are confined, indicative of quasi-periodicity---only get slightly deformed, while most do get destroyed. Increasing the chemical potential for a given magnetic field increases the intensity of our perturbations, and most regular orbits are thus destroyed to form ``dust''. However, some tori exist that do not get destroyed but only get deformed to form small fractal islands surrounded by chaos. This intriguing result undoubtedly warrants a comprehensive and meticulously detailed investigation in its own right. In any case, in QCD-like gauge theories, interesting features are expected to appear with chemical potential in the range of $\mu \lesssim 1~\text{GeV}$. The above non-trivial trend of the Lyapunov exponent that appears for $\mu \gtrsim 1~\text{GeV}$ in our model thus lies outside the physically interesting range of the chemical potential. 

We further find that the choice of the string orientation significantly affects the value of the maximum Lyapunov exponent $\lambda_{max}$. While the chemical potential enhances the chaotic nature of the string, this effect is more significant in the parallel configuration. This suggests that the anisotropic nature of the system dynamics depends on its orientation with the magnetic field. We further find that $\lambda_{max}$ increases/decreases with the magnetic field for parallel/perpendicular orientation of the string. Our Lyapunov exponent analysis, therefore, further confirms our previous expectation that the magnetic field induces anisotropy in the string dynamics, i.e.,  the chaos in the string dynamics increases/decreases with the magnetic field in parallel/perpendicular directions. This observation also makes it clear that the effect of chemical potential is dissimilar to that of the magnetic field regarding the string dynamics.

\subsection{Saddle point analysis and the MSS bound}\label{einsteinsaddle}

\begin{figure}[htbp!]
	\centering
	\subfigure[Parallel configuration]{\label{fig:potentialeinsteinframex1}	\includegraphics[width=0.4\linewidth]{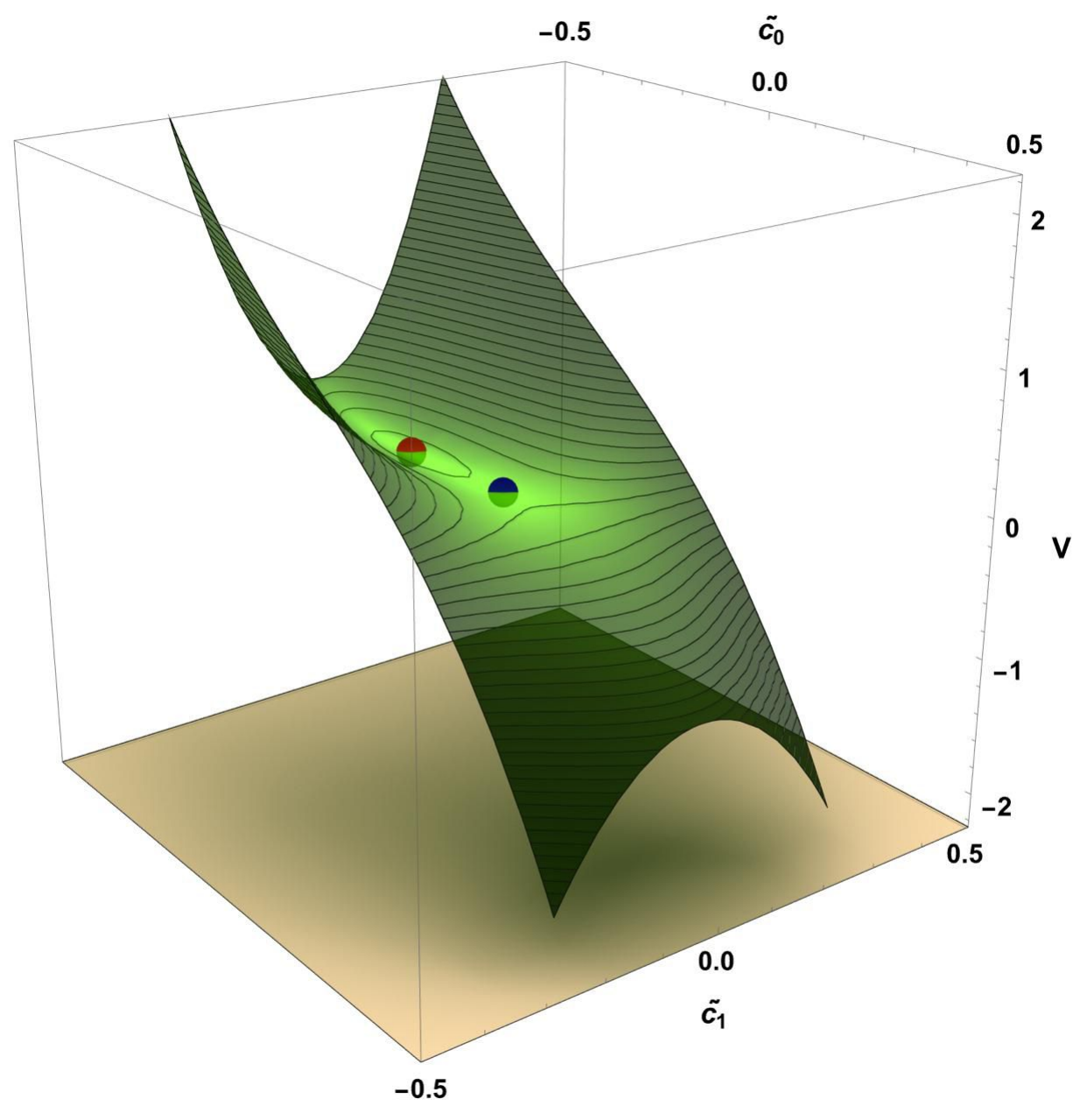}}
    \subfigure[Perpendicular configuration]{\label{fig:potentialeinsteinframex3}
	\includegraphics[width=0.4\linewidth]{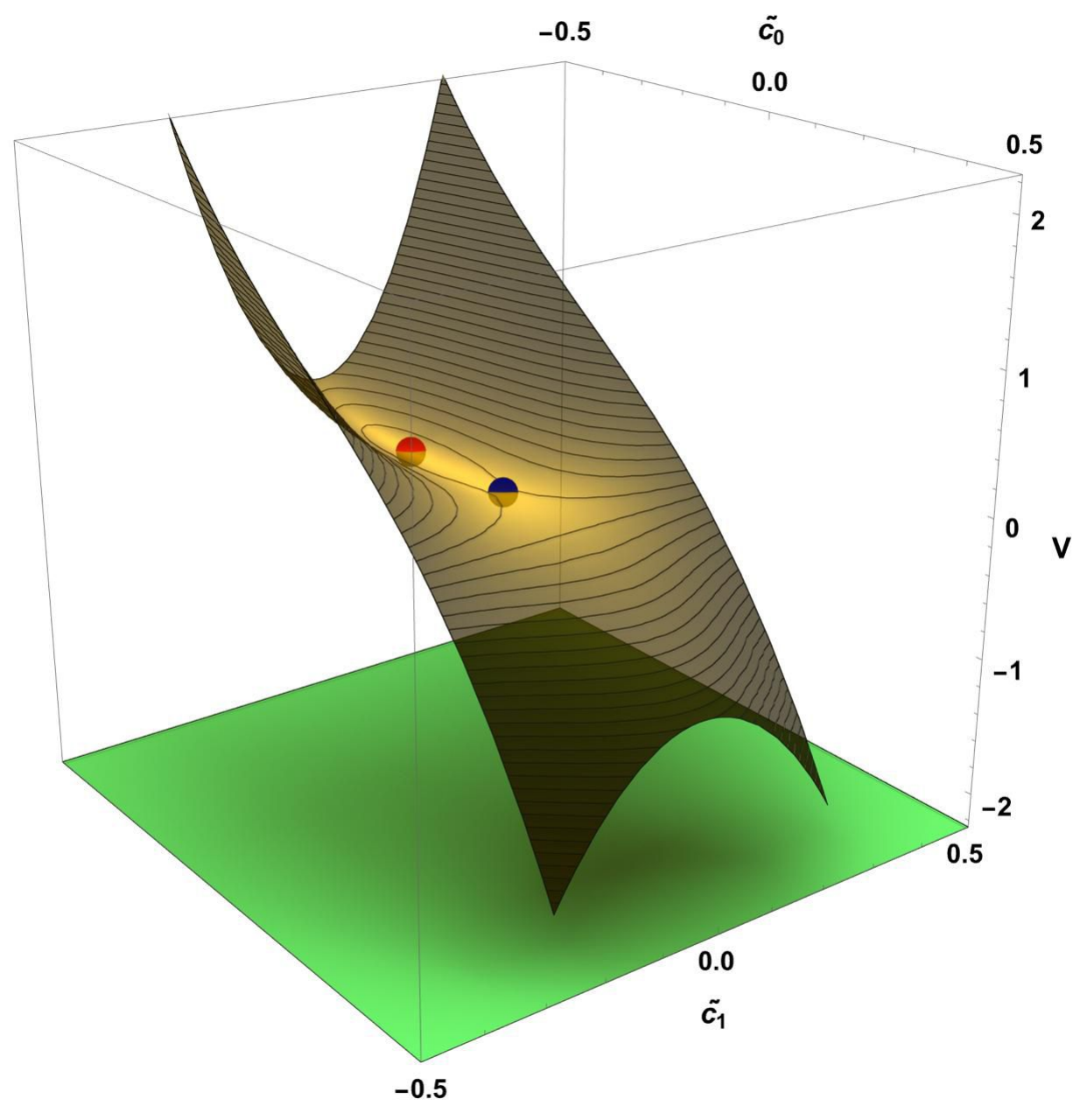}}
	\caption{\label{fig:potentialeinsteinframe}Three-dimensional plot of the potential. The red dot corresponds to the local minima and the blue dot corresponds to the saddle point. Here $B=0.2$, $\mu=0.6$, and $L=0.75$ are used. In units of GeV.}
\end{figure}

\begin{figure}[htbp!]
\centering
\includegraphics[width=0.65\textwidth]{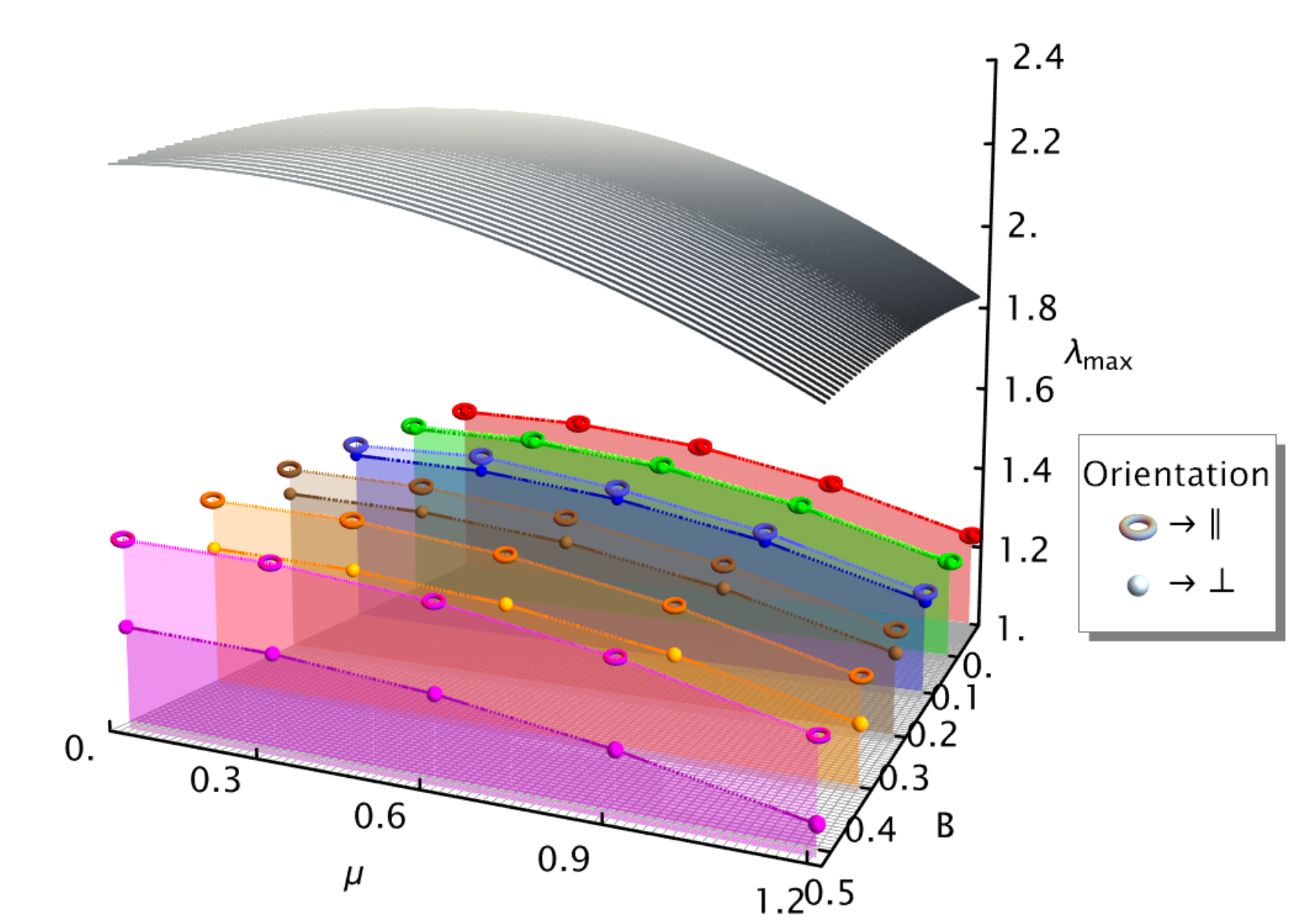}
\caption{Comparison of the maximum Lyapunov exponent $\lambda_{max}$ and the MSS bound for $x_1$ $(\parallel)$ and $x_3$ $(\perp)$ orientations for different $B$ and $\mu$ values at the unstable saddle point. Here red, green, blue, brown, orange, and magenta data points correspond to $B= 0.0$, $0.1$, $0.2$, $0.3$, $0.4$, and $0.5$, respectively. The MSS bound is shown by the grey surface above the data points.}
\label{fig:EinsteinLmaxMSS}
\end{figure}

To complete our analysis, we have again checked the validity of the MSS bound in the Einstein frame. As mentioned earlier, the perturbation around the string also generates a trapping potential, shown in Fig.~\ref{fig:potentialeinsteinframe}, with stable and unstable fixed points in the Einstein frame. We find that $\lambda_{max}$ inside the potential trap is around three orders of magnitude smaller than the MSS bound for all chemical potential and magnetic field values, thereby satisfying the MSS bound.  Similarly, like in the string frame case, the Lyapunov exponents at the unstable fixed point again asymptotically converge to $(\sqrt{-\omega_{0}^{2}}, -\sqrt{-\omega_{0}^{2}},0,0)$. The values of $\omega_0^2$ are given in Table~\ref{tab:EinsteinEigenValues}. This gives us the largest Lyapunov exponent at the saddle point $\lambda_{max}=\sqrt{-\omega_0^2}$. Although $\lambda_{max}$ at the unstable fixed point is three orders of magnitude higher than the stable point, it still remains below the MSS bound. This is explicitly shown in Fig.~\ref{fig:EinsteinLmaxMSS}. Moreover, $\lambda_{max}$ at the unstable fixed point is found to be decreasing with $\mu$ and $B$ for both orientations of the string. The chemical potential weakens the dependence of the system dynamics to the initial conditions, increasing with $\mu$ but also increasing/decreasing for parallel/perpendicular magnetic fields. This is another result different from the string frame case, where $\lambda_{max}$ exhibited similar behaviour near the unstable and stable fixed points for the string's parallel and perpendicular orientation. Our analysis, therefore, provides a curious and intriguing example where the Lyapunov exponent exhibits a different structure depending upon the fixed points involved in the system.

\bibliography{biblio.bib}

\begin{thebibliography}{130}%
\makeatletter
\providecommand \@ifxundefined [1]{%
 \@ifx{#1\undefined}
}%
\providecommand \@ifnum [1]{%
 \ifnum #1\expandafter \@firstoftwo
 \else \expandafter \@secondoftwo
 \fi
}%
\providecommand \@ifx [1]{%
 \ifx #1\expandafter \@firstoftwo
 \else \expandafter \@secondoftwo
 \fi
}%
\providecommand \natexlab [1]{#1}%
\providecommand \enquote  [1]{``#1''}%
\providecommand \bibnamefont  [1]{#1}%
\providecommand \bibfnamefont [1]{#1}%
\providecommand \citenamefont [1]{#1}%
\providecommand \href@noop [0]{\@secondoftwo}%
\providecommand \href [0]{\begingroup \@sanitize@url \@href}%
\providecommand \@href[1]{\@@startlink{#1}\@@href}%
\providecommand \@@href[1]{\endgroup#1\@@endlink}%
\providecommand \@sanitize@url [0]{\catcode `\\12\catcode `\$12\catcode
  `\&12\catcode `\#12\catcode `\^12\catcode `\_12\catcode `\%12\relax}%
\providecommand \@@startlink[1]{}%
\providecommand \@@endlink[0]{}%
\providecommand \url  [0]{\begingroup\@sanitize@url \@url }%
\providecommand \@url [1]{\endgroup\@href {#1}{\urlprefix }}%
\providecommand \urlprefix  [0]{URL }%
\providecommand \Eprint [0]{\href }%
\providecommand \doibase [0]{https://doi.org/}%
\providecommand \selectlanguage [0]{\@gobble}%
\providecommand \bibinfo  [0]{\@secondoftwo}%
\providecommand \bibfield  [0]{\@secondoftwo}%
\providecommand \translation [1]{[#1]}%
\providecommand \BibitemOpen [0]{}%
\providecommand \bibitemStop [0]{}%
\providecommand \bibitemNoStop [0]{.\EOS\space}%
\providecommand \EOS [0]{\spacefactor3000\relax}%
\providecommand \BibitemShut  [1]{\csname bibitem#1\endcsname}%
\let\auto@bib@innerbib\@empty
\bibitem [{\citenamefont {Pullirsch}\ \emph {et~al.}(1998)\citenamefont
  {Pullirsch}, \citenamefont {Rabitsch}, \citenamefont {Wettig},\ and\
  \citenamefont {Markum}}]{Pullirsch:1998ke}%
  \BibitemOpen
  \bibfield  {author} {\bibinfo {author} {\bibfnamefont {R.}~\bibnamefont
  {Pullirsch}}, \bibinfo {author} {\bibfnamefont {K.}~\bibnamefont {Rabitsch}},
  \bibinfo {author} {\bibfnamefont {T.}~\bibnamefont {Wettig}},\ and\ \bibinfo
  {author} {\bibfnamefont {H.}~\bibnamefont {Markum}},\ }\href
  {https://doi.org/10.1016/S0370-2693(98)00318-9} {\bibfield  {journal}
  {\bibinfo  {journal} {Phys. Lett. B}\ }\textbf {\bibinfo {volume} {427}},\
  \bibinfo {pages} {119} (\bibinfo {year} {1998})},\ \Eprint
  {https://arxiv.org/abs/hep-ph/9803285} {arXiv:hep-ph/9803285} \BibitemShut
  {NoStop}%
\bibitem [{\citenamefont {Asano}\ \emph {et~al.}(2015)\citenamefont {Asano},
  \citenamefont {Kawai}, \citenamefont {Kyono},\ and\ \citenamefont
  {Yoshida}}]{Asano:2015qwa}%
  \BibitemOpen
  \bibfield  {author} {\bibinfo {author} {\bibfnamefont {Y.}~\bibnamefont
  {Asano}}, \bibinfo {author} {\bibfnamefont {D.}~\bibnamefont {Kawai}},
  \bibinfo {author} {\bibfnamefont {H.}~\bibnamefont {Kyono}},\ and\ \bibinfo
  {author} {\bibfnamefont {K.}~\bibnamefont {Yoshida}},\ }\href
  {https://doi.org/10.1007/JHEP08(2015)060} {\bibfield  {journal} {\bibinfo
  {journal} {JHEP}\ }\textbf {\bibinfo {volume} {08}},\ \bibinfo {pages}
  {060}},\ \Eprint {https://arxiv.org/abs/1505.07583} {arXiv:1505.07583
  [hep-th]} \BibitemShut {NoStop}%
\bibitem [{\citenamefont {Pullirsch}\ \emph {et~al.}(1999)\citenamefont
  {Pullirsch}, \citenamefont {Markum}, \citenamefont {Rabitsch},\ and\
  \citenamefont {Wettig}}]{Pullirsch:1998wp}%
  \BibitemOpen
  \bibfield  {author} {\bibinfo {author} {\bibfnamefont {R.}~\bibnamefont
  {Pullirsch}}, \bibinfo {author} {\bibfnamefont {H.}~\bibnamefont {Markum}},
  \bibinfo {author} {\bibfnamefont {K.}~\bibnamefont {Rabitsch}},\ and\
  \bibinfo {author} {\bibfnamefont {T.}~\bibnamefont {Wettig}},\ }\href
  {https://doi.org/10.1016/S0920-5632(99)85113-5} {\bibfield  {journal}
  {\bibinfo  {journal} {Nucl. Phys. B Proc. Suppl.}\ }\textbf {\bibinfo
  {volume} {73}},\ \bibinfo {pages} {486} (\bibinfo {year} {1999})},\ \Eprint
  {https://arxiv.org/abs/hep-lat/9809057} {arXiv:hep-lat/9809057} \BibitemShut
  {NoStop}%
\bibitem [{\citenamefont {Bittner}\ \emph {et~al.}(2001)\citenamefont
  {Bittner}, \citenamefont {Markum},\ and\ \citenamefont
  {Pullirsch}}]{Bittner:2000nu}%
  \BibitemOpen
  \bibfield  {author} {\bibinfo {author} {\bibfnamefont {E.}~\bibnamefont
  {Bittner}}, \bibinfo {author} {\bibfnamefont {H.}~\bibnamefont {Markum}},\
  and\ \bibinfo {author} {\bibfnamefont {R.}~\bibnamefont {Pullirsch}},\ }\href
  {https://doi.org/10.1016/S0920-5632(01)01130-6} {\bibfield  {journal}
  {\bibinfo  {journal} {Nucl. Phys. B Proc. Suppl.}\ }\textbf {\bibinfo
  {volume} {96}},\ \bibinfo {pages} {189} (\bibinfo {year} {2001})},\ \Eprint
  {https://arxiv.org/abs/hep-lat/0009002} {arXiv:hep-lat/0009002} \BibitemShut
  {NoStop}%
\bibitem [{\citenamefont {Hashimoto}\ \emph {et~al.}(2016)\citenamefont
  {Hashimoto}, \citenamefont {Murata},\ and\ \citenamefont
  {Yoshida}}]{Hashimoto:2016wme}%
  \BibitemOpen
  \bibfield  {author} {\bibinfo {author} {\bibfnamefont {K.}~\bibnamefont
  {Hashimoto}}, \bibinfo {author} {\bibfnamefont {K.}~\bibnamefont {Murata}},\
  and\ \bibinfo {author} {\bibfnamefont {K.}~\bibnamefont {Yoshida}},\ }\href
  {https://doi.org/10.1103/PhysRevLett.117.231602} {\bibfield  {journal}
  {\bibinfo  {journal} {Phys. Rev. Lett.}\ }\textbf {\bibinfo {volume} {117}},\
  \bibinfo {pages} {231602} (\bibinfo {year} {2016})},\ \Eprint
  {https://arxiv.org/abs/1605.08124} {arXiv:1605.08124 [hep-th]} \BibitemShut
  {NoStop}%
\bibitem [{\citenamefont {Ageev}(2021)}]{Ageev:2021poy}%
  \BibitemOpen
  \bibfield  {author} {\bibinfo {author} {\bibfnamefont {D.~S.}\ \bibnamefont
  {Ageev}},\ }\href {https://doi.org/10.1103/PhysRevD.104.126013} {\bibfield
  {journal} {\bibinfo  {journal} {Phys. Rev. D}\ }\textbf {\bibinfo {volume}
  {104}},\ \bibinfo {pages} {126013} (\bibinfo {year} {2021})},\ \Eprint
  {https://arxiv.org/abs/2105.04589} {arXiv:2105.04589 [hep-th]} \BibitemShut
  {NoStop}%
\bibitem [{\citenamefont {Yadav}\ \emph {et~al.}(2024)\citenamefont {Yadav},
  \citenamefont {Kushwah},\ and\ \citenamefont {Misra}}]{Yadav:2023hyg}%
  \BibitemOpen
  \bibfield  {author} {\bibinfo {author} {\bibfnamefont {G.}~\bibnamefont
  {Yadav}}, \bibinfo {author} {\bibfnamefont {S.~S.}\ \bibnamefont {Kushwah}},\
  and\ \bibinfo {author} {\bibfnamefont {A.}~\bibnamefont {Misra}},\ }\href
  {https://doi.org/10.1007/JHEP05(2024)015} {\bibfield  {journal} {\bibinfo
  {journal} {JHEP}\ }\textbf {\bibinfo {volume} {05}},\ \bibinfo {pages}
  {015}},\ \Eprint {https://arxiv.org/abs/2311.09306} {arXiv:2311.09306
  [hep-th]} \BibitemShut {NoStop}%
\bibitem [{\citenamefont {M\"uller}\ and\ \citenamefont
  {Trayanov}(1992)}]{Muller1992Jun}%
  \BibitemOpen
  \bibfield  {author} {\bibinfo {author} {\bibfnamefont {B.}~\bibnamefont
  {M\"uller}}\ and\ \bibinfo {author} {\bibfnamefont {A.}~\bibnamefont
  {Trayanov}},\ }\href {https://doi.org/10.1103/PhysRevLett.68.3387} {\bibfield
   {journal} {\bibinfo  {journal} {Phys. Rev. Lett.}\ }\textbf {\bibinfo
  {volume} {68}},\ \bibinfo {pages} {3387} (\bibinfo {year}
  {1992})}\BibitemShut {NoStop}%
\bibitem [{\citenamefont {Bir\'o}\ \emph {et~al.}(1995)\citenamefont {Bir\'o},
  \citenamefont {Gong},\ and\ \citenamefont {M\"uller}}]{biro1995lyapunov}%
  \BibitemOpen
  \bibfield  {author} {\bibinfo {author} {\bibfnamefont {T.~S.}\ \bibnamefont
  {Bir\'o}}, \bibinfo {author} {\bibfnamefont {C.}~\bibnamefont {Gong}},\ and\
  \bibinfo {author} {\bibfnamefont {B.}~\bibnamefont {M\"uller}},\ }\href
  {https://doi.org/10.1103/PhysRevD.52.1260} {\bibfield  {journal} {\bibinfo
  {journal} {Phys. Rev. D}\ }\textbf {\bibinfo {volume} {52}},\ \bibinfo
  {pages} {1260} (\bibinfo {year} {1995})}\BibitemShut {NoStop}%
\bibitem [{\citenamefont {Tsukiji}\ \emph {et~al.}(2016)\citenamefont
  {Tsukiji}, \citenamefont {Iida}, \citenamefont {Kunihiro}, \citenamefont
  {Ohnishi},\ and\ \citenamefont {Takahashi}}]{tsukiji2016entropy}%
  \BibitemOpen
  \bibfield  {author} {\bibinfo {author} {\bibfnamefont {H.}~\bibnamefont
  {Tsukiji}}, \bibinfo {author} {\bibfnamefont {H.}~\bibnamefont {Iida}},
  \bibinfo {author} {\bibfnamefont {T.}~\bibnamefont {Kunihiro}}, \bibinfo
  {author} {\bibfnamefont {A.}~\bibnamefont {Ohnishi}},\ and\ \bibinfo {author}
  {\bibfnamefont {T.~T.}\ \bibnamefont {Takahashi}},\ }\href
  {https://doi.org/10.1103/PhysRevD.94.091502} {\bibfield  {journal} {\bibinfo
  {journal} {Phys. Rev. D}\ }\textbf {\bibinfo {volume} {94}},\ \bibinfo
  {pages} {091502} (\bibinfo {year} {2016})},\ \Eprint
  {https://arxiv.org/abs/1603.04622} {arXiv:1603.04622 [hep-ph]} \BibitemShut
  {NoStop}%
\bibitem [{\citenamefont {Kunihiro}\ \emph {et~al.}(2010)\citenamefont
  {Kunihiro}, \citenamefont {Muller}, \citenamefont {Ohnishi}, \citenamefont
  {Schafer}, \citenamefont {Takahashi},\ and\ \citenamefont
  {Yamamoto}}]{kunihiro2010chaotic}%
  \BibitemOpen
  \bibfield  {author} {\bibinfo {author} {\bibfnamefont {T.}~\bibnamefont
  {Kunihiro}}, \bibinfo {author} {\bibfnamefont {B.}~\bibnamefont {Muller}},
  \bibinfo {author} {\bibfnamefont {A.}~\bibnamefont {Ohnishi}}, \bibinfo
  {author} {\bibfnamefont {A.}~\bibnamefont {Schafer}}, \bibinfo {author}
  {\bibfnamefont {T.~T.}\ \bibnamefont {Takahashi}},\ and\ \bibinfo {author}
  {\bibfnamefont {A.}~\bibnamefont {Yamamoto}},\ }\href
  {https://doi.org/10.1103/PhysRevD.82.114015} {\bibfield  {journal} {\bibinfo
  {journal} {Phys. Rev. D}\ }\textbf {\bibinfo {volume} {82}},\ \bibinfo
  {pages} {114015} (\bibinfo {year} {2010})},\ \Eprint
  {https://arxiv.org/abs/1008.1156} {arXiv:1008.1156 [hep-ph]} \BibitemShut
  {NoStop}%
\bibitem [{\citenamefont {Maldacena}(1998)}]{Maldacena:1997re}%
  \BibitemOpen
  \bibfield  {author} {\bibinfo {author} {\bibfnamefont {J.~M.}\ \bibnamefont
  {Maldacena}},\ }\href {https://doi.org/10.4310/ATMP.1998.v2.n2.a1} {\bibfield
   {journal} {\bibinfo  {journal} {Adv. Theor. Math. Phys.}\ }\textbf {\bibinfo
  {volume} {2}},\ \bibinfo {pages} {231} (\bibinfo {year} {1998})},\ \Eprint
  {https://arxiv.org/abs/hep-th/9711200} {arXiv:hep-th/9711200} \BibitemShut
  {NoStop}%
\bibitem [{\citenamefont {Witten}(1998)}]{Witten:1998qj}%
  \BibitemOpen
  \bibfield  {author} {\bibinfo {author} {\bibfnamefont {E.}~\bibnamefont
  {Witten}},\ }\href {https://doi.org/10.4310/ATMP.1998.v2.n2.a2} {\bibfield
  {journal} {\bibinfo  {journal} {Adv. Theor. Math. Phys.}\ }\textbf {\bibinfo
  {volume} {2}},\ \bibinfo {pages} {253} (\bibinfo {year} {1998})},\ \Eprint
  {https://arxiv.org/abs/hep-th/9802150} {arXiv:hep-th/9802150} \BibitemShut
  {NoStop}%
\bibitem [{\citenamefont {Gubser}\ \emph {et~al.}(1998)\citenamefont {Gubser},
  \citenamefont {Klebanov},\ and\ \citenamefont {Polyakov}}]{Gubser:1998bc}%
  \BibitemOpen
  \bibfield  {author} {\bibinfo {author} {\bibfnamefont {S.~S.}\ \bibnamefont
  {Gubser}}, \bibinfo {author} {\bibfnamefont {I.~R.}\ \bibnamefont
  {Klebanov}},\ and\ \bibinfo {author} {\bibfnamefont {A.~M.}\ \bibnamefont
  {Polyakov}},\ }\href {https://doi.org/10.1016/S0370-2693(98)00377-3}
  {\bibfield  {journal} {\bibinfo  {journal} {Phys. Lett. B}\ }\textbf
  {\bibinfo {volume} {428}},\ \bibinfo {pages} {105} (\bibinfo {year}
  {1998})},\ \Eprint {https://arxiv.org/abs/hep-th/9802109}
  {arXiv:hep-th/9802109} \BibitemShut {NoStop}%
\bibitem [{\citenamefont {Pando~Zayas}\ and\ \citenamefont
  {Terrero-Escalante}(2010)}]{pando2010chaos}%
  \BibitemOpen
  \bibfield  {author} {\bibinfo {author} {\bibfnamefont {L.~A.}\ \bibnamefont
  {Pando~Zayas}}\ and\ \bibinfo {author} {\bibfnamefont {C.~A.}\ \bibnamefont
  {Terrero-Escalante}},\ }\href {https://doi.org/10.1007/JHEP09(2010)094}
  {\bibfield  {journal} {\bibinfo  {journal} {JHEP}\ }\textbf {\bibinfo
  {volume} {09}},\ \bibinfo {pages} {094}},\ \Eprint
  {https://arxiv.org/abs/1007.0277} {arXiv:1007.0277 [hep-th]} \BibitemShut
  {NoStop}%
\bibitem [{\citenamefont {Blake}\ and\ \citenamefont
  {Davison}(2022)}]{Blake:2021hjj}%
  \BibitemOpen
  \bibfield  {author} {\bibinfo {author} {\bibfnamefont {M.}~\bibnamefont
  {Blake}}\ and\ \bibinfo {author} {\bibfnamefont {R.~A.}\ \bibnamefont
  {Davison}},\ }\href {https://doi.org/10.1007/JHEP01(2022)013} {\bibfield
  {journal} {\bibinfo  {journal} {JHEP}\ }\textbf {\bibinfo {volume} {01}},\
  \bibinfo {pages} {013}},\ \Eprint {https://arxiv.org/abs/2111.11093}
  {arXiv:2111.11093 [hep-th]} \BibitemShut {NoStop}%
\bibitem [{\citenamefont {Giombi}\ \emph {et~al.}(2023)\citenamefont {Giombi},
  \citenamefont {Komatsu},\ and\ \citenamefont {Offertaler}}]{Giombi:2022pas}%
  \BibitemOpen
  \bibfield  {author} {\bibinfo {author} {\bibfnamefont {S.}~\bibnamefont
  {Giombi}}, \bibinfo {author} {\bibfnamefont {S.}~\bibnamefont {Komatsu}},\
  and\ \bibinfo {author} {\bibfnamefont {B.}~\bibnamefont {Offertaler}},\
  }\href {https://doi.org/10.1007/JHEP09(2023)023} {\bibfield  {journal}
  {\bibinfo  {journal} {JHEP}\ }\textbf {\bibinfo {volume} {09}},\ \bibinfo
  {pages} {023}},\ \Eprint {https://arxiv.org/abs/2212.14842} {arXiv:2212.14842
  [hep-th]} \BibitemShut {NoStop}%
\bibitem [{\citenamefont {Basu}\ \emph {et~al.}(2011)\citenamefont {Basu},
  \citenamefont {Das},\ and\ \citenamefont {Ghosh}}]{Basu:2011dg}%
  \BibitemOpen
  \bibfield  {author} {\bibinfo {author} {\bibfnamefont {P.}~\bibnamefont
  {Basu}}, \bibinfo {author} {\bibfnamefont {D.}~\bibnamefont {Das}},\ and\
  \bibinfo {author} {\bibfnamefont {A.}~\bibnamefont {Ghosh}},\ }\href
  {https://doi.org/10.1016/j.physletb.2011.04.027} {\bibfield  {journal}
  {\bibinfo  {journal} {Phys. Lett. B}\ }\textbf {\bibinfo {volume} {699}},\
  \bibinfo {pages} {388} (\bibinfo {year} {2011})},\ \Eprint
  {https://arxiv.org/abs/1103.4101} {arXiv:1103.4101 [hep-th]} \BibitemShut
  {NoStop}%
\bibitem [{\citenamefont {Basu}\ and\ \citenamefont
  {Pando~Zayas}(2011)}]{Basu:2011di}%
  \BibitemOpen
  \bibfield  {author} {\bibinfo {author} {\bibfnamefont {P.}~\bibnamefont
  {Basu}}\ and\ \bibinfo {author} {\bibfnamefont {L.~A.}\ \bibnamefont
  {Pando~Zayas}},\ }\href {https://doi.org/10.1016/j.physletb.2011.04.063}
  {\bibfield  {journal} {\bibinfo  {journal} {Phys. Lett. B}\ }\textbf
  {\bibinfo {volume} {700}},\ \bibinfo {pages} {243} (\bibinfo {year}
  {2011})},\ \Eprint {https://arxiv.org/abs/1103.4107} {arXiv:1103.4107
  [hep-th]} \BibitemShut {NoStop}%
\bibitem [{\citenamefont {Stepanchuk}\ and\ \citenamefont
  {Tseytlin}(2013)}]{Stepanchuk:2012xi}%
  \BibitemOpen
  \bibfield  {author} {\bibinfo {author} {\bibfnamefont {A.}~\bibnamefont
  {Stepanchuk}}\ and\ \bibinfo {author} {\bibfnamefont {A.~A.}\ \bibnamefont
  {Tseytlin}},\ }\href {https://doi.org/10.1088/1751-8113/46/12/125401}
  {\bibfield  {journal} {\bibinfo  {journal} {J. Phys. A}\ }\textbf {\bibinfo
  {volume} {46}},\ \bibinfo {pages} {125401} (\bibinfo {year} {2013})},\
  \Eprint {https://arxiv.org/abs/1211.3727} {arXiv:1211.3727 [hep-th]}
  \BibitemShut {NoStop}%
\bibitem [{\citenamefont {Giataganas}\ \emph {et~al.}(2014)\citenamefont
  {Giataganas}, \citenamefont {Pando~Zayas},\ and\ \citenamefont
  {Zoubos}}]{Giataganas:2013dha}%
  \BibitemOpen
  \bibfield  {author} {\bibinfo {author} {\bibfnamefont {D.}~\bibnamefont
  {Giataganas}}, \bibinfo {author} {\bibfnamefont {L.~A.}\ \bibnamefont
  {Pando~Zayas}},\ and\ \bibinfo {author} {\bibfnamefont {K.}~\bibnamefont
  {Zoubos}},\ }\href {https://doi.org/10.1007/JHEP01(2014)129} {\bibfield
  {journal} {\bibinfo  {journal} {JHEP}\ }\textbf {\bibinfo {volume} {01}},\
  \bibinfo {pages} {129}},\ \Eprint {https://arxiv.org/abs/1311.3241}
  {arXiv:1311.3241 [hep-th]} \BibitemShut {NoStop}%
\bibitem [{\citenamefont {Bai}\ \emph {et~al.}(2016)\citenamefont {Bai},
  \citenamefont {Lee}, \citenamefont {Moon},\ and\ \citenamefont
  {Chen}}]{Bai:2014wpa}%
  \BibitemOpen
  \bibfield  {author} {\bibinfo {author} {\bibfnamefont {X.}~\bibnamefont
  {Bai}}, \bibinfo {author} {\bibfnamefont {B.-H.}\ \bibnamefont {Lee}},
  \bibinfo {author} {\bibfnamefont {T.}~\bibnamefont {Moon}},\ and\ \bibinfo
  {author} {\bibfnamefont {J.}~\bibnamefont {Chen}},\ }\href
  {https://doi.org/10.3938/jkps.68.639} {\bibfield  {journal} {\bibinfo
  {journal} {J. Korean Phys. Soc.}\ }\textbf {\bibinfo {volume} {68}},\
  \bibinfo {pages} {639} (\bibinfo {year} {2016})},\ \Eprint
  {https://arxiv.org/abs/1406.5816} {arXiv:1406.5816 [hep-th]} \BibitemShut
  {NoStop}%
\bibitem [{\citenamefont {Panigrahi}\ and\ \citenamefont
  {Samal}(2016)}]{Panigrahi:2016zny}%
  \BibitemOpen
  \bibfield  {author} {\bibinfo {author} {\bibfnamefont {K.~L.}\ \bibnamefont
  {Panigrahi}}\ and\ \bibinfo {author} {\bibfnamefont {M.}~\bibnamefont
  {Samal}},\ }\href {https://doi.org/10.1016/j.physletb.2016.08.021} {\bibfield
   {journal} {\bibinfo  {journal} {Phys. Lett. B}\ }\textbf {\bibinfo {volume}
  {761}},\ \bibinfo {pages} {475} (\bibinfo {year} {2016})},\ \Eprint
  {https://arxiv.org/abs/1605.05638} {arXiv:1605.05638 [hep-th]} \BibitemShut
  {NoStop}%
\bibitem [{\citenamefont {Basu}\ \emph {et~al.}(2017)\citenamefont {Basu},
  \citenamefont {Chaturvedi},\ and\ \citenamefont {Samantray}}]{Basu:2016zkr}%
  \BibitemOpen
  \bibfield  {author} {\bibinfo {author} {\bibfnamefont {P.}~\bibnamefont
  {Basu}}, \bibinfo {author} {\bibfnamefont {P.}~\bibnamefont {Chaturvedi}},\
  and\ \bibinfo {author} {\bibfnamefont {P.}~\bibnamefont {Samantray}},\ }\href
  {https://doi.org/10.1103/PhysRevD.95.066014} {\bibfield  {journal} {\bibinfo
  {journal} {Phys. Rev. D}\ }\textbf {\bibinfo {volume} {95}},\ \bibinfo
  {pages} {066014} (\bibinfo {year} {2017})},\ \Eprint
  {https://arxiv.org/abs/1607.04466} {arXiv:1607.04466 [hep-th]} \BibitemShut
  {NoStop}%
\bibitem [{\citenamefont {Asano}\ \emph {et~al.}(2016)\citenamefont {Asano},
  \citenamefont {Kyono},\ and\ \citenamefont {Yoshida}}]{Asano:2016qsv}%
  \BibitemOpen
  \bibfield  {author} {\bibinfo {author} {\bibfnamefont {Y.}~\bibnamefont
  {Asano}}, \bibinfo {author} {\bibfnamefont {H.}~\bibnamefont {Kyono}},\ and\
  \bibinfo {author} {\bibfnamefont {K.}~\bibnamefont {Yoshida}},\ }\href
  {https://doi.org/10.1007/JHEP09(2016)103} {\bibfield  {journal} {\bibinfo
  {journal} {JHEP}\ }\textbf {\bibinfo {volume} {09}},\ \bibinfo {pages}
  {103}},\ \Eprint {https://arxiv.org/abs/1607.07302} {arXiv:1607.07302
  [hep-th]} \BibitemShut {NoStop}%
\bibitem [{\citenamefont {Ishii}\ \emph {et~al.}(2017)\citenamefont {Ishii},
  \citenamefont {Murata},\ and\ \citenamefont {Yoshida}}]{Ishii:2016rlk}%
  \BibitemOpen
  \bibfield  {author} {\bibinfo {author} {\bibfnamefont {T.}~\bibnamefont
  {Ishii}}, \bibinfo {author} {\bibfnamefont {K.}~\bibnamefont {Murata}},\ and\
  \bibinfo {author} {\bibfnamefont {K.}~\bibnamefont {Yoshida}},\ }\href
  {https://doi.org/10.1103/PhysRevD.95.066019} {\bibfield  {journal} {\bibinfo
  {journal} {Phys. Rev. D}\ }\textbf {\bibinfo {volume} {95}},\ \bibinfo
  {pages} {066019} (\bibinfo {year} {2017})},\ \Eprint
  {https://arxiv.org/abs/1610.05833} {arXiv:1610.05833 [hep-th]} \BibitemShut
  {NoStop}%
\bibitem [{\citenamefont {Rigatos}(2020)}]{Rigatos:2020hlq}%
  \BibitemOpen
  \bibfield  {author} {\bibinfo {author} {\bibfnamefont {K.~S.}\ \bibnamefont
  {Rigatos}},\ }\href {https://doi.org/10.1103/PhysRevD.102.106022} {\bibfield
  {journal} {\bibinfo  {journal} {Phys. Rev. D}\ }\textbf {\bibinfo {volume}
  {102}},\ \bibinfo {pages} {106022} (\bibinfo {year} {2020})},\ \Eprint
  {https://arxiv.org/abs/2009.11878} {arXiv:2009.11878 [hep-th]} \BibitemShut
  {NoStop}%
\bibitem [{\citenamefont {Pal}(2023)}]{Pal:2023kwc}%
  \BibitemOpen
  \bibfield  {author} {\bibinfo {author} {\bibfnamefont {J.}~\bibnamefont
  {Pal}},\ }\href {https://doi.org/10.1142/S0217732323501420} {\bibfield
  {journal} {\bibinfo  {journal} {Mod. Phys. Lett. A}\ }\textbf {\bibinfo
  {volume} {38}},\ \bibinfo {pages} {2350142} (\bibinfo {year} {2023})},\
  \Eprint {https://arxiv.org/abs/2304.10474} {arXiv:2304.10474 [hep-th]}
  \BibitemShut {NoStop}%
\bibitem [{\citenamefont {Pen\'\i{}n}\ and\ \citenamefont
  {Rigatos}(2024)}]{Penin:2024rqb}%
  \BibitemOpen
  \bibfield  {author} {\bibinfo {author} {\bibfnamefont {J.~M.}\ \bibnamefont
  {Pen\'\i{}n}}\ and\ \bibinfo {author} {\bibfnamefont {K.~C.}\ \bibnamefont
  {Rigatos}},\ }\href {https://doi.org/10.1103/PhysRevD.109.126007} {\bibfield
  {journal} {\bibinfo  {journal} {Phys. Rev. D}\ }\textbf {\bibinfo {volume}
  {109}},\ \bibinfo {pages} {126007} (\bibinfo {year} {2024})}\BibitemShut
  {NoStop}%
\bibitem [{\citenamefont {Shukla}\ \emph {et~al.}(2023)\citenamefont {Shukla},
  \citenamefont {Dudal},\ and\ \citenamefont {Mahapatra}}]{Shukla:2023pbp}%
  \BibitemOpen
  \bibfield  {author} {\bibinfo {author} {\bibfnamefont {B.}~\bibnamefont
  {Shukla}}, \bibinfo {author} {\bibfnamefont {D.}~\bibnamefont {Dudal}},\ and\
  \bibinfo {author} {\bibfnamefont {S.}~\bibnamefont {Mahapatra}},\ }\href
  {https://doi.org/10.1007/JHEP06(2023)178} {\bibfield  {journal} {\bibinfo
  {journal} {JHEP}\ }\textbf {\bibinfo {volume} {06}},\ \bibinfo {pages}
  {178}},\ \Eprint {https://arxiv.org/abs/2303.15716} {arXiv:2303.15716
  [hep-th]} \BibitemShut {NoStop}%
\bibitem [{\citenamefont {Hashimoto}\ \emph {et~al.}(2018)\citenamefont
  {Hashimoto}, \citenamefont {Murata},\ and\ \citenamefont
  {Tanahashi}}]{Hashimoto:2018fkb}%
  \BibitemOpen
  \bibfield  {author} {\bibinfo {author} {\bibfnamefont {K.}~\bibnamefont
  {Hashimoto}}, \bibinfo {author} {\bibfnamefont {K.}~\bibnamefont {Murata}},\
  and\ \bibinfo {author} {\bibfnamefont {N.}~\bibnamefont {Tanahashi}},\ }\href
  {https://doi.org/10.1103/PhysRevD.98.086007} {\bibfield  {journal} {\bibinfo
  {journal} {Phys. Rev. D}\ }\textbf {\bibinfo {volume} {98}},\ \bibinfo
  {pages} {086007} (\bibinfo {year} {2018})},\ \Eprint
  {https://arxiv.org/abs/1803.06756} {arXiv:1803.06756 [hep-th]} \BibitemShut
  {NoStop}%
\bibitem [{\citenamefont {Colangelo}\ \emph {et~al.}(2020)\citenamefont
  {Colangelo}, \citenamefont {De~Fazio},\ and\ \citenamefont
  {Losacco}}]{Colangelo:2020tpr}%
  \BibitemOpen
  \bibfield  {author} {\bibinfo {author} {\bibfnamefont {P.}~\bibnamefont
  {Colangelo}}, \bibinfo {author} {\bibfnamefont {F.}~\bibnamefont
  {De~Fazio}},\ and\ \bibinfo {author} {\bibfnamefont {N.}~\bibnamefont
  {Losacco}},\ }\href {https://doi.org/10.1103/PhysRevD.102.074016} {\bibfield
  {journal} {\bibinfo  {journal} {Phys. Rev. D}\ }\textbf {\bibinfo {volume}
  {102}},\ \bibinfo {pages} {074016} (\bibinfo {year} {2020})},\ \Eprint
  {https://arxiv.org/abs/2007.06980} {arXiv:2007.06980 [hep-ph]} \BibitemShut
  {NoStop}%
\bibitem [{\citenamefont {Colangelo}\ \emph {et~al.}(2022)\citenamefont
  {Colangelo}, \citenamefont {Giannuzzi},\ and\ \citenamefont
  {Losacco}}]{Colangelo:2021kmn}%
  \BibitemOpen
  \bibfield  {author} {\bibinfo {author} {\bibfnamefont {P.}~\bibnamefont
  {Colangelo}}, \bibinfo {author} {\bibfnamefont {F.}~\bibnamefont
  {Giannuzzi}},\ and\ \bibinfo {author} {\bibfnamefont {N.}~\bibnamefont
  {Losacco}},\ }\href {https://doi.org/10.1016/j.physletb.2022.136949}
  {\bibfield  {journal} {\bibinfo  {journal} {Phys. Lett. B}\ }\textbf
  {\bibinfo {volume} {827}},\ \bibinfo {pages} {136949} (\bibinfo {year}
  {2022})},\ \Eprint {https://arxiv.org/abs/2111.09441} {arXiv:2111.09441
  [hep-th]} \BibitemShut {NoStop}%
\bibitem [{\citenamefont {Akutagawa}\ \emph {et~al.}(2019)\citenamefont
  {Akutagawa}, \citenamefont {Hashimoto}, \citenamefont {Murata},\ and\
  \citenamefont {Ota}}]{Akutagawa:2019awh}%
  \BibitemOpen
  \bibfield  {author} {\bibinfo {author} {\bibfnamefont {T.}~\bibnamefont
  {Akutagawa}}, \bibinfo {author} {\bibfnamefont {K.}~\bibnamefont
  {Hashimoto}}, \bibinfo {author} {\bibfnamefont {K.}~\bibnamefont {Murata}},\
  and\ \bibinfo {author} {\bibfnamefont {T.}~\bibnamefont {Ota}},\ }\href
  {https://doi.org/10.1103/PhysRevD.100.046009} {\bibfield  {journal} {\bibinfo
   {journal} {Phys. Rev. D}\ }\textbf {\bibinfo {volume} {100}},\ \bibinfo
  {pages} {046009} (\bibinfo {year} {2019})},\ \Eprint
  {https://arxiv.org/abs/1903.04718} {arXiv:1903.04718 [hep-th]} \BibitemShut
  {NoStop}%
\bibitem [{\citenamefont {Giataganas}\ and\ \citenamefont
  {Zoubos}(2017)}]{Giataganas:2017guj}%
  \BibitemOpen
  \bibfield  {author} {\bibinfo {author} {\bibfnamefont {D.}~\bibnamefont
  {Giataganas}}\ and\ \bibinfo {author} {\bibfnamefont {K.}~\bibnamefont
  {Zoubos}},\ }\href {https://doi.org/10.1007/JHEP10(2017)042} {\bibfield
  {journal} {\bibinfo  {journal} {JHEP}\ }\textbf {\bibinfo {volume} {10}},\
  \bibinfo {pages} {042}},\ \Eprint {https://arxiv.org/abs/1707.04033}
  {arXiv:1707.04033 [hep-th]} \BibitemShut {NoStop}%
\bibitem [{\citenamefont {Shenker}\ and\ \citenamefont
  {Stanford}(2015)}]{Shenker:2014cwa}%
  \BibitemOpen
  \bibfield  {author} {\bibinfo {author} {\bibfnamefont {S.~H.}\ \bibnamefont
  {Shenker}}\ and\ \bibinfo {author} {\bibfnamefont {D.}~\bibnamefont
  {Stanford}},\ }\href {https://doi.org/10.1007/JHEP05(2015)132} {\bibfield
  {journal} {\bibinfo  {journal} {JHEP}\ }\textbf {\bibinfo {volume} {05}},\
  \bibinfo {pages} {132}},\ \Eprint {https://arxiv.org/abs/1412.6087}
  {arXiv:1412.6087 [hep-th]} \BibitemShut {NoStop}%
\bibitem [{\citenamefont {Maldacena}\ \emph {et~al.}(2016)\citenamefont
  {Maldacena}, \citenamefont {Shenker},\ and\ \citenamefont
  {Stanford}}]{maldacena2016bound}%
  \BibitemOpen
  \bibfield  {author} {\bibinfo {author} {\bibfnamefont {J.}~\bibnamefont
  {Maldacena}}, \bibinfo {author} {\bibfnamefont {S.~H.}\ \bibnamefont
  {Shenker}},\ and\ \bibinfo {author} {\bibfnamefont {D.}~\bibnamefont
  {Stanford}},\ }\href {https://doi.org/10.1007/JHEP08(2016)106} {\bibfield
  {journal} {\bibinfo  {journal} {JHEP}\ }\textbf {\bibinfo {volume} {08}},\
  \bibinfo {pages} {106}},\ \Eprint {https://arxiv.org/abs/1503.01409}
  {arXiv:1503.01409 [hep-th]} \BibitemShut {NoStop}%
\bibitem [{\citenamefont {Blake}\ \emph
  {et~al.}(2018{\natexlab{a}})\citenamefont {Blake}, \citenamefont {Lee},\ and\
  \citenamefont {Liu}}]{Blake:2017ris}%
  \BibitemOpen
  \bibfield  {author} {\bibinfo {author} {\bibfnamefont {M.}~\bibnamefont
  {Blake}}, \bibinfo {author} {\bibfnamefont {H.}~\bibnamefont {Lee}},\ and\
  \bibinfo {author} {\bibfnamefont {H.}~\bibnamefont {Liu}},\ }\href
  {https://doi.org/10.1007/JHEP10(2018)127} {\bibfield  {journal} {\bibinfo
  {journal} {JHEP}\ }\textbf {\bibinfo {volume} {10}},\ \bibinfo {pages}
  {127}},\ \Eprint {https://arxiv.org/abs/1801.00010} {arXiv:1801.00010
  [hep-th]} \BibitemShut {NoStop}%
\bibitem [{\citenamefont {Blake}\ \emph
  {et~al.}(2018{\natexlab{b}})\citenamefont {Blake}, \citenamefont {Davison},
  \citenamefont {Grozdanov},\ and\ \citenamefont {Liu}}]{Blake:2018leo}%
  \BibitemOpen
  \bibfield  {author} {\bibinfo {author} {\bibfnamefont {M.}~\bibnamefont
  {Blake}}, \bibinfo {author} {\bibfnamefont {R.~A.}\ \bibnamefont {Davison}},
  \bibinfo {author} {\bibfnamefont {S.}~\bibnamefont {Grozdanov}},\ and\
  \bibinfo {author} {\bibfnamefont {H.}~\bibnamefont {Liu}},\ }\href
  {https://doi.org/10.1007/JHEP10(2018)035} {\bibfield  {journal} {\bibinfo
  {journal} {JHEP}\ }\textbf {\bibinfo {volume} {10}},\ \bibinfo {pages}
  {035}},\ \Eprint {https://arxiv.org/abs/1809.01169} {arXiv:1809.01169
  [hep-th]} \BibitemShut {NoStop}%
\bibitem [{\citenamefont {Baishya}\ \emph {et~al.}(2024)\citenamefont
  {Baishya}, \citenamefont {Chakrabarti}, \citenamefont {Maity},\ and\
  \citenamefont {Nayek}}]{Baishya:2023ojl}%
  \BibitemOpen
  \bibfield  {author} {\bibinfo {author} {\bibfnamefont {B.}~\bibnamefont
  {Baishya}}, \bibinfo {author} {\bibfnamefont {S.}~\bibnamefont
  {Chakrabarti}}, \bibinfo {author} {\bibfnamefont {D.}~\bibnamefont {Maity}},\
  and\ \bibinfo {author} {\bibfnamefont {K.}~\bibnamefont {Nayek}},\ }\href
  {https://doi.org/10.1103/PhysRevD.110.086003} {\bibfield  {journal} {\bibinfo
   {journal} {Phys. Rev. D}\ }\textbf {\bibinfo {volume} {110}},\ \bibinfo
  {pages} {086003} (\bibinfo {year} {2024})},\ \Eprint
  {https://arxiv.org/abs/2312.01829} {arXiv:2312.01829 [hep-th]} \BibitemShut
  {NoStop}%
\bibitem [{\citenamefont {Pando~Zayas}\ and\ \citenamefont
  {Reichmann}(2013)}]{PandoZayas:2012ig}%
  \BibitemOpen
  \bibfield  {author} {\bibinfo {author} {\bibfnamefont {L.~A.}\ \bibnamefont
  {Pando~Zayas}}\ and\ \bibinfo {author} {\bibfnamefont {D.}~\bibnamefont
  {Reichmann}},\ }\href {https://doi.org/10.1007/JHEP04(2013)083} {\bibfield
  {journal} {\bibinfo  {journal} {JHEP}\ }\textbf {\bibinfo {volume} {04}},\
  \bibinfo {pages} {083}},\ \Eprint {https://arxiv.org/abs/1209.5902}
  {arXiv:1209.5902 [hep-th]} \BibitemShut {NoStop}%
\bibitem [{\citenamefont {Basu}\ and\ \citenamefont
  {Ghosh}(2014)}]{Basu:2013uva}%
  \BibitemOpen
  \bibfield  {author} {\bibinfo {author} {\bibfnamefont {P.}~\bibnamefont
  {Basu}}\ and\ \bibinfo {author} {\bibfnamefont {A.}~\bibnamefont {Ghosh}},\
  }\href {https://doi.org/10.1016/j.physletb.2013.12.052} {\bibfield  {journal}
  {\bibinfo  {journal} {Phys. Lett. B}\ }\textbf {\bibinfo {volume} {729}},\
  \bibinfo {pages} {50} (\bibinfo {year} {2014})},\ \Eprint
  {https://arxiv.org/abs/1304.6348} {arXiv:1304.6348 [hep-th]} \BibitemShut
  {NoStop}%
\bibitem [{\citenamefont {Shukla}\ \emph
  {et~al.}(2024{\natexlab{a}})\citenamefont {Shukla}, \citenamefont {Riyaz},\
  and\ \citenamefont {Mahapatra}}]{Shukla:2024wsu}%
  \BibitemOpen
  \bibfield  {author} {\bibinfo {author} {\bibfnamefont {B.}~\bibnamefont
  {Shukla}}, \bibinfo {author} {\bibfnamefont {O.}~\bibnamefont {Riyaz}},\ and\
  \bibinfo {author} {\bibfnamefont {S.}~\bibnamefont {Mahapatra}},\ }\href@noop
  {} {\bibfield  {journal} {\bibinfo  {journal} {arXiv}\ } (\bibinfo {year}
  {2024}{\natexlab{a}})},\ \Eprint {https://arxiv.org/abs/2411.12536}
  {arXiv:2411.12536 [hep-th]} \BibitemShut {NoStop}%
\bibitem [{\citenamefont {Jahnke}(2019)}]{Jahnke:2018off}%
  \BibitemOpen
  \bibfield  {author} {\bibinfo {author} {\bibfnamefont {V.}~\bibnamefont
  {Jahnke}},\ }\href {https://doi.org/10.1155/2019/9632708} {\bibfield
  {journal} {\bibinfo  {journal} {Adv. High Energy Phys.}\ }\textbf {\bibinfo
  {volume} {2019}},\ \bibinfo {pages} {9632708} (\bibinfo {year} {2019})},\
  \Eprint {https://arxiv.org/abs/1811.06949} {arXiv:1811.06949 [hep-th]}
  \BibitemShut {NoStop}%
\bibitem [{\citenamefont {Skokov}\ \emph {et~al.}(2009)\citenamefont {Skokov},
  \citenamefont {Illarionov},\ and\ \citenamefont {Toneev}}]{Skokov:2009qp}%
  \BibitemOpen
  \bibfield  {author} {\bibinfo {author} {\bibfnamefont {V.}~\bibnamefont
  {Skokov}}, \bibinfo {author} {\bibfnamefont {A.~Y.}\ \bibnamefont
  {Illarionov}},\ and\ \bibinfo {author} {\bibfnamefont {V.}~\bibnamefont
  {Toneev}},\ }\href {https://doi.org/10.1142/S0217751X09047570} {\bibfield
  {journal} {\bibinfo  {journal} {Int. J. Mod. Phys. A}\ }\textbf {\bibinfo
  {volume} {24}},\ \bibinfo {pages} {5925} (\bibinfo {year} {2009})},\ \Eprint
  {https://arxiv.org/abs/0907.1396} {arXiv:0907.1396 [nucl-th]} \BibitemShut
  {NoStop}%
\bibitem [{\citenamefont {Bzdak}\ and\ \citenamefont
  {Skokov}(2012)}]{Bzdak:2011yy}%
  \BibitemOpen
  \bibfield  {author} {\bibinfo {author} {\bibfnamefont {A.}~\bibnamefont
  {Bzdak}}\ and\ \bibinfo {author} {\bibfnamefont {V.}~\bibnamefont {Skokov}},\
  }\href {https://doi.org/10.1016/j.physletb.2012.02.065} {\bibfield  {journal}
  {\bibinfo  {journal} {Phys. Lett. B}\ }\textbf {\bibinfo {volume} {710}},\
  \bibinfo {pages} {171} (\bibinfo {year} {2012})},\ \Eprint
  {https://arxiv.org/abs/1111.1949} {arXiv:1111.1949 [hep-ph]} \BibitemShut
  {NoStop}%
\bibitem [{\citenamefont {D'Elia}\ \emph {et~al.}(2010)\citenamefont {D'Elia},
  \citenamefont {Mukherjee},\ and\ \citenamefont {Sanfilippo}}]{DElia:2010abb}%
  \BibitemOpen
  \bibfield  {author} {\bibinfo {author} {\bibfnamefont {M.}~\bibnamefont
  {D'Elia}}, \bibinfo {author} {\bibfnamefont {S.}~\bibnamefont {Mukherjee}},\
  and\ \bibinfo {author} {\bibfnamefont {F.}~\bibnamefont {Sanfilippo}},\
  }\href {https://doi.org/10.1103/PhysRevD.82.051501} {\bibfield  {journal}
  {\bibinfo  {journal} {Phys. Rev. D}\ }\textbf {\bibinfo {volume} {82}},\
  \bibinfo {pages} {051501} (\bibinfo {year} {2010})},\ \Eprint
  {https://arxiv.org/abs/1005.5365} {arXiv:1005.5365 [hep-lat]} \BibitemShut
  {NoStop}%
\bibitem [{\citenamefont {Deng}\ and\ \citenamefont
  {Huang}(2012)}]{Deng:2012pc}%
  \BibitemOpen
  \bibfield  {author} {\bibinfo {author} {\bibfnamefont {W.-T.}\ \bibnamefont
  {Deng}}\ and\ \bibinfo {author} {\bibfnamefont {X.-G.}\ \bibnamefont
  {Huang}},\ }\href {https://doi.org/10.1103/PhysRevC.85.044907} {\bibfield
  {journal} {\bibinfo  {journal} {Phys. Rev. C}\ }\textbf {\bibinfo {volume}
  {85}},\ \bibinfo {pages} {044907} (\bibinfo {year} {2012})},\ \Eprint
  {https://arxiv.org/abs/1201.5108} {arXiv:1201.5108 [nucl-th]} \BibitemShut
  {NoStop}%
\bibitem [{\citenamefont {Tuchin}(2013)}]{Tuchin:2013ie}%
  \BibitemOpen
  \bibfield  {author} {\bibinfo {author} {\bibfnamefont {K.}~\bibnamefont
  {Tuchin}},\ }\href {https://doi.org/10.1155/2013/490495} {\bibfield
  {journal} {\bibinfo  {journal} {Adv. High Energy Phys.}\ }\textbf {\bibinfo
  {volume} {2013}},\ \bibinfo {pages} {490495} (\bibinfo {year} {2013})},\
  \Eprint {https://arxiv.org/abs/1301.0099} {arXiv:1301.0099 [hep-ph]}
  \BibitemShut {NoStop}%
\bibitem [{\citenamefont {Voronyuk}\ \emph {et~al.}(2011)\citenamefont
  {Voronyuk}, \citenamefont {Toneev}, \citenamefont {Cassing}, \citenamefont
  {Bratkovskaya}, \citenamefont {Konchakovski},\ and\ \citenamefont
  {Voloshin}}]{Voronyuk:2011jd}%
  \BibitemOpen
  \bibfield  {author} {\bibinfo {author} {\bibfnamefont {V.}~\bibnamefont
  {Voronyuk}}, \bibinfo {author} {\bibfnamefont {V.~D.}\ \bibnamefont
  {Toneev}}, \bibinfo {author} {\bibfnamefont {W.}~\bibnamefont {Cassing}},
  \bibinfo {author} {\bibfnamefont {E.~L.}\ \bibnamefont {Bratkovskaya}},
  \bibinfo {author} {\bibfnamefont {V.~P.}\ \bibnamefont {Konchakovski}},\ and\
  \bibinfo {author} {\bibfnamefont {S.~A.}\ \bibnamefont {Voloshin}},\ }\href
  {https://doi.org/10.1103/PhysRevC.83.054911} {\bibfield  {journal} {\bibinfo
  {journal} {Phys. Rev. C}\ }\textbf {\bibinfo {volume} {83}},\ \bibinfo
  {pages} {054911} (\bibinfo {year} {2011})},\ \Eprint
  {https://arxiv.org/abs/1103.4239} {arXiv:1103.4239 [nucl-th]} \BibitemShut
  {NoStop}%
\bibitem [{\citenamefont {Gubser}\ and\ \citenamefont
  {Nellore}(2008)}]{Gubser:2008ny}%
  \BibitemOpen
  \bibfield  {author} {\bibinfo {author} {\bibfnamefont {S.~S.}\ \bibnamefont
  {Gubser}}\ and\ \bibinfo {author} {\bibfnamefont {A.}~\bibnamefont
  {Nellore}},\ }\href {https://doi.org/10.1103/PhysRevD.78.086007} {\bibfield
  {journal} {\bibinfo  {journal} {Phys. Rev. D}\ }\textbf {\bibinfo {volume}
  {78}},\ \bibinfo {pages} {086007} (\bibinfo {year} {2008})},\ \Eprint
  {https://arxiv.org/abs/0804.0434} {arXiv:0804.0434 [hep-th]} \BibitemShut
  {NoStop}%
\bibitem [{\citenamefont {Gursoy}\ and\ \citenamefont
  {Kiritsis}(2008)}]{Gursoy:2007cb}%
  \BibitemOpen
  \bibfield  {author} {\bibinfo {author} {\bibfnamefont {U.}~\bibnamefont
  {Gursoy}}\ and\ \bibinfo {author} {\bibfnamefont {E.}~\bibnamefont
  {Kiritsis}},\ }\href {https://doi.org/10.1088/1126-6708/2008/02/032}
  {\bibfield  {journal} {\bibinfo  {journal} {JHEP}\ }\textbf {\bibinfo
  {volume} {02}},\ \bibinfo {pages} {032}},\ \Eprint
  {https://arxiv.org/abs/0707.1324} {arXiv:0707.1324 [hep-th]} \BibitemShut
  {NoStop}%
\bibitem [{\citenamefont {Gursoy}\ \emph {et~al.}(2011)\citenamefont {Gursoy},
  \citenamefont {Kiritsis}, \citenamefont {Mazzanti}, \citenamefont
  {Michalogiorgakis},\ and\ \citenamefont {Nitti}}]{Gursoy:2010fj}%
  \BibitemOpen
  \bibfield  {author} {\bibinfo {author} {\bibfnamefont {U.}~\bibnamefont
  {Gursoy}}, \bibinfo {author} {\bibfnamefont {E.}~\bibnamefont {Kiritsis}},
  \bibinfo {author} {\bibfnamefont {L.}~\bibnamefont {Mazzanti}}, \bibinfo
  {author} {\bibfnamefont {G.}~\bibnamefont {Michalogiorgakis}},\ and\ \bibinfo
  {author} {\bibfnamefont {F.}~\bibnamefont {Nitti}},\ }\href
  {https://doi.org/10.1007/978-3-642-04864-7_4} {\bibfield  {journal} {\bibinfo
   {journal} {Lect. Notes Phys.}\ }\textbf {\bibinfo {volume} {828}},\ \bibinfo
  {pages} {79} (\bibinfo {year} {2011})},\ \Eprint
  {https://arxiv.org/abs/1006.5461} {arXiv:1006.5461 [hep-th]} \BibitemShut
  {NoStop}%
\bibitem [{\citenamefont {DeWolfe}\ \emph {et~al.}(2011)\citenamefont
  {DeWolfe}, \citenamefont {Gubser},\ and\ \citenamefont
  {Rosen}}]{DeWolfe:2010he}%
  \BibitemOpen
  \bibfield  {author} {\bibinfo {author} {\bibfnamefont {O.}~\bibnamefont
  {DeWolfe}}, \bibinfo {author} {\bibfnamefont {S.~S.}\ \bibnamefont
  {Gubser}},\ and\ \bibinfo {author} {\bibfnamefont {C.}~\bibnamefont
  {Rosen}},\ }\href {https://doi.org/10.1103/PhysRevD.83.086005} {\bibfield
  {journal} {\bibinfo  {journal} {Phys. Rev. D}\ }\textbf {\bibinfo {volume}
  {83}},\ \bibinfo {pages} {086005} (\bibinfo {year} {2011})},\ \Eprint
  {https://arxiv.org/abs/1012.1864} {arXiv:1012.1864 [hep-th]} \BibitemShut
  {NoStop}%
\bibitem [{\citenamefont {Dudal}\ and\ \citenamefont
  {Mahapatra}(2017{\natexlab{a}})}]{Dudal:2017max}%
  \BibitemOpen
  \bibfield  {author} {\bibinfo {author} {\bibfnamefont {D.}~\bibnamefont
  {Dudal}}\ and\ \bibinfo {author} {\bibfnamefont {S.}~\bibnamefont
  {Mahapatra}},\ }\href {https://doi.org/10.1103/PhysRevD.96.126010} {\bibfield
   {journal} {\bibinfo  {journal} {Phys. Rev. D}\ }\textbf {\bibinfo {volume}
  {96}},\ \bibinfo {pages} {126010} (\bibinfo {year} {2017}{\natexlab{a}})},\
  \Eprint {https://arxiv.org/abs/1708.06995} {arXiv:1708.06995 [hep-th]}
  \BibitemShut {NoStop}%
\bibitem [{\citenamefont {Cai}\ \emph {et~al.}(2012)\citenamefont {Cai},
  \citenamefont {He},\ and\ \citenamefont {Li}}]{Cai:2012xh}%
  \BibitemOpen
  \bibfield  {author} {\bibinfo {author} {\bibfnamefont {R.-G.}\ \bibnamefont
  {Cai}}, \bibinfo {author} {\bibfnamefont {S.}~\bibnamefont {He}},\ and\
  \bibinfo {author} {\bibfnamefont {D.}~\bibnamefont {Li}},\ }\href
  {https://doi.org/10.1007/JHEP03(2012)033} {\bibfield  {journal} {\bibinfo
  {journal} {JHEP}\ }\textbf {\bibinfo {volume} {03}},\ \bibinfo {pages}
  {033}},\ \Eprint {https://arxiv.org/abs/1201.0820} {arXiv:1201.0820 [hep-th]}
  \BibitemShut {NoStop}%
\bibitem [{\citenamefont {He}\ \emph {et~al.}(2013{\natexlab{a}})\citenamefont
  {He}, \citenamefont {Wu}, \citenamefont {Yang},\ and\ \citenamefont
  {Yuan}}]{He:2013qq}%
  \BibitemOpen
  \bibfield  {author} {\bibinfo {author} {\bibfnamefont {S.}~\bibnamefont
  {He}}, \bibinfo {author} {\bibfnamefont {S.-Y.}\ \bibnamefont {Wu}}, \bibinfo
  {author} {\bibfnamefont {Y.}~\bibnamefont {Yang}},\ and\ \bibinfo {author}
  {\bibfnamefont {P.-H.}\ \bibnamefont {Yuan}},\ }\href
  {https://doi.org/10.1007/JHEP04(2013)093} {\bibfield  {journal} {\bibinfo
  {journal} {JHEP}\ }\textbf {\bibinfo {volume} {04}},\ \bibinfo {pages}
  {093}},\ \Eprint {https://arxiv.org/abs/1301.0385} {arXiv:1301.0385 [hep-th]}
  \BibitemShut {NoStop}%
\bibitem [{\citenamefont {Bohra}\ \emph {et~al.}(2021)\citenamefont {Bohra},
  \citenamefont {Dudal}, \citenamefont {Hajilou},\ and\ \citenamefont
  {Mahapatra}}]{bohra2021chiral}%
  \BibitemOpen
  \bibfield  {author} {\bibinfo {author} {\bibfnamefont {H.}~\bibnamefont
  {Bohra}}, \bibinfo {author} {\bibfnamefont {D.}~\bibnamefont {Dudal}},
  \bibinfo {author} {\bibfnamefont {A.}~\bibnamefont {Hajilou}},\ and\ \bibinfo
  {author} {\bibfnamefont {S.}~\bibnamefont {Mahapatra}},\ }\href
  {https://doi.org/10.1103/PhysRevD.103.086021} {\bibfield  {journal} {\bibinfo
   {journal} {Phys. Rev. D}\ }\textbf {\bibinfo {volume} {103}},\ \bibinfo
  {pages} {086021} (\bibinfo {year} {2021})},\ \Eprint
  {https://arxiv.org/abs/2010.04578} {arXiv:2010.04578 [hep-th]} \BibitemShut
  {NoStop}%
\bibitem [{\citenamefont {Dudal}\ \emph {et~al.}(2021)\citenamefont {Dudal},
  \citenamefont {Hajilou},\ and\ \citenamefont {Mahapatra}}]{Dudal:2021jav}%
  \BibitemOpen
  \bibfield  {author} {\bibinfo {author} {\bibfnamefont {D.}~\bibnamefont
  {Dudal}}, \bibinfo {author} {\bibfnamefont {A.}~\bibnamefont {Hajilou}},\
  and\ \bibinfo {author} {\bibfnamefont {S.}~\bibnamefont {Mahapatra}},\ }\href
  {https://doi.org/10.1140/epja/s10050-021-00461-4} {\bibfield  {journal}
  {\bibinfo  {journal} {Eur. Phys. J. A}\ }\textbf {\bibinfo {volume} {57}},\
  \bibinfo {pages} {142} (\bibinfo {year} {2021})},\ \Eprint
  {https://arxiv.org/abs/2103.01185} {arXiv:2103.01185 [hep-th]} \BibitemShut
  {NoStop}%
\bibitem [{\citenamefont {Rougemont}\ \emph {et~al.}(2016)\citenamefont
  {Rougemont}, \citenamefont {Critelli},\ and\ \citenamefont
  {Noronha}}]{Rougemont:2015oea}%
  \BibitemOpen
  \bibfield  {author} {\bibinfo {author} {\bibfnamefont {R.}~\bibnamefont
  {Rougemont}}, \bibinfo {author} {\bibfnamefont {R.}~\bibnamefont
  {Critelli}},\ and\ \bibinfo {author} {\bibfnamefont {J.}~\bibnamefont
  {Noronha}},\ }\href {https://doi.org/10.1103/PhysRevD.93.045013} {\bibfield
  {journal} {\bibinfo  {journal} {Phys. Rev. D}\ }\textbf {\bibinfo {volume}
  {93}},\ \bibinfo {pages} {045013} (\bibinfo {year} {2016})},\ \Eprint
  {https://arxiv.org/abs/1505.07894} {arXiv:1505.07894 [hep-th]} \BibitemShut
  {NoStop}%
\bibitem [{\citenamefont {Finazzo}\ \emph {et~al.}(2016)\citenamefont
  {Finazzo}, \citenamefont {Critelli}, \citenamefont {Rougemont},\ and\
  \citenamefont {Noronha}}]{Finazzo:2016mhm}%
  \BibitemOpen
  \bibfield  {author} {\bibinfo {author} {\bibfnamefont {S.~I.}\ \bibnamefont
  {Finazzo}}, \bibinfo {author} {\bibfnamefont {R.}~\bibnamefont {Critelli}},
  \bibinfo {author} {\bibfnamefont {R.}~\bibnamefont {Rougemont}},\ and\
  \bibinfo {author} {\bibfnamefont {J.}~\bibnamefont {Noronha}},\ }\href
  {https://doi.org/10.1103/PhysRevD.94.054020} {\bibfield  {journal} {\bibinfo
  {journal} {Phys. Rev. D}\ }\textbf {\bibinfo {volume} {94}},\ \bibinfo
  {pages} {054020} (\bibinfo {year} {2016})},\ \bibinfo {note} {[Erratum:
  Phys.Rev.D 96, 019903 (2017)]},\ \Eprint {https://arxiv.org/abs/1605.06061}
  {arXiv:1605.06061 [hep-ph]} \BibitemShut {NoStop}%
\bibitem [{\citenamefont {Aref'eva}\ \emph
  {et~al.}(2024{\natexlab{a}})\citenamefont {Aref'eva}, \citenamefont
  {Hajilou}, \citenamefont {Nikolaev},\ and\ \citenamefont
  {Slepov}}]{Arefeva:2024xmg}%
  \BibitemOpen
  \bibfield  {author} {\bibinfo {author} {\bibfnamefont {I.~Y.}\ \bibnamefont
  {Aref'eva}}, \bibinfo {author} {\bibfnamefont {A.}~\bibnamefont {Hajilou}},
  \bibinfo {author} {\bibfnamefont {A.}~\bibnamefont {Nikolaev}},\ and\
  \bibinfo {author} {\bibfnamefont {P.}~\bibnamefont {Slepov}},\ }\href
  {https://doi.org/10.1103/PhysRevD.110.086021} {\bibfield  {journal} {\bibinfo
   {journal} {Phys. Rev. D}\ }\textbf {\bibinfo {volume} {110}},\ \bibinfo
  {pages} {086021} (\bibinfo {year} {2024}{\natexlab{a}})},\ \Eprint
  {https://arxiv.org/abs/2407.11924} {arXiv:2407.11924 [hep-th]} \BibitemShut
  {NoStop}%
\bibitem [{\citenamefont {Bohra}\ \emph {et~al.}(2020)\citenamefont {Bohra},
  \citenamefont {Dudal}, \citenamefont {Hajilou},\ and\ \citenamefont
  {Mahapatra}}]{Bohra:2019ebj}%
  \BibitemOpen
  \bibfield  {author} {\bibinfo {author} {\bibfnamefont {H.}~\bibnamefont
  {Bohra}}, \bibinfo {author} {\bibfnamefont {D.}~\bibnamefont {Dudal}},
  \bibinfo {author} {\bibfnamefont {A.}~\bibnamefont {Hajilou}},\ and\ \bibinfo
  {author} {\bibfnamefont {S.}~\bibnamefont {Mahapatra}},\ }\href
  {https://doi.org/10.1016/j.physletb.2019.135184} {\bibfield  {journal}
  {\bibinfo  {journal} {Phys. Lett. B}\ }\textbf {\bibinfo {volume} {801}},\
  \bibinfo {pages} {135184} (\bibinfo {year} {2020})},\ \Eprint
  {https://arxiv.org/abs/1907.01852} {arXiv:1907.01852 [hep-th]} \BibitemShut
  {NoStop}%
\bibitem [{\citenamefont {Gursoy}\ \emph {et~al.}(2018)\citenamefont {Gursoy},
  \citenamefont {Jarvinen},\ and\ \citenamefont {Nijs}}]{Gursoy:2017wzz}%
  \BibitemOpen
  \bibfield  {author} {\bibinfo {author} {\bibfnamefont {U.}~\bibnamefont
  {Gursoy}}, \bibinfo {author} {\bibfnamefont {M.}~\bibnamefont {Jarvinen}},\
  and\ \bibinfo {author} {\bibfnamefont {G.}~\bibnamefont {Nijs}},\ }\href
  {https://doi.org/10.1103/PhysRevLett.120.242002} {\bibfield  {journal}
  {\bibinfo  {journal} {Phys. Rev. Lett.}\ }\textbf {\bibinfo {volume} {120}},\
  \bibinfo {pages} {242002} (\bibinfo {year} {2018})},\ \Eprint
  {https://arxiv.org/abs/1707.00872} {arXiv:1707.00872 [hep-th]} \BibitemShut
  {NoStop}%
\bibitem [{\citenamefont {Aref'eva}\ \emph
  {et~al.}(2021{\natexlab{a}})\citenamefont {Aref'eva}, \citenamefont {Rannu},\
  and\ \citenamefont {Slepov}}]{Arefeva:2020vae}%
  \BibitemOpen
  \bibfield  {author} {\bibinfo {author} {\bibfnamefont {I.~Y.}\ \bibnamefont
  {Aref'eva}}, \bibinfo {author} {\bibfnamefont {K.}~\bibnamefont {Rannu}},\
  and\ \bibinfo {author} {\bibfnamefont {P.}~\bibnamefont {Slepov}},\ }\href
  {https://doi.org/10.1007/JHEP07(2021)161} {\bibfield  {journal} {\bibinfo
  {journal} {JHEP}\ }\textbf {\bibinfo {volume} {07}},\ \bibinfo {pages}
  {161}},\ \Eprint {https://arxiv.org/abs/2011.07023} {arXiv:2011.07023
  [hep-th]} \BibitemShut {NoStop}%
\bibitem [{\citenamefont {Aref'eva}\ \emph
  {et~al.}(2024{\natexlab{b}})\citenamefont {Aref'eva}, \citenamefont {Rannu},\
  and\ \citenamefont {Slepov}}]{Arefeva:2024mtl}%
  \BibitemOpen
  \bibfield  {author} {\bibinfo {author} {\bibfnamefont {I.~Y.}\ \bibnamefont
  {Aref'eva}}, \bibinfo {author} {\bibfnamefont {K.}~\bibnamefont {Rannu}},\
  and\ \bibinfo {author} {\bibfnamefont {P.}~\bibnamefont {Slepov}},\
  }\href@noop {} {\bibfield  {journal} {\bibinfo  {journal} {arXiv}\ }
  (\bibinfo {year} {2024}{\natexlab{b}})},\ \Eprint
  {https://arxiv.org/abs/2409.12131} {arXiv:2409.12131 [hep-th]} \BibitemShut
  {NoStop}%
\bibitem [{\citenamefont {Dudal}\ and\ \citenamefont
  {Mahapatra}(2018)}]{Dudal:2018ztm}%
  \BibitemOpen
  \bibfield  {author} {\bibinfo {author} {\bibfnamefont {D.}~\bibnamefont
  {Dudal}}\ and\ \bibinfo {author} {\bibfnamefont {S.}~\bibnamefont
  {Mahapatra}},\ }\href {https://doi.org/10.1007/JHEP07(2018)120} {\bibfield
  {journal} {\bibinfo  {journal} {JHEP}\ }\textbf {\bibinfo {volume} {07}},\
  \bibinfo {pages} {120}},\ \Eprint {https://arxiv.org/abs/1805.02938}
  {arXiv:1805.02938 [hep-th]} \BibitemShut {NoStop}%
\bibitem [{\citenamefont {Mahapatra}(2019)}]{Mahapatra:2019uql}%
  \BibitemOpen
  \bibfield  {author} {\bibinfo {author} {\bibfnamefont {S.}~\bibnamefont
  {Mahapatra}},\ }\href {https://doi.org/10.1007/JHEP04(2019)137} {\bibfield
  {journal} {\bibinfo  {journal} {JHEP}\ }\textbf {\bibinfo {volume} {04}},\
  \bibinfo {pages} {137}},\ \Eprint {https://arxiv.org/abs/1903.05927}
  {arXiv:1903.05927 [hep-th]} \BibitemShut {NoStop}%
\bibitem [{\citenamefont {Dudal}\ and\ \citenamefont
  {Mahapatra}(2017{\natexlab{b}})}]{Dudal:2016joz}%
  \BibitemOpen
  \bibfield  {author} {\bibinfo {author} {\bibfnamefont {D.}~\bibnamefont
  {Dudal}}\ and\ \bibinfo {author} {\bibfnamefont {S.}~\bibnamefont
  {Mahapatra}},\ }\href {https://doi.org/10.1007/JHEP04(2017)031} {\bibfield
  {journal} {\bibinfo  {journal} {JHEP}\ }\textbf {\bibinfo {volume} {04}},\
  \bibinfo {pages} {031}},\ \Eprint {https://arxiv.org/abs/1612.06248}
  {arXiv:1612.06248 [hep-th]} \BibitemShut {NoStop}%
\bibitem [{\citenamefont {Rougemont}\ \emph {et~al.}(2015)\citenamefont
  {Rougemont}, \citenamefont {Critelli},\ and\ \citenamefont
  {Noronha}}]{Rougemont:2014efa}%
  \BibitemOpen
  \bibfield  {author} {\bibinfo {author} {\bibfnamefont {R.}~\bibnamefont
  {Rougemont}}, \bibinfo {author} {\bibfnamefont {R.}~\bibnamefont
  {Critelli}},\ and\ \bibinfo {author} {\bibfnamefont {J.}~\bibnamefont
  {Noronha}},\ }\href {https://doi.org/10.1103/PhysRevD.91.066001} {\bibfield
  {journal} {\bibinfo  {journal} {Phys. Rev. D}\ }\textbf {\bibinfo {volume}
  {91}},\ \bibinfo {pages} {066001} (\bibinfo {year} {2015})},\ \Eprint
  {https://arxiv.org/abs/1409.0556} {arXiv:1409.0556 [hep-th]} \BibitemShut
  {NoStop}%
\bibitem [{\citenamefont {Fuini}\ and\ \citenamefont
  {Yaffe}(2015)}]{Fuini:2015hba}%
  \BibitemOpen
  \bibfield  {author} {\bibinfo {author} {\bibfnamefont {J.~F.}\ \bibnamefont
  {Fuini}}\ and\ \bibinfo {author} {\bibfnamefont {L.~G.}\ \bibnamefont
  {Yaffe}},\ }\href {https://doi.org/10.1007/JHEP07(2015)116} {\bibfield
  {journal} {\bibinfo  {journal} {JHEP}\ }\textbf {\bibinfo {volume} {07}},\
  \bibinfo {pages} {116}},\ \Eprint {https://arxiv.org/abs/1503.07148}
  {arXiv:1503.07148 [hep-th]} \BibitemShut {NoStop}%
\bibitem [{\citenamefont {Cartwright}\ and\ \citenamefont
  {Kaminski}(2019)}]{Cartwright:2019opv}%
  \BibitemOpen
  \bibfield  {author} {\bibinfo {author} {\bibfnamefont {C.}~\bibnamefont
  {Cartwright}}\ and\ \bibinfo {author} {\bibfnamefont {M.}~\bibnamefont
  {Kaminski}},\ }\href {https://doi.org/10.1007/JHEP09(2019)072} {\bibfield
  {journal} {\bibinfo  {journal} {JHEP}\ }\textbf {\bibinfo {volume} {09}},\
  \bibinfo {pages} {072}},\ \Eprint {https://arxiv.org/abs/1904.11507}
  {arXiv:1904.11507 [hep-th]} \BibitemShut {NoStop}%
\bibitem [{\citenamefont {Fukushima}\ and\ \citenamefont
  {Okutsu}(2022)}]{Fukushima:2021got}%
  \BibitemOpen
  \bibfield  {author} {\bibinfo {author} {\bibfnamefont {K.}~\bibnamefont
  {Fukushima}}\ and\ \bibinfo {author} {\bibfnamefont {A.}~\bibnamefont
  {Okutsu}},\ }\href {https://doi.org/10.1103/PhysRevD.105.054016} {\bibfield
  {journal} {\bibinfo  {journal} {Phys. Rev. D}\ }\textbf {\bibinfo {volume}
  {105}},\ \bibinfo {pages} {054016} (\bibinfo {year} {2022})},\ \Eprint
  {https://arxiv.org/abs/2106.07968} {arXiv:2106.07968 [hep-ph]} \BibitemShut
  {NoStop}%
\bibitem [{\citenamefont {Ballon-Bayona}\ \emph {et~al.}(2022)\citenamefont
  {Ballon-Bayona}, \citenamefont {Shock},\ and\ \citenamefont
  {Zoakos}}]{Ballon-Bayona:2022uyy}%
  \BibitemOpen
  \bibfield  {author} {\bibinfo {author} {\bibfnamefont {A.}~\bibnamefont
  {Ballon-Bayona}}, \bibinfo {author} {\bibfnamefont {J.~P.}\ \bibnamefont
  {Shock}},\ and\ \bibinfo {author} {\bibfnamefont {D.}~\bibnamefont
  {Zoakos}},\ }\href {https://doi.org/10.1007/JHEP06(2022)154} {\bibfield
  {journal} {\bibinfo  {journal} {JHEP}\ }\textbf {\bibinfo {volume} {06}},\
  \bibinfo {pages} {154}},\ \Eprint {https://arxiv.org/abs/2203.00050}
  {arXiv:2203.00050 [hep-th]} \BibitemShut {NoStop}%
\bibitem [{\citenamefont {Rodrigues}\ \emph {et~al.}(2018)\citenamefont
  {Rodrigues}, \citenamefont {Folco~Capossoli},\ and\ \citenamefont
  {Boschi-Filho}}]{Rodrigues:2017cha}%
  \BibitemOpen
  \bibfield  {author} {\bibinfo {author} {\bibfnamefont {D.~M.}\ \bibnamefont
  {Rodrigues}}, \bibinfo {author} {\bibfnamefont {E.}~\bibnamefont
  {Folco~Capossoli}},\ and\ \bibinfo {author} {\bibfnamefont {H.}~\bibnamefont
  {Boschi-Filho}},\ }\href {https://doi.org/10.1016/j.physletb.2018.02.049}
  {\bibfield  {journal} {\bibinfo  {journal} {Phys. Lett. B}\ }\textbf
  {\bibinfo {volume} {780}},\ \bibinfo {pages} {37} (\bibinfo {year} {2018})},\
  \Eprint {https://arxiv.org/abs/1709.09258} {arXiv:1709.09258 [hep-th]}
  \BibitemShut {NoStop}%
\bibitem [{\citenamefont {Aref'eva}\ \emph
  {et~al.}(2023{\natexlab{a}})\citenamefont {Aref'eva}, \citenamefont
  {Hajilou}, \citenamefont {Rannu},\ and\ \citenamefont
  {Slepov}}]{Arefeva:2023jjh}%
  \BibitemOpen
  \bibfield  {author} {\bibinfo {author} {\bibfnamefont {I.~Y.}\ \bibnamefont
  {Aref'eva}}, \bibinfo {author} {\bibfnamefont {A.}~\bibnamefont {Hajilou}},
  \bibinfo {author} {\bibfnamefont {K.}~\bibnamefont {Rannu}},\ and\ \bibinfo
  {author} {\bibfnamefont {P.}~\bibnamefont {Slepov}},\ }\href
  {https://doi.org/10.1140/epjc/s10052-023-12309-w} {\bibfield  {journal}
  {\bibinfo  {journal} {Eur. Phys. J. C}\ }\textbf {\bibinfo {volume} {83}},\
  \bibinfo {pages} {1143} (\bibinfo {year} {2023}{\natexlab{a}})},\ \Eprint
  {https://arxiv.org/abs/2305.06345} {arXiv:2305.06345 [hep-th]} \BibitemShut
  {NoStop}%
\bibitem [{\citenamefont {Aref'eva}\ \emph {et~al.}(2022)\citenamefont
  {Aref'eva}, \citenamefont {Ermakov},\ and\ \citenamefont
  {Slepov}}]{Arefeva:2021jpa}%
  \BibitemOpen
  \bibfield  {author} {\bibinfo {author} {\bibfnamefont {I.~Y.}\ \bibnamefont
  {Aref'eva}}, \bibinfo {author} {\bibfnamefont {A.}~\bibnamefont {Ermakov}},\
  and\ \bibinfo {author} {\bibfnamefont {P.}~\bibnamefont {Slepov}},\ }\href
  {https://doi.org/10.1140/epjc/s10052-022-10025-5} {\bibfield  {journal}
  {\bibinfo  {journal} {Eur. Phys. J. C}\ }\textbf {\bibinfo {volume} {82}},\
  \bibinfo {pages} {85} (\bibinfo {year} {2022})},\ \Eprint
  {https://arxiv.org/abs/2104.14582} {arXiv:2104.14582 [hep-th]} \BibitemShut
  {NoStop}%
\bibitem [{\citenamefont {Jena}\ \emph {et~al.}(2022)\citenamefont {Jena},
  \citenamefont {Shukla}, \citenamefont {Dudal},\ and\ \citenamefont
  {Mahapatra}}]{Jena:2022nzw}%
  \BibitemOpen
  \bibfield  {author} {\bibinfo {author} {\bibfnamefont {S.~S.}\ \bibnamefont
  {Jena}}, \bibinfo {author} {\bibfnamefont {B.}~\bibnamefont {Shukla}},
  \bibinfo {author} {\bibfnamefont {D.}~\bibnamefont {Dudal}},\ and\ \bibinfo
  {author} {\bibfnamefont {S.}~\bibnamefont {Mahapatra}},\ }\href
  {https://doi.org/10.1103/PhysRevD.105.086011} {\bibfield  {journal} {\bibinfo
   {journal} {Phys. Rev. D}\ }\textbf {\bibinfo {volume} {105}},\ \bibinfo
  {pages} {086011} (\bibinfo {year} {2022})},\ \Eprint
  {https://arxiv.org/abs/2202.01486} {arXiv:2202.01486 [hep-th]} \BibitemShut
  {NoStop}%
\bibitem [{\citenamefont {Jain}\ \emph {et~al.}(2023)\citenamefont {Jain},
  \citenamefont {Jena},\ and\ \citenamefont {Mahapatra}}]{Jain:2022hxl}%
  \BibitemOpen
  \bibfield  {author} {\bibinfo {author} {\bibfnamefont {P.}~\bibnamefont
  {Jain}}, \bibinfo {author} {\bibfnamefont {S.~S.}\ \bibnamefont {Jena}},\
  and\ \bibinfo {author} {\bibfnamefont {S.}~\bibnamefont {Mahapatra}},\ }\href
  {https://doi.org/10.1103/PhysRevD.107.086016} {\bibfield  {journal} {\bibinfo
   {journal} {Phys. Rev. D}\ }\textbf {\bibinfo {volume} {107}},\ \bibinfo
  {pages} {086016} (\bibinfo {year} {2023})},\ \Eprint
  {https://arxiv.org/abs/2209.15355} {arXiv:2209.15355 [hep-th]} \BibitemShut
  {NoStop}%
\bibitem [{\citenamefont {Aref'eva}\ \emph
  {et~al.}(2023{\natexlab{b}})\citenamefont {Aref'eva}, \citenamefont
  {Ermakov}, \citenamefont {Rannu},\ and\ \citenamefont
  {Slepov}}]{Arefeva:2022avn}%
  \BibitemOpen
  \bibfield  {author} {\bibinfo {author} {\bibfnamefont {I.~Y.}\ \bibnamefont
  {Aref'eva}}, \bibinfo {author} {\bibfnamefont {A.}~\bibnamefont {Ermakov}},
  \bibinfo {author} {\bibfnamefont {K.}~\bibnamefont {Rannu}},\ and\ \bibinfo
  {author} {\bibfnamefont {P.}~\bibnamefont {Slepov}},\ }\href
  {https://doi.org/10.1140/epjc/s10052-022-11166-3} {\bibfield  {journal}
  {\bibinfo  {journal} {Eur. Phys. J. C}\ }\textbf {\bibinfo {volume} {83}},\
  \bibinfo {pages} {79} (\bibinfo {year} {2023}{\natexlab{b}})},\ \Eprint
  {https://arxiv.org/abs/2203.12539} {arXiv:2203.12539 [hep-th]} \BibitemShut
  {NoStop}%
\bibitem [{\citenamefont {Chen}\ \emph {et~al.}(2022)\citenamefont {Chen},
  \citenamefont {Zhang},\ and\ \citenamefont {Hou}}]{Chen:2021gop}%
  \BibitemOpen
  \bibfield  {author} {\bibinfo {author} {\bibfnamefont {X.}~\bibnamefont
  {Chen}}, \bibinfo {author} {\bibfnamefont {L.}~\bibnamefont {Zhang}},\ and\
  \bibinfo {author} {\bibfnamefont {D.}~\bibnamefont {Hou}},\ }\href
  {https://doi.org/10.1088/1674-1137/ac5c2d} {\bibfield  {journal} {\bibinfo
  {journal} {Chin. Phys. C}\ }\textbf {\bibinfo {volume} {46}},\ \bibinfo
  {pages} {073101} (\bibinfo {year} {2022})},\ \Eprint
  {https://arxiv.org/abs/2108.03840} {arXiv:2108.03840 [hep-ph]} \BibitemShut
  {NoStop}%
\bibitem [{\citenamefont {Braga}\ and\ \citenamefont
  {da~Mata}(2020)}]{Braga:2020hhs}%
  \BibitemOpen
  \bibfield  {author} {\bibinfo {author} {\bibfnamefont {N.~R.~F.}\
  \bibnamefont {Braga}}\ and\ \bibinfo {author} {\bibfnamefont
  {R.}~\bibnamefont {da~Mata}},\ }\href
  {https://doi.org/10.1016/j.physletb.2020.135918} {\bibfield  {journal}
  {\bibinfo  {journal} {Phys. Lett. B}\ }\textbf {\bibinfo {volume} {811}},\
  \bibinfo {pages} {135918} (\bibinfo {year} {2020})},\ \Eprint
  {https://arxiv.org/abs/2008.10457} {arXiv:2008.10457 [hep-th]} \BibitemShut
  {NoStop}%
\bibitem [{\citenamefont {Zhou}\ \emph {et~al.}(2020)\citenamefont {Zhou},
  \citenamefont {Chen}, \citenamefont {Zhao},\ and\ \citenamefont
  {Ping}}]{Zhou:2020ssi}%
  \BibitemOpen
  \bibfield  {author} {\bibinfo {author} {\bibfnamefont {J.}~\bibnamefont
  {Zhou}}, \bibinfo {author} {\bibfnamefont {X.}~\bibnamefont {Chen}}, \bibinfo
  {author} {\bibfnamefont {Y.-Q.}\ \bibnamefont {Zhao}},\ and\ \bibinfo
  {author} {\bibfnamefont {J.}~\bibnamefont {Ping}},\ }\href
  {https://doi.org/10.1103/PhysRevD.102.086020} {\bibfield  {journal} {\bibinfo
   {journal} {Phys. Rev. D}\ }\textbf {\bibinfo {volume} {102}},\ \bibinfo
  {pages} {086020} (\bibinfo {year} {2020})},\ \Eprint
  {https://arxiv.org/abs/2006.09062} {arXiv:2006.09062 [hep-ph]} \BibitemShut
  {NoStop}%
\bibitem [{\citenamefont {Ballon-Bayona}\ \emph {et~al.}(2020)\citenamefont
  {Ballon-Bayona}, \citenamefont {Shock},\ and\ \citenamefont
  {Zoakos}}]{Ballon-Bayona:2020xtf}%
  \BibitemOpen
  \bibfield  {author} {\bibinfo {author} {\bibfnamefont {A.}~\bibnamefont
  {Ballon-Bayona}}, \bibinfo {author} {\bibfnamefont {J.~P.}\ \bibnamefont
  {Shock}},\ and\ \bibinfo {author} {\bibfnamefont {D.}~\bibnamefont
  {Zoakos}},\ }\href {https://doi.org/10.1007/JHEP10(2020)193} {\bibfield
  {journal} {\bibinfo  {journal} {JHEP}\ }\textbf {\bibinfo {volume} {10}},\
  \bibinfo {pages} {193}},\ \Eprint {https://arxiv.org/abs/2005.00500}
  {arXiv:2005.00500 [hep-th]} \BibitemShut {NoStop}%
\bibitem [{\citenamefont {Zhao}\ and\ \citenamefont
  {Hou}(2022)}]{Zhao:2021ogc}%
  \BibitemOpen
  \bibfield  {author} {\bibinfo {author} {\bibfnamefont {Y.-Q.}\ \bibnamefont
  {Zhao}}\ and\ \bibinfo {author} {\bibfnamefont {D.}~\bibnamefont {Hou}},\
  }\href {https://doi.org/10.1140/epjc/s10052-022-11065-7} {\bibfield
  {journal} {\bibinfo  {journal} {Eur. Phys. J. C}\ }\textbf {\bibinfo {volume}
  {82}},\ \bibinfo {pages} {1102} (\bibinfo {year} {2022})},\ \Eprint
  {https://arxiv.org/abs/2108.08479} {arXiv:2108.08479 [hep-ph]} \BibitemShut
  {NoStop}%
\bibitem [{\citenamefont {Dudal}\ and\ \citenamefont
  {Mertens}(2018)}]{Dudal:2018rki}%
  \BibitemOpen
  \bibfield  {author} {\bibinfo {author} {\bibfnamefont {D.}~\bibnamefont
  {Dudal}}\ and\ \bibinfo {author} {\bibfnamefont {T.~G.}\ \bibnamefont
  {Mertens}},\ }\href {https://doi.org/10.1103/PhysRevD.97.054035} {\bibfield
  {journal} {\bibinfo  {journal} {Phys. Rev. D}\ }\textbf {\bibinfo {volume}
  {97}},\ \bibinfo {pages} {054035} (\bibinfo {year} {2018})},\ \Eprint
  {https://arxiv.org/abs/1802.02805} {arXiv:1802.02805 [hep-th]} \BibitemShut
  {NoStop}%
\bibitem [{\citenamefont {Jena}\ \emph {et~al.}(2024)\citenamefont {Jena},
  \citenamefont {Barman}, \citenamefont {Toniato}, \citenamefont {Dudal},\ and\
  \citenamefont {Mahapatra}}]{Jena:2024cqs}%
  \BibitemOpen
  \bibfield  {author} {\bibinfo {author} {\bibfnamefont {S.~S.}\ \bibnamefont
  {Jena}}, \bibinfo {author} {\bibfnamefont {J.}~\bibnamefont {Barman}},
  \bibinfo {author} {\bibfnamefont {B.}~\bibnamefont {Toniato}}, \bibinfo
  {author} {\bibfnamefont {D.}~\bibnamefont {Dudal}},\ and\ \bibinfo {author}
  {\bibfnamefont {S.}~\bibnamefont {Mahapatra}},\ }\href
  {https://doi.org/10.1007/JHEP12(2024)096} {\bibfield  {journal} {\bibinfo
  {journal} {JHEP}\ }\textbf {\bibinfo {volume} {12}},\ \bibinfo {pages}
  {096}},\ \Eprint {https://arxiv.org/abs/2408.14813} {arXiv:2408.14813
  [hep-th]} \BibitemShut {NoStop}%
\bibitem [{\citenamefont {Ballon-Bayona}\ \emph {et~al.}(2024)\citenamefont
  {Ballon-Bayona}, \citenamefont {Bartz}, \citenamefont {Mamani},\ and\
  \citenamefont {Rodrigues}}]{Ballon-Bayona:2024twa}%
  \BibitemOpen
  \bibfield  {author} {\bibinfo {author} {\bibfnamefont {A.}~\bibnamefont
  {Ballon-Bayona}}, \bibinfo {author} {\bibfnamefont {S.}~\bibnamefont
  {Bartz}}, \bibinfo {author} {\bibfnamefont {L.~A.~H.}\ \bibnamefont
  {Mamani}},\ and\ \bibinfo {author} {\bibfnamefont {D.~M.}\ \bibnamefont
  {Rodrigues}},\ }\href@noop {} {\bibfield  {journal} {\bibinfo  {journal}
  {arXiv}\ } (\bibinfo {year} {2024})},\ \Eprint
  {https://arxiv.org/abs/2410.23471} {arXiv:2410.23471 [hep-ph]} \BibitemShut
  {NoStop}%
\bibitem [{\citenamefont {Braga}\ \emph {et~al.}(2022)\citenamefont {Braga},
  \citenamefont {Ferreira},\ and\ \citenamefont {Ferreira}}]{Braga:2021fey}%
  \BibitemOpen
  \bibfield  {author} {\bibinfo {author} {\bibfnamefont {N.~R.~F.}\
  \bibnamefont {Braga}}, \bibinfo {author} {\bibfnamefont {Y.~F.}\ \bibnamefont
  {Ferreira}},\ and\ \bibinfo {author} {\bibfnamefont {L.~F.}\ \bibnamefont
  {Ferreira}},\ }\href {https://doi.org/10.1103/PhysRevD.105.114044} {\bibfield
   {journal} {\bibinfo  {journal} {Phys. Rev. D}\ }\textbf {\bibinfo {volume}
  {105}},\ \bibinfo {pages} {114044} (\bibinfo {year} {2022})},\ \Eprint
  {https://arxiv.org/abs/2110.04560} {arXiv:2110.04560 [hep-th]} \BibitemShut
  {NoStop}%
\bibitem [{\citenamefont {Rougemont}\ \emph {et~al.}(2024)\citenamefont
  {Rougemont}, \citenamefont {Grefa}, \citenamefont {Hippert}, \citenamefont
  {Noronha}, \citenamefont {Noronha-Hostler}, \citenamefont {Portillo},\ and\
  \citenamefont {Ratti}}]{Rougemont:2023gfz}%
  \BibitemOpen
  \bibfield  {author} {\bibinfo {author} {\bibfnamefont {R.}~\bibnamefont
  {Rougemont}}, \bibinfo {author} {\bibfnamefont {J.}~\bibnamefont {Grefa}},
  \bibinfo {author} {\bibfnamefont {M.}~\bibnamefont {Hippert}}, \bibinfo
  {author} {\bibfnamefont {J.}~\bibnamefont {Noronha}}, \bibinfo {author}
  {\bibfnamefont {J.}~\bibnamefont {Noronha-Hostler}}, \bibinfo {author}
  {\bibfnamefont {I.}~\bibnamefont {Portillo}},\ and\ \bibinfo {author}
  {\bibfnamefont {C.}~\bibnamefont {Ratti}},\ }\href
  {https://doi.org/10.1016/j.ppnp.2023.104093} {\bibfield  {journal} {\bibinfo
  {journal} {Prog. Part. Nucl. Phys.}\ }\textbf {\bibinfo {volume} {135}},\
  \bibinfo {pages} {104093} (\bibinfo {year} {2024})},\ \Eprint
  {https://arxiv.org/abs/2307.03885} {arXiv:2307.03885 [nucl-th]} \BibitemShut
  {NoStop}%
\bibitem [{\citenamefont {Hoyos}\ \emph {et~al.}(2022)\citenamefont {Hoyos},
  \citenamefont {Jokela},\ and\ \citenamefont {Vuorinen}}]{Hoyos:2021uff}%
  \BibitemOpen
  \bibfield  {author} {\bibinfo {author} {\bibfnamefont {C.}~\bibnamefont
  {Hoyos}}, \bibinfo {author} {\bibfnamefont {N.}~\bibnamefont {Jokela}},\ and\
  \bibinfo {author} {\bibfnamefont {A.}~\bibnamefont {Vuorinen}},\ }\href
  {https://doi.org/10.1016/j.ppnp.2022.103972} {\bibfield  {journal} {\bibinfo
  {journal} {Prog. Part. Nucl. Phys.}\ }\textbf {\bibinfo {volume} {126}},\
  \bibinfo {pages} {103972} (\bibinfo {year} {2022})},\ \Eprint
  {https://arxiv.org/abs/2112.08422} {arXiv:2112.08422 [hep-th]} \BibitemShut
  {NoStop}%
\bibitem [{\citenamefont {J\"arvinen}(2022)}]{Jarvinen:2021jbd}%
  \BibitemOpen
  \bibfield  {author} {\bibinfo {author} {\bibfnamefont {M.}~\bibnamefont
  {J\"arvinen}},\ }\href {https://doi.org/10.1140/epjc/s10052-022-10227-x}
  {\bibfield  {journal} {\bibinfo  {journal} {Eur. Phys. J. C}\ }\textbf
  {\bibinfo {volume} {82}},\ \bibinfo {pages} {282} (\bibinfo {year} {2022})},\
  \Eprint {https://arxiv.org/abs/2110.08281} {arXiv:2110.08281 [hep-ph]}
  \BibitemShut {NoStop}%
\bibitem [{\citenamefont {Hashimoto}\ and\ \citenamefont
  {Tanahashi}(2017)}]{Hashimoto:2016dfz}%
  \BibitemOpen
  \bibfield  {author} {\bibinfo {author} {\bibfnamefont {K.}~\bibnamefont
  {Hashimoto}}\ and\ \bibinfo {author} {\bibfnamefont {N.}~\bibnamefont
  {Tanahashi}},\ }\href {https://doi.org/10.1103/PhysRevD.95.024007} {\bibfield
   {journal} {\bibinfo  {journal} {Phys. Rev. D}\ }\textbf {\bibinfo {volume}
  {95}},\ \bibinfo {pages} {024007} (\bibinfo {year} {2017})},\ \Eprint
  {https://arxiv.org/abs/1610.06070} {arXiv:1610.06070 [hep-th]} \BibitemShut
  {NoStop}%
\bibitem [{\citenamefont {Giataganas}(2022)}]{Giataganas:2021ghs}%
  \BibitemOpen
  \bibfield  {author} {\bibinfo {author} {\bibfnamefont {D.}~\bibnamefont
  {Giataganas}},\ }\href {https://doi.org/10.1002/prop.202200001} {\bibfield
  {journal} {\bibinfo  {journal} {Fortsch. Phys.}\ }\textbf {\bibinfo {volume}
  {70}},\ \bibinfo {pages} {2200001} (\bibinfo {year} {2022})},\ \Eprint
  {https://arxiv.org/abs/2112.02081} {arXiv:2112.02081 [hep-th]} \BibitemShut
  {NoStop}%
\bibitem [{\citenamefont {Djuki\'c}\ and\ \citenamefont
  {\v{C}ubrovi\'c}(2024)}]{Djukic:2023dgk}%
  \BibitemOpen
  \bibfield  {author} {\bibinfo {author} {\bibfnamefont {V.}~\bibnamefont
  {Djuki\'c}}\ and\ \bibinfo {author} {\bibfnamefont {M.}~\bibnamefont
  {\v{C}ubrovi\'c}},\ }\href {https://doi.org/10.1007/JHEP04(2024)025}
  {\bibfield  {journal} {\bibinfo  {journal} {JHEP}\ }\textbf {\bibinfo
  {volume} {04}},\ \bibinfo {pages} {025}},\ \Eprint
  {https://arxiv.org/abs/2310.15697} {arXiv:2310.15697 [hep-th]} \BibitemShut
  {NoStop}%
\bibitem [{\citenamefont {Ishii}\ and\ \citenamefont
  {Murata}(2015)}]{Ishii:2015wua}%
  \BibitemOpen
  \bibfield  {author} {\bibinfo {author} {\bibfnamefont {T.}~\bibnamefont
  {Ishii}}\ and\ \bibinfo {author} {\bibfnamefont {K.}~\bibnamefont {Murata}},\
  }\href {https://doi.org/10.1007/JHEP06(2015)086} {\bibfield  {journal}
  {\bibinfo  {journal} {JHEP}\ }\textbf {\bibinfo {volume} {06}},\ \bibinfo
  {pages} {086}},\ \Eprint {https://arxiv.org/abs/1504.02190} {arXiv:1504.02190
  [hep-th]} \BibitemShut {NoStop}%
\bibitem [{\citenamefont {Faraoni}\ and\ \citenamefont
  {Gunzig}(1999)}]{faraoni1999einstein}%
  \BibitemOpen
  \bibfield  {author} {\bibinfo {author} {\bibfnamefont {V.}~\bibnamefont
  {Faraoni}}\ and\ \bibinfo {author} {\bibfnamefont {E.}~\bibnamefont
  {Gunzig}},\ }\href {https://doi.org/10.1023/A:1026645510351} {\bibfield
  {journal} {\bibinfo  {journal} {Int. J. Theor. Phys.}\ }\textbf {\bibinfo
  {volume} {38}},\ \bibinfo {pages} {217} (\bibinfo {year} {1999})},\ \Eprint
  {https://arxiv.org/abs/astro-ph/9910176} {arXiv:astro-ph/9910176}
  \BibitemShut {NoStop}%
\bibitem [{\citenamefont {Cho}(1992)}]{cho1992reinterpretation}%
  \BibitemOpen
  \bibfield  {author} {\bibinfo {author} {\bibfnamefont {Y.~M.}\ \bibnamefont
  {Cho}},\ }\href {https://doi.org/10.1103/PhysRevLett.68.3133} {\bibfield
  {journal} {\bibinfo  {journal} {Phys. Rev. Lett.}\ }\textbf {\bibinfo
  {volume} {68}},\ \bibinfo {pages} {3133} (\bibinfo {year}
  {1992})}\BibitemShut {NoStop}%
\bibitem [{\citenamefont {Sk}\ and\ \citenamefont
  {Sanyal}(2017)}]{sk2017scalar}%
  \BibitemOpen
  \bibfield  {author} {\bibinfo {author} {\bibfnamefont {N.}~\bibnamefont
  {Sk}}\ and\ \bibinfo {author} {\bibfnamefont {A.~K.}\ \bibnamefont
  {Sanyal}},\ }\href {https://doi.org/10.1142/S0218271817501620} {\bibfield
  {journal} {\bibinfo  {journal} {Int. J. Mod. Phys. D}\ }\textbf {\bibinfo
  {volume} {26}},\ \bibinfo {pages} {1750162} (\bibinfo {year} {2017})},\
  \Eprint {https://arxiv.org/abs/1609.01824} {arXiv:1609.01824 [gr-qc]}
  \BibitemShut {NoStop}%
\bibitem [{\citenamefont {Capozziello}\ \emph {et~al.}(2010)\citenamefont
  {Capozziello}, \citenamefont {Martin-Moruno},\ and\ \citenamefont
  {Rubano}}]{capozziello2010physical}%
  \BibitemOpen
  \bibfield  {author} {\bibinfo {author} {\bibfnamefont {S.}~\bibnamefont
  {Capozziello}}, \bibinfo {author} {\bibfnamefont {P.}~\bibnamefont
  {Martin-Moruno}},\ and\ \bibinfo {author} {\bibfnamefont {C.}~\bibnamefont
  {Rubano}},\ }\href {https://doi.org/10.1016/j.physletb.2010.04.058}
  {\bibfield  {journal} {\bibinfo  {journal} {Phys. Lett. B}\ }\textbf
  {\bibinfo {volume} {689}},\ \bibinfo {pages} {117} (\bibinfo {year}
  {2010})},\ \Eprint {https://arxiv.org/abs/1003.5394} {arXiv:1003.5394
  [gr-qc]} \BibitemShut {NoStop}%
\bibitem [{\citenamefont {Corda}(2011)}]{corda2011gravitational}%
  \BibitemOpen
  \bibfield  {author} {\bibinfo {author} {\bibfnamefont {C.}~\bibnamefont
  {Corda}},\ }\href {https://doi.org/10.1016/j.astropartphys.2010.10.006}
  {\bibfield  {journal} {\bibinfo  {journal} {Astropart. Phys.}\ }\textbf
  {\bibinfo {volume} {34}},\ \bibinfo {pages} {412} (\bibinfo {year} {2011})},\
  \Eprint {https://arxiv.org/abs/1010.2086} {arXiv:1010.2086 [gr-qc]}
  \BibitemShut {NoStop}%
\bibitem [{\citenamefont {Quiros}\ \emph {et~al.}(2013)\citenamefont {Quiros},
  \citenamefont {Garcia-Salcedo}, \citenamefont {Madriz~Aguilar},\ and\
  \citenamefont {Matos}}]{quiros2013conformal}%
  \BibitemOpen
  \bibfield  {author} {\bibinfo {author} {\bibfnamefont {I.}~\bibnamefont
  {Quiros}}, \bibinfo {author} {\bibfnamefont {R.}~\bibnamefont
  {Garcia-Salcedo}}, \bibinfo {author} {\bibfnamefont {J.~E.}\ \bibnamefont
  {Madriz~Aguilar}},\ and\ \bibinfo {author} {\bibfnamefont {T.}~\bibnamefont
  {Matos}},\ }\href {https://doi.org/10.1007/s10714-012-1484-7} {\bibfield
  {journal} {\bibinfo  {journal} {Gen. Rel. Grav.}\ }\textbf {\bibinfo {volume}
  {45}},\ \bibinfo {pages} {489} (\bibinfo {year} {2013})},\ \Eprint
  {https://arxiv.org/abs/1108.5857} {arXiv:1108.5857 [gr-qc]} \BibitemShut
  {NoStop}%
\bibitem [{\citenamefont {Maci\'\i{}as}\ and\ \citenamefont
  {Garc\'\i{}a}(2001)}]{macias2001jordan}%
  \BibitemOpen
  \bibfield  {author} {\bibinfo {author} {\bibfnamefont {A.}~\bibnamefont
  {Maci\'\i{}as}}\ and\ \bibinfo {author} {\bibfnamefont {A.}~\bibnamefont
  {Garc\'\i{}a}},\ }\href {https://doi.org/10.1023/A:1010212025682} {\bibfield
  {journal} {\bibinfo  {journal} {Gen. Relativ. Gravitation}\ }\textbf
  {\bibinfo {volume} {33}},\ \bibinfo {pages} {889} (\bibinfo {year}
  {2001})}\BibitemShut {NoStop}%
\bibitem [{\citenamefont {Galaverni}\ and\ \citenamefont
  {S.~J.}(2022)}]{galaverni2022jordan}%
  \BibitemOpen
  \bibfield  {author} {\bibinfo {author} {\bibfnamefont {M.}~\bibnamefont
  {Galaverni}}\ and\ \bibinfo {author} {\bibfnamefont {G.~G.}\ \bibnamefont
  {S.~J.}},\ }\href {https://doi.org/10.1103/PhysRevD.105.084008} {\bibfield
  {journal} {\bibinfo  {journal} {Phys. Rev. D}\ }\textbf {\bibinfo {volume}
  {105}},\ \bibinfo {pages} {084008} (\bibinfo {year} {2022})},\ \Eprint
  {https://arxiv.org/abs/2110.12222} {arXiv:2110.12222 [gr-qc]} \BibitemShut
  {NoStop}%
\bibitem [{\citenamefont {Nojiri}\ \emph {et~al.}(2001)\citenamefont {Nojiri},
  \citenamefont {Obregon}, \citenamefont {Odintsov},\ and\ \citenamefont
  {Tkach}}]{nojiri2001string}%
  \BibitemOpen
  \bibfield  {author} {\bibinfo {author} {\bibfnamefont {S.}~\bibnamefont
  {Nojiri}}, \bibinfo {author} {\bibfnamefont {O.}~\bibnamefont {Obregon}},
  \bibinfo {author} {\bibfnamefont {S.~D.}\ \bibnamefont {Odintsov}},\ and\
  \bibinfo {author} {\bibfnamefont {V.~I.}\ \bibnamefont {Tkach}},\ }\href
  {https://doi.org/10.1103/PhysRevD.64.043505} {\bibfield  {journal} {\bibinfo
  {journal} {Phys. Rev. D}\ }\textbf {\bibinfo {volume} {64}},\ \bibinfo
  {pages} {043505} (\bibinfo {year} {2001})}\BibitemShut {NoStop}%
\bibitem [{\citenamefont {Jarv}\ \emph {et~al.}(2007)\citenamefont {Jarv},
  \citenamefont {Kuusk},\ and\ \citenamefont {Saal}}]{jarv2007scalar}%
  \BibitemOpen
  \bibfield  {author} {\bibinfo {author} {\bibfnamefont {L.}~\bibnamefont
  {Jarv}}, \bibinfo {author} {\bibfnamefont {P.}~\bibnamefont {Kuusk}},\ and\
  \bibinfo {author} {\bibfnamefont {M.}~\bibnamefont {Saal}},\ }\href
  {https://doi.org/10.1103/PhysRevD.76.103506} {\bibfield  {journal} {\bibinfo
  {journal} {Phys. Rev. D}\ }\textbf {\bibinfo {volume} {76}},\ \bibinfo
  {pages} {103506} (\bibinfo {year} {2007})},\ \Eprint
  {https://arxiv.org/abs/0705.4644} {arXiv:0705.4644 [gr-qc]} \BibitemShut
  {NoStop}%
\bibitem [{\citenamefont {Girardello}\ \emph {et~al.}(1999)\citenamefont
  {Girardello}, \citenamefont {Petrini}, \citenamefont {Porrati},\ and\
  \citenamefont {Zaffaroni}}]{Girardello:1999hj}%
  \BibitemOpen
  \bibfield  {author} {\bibinfo {author} {\bibfnamefont {L.}~\bibnamefont
  {Girardello}}, \bibinfo {author} {\bibfnamefont {M.}~\bibnamefont {Petrini}},
  \bibinfo {author} {\bibfnamefont {M.}~\bibnamefont {Porrati}},\ and\ \bibinfo
  {author} {\bibfnamefont {A.}~\bibnamefont {Zaffaroni}},\ }\href
  {https://doi.org/10.1088/1126-6708/1999/05/026} {\bibfield  {journal}
  {\bibinfo  {journal} {JHEP}\ }\textbf {\bibinfo {volume} {05}},\ \bibinfo
  {pages} {026}},\ \Eprint {https://arxiv.org/abs/hep-th/9903026}
  {arXiv:hep-th/9903026} \BibitemShut {NoStop}%
\bibitem [{\citenamefont {Mahapatra}\ and\ \citenamefont
  {Roy}(2018)}]{Mahapatra:2018gig}%
  \BibitemOpen
  \bibfield  {author} {\bibinfo {author} {\bibfnamefont {S.}~\bibnamefont
  {Mahapatra}}\ and\ \bibinfo {author} {\bibfnamefont {P.}~\bibnamefont
  {Roy}},\ }\href {https://doi.org/10.1007/JHEP11(2018)138} {\bibfield
  {journal} {\bibinfo  {journal} {JHEP}\ }\textbf {\bibinfo {volume} {11}},\
  \bibinfo {pages} {138}},\ \Eprint {https://arxiv.org/abs/1808.09917}
  {arXiv:1808.09917 [hep-th]} \BibitemShut {NoStop}%
\bibitem [{\citenamefont {Aref'eva}\ and\ \citenamefont
  {Rannu}(2018)}]{Arefeva:2018hyo}%
  \BibitemOpen
  \bibfield  {author} {\bibinfo {author} {\bibfnamefont {I.}~\bibnamefont
  {Aref'eva}}\ and\ \bibinfo {author} {\bibfnamefont {K.}~\bibnamefont
  {Rannu}},\ }\href {https://doi.org/10.1007/JHEP05(2018)206} {\bibfield
  {journal} {\bibinfo  {journal} {JHEP}\ }\textbf {\bibinfo {volume} {05}},\
  \bibinfo {pages} {206}},\ \Eprint {https://arxiv.org/abs/1802.05652}
  {arXiv:1802.05652 [hep-th]} \BibitemShut {NoStop}%
\bibitem [{\citenamefont {Aref'eva}\ \emph
  {et~al.}(2021{\natexlab{b}})\citenamefont {Aref'eva}, \citenamefont {Rannu},\
  and\ \citenamefont {Slepov}}]{Arefeva:2020byn}%
  \BibitemOpen
  \bibfield  {author} {\bibinfo {author} {\bibfnamefont {I.~Y.}\ \bibnamefont
  {Aref'eva}}, \bibinfo {author} {\bibfnamefont {K.}~\bibnamefont {Rannu}},\
  and\ \bibinfo {author} {\bibfnamefont {P.}~\bibnamefont {Slepov}},\ }\href
  {https://doi.org/10.1007/JHEP06(2021)090} {\bibfield  {journal} {\bibinfo
  {journal} {JHEP}\ }\textbf {\bibinfo {volume} {06}},\ \bibinfo {pages}
  {090}},\ \Eprint {https://arxiv.org/abs/2009.05562} {arXiv:2009.05562
  [hep-th]} \BibitemShut {NoStop}%
\bibitem [{\citenamefont {Alanen}\ \emph {et~al.}(2009)\citenamefont {Alanen},
  \citenamefont {Kajantie},\ and\ \citenamefont {Suur-Uski}}]{Alanen:2009xs}%
  \BibitemOpen
  \bibfield  {author} {\bibinfo {author} {\bibfnamefont {J.}~\bibnamefont
  {Alanen}}, \bibinfo {author} {\bibfnamefont {K.}~\bibnamefont {Kajantie}},\
  and\ \bibinfo {author} {\bibfnamefont {V.}~\bibnamefont {Suur-Uski}},\ }\href
  {https://doi.org/10.1103/PhysRevD.80.126008} {\bibfield  {journal} {\bibinfo
  {journal} {Phys. Rev. D}\ }\textbf {\bibinfo {volume} {80}},\ \bibinfo
  {pages} {126008} (\bibinfo {year} {2009})},\ \Eprint
  {https://arxiv.org/abs/0911.2114} {arXiv:0911.2114 [hep-ph]} \BibitemShut
  {NoStop}%
\bibitem [{\citenamefont {Mahapatra}\ \emph {et~al.}(2020)\citenamefont
  {Mahapatra}, \citenamefont {Priyadarshinee}, \citenamefont {Reddy},\ and\
  \citenamefont {Shukla}}]{Mahapatra:2020wym}%
  \BibitemOpen
  \bibfield  {author} {\bibinfo {author} {\bibfnamefont {S.}~\bibnamefont
  {Mahapatra}}, \bibinfo {author} {\bibfnamefont {S.}~\bibnamefont
  {Priyadarshinee}}, \bibinfo {author} {\bibfnamefont {G.~N.}\ \bibnamefont
  {Reddy}},\ and\ \bibinfo {author} {\bibfnamefont {B.}~\bibnamefont
  {Shukla}},\ }\href {https://doi.org/10.1103/PhysRevD.102.024042} {\bibfield
  {journal} {\bibinfo  {journal} {Phys. Rev. D}\ }\textbf {\bibinfo {volume}
  {102}},\ \bibinfo {pages} {024042} (\bibinfo {year} {2020})},\ \Eprint
  {https://arxiv.org/abs/2004.00921} {arXiv:2004.00921 [hep-th]} \BibitemShut
  {NoStop}%
\bibitem [{\citenamefont {Priyadarshinee}\ \emph {et~al.}(2021)\citenamefont
  {Priyadarshinee}, \citenamefont {Mahapatra},\ and\ \citenamefont
  {Banerjee}}]{Priyadarshinee:2021rch}%
  \BibitemOpen
  \bibfield  {author} {\bibinfo {author} {\bibfnamefont {S.}~\bibnamefont
  {Priyadarshinee}}, \bibinfo {author} {\bibfnamefont {S.}~\bibnamefont
  {Mahapatra}},\ and\ \bibinfo {author} {\bibfnamefont {I.}~\bibnamefont
  {Banerjee}},\ }\href {https://doi.org/10.1103/PhysRevD.104.084023} {\bibfield
   {journal} {\bibinfo  {journal} {Phys. Rev. D}\ }\textbf {\bibinfo {volume}
  {104}},\ \bibinfo {pages} {084023} (\bibinfo {year} {2021})},\ \Eprint
  {https://arxiv.org/abs/2108.02514} {arXiv:2108.02514 [hep-th]} \BibitemShut
  {NoStop}%
\bibitem [{\citenamefont {Priyadarshinee}\ and\ \citenamefont
  {Mahapatra}(2023)}]{Priyadarshinee:2023cmi}%
  \BibitemOpen
  \bibfield  {author} {\bibinfo {author} {\bibfnamefont {S.}~\bibnamefont
  {Priyadarshinee}}\ and\ \bibinfo {author} {\bibfnamefont {S.}~\bibnamefont
  {Mahapatra}},\ }\href {https://doi.org/10.1103/PhysRevD.108.044017}
  {\bibfield  {journal} {\bibinfo  {journal} {Phys. Rev. D}\ }\textbf {\bibinfo
  {volume} {108}},\ \bibinfo {pages} {044017} (\bibinfo {year} {2023})},\
  \Eprint {https://arxiv.org/abs/2305.09172} {arXiv:2305.09172 [gr-qc]}
  \BibitemShut {NoStop}%
\bibitem [{\citenamefont {Daripa}\ and\ \citenamefont
  {Mahapatra}(2024)}]{Daripa:2024ksg}%
  \BibitemOpen
  \bibfield  {author} {\bibinfo {author} {\bibfnamefont {A.}~\bibnamefont
  {Daripa}}\ and\ \bibinfo {author} {\bibfnamefont {S.}~\bibnamefont
  {Mahapatra}},\ }\href {https://doi.org/10.1103/PhysRevD.109.124039}
  {\bibfield  {journal} {\bibinfo  {journal} {Phys. Rev. D}\ }\textbf {\bibinfo
  {volume} {109}},\ \bibinfo {pages} {124039} (\bibinfo {year} {2024})},\
  \Eprint {https://arxiv.org/abs/2401.04561} {arXiv:2401.04561 [gr-qc]}
  \BibitemShut {NoStop}%
\bibitem [{\citenamefont {He}\ \emph {et~al.}(2013{\natexlab{b}})\citenamefont
  {He}, \citenamefont {Wu}, \citenamefont {Yang},\ and\ \citenamefont
  {Yuan}}]{he2013phase}%
  \BibitemOpen
  \bibfield  {author} {\bibinfo {author} {\bibfnamefont {S.}~\bibnamefont
  {He}}, \bibinfo {author} {\bibfnamefont {S.-Y.}\ \bibnamefont {Wu}}, \bibinfo
  {author} {\bibfnamefont {Y.}~\bibnamefont {Yang}},\ and\ \bibinfo {author}
  {\bibfnamefont {P.-H.}\ \bibnamefont {Yuan}},\ }\href
  {https://doi.org/10.1007/JHEP04(2013)093} {\bibfield  {journal} {\bibinfo
  {journal} {JHEP}\ }\textbf {\bibinfo {volume} {04}},\ \bibinfo {pages}
  {093}},\ \Eprint {https://arxiv.org/abs/1301.0385} {arXiv:1301.0385 [hep-th]}
  \BibitemShut {NoStop}%
\bibitem [{\citenamefont {Gubser}(2000)}]{gubser2000curvature}%
  \BibitemOpen
  \bibfield  {author} {\bibinfo {author} {\bibfnamefont {S.~S.}\ \bibnamefont
  {Gubser}},\ }\href {https://doi.org/10.4310/ATMP.2000.v4.n3.a6} {\bibfield
  {journal} {\bibinfo  {journal} {Adv. Theor. Math. Phys.}\ }\textbf {\bibinfo
  {volume} {4}},\ \bibinfo {pages} {679} (\bibinfo {year} {2000})},\ \Eprint
  {https://arxiv.org/abs/hep-th/0002160} {arXiv:hep-th/0002160} \BibitemShut
  {NoStop}%
\bibitem [{\citenamefont {Breitenlohner}\ and\ \citenamefont
  {Freedman}(1982)}]{Breitenlohner:1982bm}%
  \BibitemOpen
  \bibfield  {author} {\bibinfo {author} {\bibfnamefont {P.}~\bibnamefont
  {Breitenlohner}}\ and\ \bibinfo {author} {\bibfnamefont {D.~Z.}\ \bibnamefont
  {Freedman}},\ }\href {https://doi.org/10.1016/0370-2693(82)90643-8}
  {\bibfield  {journal} {\bibinfo  {journal} {Phys. Lett. B}\ }\textbf
  {\bibinfo {volume} {115}},\ \bibinfo {pages} {197} (\bibinfo {year}
  {1982})}\BibitemShut {NoStop}%
\bibitem [{\citenamefont {Shukla}\ \emph
  {et~al.}(2024{\natexlab{b}})\citenamefont {Shukla}, \citenamefont {Das},
  \citenamefont {Dudal},\ and\ \citenamefont {Mahapatra}}]{Shukla:2024tkw}%
  \BibitemOpen
  \bibfield  {author} {\bibinfo {author} {\bibfnamefont {B.}~\bibnamefont
  {Shukla}}, \bibinfo {author} {\bibfnamefont {P.~P.}\ \bibnamefont {Das}},
  \bibinfo {author} {\bibfnamefont {D.}~\bibnamefont {Dudal}},\ and\ \bibinfo
  {author} {\bibfnamefont {S.}~\bibnamefont {Mahapatra}},\ }\href
  {https://doi.org/10.1103/PhysRevD.110.024068} {\bibfield  {journal} {\bibinfo
   {journal} {Phys. Rev. D}\ }\textbf {\bibinfo {volume} {110}},\ \bibinfo
  {pages} {024068} (\bibinfo {year} {2024}{\natexlab{b}})},\ \Eprint
  {https://arxiv.org/abs/2404.02095} {arXiv:2404.02095 [hep-th]} \BibitemShut
  {NoStop}%
\bibitem [{\citenamefont {Guo}\ \emph {et~al.}(2022)\citenamefont {Guo},
  \citenamefont {Lu}, \citenamefont {Mu},\ and\ \citenamefont
  {Wang}}]{guo2022probing}%
  \BibitemOpen
  \bibfield  {author} {\bibinfo {author} {\bibfnamefont {X.}~\bibnamefont
  {Guo}}, \bibinfo {author} {\bibfnamefont {Y.}~\bibnamefont {Lu}}, \bibinfo
  {author} {\bibfnamefont {B.}~\bibnamefont {Mu}},\ and\ \bibinfo {author}
  {\bibfnamefont {P.}~\bibnamefont {Wang}},\ }\href
  {https://doi.org/10.1007/JHEP08(2022)153} {\bibfield  {journal} {\bibinfo
  {journal} {JHEP}\ }\textbf {\bibinfo {volume} {08}},\ \bibinfo {pages}
  {153}},\ \Eprint {https://arxiv.org/abs/2205.02122} {arXiv:2205.02122
  [gr-qc]} \BibitemShut {NoStop}%
\bibitem [{\citenamefont {Yang}\ \emph {et~al.}(2023)\citenamefont {Yang},
  \citenamefont {Tao}, \citenamefont {Mu},\ and\ \citenamefont
  {He}}]{yang2023lyapunov}%
  \BibitemOpen
  \bibfield  {author} {\bibinfo {author} {\bibfnamefont {S.}~\bibnamefont
  {Yang}}, \bibinfo {author} {\bibfnamefont {J.}~\bibnamefont {Tao}}, \bibinfo
  {author} {\bibfnamefont {B.}~\bibnamefont {Mu}},\ and\ \bibinfo {author}
  {\bibfnamefont {A.}~\bibnamefont {He}},\ }\href
  {https://doi.org/10.1088/1475-7516/2023/07/045} {\bibfield  {journal}
  {\bibinfo  {journal} {JCAP}\ }\textbf {\bibinfo {volume} {07}},\ \bibinfo
  {pages} {045}},\ \Eprint {https://arxiv.org/abs/2304.01877} {arXiv:2304.01877
  [gr-qc]} \BibitemShut {NoStop}%
\bibitem [{\citenamefont {Lyu}\ \emph {et~al.}(2024)\citenamefont {Lyu},
  \citenamefont {Tao},\ and\ \citenamefont {Wang}}]{lyu2023probing}%
  \BibitemOpen
  \bibfield  {author} {\bibinfo {author} {\bibfnamefont {X.}~\bibnamefont
  {Lyu}}, \bibinfo {author} {\bibfnamefont {J.}~\bibnamefont {Tao}},\ and\
  \bibinfo {author} {\bibfnamefont {P.}~\bibnamefont {Wang}},\ }\href
  {https://doi.org/10.1140/epjc/s10052-024-13354-9} {\bibfield  {journal}
  {\bibinfo  {journal} {Eur. Phys. J. C}\ }\textbf {\bibinfo {volume} {84}},\
  \bibinfo {pages} {974} (\bibinfo {year} {2024})},\ \Eprint
  {https://arxiv.org/abs/2312.11912} {arXiv:2312.11912 [gr-qc]} \BibitemShut
  {NoStop}%
\bibitem [{\citenamefont {Kumara}\ \emph {et~al.}(2024)\citenamefont {Kumara},
  \citenamefont {Punacha},\ and\ \citenamefont {Ali}}]{kumara2024lyapunov}%
  \BibitemOpen
  \bibfield  {author} {\bibinfo {author} {\bibfnamefont {A.~N.}\ \bibnamefont
  {Kumara}}, \bibinfo {author} {\bibfnamefont {S.}~\bibnamefont {Punacha}},\
  and\ \bibinfo {author} {\bibfnamefont {M.~S.}\ \bibnamefont {Ali}},\ }\href
  {https://doi.org/10.1088/1475-7516/2024/07/061} {\bibfield  {journal}
  {\bibinfo  {journal} {JCAP}\ }\textbf {\bibinfo {volume} {07}},\ \bibinfo
  {pages} {061}},\ \Eprint {https://arxiv.org/abs/2401.05181} {arXiv:2401.05181
  [gr-qc]} \BibitemShut {NoStop}%
\bibitem [{\citenamefont {Du}\ \emph {et~al.}(2025)\citenamefont {Du},
  \citenamefont {Li}, \citenamefont {Ma},\ and\ \citenamefont
  {Gu}}]{Du:2024uhd}%
  \BibitemOpen
  \bibfield  {author} {\bibinfo {author} {\bibfnamefont {Y.-Z.}\ \bibnamefont
  {Du}}, \bibinfo {author} {\bibfnamefont {H.-F.}\ \bibnamefont {Li}}, \bibinfo
  {author} {\bibfnamefont {Y.-B.}\ \bibnamefont {Ma}},\ and\ \bibinfo {author}
  {\bibfnamefont {Q.}~\bibnamefont {Gu}},\ }\href
  {https://doi.org/10.1140/epjc/s10052-025-13809-7} {\bibfield  {journal}
  {\bibinfo  {journal} {Eur. Phys. J. C}\ }\textbf {\bibinfo {volume} {85}},\
  \bibinfo {pages} {78} (\bibinfo {year} {2025})},\ \Eprint
  {https://arxiv.org/abs/2403.20083} {arXiv:2403.20083 [hep-th]} \BibitemShut
  {NoStop}%
\bibitem [{\citenamefont {Gogoi}\ \emph {et~al.}(2024)\citenamefont {Gogoi},
  \citenamefont {Acharjee},\ and\ \citenamefont {Phukon}}]{Gogoi:2024akv}%
  \BibitemOpen
  \bibfield  {author} {\bibinfo {author} {\bibfnamefont {N.~J.}\ \bibnamefont
  {Gogoi}}, \bibinfo {author} {\bibfnamefont {S.}~\bibnamefont {Acharjee}},\
  and\ \bibinfo {author} {\bibfnamefont {P.}~\bibnamefont {Phukon}},\ }\href
  {https://doi.org/10.1140/epjc/s10052-024-13520-z} {\bibfield  {journal}
  {\bibinfo  {journal} {Eur. Phys. J. C}\ }\textbf {\bibinfo {volume} {84}},\
  \bibinfo {pages} {1144} (\bibinfo {year} {2024})},\ \Eprint
  {https://arxiv.org/abs/2404.03947} {arXiv:2404.03947 [hep-th]} \BibitemShut
  {NoStop}%
\bibitem [{\citenamefont {Sandri}(1996)}]{sandri1996numerical}%
  \BibitemOpen
  \bibfield  {author} {\bibinfo {author} {\bibfnamefont {M.}~\bibnamefont
  {Sandri}},\ }\href@noop {} {\bibfield  {journal} {\bibinfo  {journal}
  {Mathematica Journal}\ }\textbf {\bibinfo {volume} {6}},\ \bibinfo {pages}
  {78} (\bibinfo {year} {1996})}\BibitemShut {NoStop}%
\bibitem [{\citenamefont {Wolf}\ \emph {et~al.}(1985)\citenamefont {Wolf},
  \citenamefont {Swift}, \citenamefont {Swinney},\ and\ \citenamefont
  {Vastano}}]{wolf1985determining}%
  \BibitemOpen
  \bibfield  {author} {\bibinfo {author} {\bibfnamefont {A.}~\bibnamefont
  {Wolf}}, \bibinfo {author} {\bibfnamefont {J.~B.}\ \bibnamefont {Swift}},
  \bibinfo {author} {\bibfnamefont {H.~L.}\ \bibnamefont {Swinney}},\ and\
  \bibinfo {author} {\bibfnamefont {J.~A.}\ \bibnamefont {Vastano}},\ }\href
  {https://doi.org/10.1016/0167-2789(85)90011-9} {\bibfield  {journal}
  {\bibinfo  {journal} {Physica D: nonlinear phenomena}\ }\textbf {\bibinfo
  {volume} {16}},\ \bibinfo {pages} {285} (\bibinfo {year} {1985})}\BibitemShut
  {NoStop}%
\bibitem [{\citenamefont {Bali}\ \emph {et~al.}(2012)\citenamefont {Bali},
  \citenamefont {Bruckmann}, \citenamefont {Endrodi}, \citenamefont {Fodor},
  \citenamefont {Katz}, \citenamefont {Krieg}, \citenamefont {Schafer},\ and\
  \citenamefont {Szabo}}]{Bali:2011qj}%
  \BibitemOpen
  \bibfield  {author} {\bibinfo {author} {\bibfnamefont {G.~S.}\ \bibnamefont
  {Bali}}, \bibinfo {author} {\bibfnamefont {F.}~\bibnamefont {Bruckmann}},
  \bibinfo {author} {\bibfnamefont {G.}~\bibnamefont {Endrodi}}, \bibinfo
  {author} {\bibfnamefont {Z.}~\bibnamefont {Fodor}}, \bibinfo {author}
  {\bibfnamefont {S.~D.}\ \bibnamefont {Katz}}, \bibinfo {author}
  {\bibfnamefont {S.}~\bibnamefont {Krieg}}, \bibinfo {author} {\bibfnamefont
  {A.}~\bibnamefont {Schafer}},\ and\ \bibinfo {author} {\bibfnamefont {K.~K.}\
  \bibnamefont {Szabo}},\ }\href {https://doi.org/10.1007/JHEP02(2012)044}
  {\bibfield  {journal} {\bibinfo  {journal} {JHEP}\ }\textbf {\bibinfo
  {volume} {02}},\ \bibinfo {pages} {044}},\ \Eprint
  {https://arxiv.org/abs/1111.4956} {arXiv:1111.4956 [hep-lat]} \BibitemShut
  {NoStop}%
\bibitem [{\citenamefont {D'Elia}\ \emph {et~al.}(2021)\citenamefont {D'Elia},
  \citenamefont {Maio}, \citenamefont {Sanfilippo},\ and\ \citenamefont
  {Stanzione}}]{DElia:2021tfb}%
  \BibitemOpen
  \bibfield  {author} {\bibinfo {author} {\bibfnamefont {M.}~\bibnamefont
  {D'Elia}}, \bibinfo {author} {\bibfnamefont {L.}~\bibnamefont {Maio}},
  \bibinfo {author} {\bibfnamefont {F.}~\bibnamefont {Sanfilippo}},\ and\
  \bibinfo {author} {\bibfnamefont {A.}~\bibnamefont {Stanzione}},\ }\href
  {https://doi.org/10.1103/PhysRevD.104.114512} {\bibfield  {journal} {\bibinfo
   {journal} {Phys. Rev. D}\ }\textbf {\bibinfo {volume} {104}},\ \bibinfo
  {pages} {114512} (\bibinfo {year} {2021})},\ \Eprint
  {https://arxiv.org/abs/2109.07456} {arXiv:2109.07456 [hep-lat]} \BibitemShut
  {NoStop}%
\bibitem [{\citenamefont {Gutzwiller}(1990)}]{Gutzwiller1990}%
  \BibitemOpen
  \bibfield  {author} {\bibinfo {author} {\bibfnamefont {M.~C.}\ \bibnamefont
  {Gutzwiller}},\ }\bibinfo {title} {Soft chaos and the kam theorem},\ in\
  \href {https://doi.org/10.1007/978-1-4612-0983-6_10} {\emph {\bibinfo
  {booktitle} {Chaos in Classical and Quantum Mechanics}}}\ (\bibinfo
  {publisher} {Springer New York},\ \bibinfo {address} {New York, NY},\
  \bibinfo {year} {1990})\ pp.\ \bibinfo {pages} {116--141}\BibitemShut
  {NoStop}%
\end{thebibliography}%
\bibliographystyle{apsrev4-2}

\end{document}